\documentclass[12pt,a4paper,final]{article}

\usepackage{lmodern} 
\usepackage[utf8]{inputenc}
\usepackage[T1]{fontenc}

\usepackage[english]{babel}

\usepackage{tabularx}
\usepackage{graphicx} 
\usepackage{adjustbox}

\usepackage{caption}
\usepackage{booktabs} 
\usepackage{siunitx}

\usepackage{subfig}

\usepackage{csquotes}
\usepackage[
    giveninits=true, 
    backend=biber, 
    bibstyle=authoryear, 
    citestyle=authoryear-comp, 
    dashed=false, 
    doi=true, 
    maxbibnames=99, 
    maxcitenames=99, 
    uniquename=init 
]{biblatex} 
\DeclareFieldFormat[article]{volume}{\bibstring{jourvol}\addnbspace #1} 
\DeclareFieldFormat[article]{number}{\bibstring{number}\addnbspace #1} 
\renewbibmacro*{volume+number+eid}{
  \printfield{volume}%
  \setunit{\addcomma\space}%
  \printfield{number}%
  \setunit{\addcomma\space}%
  \printfield{eid}
}
\renewbibmacro{in:}{} 

\usepackage{lscape}
\usepackage{enumitem}

\usepackage{amsmath} 
\usepackage{amssymb} 
\usepackage{amsthm} 
\usepackage{mathrsfs}

\usepackage{fullpage} 

\usepackage{float}

\allowdisplaybreaks

\newtheorem{assumption}{Assumption}[section]
\newtheorem{theorem}{Theorem}[section]
\newtheorem{corollary}{Corollary}[section]
\newtheorem{lemma}{Lemma}[section]
\newtheoremstyle{boldremark}
    {\dimexpr\topsep/2\relax} 
    {\dimexpr\topsep/2\relax} 
    {}          
    {}          
    {\bfseries} 
    {.}         
    {.5em}      
    {}          
\theoremstyle{boldremark}

\graphicspath{ {./images/} }
\usepackage{dcolumn}

\usepackage{array}
\newcolumntype{E}{D{.}{.}{2,3}}
\newcolumntype{C}{>{$}c<{$}} 
\newcolumntype{R}{>{$}r<{$}} 
\usepackage{multirow} 

\usepackage{authblk}

\usepackage[margin=2.5cm]{geometry} 
\usepackage{afterpage} 

\usepackage[nodayofweek]{datetime} 

\usepackage[UKenglish]{isodate} 

\usepackage[colorlinks=false,urlcolor=blue,citecolor=blue,linkcolor=blue,bookmarks,bookmarksopen=true,bookmarksnumbered=true]{hyperref}

\usepackage[notcite]{showkeys} 
\usepackage{soul} 

\newcommand{\DL}{\mathit{DL}}
\newcommand{\DLd}{\mathit{D_dL}} 
\newcommand{\DLvp}{\mathit{D_{\varphi}L}} 
\newcommand{\DLvt}{\mathit{D_{\vartheta}L}}

\usepackage{fancyhdr}

\title{The modified conditional sum-of-squares estimator for fractionally integrated models\thanks{We are grateful to Richard Baillie, J\"{o}rg Breitung, Alexander Mayer, Sven Otto, Marc-Oliver Pohle, Philipp Sibbertsen and
    Dominik Wied for illuminating discussions and helpful suggestions. Valuable comments were also made by participants at the econometrics research seminars at Cologne University and at WHU -- Otto Beisheim School of Management,
    at the European Meeting of the Econometric Society in Milan, and at Statistische Woche in Dortmund. The financial support of Deutsche Forschungsgemeinschaft (project number 258395632) is gratefully acknowledged.}}
\author[1]{Mustafa R. K{\i}l{\i}n\c{c}} \affil[1]{WHU -- Otto Beisheim School of Management,\linebreak Chair of Econometrics and Statistics, Vallendar, Germany\vspace{1ex}} \author[1,2]{Michael Massmann\footnote{corresponding
    author: \href{mailto:michael.massmann@whu.edu}{\texttt{michael.massmann@whu.edu}} }$^,$} \affil[2]{Vrije Universiteit, Department of Econometrics and Data Science, Amsterdam, The Netherlands}
\date{12th March 2026}

\begin{document}

\fancyhf{}
\fancyhead[L]{\small \sf This is the Manuscript Accepted by the Journal of Econometrics. \newline \textcopyright \ 2026. This manuscript version is made available under the \href{https://creativecommons.org/licenses/by-nc-nd/4.0/}{CC-BY-NC-ND 4.0 license}.}
\renewcommand{\headrulewidth}{0pt}
\maketitle
\thispagestyle{fancy}

\begin{abstract}
  \noindent In this paper, we analyse the influence of estimating a constant term on the bias of the conditional sum-of-squares (CSS) estimator in a stationary or non-stationary type-II ARFIMA ($p_1$,$d$,$p_2$) model.  We derive expressions
  for the estimator's bias and show that the leading term can be easily removed by a simple modification of the CSS objective function. We call this new estimator the modified conditional sum-of-squares (MCSS) estimator. We show
  theoretically and by means of Monte Carlo simulations that its performance relative to that of the CSS estimator is markedly improved even for small sample sizes. Finally, we revisit three classical short datasets that have in
  the past been described by ARFIMA($p_1$,$d$,$p_2$) models with constant term, namely the post-second World War real GNP data, the extended Nelson-Plosser data, and the Nile data.



\end{abstract}

\clearpage

\newpage

\pagestyle{plain} 

\section{Introduction}

Fractionally integrated autoregressive moving average (ARFIMA) models are applied in a wide range of fields for describing long-memory phenomena, witness inter alia the economic and political as well as the natural sciences; see
\textcite{hassler2019time} and \textcite{hualde2021frac} for general treatments. One particular variant of this model class that has recently gained popularity is the so-called type-II ARFIMA model, which
truncates the fractional integration operator and allows both stationary and non-stationary processes to be described, see for example \textcite{nielsen2004efficient}, \textcite{ robinson2005distance} and
\textcite{johansen2008representation}. A popular choice for estimating this model is the conditional sum-of-squares (CSS) estimator whose main appealing features are that it is computationally straightforward and that the memory
parameter can be estimated consistently as long as it lies in an arbitrary compact interval on the real line. It was introduced by \textcite{li1986fractional} in the context of stationary fractionally integrated
models. Subsequent papers allowed for non-stationary models, see for instance \textcite{beran1995maximum} and \textcite{velasco2000whittle}. Local consistency proofs were provided by \textcite{tanaka1999nonstationary},
\textcite{nielsen2004efficient} and \textcite{robinson2006conditional}. Global consistency was proved by \textcite{hualde2011gaussian} and \textcite{nielsen2015asymptotics} in a model without deterministic components.  Only
recently, \textcite{hualde2020truncated,hualde2021truncated} derived global consistency and the asymptotic normality of the CSS estimator in a model with deterministic components, such as a constant or a trending term. Empirical
applications include \textcite{hualde2011gaussian} for aggregate income and consumption data and \textcite{johansen2016role} for opinion poll data.

While the literature dealing with asymptotic inferences in the context of parametric ARFIMA models is well-developed, some issues still require attention. One such concern pertains to the small sample performance of the CSS
estimator. Despite the widespread use of the CSS estimator little is currently known about the impact deterministic terms have on the properties of the estimator of the memory parameter in small samples.  Early on,
\textcite{chung1993small} and \textcite{cheung1994maximum} conducted simulation studies and found that the inclusion of a constant term in the model can substantially increase the small-sample bias and mean squared error (MSE) of the
estimated memory parameter. \textcite{lieberman2005expansions} and \textcite{johansen2016role} are among the few theoretical contributions to shed light on the issue. \textcite{lieberman2005expansions}
derive the Edgeworth expansion of the memory parameter for the Gaussian maximum likelihood estimator in stationary fractional time series model. \textcite{johansen2016role} investigate the impact of observed and unobserved initial
values on the bias of the memory parameter estimator in a non-stationary fractional time series model. Neither paper, however, includes short-run dynamics in its model. In addition, we are not aware of any related work that
simultaneously tackles both stationary and non-stationary processes.

The purpose of the present paper is therefore to add to this literature and analyse the small-sample bias of the CSS estimator in a type-II fractional model with short-run dynamics and constant term from an analytical, empirical
and simulation point of view. In particular, our analysis reveals that incorporating the level parameter into the model introduces an additional bias in the CSS estimator. This bias is due to a biased score which is particularly
pronounced when the data is stationary. We will suggest what we call the modified conditional sum-of-squares (MCSS) estimator which (i) is easy to compute, (ii) removes the leading bias term and (iii) allows much more accurate small-sample
inference.

To do so, we will interpret the constant term as nuisance parameter and draw on a large literature on bias correction. \textcite{laskar1998modified} provide an overview of this literature. We build on the approach to dealing with
nuisance parameters initiated by \textcite{conniffe1987expected} and \textcite{mccullagh1990simple} and recently applied by \textcite{bartolucci2016modified} and by \textcite{martellosio2020adjusted}, i.e.\ we adjust the score
function so that its expectation equals zero. The idea is as follows: We find a stochastic higher-order expansion of the estimator as a function of the derivatives of the profile likelihood, cf.\
\textcite{johansen2016role,lawley1956general}. The expansion is simplified by approximating the derivatives by their leading terms. This allows the expectation of the estimator to be evaluated explicitly. We show that a leading
component of the bias is attributable to the nonzero expectation of the score. By premultiplying the objective function by a suitable modification term then results in the expected score evaluated at the true parameter to be equal
to zero, thereby mitigating the bias of the estimator. 

It is important to emphasise that this paper tackles the correction of the CSS estimator's bias that is due to the estimation of the unknown constant term in the model, henceforth referred to as ``unknown-level bias''.  Of
course, there may be other, additional, sources of bias, yet they are not the focus of the present investigation. In particular, as we will discuss below, the estimator is also subject to what we call ``intrinsic bias'', i.e.\ the bias
also present if the constant term in the model is known. Moreover, what we will refer to as ``misspecification bias'' is due to conditioning on unobserved pre-sample values, as considered by \textcite{johansen2016role} and
\textcite{hualde2020truncated}. While there is no way around the intrinsic bias in our model, we abstract initially from the misspecification bias for expositional clarity by making a suitable simplifying assumption. Subsequently,
our results are extended to a more general model setup that includes unobserved pre-sample values.  

The main contributions of this paper to the literature are threefold: First, we examine our MCSS estimator in type-II ARFIMA($p_1$,$d$,$p_2$) models with constant term and compare it to the standard CSS estimator. In particular,
we derive its exact bias and we show that it is consistent and asymptotically normally distributed. The results generate new insights into the sources of bias and into bias correction of other models nested in our setup, such as purely fractional ARFIMA models and stationary ARMA($p_1$,$p_2$) models. Secondly, we re-visit three classical datasets
that have in the past been described by ARFIMA($p_1$,$d$,$p_2$) models with constant term, namely the post-second World War real GNP data, the extended Nelson-Plosser dataset, and the Nile data, by applying our MCSS estimator to
estimate the long-memory parameter and the short-run dynamics. All three time series are short and therefore warrant the use of small-sample bias corrections. Our conclusion sheds new light on the interpretation of these
datasets. Thirdly, this paper paves the way to extending the analysis of small-sample bias from univariate type-II ARFIMA modes to panel settings, see also the contributions of \textcite{robinson2015efficient} and
\textcite{schumann2023role}.

The remainder of the paper is organised as follows. In Section \ref{secgen} we present the MCSS estimator for a parametric
fractional time series model. In Section \ref{Ssimgen} we conduct a simulation study to examine the small sample properties of the estimators.
Section \ref{illustrations} presents
the empirical illustrations. Section \ref{S5} contains  concluding remarks. All proofs are relegated to the appendix.

\section{The modified conditional sum-of-squares estimator} \label{secgen}

\subsection{The model}
\label{model}

Consider a so-called type-II fractional process $z_t$, $t = 0,\pm 1,\pm 2,\ldots$, generated by the model
\begin{align}
    z_t = \Delta_+^{-d} \epsilon_t, \label{qq11} 
\end{align} 
with $\epsilon_t \sim \textit{IID}(0,\sigma^2)$, where $0 < \sigma^2 < \infty $. Here, $\Delta = 1 - L$ and $L$ are the difference and lag operators, respectively, and $d$ can take any value in $\mathbb{R}$. For any series $v_t$, real number $\zeta$ and time index $t \geq 1$, the so-called truncation operator $\Delta_+^{\zeta}$ is defined by $\Delta_+^{\zeta} v_t  = \Delta^{\zeta} \{ v_t I(t \geq 1)  \} = \sum_{i = 0}^{t-1} \pi_{i}(-\zeta) v_{t-i},$ with $I(\cdot)$ being the indicator function and with $\pi_{i}(a)$ denoting the coefficients in the usual binomial expansion $\Delta^{-a} = \sum_{i = 0}^{\infty} \pi_i(a) L^i$, where $\pi_{0}(a) = 1$ and $\pi_{i}(a) = (i - 1 + a) \pi_{i-1}(a) / i,$ for $i = 1, 2, \ldots$, see e.g.\ \textcite{hassler2019time}. The parameter $d$ in \eqref{qq11} is known as the memory parameter or the fractional parameter. A consequence of the indicator function in the definition of the truncation operator is that $z_t = 0$ for all $t \leq 0$. The process $z_{t}$ has been widely applied in the literature, see \textcite{marinucci2000weak,marinucci2001semiparametric},
\textcite{robinson2003cointegration}, \textcite{nielsen2004efficient}, \textcite{shimotsu2005exact}, \textcite{robinson2005distance} and \textcite{johansen2008representation}, among others.
An extension of this setup is introduced by \textcite{johansen2016role} who allow for $N_0$ unobserved pre-sample values of $z_t$ by defining
$ \Delta_{-N_0}^{\zeta} v_t = \sum_{i = 0}^{t+N_0-1} \pi_i(-\zeta) v_{t-i}$. As a result, \eqref{qq11} would become $z_t = \Delta_{-N_0}^{-d} \epsilon_t$. Setting $N_0 = 0$, the original truncation operator is recovered; the greater the value of $N_0$, the longer the burn-in period before the process is observed. 

Two comments on the memory parameter are of interest: First, its range is commonly divided into a ``stationary'' and a ``non-stationary'' region: $d < 1/2$ and $d \geq 1/2$, respectively. Since  the definition of the truncation operator implies
that $z_t = 0$ for $t \leq 0$ the process $z_t$ is in fact not covariance stationary when $d < 1/2$ and $d \neq 0$. However, it may be considered asymptotically stationary for any such $d$. To see this, consider the
so-called type-I fractional process $\tilde{z}_t = \Delta^{-d} \epsilon_t$ which is known to be covariance stationary for any $d < 1/2$. \textcite{marinucci1999alternative} observe that for $|d| < 1/2$,
$ E\left(  z_t - \tilde{z}_t \right)^2 = O(t^{2d -1}) $ 
as $t \rightarrow \infty$, and hence the difference between $\tilde{z}_t $ to $z_t$ vanishes. Although \textcite{marinucci1999alternative} consider only $|d| < 1/2$, their result actually holds for any $d < 1/2$. This follows from
Stirling's approximation and \textcite[Lemma A.1]{johansen2016role}. This asymptotic equivalence prompts us to retain the terminological dichotomy between stationarity and non-stationarity.  Secondly, it is worth noting that even
for $d \geq 1/2$, i.e.\ in the non-stationary region, the truncation operator ensures that the process $z_{t}$ is well-defined in the mean-square sense, see \textcite[Section A.4]{johansen2008representation} and
\textcite{hualde2011gaussian}.
    
While the model in \eqref{qq11} covers a wide range of dynamics, it seems unsuitable for many empirical applications for two key reasons. First, it implies that $E(z_t) = 0$. Secondly, the simple \textit{IID} errors assumed in model \eqref{qq11} are overly restrictive. In order to make our model more widely applicable, we therefore complement the model in \eqref{qq11} by a constant term $\mu$ and replace $\epsilon_t$ by $u_t$ to yield
\begin{align}
    x_t &= \mu I(t \geq 1) + \Delta_+^{-d} u_t, \label{genq1}\\
    u_t &= \omega(L;\varphi) \epsilon_t, \label{genq2}
\end{align}
where  $\omega$ is a lag polynomial and captures the short-run dependence structure parametrically, given by 
\begin{align}
     \omega(L;\varphi) = \sum_{j = 0}^{\infty} \omega_j(\varphi) L^j, \label{repmainf}
\end{align}
with $\varphi$ being an unknown $p\times1$ vector, $\omega_0(\varphi) = 1$, $|\omega(s;\varphi)| \neq 0$ for $|s| \leq 1$, and $\sum_{j = 0}^{\infty} |\omega_j(\varphi)| < \infty$. More precise conditions on $\omega$ will be
specified below.  The representation of $u_t$ in \eqref{genq2} as a MA$(\infty)$ model is common in the literature and considered by, among others, \textcite{hualde2011gaussian} and
\textcite{hualde2020truncated,hualde2021truncated}.  As a consequence of the constant term, $E(x_t) = \mu$ for $t \geq 1$. Note that equation \eqref{genq1} implies that $x_t = 0$ for $t \leq 0$ although equation \eqref{genq2} allows for the process $u_t$ to have started in the infinite past. Indeed, the literature has so far considered several different specifications of pre-sample values, as we will discuss further below. 

One well-known special case of $u_t$ in \eqref{genq2} is an ARMA$(p_1,p_2)$ model which is given by 
\begin{align}
    \omega(L;\varphi) = \frac{\alpha(L;\varphi)}{\beta(L;\varphi)}, \label{arma}
\end{align}
where $\beta(L;\varphi)$ is the AR polynomial of order $p_1$ and $\alpha(L;\varphi)$ is the MA polynomial of order $p_2$. It is assumed that the polynomials do not have common roots and that their roots lie outside the unit
circle. Then \eqref{genq1}, \eqref{genq2} and \eqref{arma} is an ARFIMA$(p_1,d,p_2)$ model. Another special case of $u_t$ in \eqref{genq2} is \citeauthor{bloomfield1973exponential}'s (1973) exponential spectrum model, see
\textcite{robinson1994efficient} and \textcite{hassler2019time}.

Following \textcite{hualde2020truncated}, we make the following assumptions on the model's short-run dynamics and on the admissible parameter space. We use the notation $\vartheta = (d,\varphi')'$. True parameter values are
denoted by the subscript 0, i.e.\ $\vartheta_0 = (d_0,\varphi_0')'$ and $\mu_0$.

\begin{assumption}\label{A2}
The errors $\epsilon_t$ are \textit{IID}(0,$\sigma_0^2$) with finite fourth moments.
\end{assumption}

\begin{assumption}\label{A3}
The parameter space for $\vartheta = (d,\varphi')'$ is given by $\Theta  = [\nabla_1,\nabla_2] \times \Phi$, with $-\infty < \nabla_1 < \nabla_2 < \infty$ and $\Phi$ being a compact and convex subset of $\mathbb{R}^p$. The parameter space for $\mu$ is $\mathbb{R}$. The true value $\vartheta_0 = (d_0,\varphi_0')' \in \Theta $, with $d_0$ not equal to 1/2, and $\mu_0 \in \mathbb{R}$. 
\end{assumption}

\begin{assumption}\label{A1}
\begin{enumerate}[label=(\roman*)]
    \item For all $\varphi \in \Phi \backslash \{\varphi_0 \}$, $|\omega(s;\varphi)|\neq|\omega(s;\varphi_0)|$  on a set $S \subset \{ s : |s| = 1 \}$ of positive Lebesgue measure. \label{i}
    \item For all $\varphi \in \Phi$, $\omega(e^{i\lambda};\varphi)$ is differentiable in $\lambda$ with derivative in Lip$(\varsigma )$ for $1/2 < \varsigma  \leq 1$. 
    \item For all $\lambda$, $\omega(e^{i\lambda};\varphi)$ is continuous in $\varphi$.
    \item For all $\varphi \in \Phi $, $|\omega(s;\varphi)|\neq 0$, $|s| \leq 1$. 
    \item The true value $\vartheta_0$ is in the interior of $\Theta$.
    \item For all $\lambda$,  $\omega(e^{i\lambda};\varphi)$ is thrice continuously differentiable in $\varphi$ in a closed neighbourhood $\mathcal{N}_{\varepsilon}(\varphi_0)$ of radius $\varepsilon \in (0,1/2)$ about $\varphi_0$. For all $\varphi \in \mathcal{N}_{\varepsilon}(\varphi_0)$ these partial derivatives with respect to $\varphi$ are themselves differentiable in $\lambda$ with derivative in Lip$(\varsigma )$ for $1/2 < \varsigma  \leq 1$. 
    \item The matrix 
    \begin{align}
    A = \begin{pmatrix}
\pi^2/6 & - \sum_{j = 1}^{\infty} b_{\varphi' j}(\varphi_0)/j \\
- \sum_{j = 1}^{\infty} b_{\varphi j}(\varphi_0)/j  & \sum_{j = 1}^{\infty} b_{\varphi j}(\varphi_0) b_{\varphi' j}(\varphi_0)
\end{pmatrix}  \label{genA}
\end{align}
is nonsingular, where $  b_{\varphi j}(\varphi_0) = \sum_{k = 0}^{j-1} \omega_k(\varphi_0) \partial \phi_{j-k}(\varphi_0)/\partial \varphi $
and where $\phi_{j}$ is defined by $ \phi(s;\varphi) = \omega^{-1}(s;\varphi) = \sum_{j = 0}^{\infty} \phi_j(\varphi) s^j.$ 
\end{enumerate}
\end{assumption}

Assumption \ref{A2} states that the errors $\epsilon_t$ are \textit{IID}, but it does not restrict them to normality. In fact, the \textit{IID} assumption can be weakened to martingale difference series as in
\textcite{hualde2020truncated,hualde2021truncated} but for the sake of convenience we keep this condition simple. Assumption \ref{A3} covers the admissable parameter spaces and the true parameter values. Note that we exclude
the boundary case of $d_0 = 1/2$. The reason is that our analysis of the CSS estimator's bias takes its consistency for granted. To the best of our knowledge, \textcite{hualde2020truncated,hualde2021truncated} are the only papers
to show consistency in a model with constant term, yet they exclude the value of $d_0 = 1/2$.\footnote{Specifically, \textcite{hualde2020truncated,hualde2021truncated} study CSS estimation in models with generalised
  polynomial or power-law trends whose coefficients and exponents are unknown, and they exclude the boundary case $d_0 = 1/2$ for technical reasons. A constant term is, of course, a known deterministic structure, yet proving
  consistency of the CSS estimator in this model is beyond the scope of our analysis.}  Assumption \ref{A1}$(i)$-$(iv)$, which ensures the identification of the short-term dynamics, is standard in the literature on parametric
short-memory models since its introduction by \textcite{hannan1973asymptotic}. Assumption \ref{A1}$(v)$-$(vii)$ serve as additional regulatory conditions necessary to establish the asymptotic distribution theory. We refer to the
papers by \textcite{hualde2011gaussian}, \textcite{nielsen2015asymptotics}, \textcite{hualde2020truncated,hualde2021truncated} for a detailed discussion of Assumption \ref{A1}. Importantly, it is satisfied for the stationary and
invertible ARMA model and also the exponential spectrum model of \textcite{bloomfield1973exponential}.

For the purpose of expositional clarity we also make the following assumption in our initial analysis. Subsequently, our main results will be generalised to a model setting in which this restriction is ignored.
\begin{assumption}\label{A5}
 For all $t \leq 0$, we assume that $\epsilon_t = 0$  in \eqref{genq2} while, for $t > 1$, $\epsilon_t$ satisfies Assumption \ref{A2}.
\end{assumption}
Assumption \ref{A5} boils down to not only the obervations $x_t$ in \eqref{genq1} but also the unobserved error terms $u_t$ and $\varepsilon_t$ in \eqref{genq2} being zero before $t = 1$.
Assumption \ref{A5} reflects the definition of the CSS estimator in that the latter conditions on all pre-sample information. Importantly, it is not required for the consistency or asymptotic normality of our MCSS estimator, nor
does it have an impact on the correction of the unknown-level bias. Instead, as will become plain in the generalisation presented in Section \ref{sc:initial}, it helps to disentangle the unknown-level bias from the model
misspecification bias, i.e.\ from the bias that is due to the CSS estimator not making use of pre-sample information. Several papers in the literature have employed an assumption similar to Assumption \ref{A5}, see e.g.\
\textcite{SibbertsenEtAl18} and \textcite{robinson2020estimation}.  Alternative initialisation schemes also exist: the paper by \textcite{johansen2016role} allows for a finite
number of unobserved pre-sample values, while \textcite{hualde2011gaussian} and \textcite{hualde2020truncated} assume that both $u_t$ and $\varepsilon_t$ have started in the infinite past. Indeed, this latter generalisation will
be used in Section \ref{sc:initial} below to dispense with Assumption \ref{A5} and to extend the main results of our analysis.

\subsection{The conditional sum-of-squares estimator} \label{subsect22}

We now discuss the conditional sum-of-squares (CSS) estimator of the parameters in model \eqref{genq1}-\eqref{repmainf}. This is the estimator considered by e.g.\ \textcite{hualde2011gaussian} who, however, look at a model without the constant term. We
distinguish the case in which $\mu$ is unknown from that in which it is known. As will be seen in Section \ref{sectionmod} below, the CSS estimator may also be motivated as a conditional maximum likelihood estimator under the
assumption of Gaussian innovation terms $\epsilon_t$, as in \textcite{johansen2016role} and \textcite{hualde2020truncated}.

Consider a sample of observations for $t = 1, \ldots, T$. For any $(\vartheta,\mu)$ in the admissible parameter space, define the residuals $\epsilon_t(\vartheta,\mu) = \phi(L;\varphi) \Delta_+^{d}(x_t-\mu)$. The CSS objective
function is then given by
\begin{align}
    L(\vartheta,\mu) = \frac{1}{2} \sum_{t = 1}^T \epsilon^2_t(\vartheta,\mu) =   \frac{1}{2} \sum_{t = 1}^T \left(  \phi(L;\varphi)\Delta_{+}^{d} x_t- \mu c_t(\vartheta)  \right)^2, \label{genlikmu1}
\end{align}
where we define the convoluted coefficient
\begin{align}
    c_t(\vartheta) = \phi(L;\varphi)\Delta_{+}^{d}  I(t \geq 1) = \sum_{j = 0}^{t-1} \phi_j(\varphi) \kappa_{0(t-j)}(d), \label{convcoef}
\end{align}
with $ \kappa_{0t}(d) = \Delta_{+}^{d} I(t \geq 1)=  \sum_{n = 0}^{t-1} \pi_n(-d) = \pi_{t-1}(1-d), $ 
cf.\ \textcite[Lemma A.4]{johansen2016role}.

Since $L(\vartheta,\mu)$ in \eqref{genlikmu1} is quadratic in $\mu$ we can concentrate it. Unsurprisingly, the CSS estimator of $\mu$ for fixed $\vartheta$ is given by
\begin{align}
    \hat{\mu}(\vartheta) = \frac{\sum_{t = 1}^T ( \phi(L;\varphi)\Delta_{+}^{d} x_t)c_t(\vartheta)}{\sum_{t = 1}^T c^2_t(\vartheta)}. \label{genmu1}
\end{align}
Substituting $\hat \mu (\vartheta)$ into \eqref{genlikmu1} yields the profile (or concentrated) CSS function
\begin{align} 
    L^*(\vartheta) = \frac{1}{2} \sum_{t = 1}^T \left(  \phi(L;\varphi)\Delta_{+}^{d} x_t- \hat{\mu}(\vartheta) c_t(\vartheta)  \right)^2. \label{genL1}
\end{align} 
Note that we use asterisks to emphasise that we are dealing with a \textit{profile} objective function. The resulting CSS estimator of $\vartheta = (d,\varphi')'$ is given by
\begin{align}
    \hat{\vartheta} = \operatorname*{argmin}_{\vartheta \in \Theta}  L^*(\vartheta). \label{genCSS}
\end{align}
As discussed in Section \ref{model}, the model effectively conditions on $x_t = 0$, for $t \leq 0$. For this reason, \textcite{hualde2011gaussian} and \textcite{hualde2020truncated} prefer to call the estimator in \eqref{genCSS}
the truncated sum-of-squares estimator.  \textcite{hualde2020truncated} show that if $x_t$ is generated by \eqref{genq1}-\eqref{repmainf} and if Assumptions \ref{A2} to \ref{A1} hold, then, as $T \rightarrow \infty$,
$\hat{\vartheta} \overset{P}{\rightarrow} \vartheta_0$ and $ \sqrt{T} (\hat{\vartheta} - \vartheta_0 ) \xrightarrow{D} N(0_{p+1},A^{-1}), $ 
where $A$ is given in \eqref{genA}. Note that the authors exclude the boundary case of $d_0 = 1/2$ from their analysis, see also the discussion of our Assumption \ref{A3} above. We take this consistency of the CSS estimator as
basis for the following analysis of its bias.

A few remarks about the estimator $\hat{\mu}(\vartheta)$ in \eqref{genmu1} are instructive. For $\vartheta = \vartheta_0$ we have that $\hat{\mu}(\vartheta_0)-\mu_0 = \sum_{t = 1}^T \epsilon_t c_t(\vartheta_0) / \sum_{t = 1}^T c^2_t(\vartheta_0) ,$
which has mean zero and variance $\sigma_0^2 (\sum_{t = 1}^T c^2_t(\vartheta_0))^{-1}$. In the stationary region, i.e.\ when $d_0 < 1/2$, this variance goes to zero because then $\sum_{t = 1}^T  c^2_t(\vartheta_0)$ diverges in $T$,
see Lemma \ref{genlemmaaaa2s}. As opposed to that, in the non-stationary region, i.e.\ when $d_0 > 1/2$, this variance does not converge to zero because then $\sum_{t = 1}^T  c^2_t(\vartheta_0)$ is bounded in $T$, see Lemma \ref{genlemmaaaa2n}. This
is the reason why $\hat{\mu}(\hat{\vartheta}) \xrightarrow{P} \mu_0$ only if $d_0 < 1/2$, see \textcite[Corollary 1]{hualde2020truncated} for the proof.

For comparison, we also analyse the situation where the true $\mu_0$ is known. The CSS estimator for this model can be derived by substituting $\mu_0$ into \eqref{genlikmu1} to have 
\begin{align}
    L^*_{\mu_0}(\vartheta)   &=   \frac{1}{2} \sum_{t = 1}^T \left(  \phi(L;\varphi)\Delta_{+}^{d} x_t- \mu_0 c_t(\vartheta)  \right)^2, \label{genlikmu1known}
\end{align}
such that 
\begin{align}
    \hat{\vartheta}_{\mu_0} = \operatorname*{argmin}_{\vartheta \in \Theta} L^*_{\mu_0}(\vartheta). \label{genCSSmuknown}
\end{align}
This estimator is considered by \textcite{hualde2011gaussian} and \textcite{nielsen2015asymptotics} who show that if $x_t$ is generated by \eqref{genq1}-\eqref{repmainf} and if Assumptions \ref{A2} to \ref{A1} hold, then
$\hat \vartheta_{\mu_0}$ is consistent, too, attaining the same limiting distribution as $\hat{\vartheta}$. In other words, the distribution does not depend on whether $\mu$ is known or needs to be
estimated.

\subsection{The modified profile likelihood} \label{sectionmod}

A central concern in this paper is to investigate the bias of $\hat{\vartheta}$ in \eqref{genCSS} and of $\hat{\vartheta}_{\mu_0}$ in \eqref{genCSSmuknown}. This will be done in Section \ref{S3} below. It will turn out that the expectation of the CSS estimators is a function of the expectation of the score functions, or first derivatives, of $L^*(\vartheta)$ and $L_{\mu_0}^*(\vartheta)$ evaluated at $\vartheta = \vartheta_0$, respectively. The present section therefore examines the bias of the two scores
and builds on an approach by \textcite{mccullagh1990simple} to correct for it.

To that end, it will be instructive to interpret the CSS objective in \eqref{genlikmu1} as a log-likelihood function, as do  \textcite{johansen2016role} and \textcite{hualde2020truncated}. Assuming for the moment that
$\epsilon_t \sim \textit{NID}(0,\sigma^2)$, the Gaussian log-likelihood of $x_t$ in \eqref{genq1}, conditional on $x_t$ = $0$ for $t\leq 0$, is given by $l (\vartheta,\mu,\sigma^2) = - T \log (\sigma^2 ) /2 - \sum_{t = 1}^T (  \phi(L;\varphi) \Delta_+^{d} (x_t-\mu ) )^2 / (2\sigma^2). $ Throughout the paper, we omit additive constants in log-likelihood functions for notational simplicity. 
Maximising $l (\vartheta,\mu,\sigma^2)$ with respect to $\sigma^2$ yields $    \hat{\sigma}^2(\vartheta,\mu) = T^{-1} \sum_{t = 1}^T ( \phi(L;\varphi) \Delta_+^{d} (x_t-\mu ) )^2 $
and the resulting profile log-likelihood is
\begin{align}
     \ell(\vartheta,\mu) = l (\vartheta,\mu,\hat{\sigma}^2(\vartheta,\mu)) = -\frac{T}{2} \log\left( \frac{1}{T} \sum_{t = 1}^T \left(   \phi(L;\varphi) \Delta_+^{d} (x_t-\mu ) \right)^2\right). \label{lq2}
\end{align}  
Maximising $\ell (\vartheta,\mu)$ further with respect to $\mu$ results in $\hat \mu (\vartheta)$ in \eqref{genmu1} and the profile log-likelihood function becomes
\begin{align}
    \ell^*(\vartheta) = \ell (\vartheta,\hat{\mu}(\vartheta) )  = -\frac{T}{2} \log\left( \frac{1}{T} \sum_{t = 1}^T \left(   \phi(L;\varphi) \Delta_+^{d} (x_t-\hat{\mu}(\vartheta) ) \right)^2\right).  \label{lq3}
\end{align}
Clearly, the estimator of $\vartheta$ resulting from maximising \eqref{lq3} is identical to that obtained by minimising \eqref{genL1} since
\begin{align}
     L^*(\vartheta) &=  \frac{T}{2}  \exp \left( -\frac{2}{T} \ell^*(\vartheta) \right).  \label{Lstar}
\end{align}
So, the CSS objective $L^*(\vartheta)$ can be seen as a negative non-logged profile likelihood even though we do not impose Normality on the error term $\epsilon_t$ in Assumption \ref{A2}. As the maximum likelihood estimator of $\vartheta$ is asymptotically efficient, see \textcite{hualde2020truncated}, so is the CSS estimator $\hat \vartheta$ in
\eqref{genCSS}. The same can of course be said of $\hat \vartheta_{\mu_0}$ in \eqref{genCSSmuknown} since the profile CSS objective $ L^*_{\mu_0}(d)$ in \eqref{genlikmu1known} can be obtained from \eqref{lq2} by replacing $\mu$ by its known value
$\mu_0$ such that $ L^*_{\mu_0}(\vartheta) = T  \exp ( - 2 \ell(\vartheta,\mu_0)/T ) /2 . $ 

We will in the present section therefore interpret $L^*(\vartheta)$ in \eqref{genL1} as a profile likelihood. As such, it is not a genuine likelihood, for it is not directly based on observable quantities, see
\textcite{barndorff1983formula} and \textcite{severini2000likelihood}. Instead, it is a function of the maximum likelihood estimators of $\mu$ and $\sigma^2$ which are treated as if they were the true parameter values. In large
samples, the concentration procedure has relatively minor effects, yet \textcite{chung1993small} showed in Monte Carlo simulations that in small samples it leads to a strong bias in $\hat{\vartheta}$. This is because profile
likelihoods do not necessarily possess the same properties as genuine likelihoods. It is well-known that, under classical regularity conditions and with a fixed number of regressors, the score of the profile likelihood is
biased. In particular, its expectation is $O(1)$, see \textcite{kalbfleisch1973marginal}, \textcite{mccullagh1990simple} and \textcite{liang1995inference}. The following theorem derives the bias of the scores of $L^*(\vartheta)$
and $L^*_{\mu_0}(\vartheta)$. The proof will be given in Appendix \ref{generalizationsappendix} and a generalisation that dispenses with Assumption \ref{A5} in Theorem \ref{genlemmaexpectations-} of Appendix
\ref{Initschemes}. We use the notation $D_i f(\vartheta) = D_i f(d, \varphi)$ to denote the first derivative of a function $f (\vartheta) = f(d, \varphi)$ with respect to parameter $i \in \{\vartheta, d, \varphi\}$.
\begin{theorem}\label{t52} 
Let $x_t$, $t$ = 1,$\ldots$,$T$, be given by \eqref{genq1}-\eqref{repmainf} and let Assumptions \ref{A2} to \ref{A5} be satisfied. Then, the expected scores of $L^* (d,\varphi)$, evaluated at the true parameters $d_0$ and $\varphi_0$, are given by
\begin{align}
    E( \DLd^*(d_0,\varphi_0)) &= O(\log(T)I(d_0 < 1/2) + I(d_0 > 1/2) ), \label{genscored}\\
    E( \DLvp^*(d_0,\varphi_0)) &=  O(1) \label{genscorevarphi},
\end{align}
when $T \rightarrow \infty$. The expected scores of $L_{\mu_0}^* (d,\varphi)$, evaluated at the true parameters $d_0$ and $\varphi_0$, are given by
\begin{align}
    E( \DLd_{\mu_0}^*(d_0,\varphi_0)) &= 0, \label{genscoredknown}\\
    E( \DLvp_{\mu_0}^*(d_0,\varphi_0)) &= 0_p. \label{genscorevarphiknown}
\end{align}

\end{theorem} 

Clearly, the scores $\DLd^*(d_0,\varphi_0)$ and $\DLvp^*(d_0,\varphi_0)$ are biased. In addition, the expectation of the score in \eqref{genscored}, i.e.\ the score with respect to $d$, is not uniform in $d_0$. For $d_0 < 1/2$ it
diverges at the rate of $\log(T)$, while it is $O(1)$ for $d_0 > 1/2$. This is not true for the expectation of the score in \eqref{genscorevarphi}, i.e.\ the score with respect to $\varphi$, which is $O(1)$ uniformly in $d_0$.
The rationale behind this is that the score bias measures the relative strengths of the level parameter and the stochastic component. The score bias with respect to $d$ gauges the strength of the level parameter relative to the
fractional dynamics, whereas the score bias with respect to $\varphi$ evaluates the strength of the level parameter in relation to short-run dynamics. In the non-stationary region, i.e.\ when $d_0 > 1/2$, we recall that $\mu$
is not consistently estimated, see the discussion in Section \ref{subsect22}. The reason is that the stochastic component $\Delta_+^{-d} u_t$ in \eqref{genq1} dominates the deterministic component
$\mu$. Hence, the bias in the score is less influenced by $\hat{\mu}(\hat{\vartheta})$, resulting in the expected scores being $O(1)$ for such $d_0$. On the other hand, if $d_0 < 1/2$, $\mu$ is consistently
estimated and $\hat{\mu}(\hat{\vartheta})$ plays a more important role in the bias of the scores and especially for the score with respect to the fractional dynamics. This is reflected by the expected score in \eqref{genscored}
being $O(\log(T))$ for such $d_0$. Recall that, at the beginning of this section, we mentioned that the bias of the CSS estimator is a function of the score bias. One might be tempted to think, from \eqref{genscored} and
\eqref{genscorevarphi}, that in the stationary region the bias of $\hat{d}$ will be of a larger order of magnitude than the bias of $\hat{\varphi}$.  Yet this turns out not to be true. As will be shown below, the biases of
$\hat{d}$ and $\hat{\varphi}$ are functions of not only their own score biases but, instead, of a weighted sum of both score biases. This will lead to the order of the bias of the short-run dynamics to be the same as that of the
memory parameter. The situation for $L_{\mu_0}^*(d,\varphi)$ in \eqref{genlikmu1known} is somewhat different. Although, technically speaking, $L_{\mu_0}^*(d,\varphi)$ is also a profile likelihood due to the substitution of
$\hat{\sigma}^2$ for $\sigma^2$, its scores are unbiased, as shown in \eqref{genscoredknown} and \eqref{genscorevarphiknown}.

This discussion highlights the need for a modification of the profile likelihood function such that it behaves more like a genuine likelihood in terms of score unbiasedness. This modification will eliminate the bias of
the CSS estimator $\hat \vartheta$ stemming from the presence of the unknown nuisance parameter $\mu$, as will be seen in Section \ref{S3}. The idea of modifying the profile likelihood to obtain score unbiasedness is in fact not new
and was previously discussed by \textcite{mccullagh1990simple}. \textcite{martellosio2020adjusted}, for instance, implement this idea for a spatial model.

To develop the idea of an unbiased score we follow \textcite{mccullagh1990simple} by considering for simplicity a profile log-likelihood in a scalar parameter $\vartheta$. Recentering its score yields, say, $ D  \ell^*_{a}(\vartheta) = D  \ell^*(\vartheta) - a(\vartheta), $ where $ \ell^*(\vartheta)$ denotes, as before, the profile log-likelihood function and where $a(\vartheta)$ is an adjustment function only depending on $\vartheta$. Then \textcite{mccullagh1990simple} require that  $   E \left( D  \ell^*_{a}(\vartheta_0) \right) = 0, $ which implies that 
\begin{align}
    a(\vartheta_0) = E \left(  D  \ell^*(\vartheta_0)  \right), \label{tm3}
\end{align}
for all $\vartheta_0$.
Finally, they call $\ell^*_{a}(\vartheta) = \int_{\Theta}  D  \ell^*_{a}(t) dt$  the adjusted profile log-likelihood, which is subsequently maximised w.r.t.\ $\vartheta$.

Note that \textcite{mccullagh1990simple} further adjust $D\ell^*_{a}(\vartheta)$ to make it information unbiased, i.e.\ by making its variance equal the negative expectation of the derivative of the score. While these adjustments
may improve the efficiency of the estimator, they are not addressed in the present paper because they do not affect the location of the zeros of $D\ell^*_{a}(\vartheta)$.

The adjustment function $a(\vartheta)$ can in principle be computed from \eqref{tm3}. Yet this calculation is challenging as can be seen by rewriting \eqref{tm3} as 
\begin{align}
      a(\vartheta_0)  &= E \left( - \frac{1}{2}\frac{\DL^*(\vartheta_0)}{T^{-1} L^*(\vartheta_0)}  \right). \label{diffadj}
\end{align}
In special cases, the computation of $a(\vartheta)$ may indeed be feasible, see for instance the spatial model with Gaussian errors considered by \textcite{martellosio2020adjusted}. 
In general, however, the evaluation of \eqref{diffadj} is not straightforward. In the sequel, we therefore present an argument of how the expectation of the fraction can be circumvented.
To that end, we consider the profile CSS objective function $L^*(\vartheta)$ as basis for the adjustment. Indeed, \textcite{mccullagh1990simple} in their Remark 3 allude to the possibility of using an
objective function other than the profile log-likelihood $\ell^*(\vartheta)$ for deriving an adjustment.

With a view to framing the approach of \textcite{mccullagh1990simple} in terms of $L^*(\vartheta)$, note that the adjusted profile log-likelihood can be written as
\begin{align}
      \ell^*_{a}(\vartheta) = \int_{\Theta} \left(D  \ell^*(t) - a(t)\right) dt =  \ell^*(\vartheta) - A(\vartheta), \label{tbs51}
\end{align}
with $A(\vartheta) = \int_{\Theta} a(t) dt$. Based on the relationship between $ \ell^*(\vartheta)$ and $L^*(\vartheta)$ in \eqref{Lstar}, we can write \eqref{tbs51} as
\begin{align}
      \ell^*_{a}(\vartheta) &= -\frac{T}{2} \log\left( \frac{2}{T}  L^*(\vartheta)  \right) - A(\vartheta). \label{tbs1}
\end{align}
Clearly, maximising the adjusted profile log-likelihood $ \ell^*_a (\vartheta)$ in \eqref{tbs1} is equivalent to minimising the adjusted profile CSS objective $L^*_{a}(\vartheta) =  T \exp ( -2 \ell^*_{a}(\vartheta)/ T) / 2 $  
which, upon substituting in $ \ell^*_a (\vartheta)$, is
\begin{align}
     L^*_{a}(\vartheta) &= \exp{\left(\frac{2}{T} A(\vartheta)\right)}L^*(\vartheta). \label{tbs3}
\end{align}
It is important to note that while the adjustment in \eqref{tbs1} is additive, it is multiplicative in \eqref{tbs3}. Recall that $a(\vartheta)$ in \eqref{diffadj}, and thus $A(\vartheta)$ in \eqref{tbs3}, is difficult to compute. We therefore define, as an alternative, the modified profile CSS objective function
\begin{align}
    L^*_{m}(\vartheta) &= m(\vartheta) L^*(\vartheta),  \label{genmlik}
\end{align}
where the multiplicative modification term $m(\vartheta) > 0$ depends only on $\vartheta$. The corresponding score function is the first derivative of \eqref{genmlik}:
\begin{align}
    \DL^*_{m}(\vartheta) = m(\vartheta) \DL^*(\vartheta) +  D m(\vartheta)   L^*(\vartheta). \label{gentibmc12}
\end{align}
As do \textcite{mccullagh1990simple}, we now require that our objective function is score unbiased, i.e.\ that the score function in \eqref{gentibmc12} satisfies $E \left( \DL^*_{m}(\vartheta_0)  \right) = 0. $  
Using \eqref{gentibmc12} and the fact that $ D \log (m(\vartheta)) = D m(\vartheta) /m(\vartheta)$ as well as $m(\vartheta) > 0$, it follows that this condition is equivalent to
\begin{align}
     D \log\left( m(\vartheta_0) \right) = -\frac{E\left( \DL^*(\vartheta_0) \right)}{E\left(L^*(\vartheta_0) \right)}. \label{solveformgen}
\end{align}
It will be seen in the next section that the evaluation of \eqref{solveformgen} is straightforward, as opposed to the evaluation of \eqref{diffadj}.

\subsection{The modified conditional sum-of-squares estimator} \label{Smcss}

Re-write now the condition in \eqref{solveformgen} in terms of a parameter vector $\vartheta$ so $D_{\vartheta}L^*(\vartheta)$ denotes the vector of derivatives of $L^* (\vartheta)$ w.r.t.\ $\vartheta$.  The condition can now be
used for finding the modification term $m(\vartheta)$ for the modified profile CSS objective in \eqref{genmlik}: First, it is shown in Lemma \ref{genlemmaexpectations} that the expectation of $D_{\vartheta}L^*(\vartheta_0)$ equals
\begin{align*}
  E\left( D_{\vartheta}L^*(\vartheta_0) \right) =  -\sigma^2_0 \frac{\sum_{t = 1}^T c_{t}(\vartheta_0)  D_{\vartheta}c_t(\vartheta_0) }{\sum_{t = 1}^T c^2_{t}(\vartheta_0)},
\end{align*}
where $c_{t}(\vartheta_0)$ is given in  \eqref{convcoef}. It is also shown in Lemma \ref{genlemmaexpectations} that $E\left( L^*(\vartheta_0) \right) =  \sigma^2_0 (T-1)/2$. Consequently, from \eqref{solveformgen}, we have
\begin{align}
  D_{\vartheta} \log\left( m(\vartheta_0) \right) = \frac{2}{T-1} \frac{ \sum_{t = 1}^T c_t(\vartheta_0)  D_{\vartheta}c_t(\vartheta_0)   }{  \sum_{t = 1}^T  c^2_t(\vartheta_0) }. \label{genanitd}
\end{align}
Integrating and exponentiating \eqref{genanitd}  yields $ m(\vartheta) = e^k ( \sum_{t = 1}^T c^2_t(\vartheta)  )^{1/\left(T-1 \right)}, $
where $k$ is the constant of integration.

The modified profile CSS objective function in \eqref{genmlik} is thus given by the product of $ m(\vartheta)$ and $ L^*(\vartheta)$. Clearly the argument that minimises the likelihood is invariant to the choice of $k$ so
that, in the sequel, we can set $k = 0$ without loss of generality. The resulting modified profile CSS objective function is then given by
\begin{align*}
   L_m^*(\vartheta) = \left( \sum_{t = 1}^T c^2_t(\vartheta)  \right)^{\frac{1}{T-1}}  \frac{1}{2} \sum_{t = 1}^T \left(  \phi(L;\varphi)\Delta_{+}^{d} x_t- \hat{\mu}(\vartheta) c_t(\vartheta)  \right)^2 .
\end{align*}
We call the argument that minimises $L_m^*(\vartheta)$ the modified conditional sum-of-squares (MCSS) estimator and denote it by $\hat{\vartheta}_{m}$, i.e.\,
\begin{align}
    \hat{\vartheta}_{m} = \operatorname*{argmin}_{\vartheta \in \Theta} L^*_{m}(\vartheta).  \label{MCSSgen1}
\end{align}
The modification term, with $k = 0$ imposed, is
\begin{align}
     m(\vartheta) &= \left( \sum_{t = 1}^T c^2_t(\vartheta)  \right)^{\frac{1}{T-1}}. \label{genmodificationterm}
\end{align}
Two important properties of the modification term $m(\vartheta)$ are stated in the following lemma. See Appendix \ref{lemma31} for the proof. 
\begin{lemma} \label{gen:lm:md}
  For all $d \in \mathbb{R} $ and $\varphi \in \Phi$,
  \begin{align}
    m(\vartheta) &\geq 1. \label{genmqeq1}
    \end{align}
Here, equality holds if $d$ = 1 and $\omega(L;\varphi) = 1$.
Also, it holds that, for $T \rightarrow \infty$,
    {\small
    \begin{align}
    m(\vartheta) &= 1 + O\Big(T^{-1}\log(T) I(d < 1/2) + T^{-1}\log(\log(T)) I(d = 1/2) + T^{-1} I(d > 1/2) \Big).  \label{geneq111} 
  \end{align}
  }
\end{lemma}
The property in \eqref{genmqeq1} implies that the modification term $m(\vartheta)$ acts as a penalty in the minimisation of the modified profile likelihood $L_m^*(\vartheta)$ through inflating $L^*(\vartheta)$ by
$m(\vartheta)$. The property in \eqref{geneq111} ensures that $m(\vartheta) \rightarrow 1$ such that the asymptotic properties of the MCSS estimator $\hat \vartheta_m$ are the same as those of the CSS estimator $\hat \vartheta$ in
\eqref{genCSS}. This is desirable because the CSS estimator is efficient under Gaussianity, as argued in Section \ref{sectionmod}. The asymptotic properties of $\hat \vartheta_m$ are summarised for completeness in the following
theorem and are proved in Appendix \ref{generalizationsappendix}. Note that no use is made of Assumption \ref{A5}.
\begin{theorem}\label{t51} 
Let $x_t$, $t$ = 1,$\ldots$,$T$, be given by \eqref{genq1}-\eqref{repmainf} and let Assumptions \ref{A2} to \ref{A1} be satisfied. Then, as $T \rightarrow \infty$, $\hat{\vartheta}_{m} \overset{P}{\rightarrow} \vartheta_0$ and $\sqrt{T} (\hat{\vartheta}_{m} - \vartheta_0 ) \xrightarrow{D} N(0_{p+1},A^{-1})$ 
where $A$ is given in \eqref{genA}.
\end{theorem} 

The intuitive explanation of Theorem \ref{t51} follows from noticing that
\begin{align}
\label{ObjecApprox}
\begin{split}
    L^*_{m}(\vartheta_0) &=  L^*(\vartheta_0) + O_P(1) \  \ \  \ \ \ \ \text{ for }d_0 > 1/2, \\
    L^*_{m}(\vartheta_0) &=  L^*(\vartheta_0) + O_P(\log(T))\text{ for }d_0 < 1/2, 
\end{split}
\end{align}
where use was made of the definition $L^*_{m}(\vartheta_0)$ in \eqref{genmlik} and the asymptotic behaviour of $m(\vartheta)$ in \eqref{geneq111} of Lemma \ref{gen:lm:md}. Since $ L^*(\vartheta_0)$ in \eqref{ObjecApprox} is $O_P(T)$, the second summands have no influence on the asymptotic distribution of $\hat{\vartheta}_{m}$. For the bias, however, the latter terms require further analysis, which is carried out below in Section \ref{S3}.

\begin{figure}[H]
  \centering
  \hspace{-1.25ex}\subfloat[modification term]{
    \includegraphics[width=0.35\textwidth]{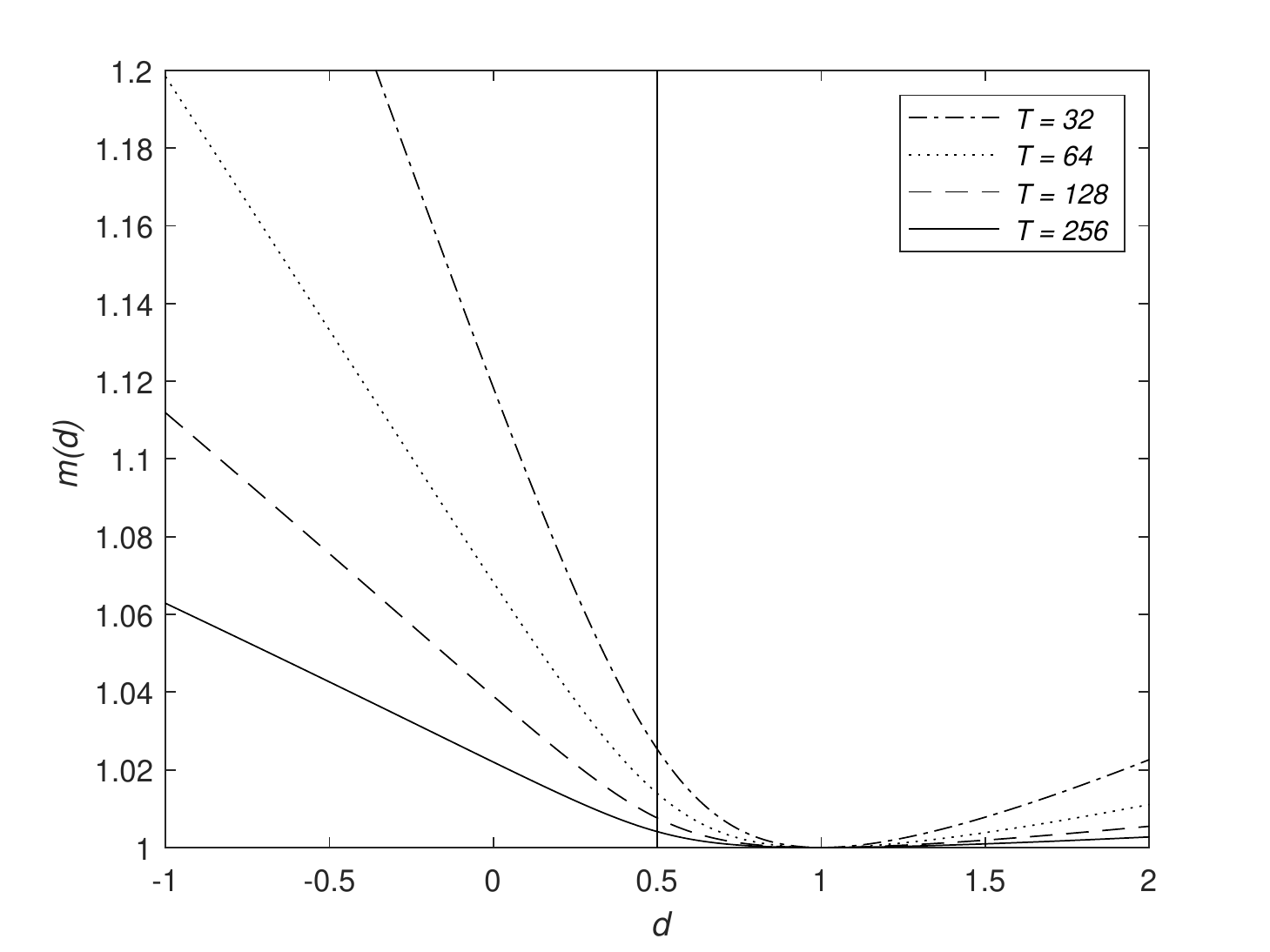}
  }
  \subfloat[likelihoods]{
    \includegraphics[width=0.35\textwidth]{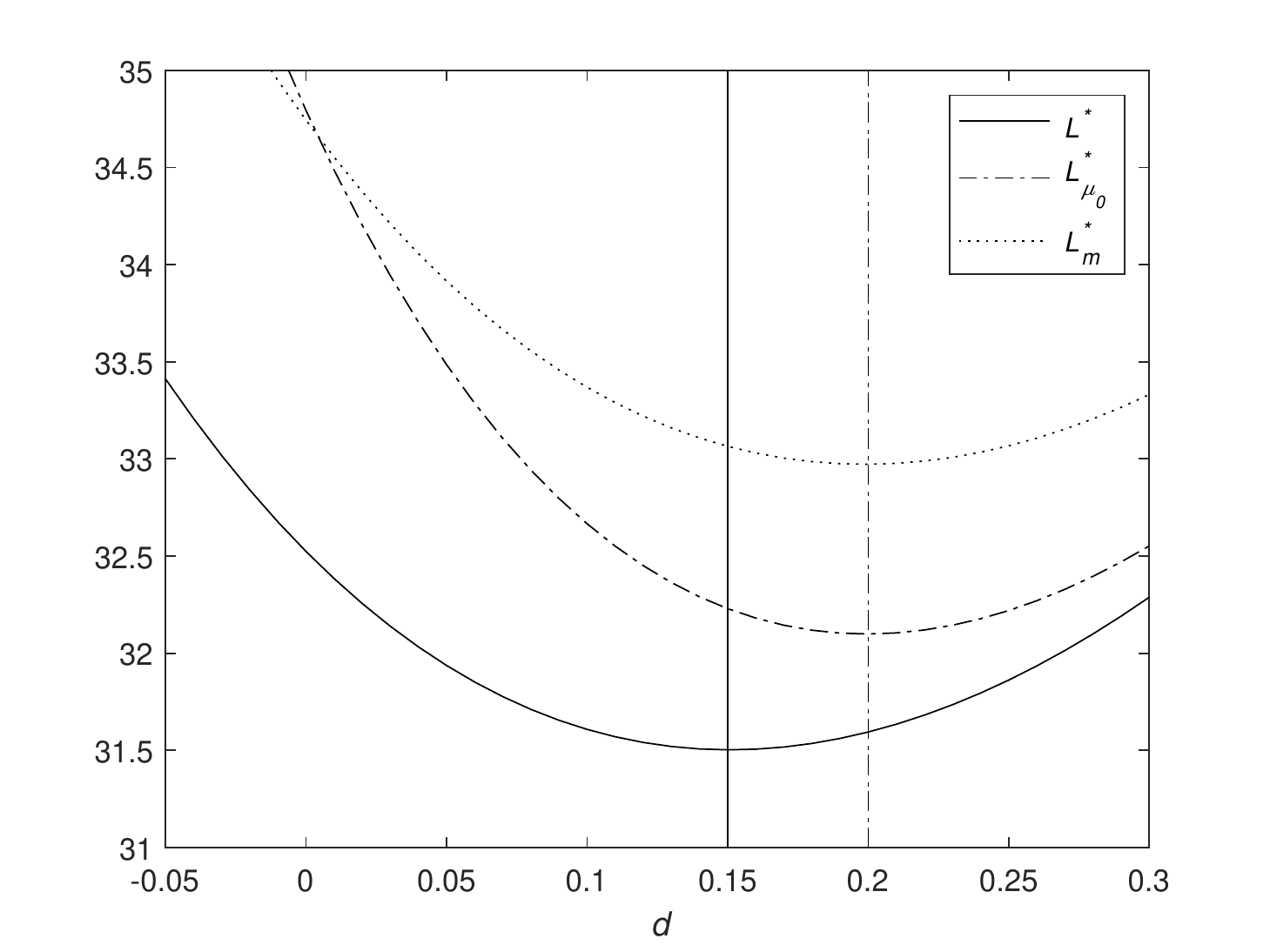}
  }
  \caption{Panel (a) plots the modification term $m(d)$ in \eqref{genmodificationterm} for $d$ between $-1$ and 2, and $T$ = 32, 64, 128, 256, and without short-run dynamics, i.e.\ $\phi(L;\varphi) = 1$. The value of $d = 1/2$ is
    added as a vertical line for clarity.  Panel (b) shows the Monte Carlo average over 10,000 replications of $L^*(d)$, $L_{\mu_0}^*(d)$ and $L_{m}^*(d)$. The DGP is given in \eqref{genq1}-\eqref{genq2} with
    $\epsilon_t \sim \textit{NID}(0,1)$, $\omega(L;\varphi_0) = 1$, $d_0 = 0.2$, $\mu_0 = 0$ and $T$ = 64.}
  \label{fig1}%
\end{figure}

We now provide an illustration of how the modification term behaves for some selected parameter values. For simplicity, we consider a purely fractional model, i.e.\ we set $\omega(L;\varphi) = 1$ in model \eqref{genq1}-\eqref{genq2} so that, effectively, $\vartheta = d$. The corresponding modification term $m(d)$ is then given by \eqref{genmodificationterm} with $\phi(L;\varphi) = 1$.
This modification term is plotted in panel (a) of Figure \ref{fig1} for some illustrative values of $d$ and $T$. Four important observations can be made: First, recall from \eqref{genmqeq1} that the modification term $m(d)$ penalises
the CSS objective $L^*(d)$ through inflating it by the factor $m(d)$. It appears from the plot that in the stationary region, i.e.\ when $d < 1/2$, $m(d)$ inflates $L^*(d)$ more than in the non-stationary region, i.e.\ when
$d \geq 1/2$. This is a reflection of the fact that the order of the bias in the score is larger in the stationary region, as was argued in \eqref{genscored} of Theorem \ref{t52}.  Secondly, it is plain that when $d = 1$ the bias caused by estimating the
constant term $\mu$ is the smallest, as predicted in Lemma \ref{gen:lm:md}. Thirdly, even for a moderately large sample of size $T = 256$, $m(d)$ still turns out to be substantial in the stationary region, implying that the
corresponding bias in the score is large. Fourthly, the negative slope of the modification term $m(d)$ for $d < 1$ implies that the minimum of $L^*_m(d)$ is shifted to the right of that of $L^*(d)$. This is illustrated in panel
(b) of Figure \ref{fig1} which displays a Monte Carlo simulation of the CSS and MCSS objective functions. The DGP is stationary and corresponds to the model in \eqref{genq1}-\eqref{genq2} with $\epsilon_t \sim \textit{NID}(0,1)$, $\omega(L;\varphi_0) = 1$, $d_0 = 0.2$ and
$\mu_0 = 0$. The sample size is $T = 64$ and the number of replications is 10,000. On display is the Monte Carlo average of the simulated $L^*(d)$, $L^*_{\mu_0}(d)$ and $L_{m}^*(d)$. The solid line represents the Monte Carlo
average of $L^*(d)$: it can be seen that the CSS estimator underestimates the true $d_0 = 0.2$ on average. The dash-dotted line represents the Monte Carlo average of $L_{\mu_0}^*(d)$, which takes the constant term as known. This
estimator is, on average, close to $d_0$. The dotted line represents the Monte Carlo average of $L_{m}^*(d)$, whose minimum is shifted to the right of that of $L^*(d)$. It therefore corrects for the distortion in $L^*(d)$ caused
by estimating $\mu$.

\subsection{Relationship with other modifications} \label{relwother}

There is a large literature on correcting the bias of maximum likelihood caused by the presence of unknown nuisance parameters. Seminal contributions include \textcite{barndorff1983formula} who proposed the modified likelihood
function, and \textcite{cox1987parameter} who contributed the idea of the conditional profile likelihood by approximating the modified likelihood function. Both modifications result in modified profile likelihoods that are
approximately score unbiased, see \textcite{liang1987estimating} and \textcite{cox1994inference}. It is therefore illuminating to investigate how our MCSS objective, with an expected score exactly equal to zero, relates to alternative approaches to bias-reduction, or how our modification term $m(\vartheta)$ compares to alternative adjustments. This
section discusses two such ideas.

First, reconsider the adjusted profile log-likelihood $\ell^*_a (\vartheta)$ proposed by \textcite{mccullagh1990simple} and derived in Section \ref{sectionmod}. Denote the corresponding estimator by $    \hat{\vartheta}_{a} = \operatorname*{argmax}_{\vartheta \in \Theta}  \ell^*_{a}(\vartheta)$. The following corollary establishes the condition  under which the adjusted profile log-likelihood estimator $\hat \vartheta_a$ is identical to our MCSS estimator $\hat \vartheta_m$. The
proof follows from \eqref{diffadj} and \eqref{solveformgen} and is omitted.
\begin{corollary} \label{corrrelationship}
Let $x_t$, $t$ = 1,$\ldots$,$T$, be given by \eqref{genq1}-\eqref{repmainf} and let Assumptions \ref{A2} to \ref{A5} be satisfied. Then, $\hat{\vartheta}_{m} = \hat{\vartheta}_{a}$ if and only if
\begin{align}
    E \left( \frac{ D_{\vartheta}L^*(\vartheta_0) }{L^*(\vartheta_0) } \right)  = \frac{E \left( D_{\vartheta}L^*(\vartheta_0) \right)}{E \left( L^*(\vartheta_0)  \right)}. \label{relatlL}
\end{align}
\end{corollary}
However, there is no particular reason to expect \eqref{relatlL} to hold in general. Therefore, our modified estimator $\hat \vartheta_m$ will in general be different from the adjusted estimator $\hat{\vartheta}_{a}$ suggested by \textcite{mccullagh1990simple}.  

Secondly, a modification term closely related to $m(\vartheta)$ in \eqref{genmodificationterm} is the one discussed in \textcite{an1993cox} who implement the idea of \textcite{cox1987parameter} to adjust the log-likelihood function.  The setup in
\textcite{an1993cox} is different from ours, however: They consider a stationary Gaussian type-I ARFIMA$(p_1,d,p_2)$ process $\tilde{x}_t$ generated by the model $\tilde{x}_t = \mu + \Delta^{-d} u_t$ where $u_t$ is defined in \eqref{genq2} and \eqref{arma} with $\epsilon_t\sim \textit{NID}(0,\sigma^2)$ and $|d| < 1/2$. This contrasts to our type-II process whose $d$ is also allowed to lie in the non-stationary region and whose error term
$\epsilon_t$ is not assumed to be Normally distributed.

It will prove helpful to phrase the approach by \textcite{an1993cox} in matrix notation: Define the $T \times 1$ vector $\tilde{x} = (\tilde{x}_1,\ldots,\tilde{x}_T)'$ such that $\tilde{x} \sim N(\mu \iota, \sigma^2 \Sigma(\vartheta))$ where
$\iota$ is a $T\times1$ vector of ones and $ \sigma^2 \Sigma(\vartheta)$ is the $T \times T$ variance-covariance matrix of $\tilde{x}$, see for instance \textcite{hoskingdiff1981} for the
elements of $\Sigma(\vartheta)$. The log-likelihood function is then given by $\tilde{\ell}(\vartheta,\mu,\sigma^2) = - \log( |\Sigma(\vartheta)| ) /2 - T  \log( \sigma^2)/2  -  (  \tilde{x}  - \iota \mu )' \Sigma(\vartheta)^{-1} (  \tilde{x}  - \iota \mu ) / (2 \sigma^2). $ 
Substituting in the maximum likelihood estimators $\hat{\mu}(\vartheta)$ and $\hat{\sigma}^2(\vartheta,\mu)$ yields the profile log-likelihood function
\begin{align*}
     \tilde{\ell}^*(\vartheta) = -\frac{1}{2}  \log( |\Sigma(\vartheta)| ) - \frac{T}{2}  \log \left( \frac{1}{T} \left(  \tilde{x}  - \iota \hat \mu (\vartheta) \right)' \Sigma(\vartheta)^{-1} \left(  \tilde{x}  - \iota \hat \mu (\vartheta) \right) \right).
\end{align*}
The modified profile log-likelihood function that \textcite{an1993cox} find is
\begin{align}
    \tilde{\ell}^*_m(\vartheta) &=  \tilde{\ell}^*(\vartheta)  + \frac{3}{2} \log \left( \hat{\sigma}^2(\vartheta,\hat{\mu}(\vartheta)) \right) + \frac{1}{T} \log( |\Sigma(\vartheta)|) - \frac{1}{2} \log\left(   \iota' \Sigma(\vartheta)^{-1} \iota \right). \label{expcox}
\end{align}
It can be shown that the orders of magnitude of the last three summands on the right-hand side of \eqref{expcox} are $O_P(1)$, $o(1)$ and $ O(\log(T))$, respectively. For the proof of the second summand, we refer to
\textcite{dahlhaus1989efficient}. As for the third summand, denote by $f_{\vartheta}(\omega)$ the spectral density of $\tilde{x}_t$ such that 
$T^{-1} \log |\Sigma(\vartheta)|  \rightarrow \frac{1}{2 \pi} \int_{-\pi}^{\pi} \log( f_{\vartheta}(\omega)/\sigma^2) d\omega $ which is equal to 0 by virtue of the well-known Szeg\H o--Kolmogorov formula, see \textcite{Chan2006}. The order of magnitude of the fourth summand follows from $\iota' \Sigma(\vartheta)^{-1} \iota = O(T^{1-2d+\varepsilon})$ for each $\varepsilon > 0$, cf.\ \textcite[Theorem 5.2]{adenstedt1974large}. The leading of the three summands is therefore the last one, its  order of magnitude being $O(\log(T))$. 

In order to compare $\tilde{\ell}^*_m (\vartheta)$ to our MCSS function $L_m^*(\vartheta)$ in \eqref{genmlik} we transform the latter again in a fashion similar to that in \eqref{Lstar} and define $     \ell^*_{m}(\vartheta) = - T \log ( 2 L_{m}^*(\vartheta) /T ) /2, $ such that $\ell^*_{m}(\vartheta) =  \ell^*(\vartheta)- T \log (  m(\vartheta) ) /2, $ with $\ell^* (\vartheta)$ given in \eqref{lq3}. Using the definition of $m(\vartheta)$ in \eqref{genmodificationterm} yields
\begin{align*}
     \ell^*_{m}(\vartheta)  =  \ell^*(\vartheta)-\frac{T}{(T-1)} \frac{1}{2} \log\left(  \sum_{t = 1}^T c^2_{t}(\vartheta)   \right) = \ell^*(\vartheta)-\frac{1}{2} \log\left(  \sum_{t = 1}^T  c^2_{t}(\vartheta)   \right) + O_P(T^{-1}\log(T)), 
\end{align*}
since $T/(T-1) = 1 + 1/(T-1)$. As is shown in Lemma \ref{genlemmaaaa2s}, the second summand is of order $O (\log T)$.

Two observations are now instructive. First, the leading modification term in $\ell^*_m (\vartheta)$ is of the same order of magnitude as that in $\tilde \ell^*_m$ in \eqref{expcox}, namely $O(\log(T))$. Second, we note that the Cholesky factor of $\Sigma(\vartheta)^{-1}$ in \eqref{expcox} is
the GLS transformation matrix that filters out the correlation structure of the type-I error term $\Delta^{-d} u_t$. Similarly, in our setting, $\phi(L;\varphi)\Delta_{+}^{d}$ filters out the correlation structure of the type-II error
$\Delta_+^{-d} u_t$. Indeed, if it were possible to replace $\Sigma(\vartheta)^{-1/2} \iota$ in \eqref{expcox} by $\phi(L;\varphi)\Delta_{+}^{d} \iota$ we would obtain
\begin{align*}
  \frac{1}{2} \log \left( (\phi(L;\varphi)\Delta_{+}^{d} \iota )' (\phi(L;\varphi)\Delta_{+}^{d} \iota) \right) = \frac{1}{2} \log \left( \sum_{t = 1}^T c^2_{t}(\vartheta)   \right) .
\end{align*}
using the definition of $c_{t}(\vartheta)$ in \eqref{convcoef}.
Let us emphasise again, however, that the approach by \textcite{an1993cox}, although asymptotically equivalent to ours, is based on a model that assumes stationary and Normally distributed data. In addition, it
necessitates the computation of the $T \times T$ variance-covariance matrix, or its Cholesky factor, which is often onerous computationally.

\subsection{Asymptotic biases}  \label{S3}

This section investigates the asymptotic biases of the estimators of $\vartheta$. Two questions are of central interest. First, by how much does the MCSS estimator $\hat{\vartheta}_m$ reduce the bias of the CSS estimator $\hat{\vartheta}$? Second, is the bias of the MCSS estimator $\hat \vartheta_m$ comparable
to that of the CSS estimator with known $\mu_0$? To address both questions, we proceed in a similar fashion as do \textcite{johansen2016role}, involving two steps: first, we find asymptotic expansions of the estimators and, secondly, we approximate these expansions. Since there would be a risk of
overloading the exposition with bulky notation, the treatment is here mainly of verbal-descriptive type. Technical details are contained in Appendix \ref{generalizationsappendix}. Additional intuition can also be
gained from the discussion of the asymptotic biases in the simple model without short-run dynamics, presented in Section \ref{arfima0d}.

With a view to deriving the asymptotic bias of the CSS estimator $\hat{\vartheta}$, take a second-order Taylor series expansion of $D_\vartheta L^* ( \hat \vartheta) = 0$ around $\vartheta_0$, which results in an expression
involving the first three derivatives of $L^* ( \vartheta)$, denoted by $D_\vartheta L^* ( \vartheta)$, $D_{\vartheta\vartheta'} L^* (\vartheta) $, and $D_{\vartheta_i \vartheta \vartheta'} L^* (\vartheta) $ for
$i = 1,\ldots, p+1$.  Since the representation of the third derivative in particular is unwieldy, we refer the reader to equation \eqref{gentaylorexp} for an explicit formula. We demonstrate in Lemmata \ref{asyappgennon1} and
\ref{asyappgenstat1} that the derivatives satisfy $D_\vartheta L^* ( \vartheta_0) = O_P(T^{1/2})$, $D_{\vartheta'\vartheta} L^* (\vartheta_0) = O_P(T)$, and $D_{\vartheta_i \vartheta \vartheta'} L^* (\vartheta_0) = O_P(T)$ for
$i = 1,\ldots, p+1$. Importantly, the orders of magnitude hold uniformly in $d_0$, allowing us to treat the stationary and non-stationary region jointly. Using the Taylor expansion of $D_\vartheta L^* ( \hat \vartheta) = 0$ and
the expressions of the derivatives of $L^* ( \vartheta)$ allows us to find $G_{1T}$ and $G_{2T}$ in $\hat{\vartheta}-\vartheta_0 = T^{-1/2} G_{1T} + T^{-1} G_{2T} + O_p(T^{-3/2})$ and, thence,
\begin{align}
    E\left( \hat{\vartheta} - \vartheta_0 \right) = S_T(\vartheta_0) + B_T(\varphi_0) + o(T^{-1}), \label{trian}
\end{align}
where $S_T(\vartheta_0 ) = - A^{-1} T^{-1} [ \sigma^{-2}_0 E \left( \DLvt^*(\vartheta_0) \right) ]$ and $B_T (\varphi_0)$ are  given in equations \eqref{exactSappend} and \eqref{exactBappend}, respectively. The purpose of the
decomposition in \eqref{trian} is to separate the expectation of the first derivative of $L^* (\vartheta)$ from the correlations between the derivatives: the former enters $S_T (\vartheta_0)$ only, while the latter are gathered in
$B_T (\varphi_0)$.  We therefore call $S_T (\vartheta_0)$ the score bias of $\hat \vartheta$ and $B_T(\varphi_0)$ the intrinsic bias.

Importantly, we refer to $S_T(\vartheta_0)$ and $B_T(\varphi_0)$ as ``exact'' biases as we evaluate the sample sums inside these expressions for a fixed $T$. Following the literature, we also derive simplified expressions by
replacing the sample sums, suitably scaled, by their asymptotic counterparts. The resulting bias expressions, denoted by $\mathcal{S}_T(\vartheta_0)$ and $\mathcal{B}_T(\varphi_0)$, are referred to as ``approximate'' and will make
the representations in Theorem \ref{t53} and Theorem \ref{t1} more intelligible. The simplification of the exact intrinsic bias $B_T(\varphi_0)$ is based on finding
$\mathcal{B}(\varphi_0) = \lim_{T \rightarrow \infty} T B_T(\varphi_0)$ and defining $\mathcal{B}_T(\varphi_0) = \mathcal{B}(\varphi_0)/T$. The simplified expression of the exact score bias $S_T(\vartheta_0)$ is more intricate
because its order of magnitude depends on $d_0$. From Theorem \ref{t52}, $T S_T (\vartheta_0) = O(1)$ when $d_0 > 1/2$, such that the approximate bias can also be based on finding
$\mathcal{S}(\vartheta_0) = \lim_{T \rightarrow \infty} T S_T(\vartheta_0)$ and writing $\mathcal{S}_T(\vartheta_0) = \mathcal{S}(\vartheta_0)/T$. However, when $d_0 < 1/2$, it is shown that $T S_T(\vartheta_0) = O(\log(T))$. Yet,
we derive in \eqref{niceS_T} the asymptotic expansion $T S_T(\vartheta_0) =\mathscr{S}_T(\vartheta_0) + o(1)$ that allows us to define $\mathcal{S}_T(\vartheta_0) = \mathscr{S}_T(\vartheta_0)/T$. Note that the exact and the approximate biases
are potentially very different from each other. For instance, the number of sample sums in $B_T(\varphi_0)$ that are approximated by their limiting value in $\mathcal{B}_T(\varphi_0)$ equals $3 (p+1)^3 + (p+1)^2$. If $T$ is
relatively small, the performance of $\mathcal{B}_T(\varphi_0)$ can therefore deteriorate substantially relative to that of $B_T(\varphi_0)$. \textcite{lieberman2005expansions} present a similar argument in the context of Edgeworth
expansions of the memory parameter.

It follows from Lemmata \ref{asyappgennon2}, \ref{asyappgennon3}, \ref{asyappgenstat2} and \ref{asyappgenstat3} that analogues of \eqref{trian} also hold for $\hat{\vartheta}_{\mu_0}$ and $\hat{\vartheta}_m$. In particular,
analogous derivations result in expressions of the score bias $S_T (\vartheta_0)$ involving $ E ( \DLvt_{\mu_0}^*(\vartheta_0) )$ and $ E (\DLvt_{m}^*(\vartheta_0))$, respectively. Clearly, both expectations are equal to zero, the
former due to Theorem \ref{t52} and the latter by construction. The following theorem is the main result of this paper and presents the approximate biases of $\hat{\vartheta}$, $\hat{\vartheta}_{\mu_0}$ and
$\hat{\vartheta}_{m}$. The proof of the theorem is given in Appendix \ref{generalizationsappendix} and an extension that discards Assumption \ref{A5} in Theorem \ref{th:corinitial} of Section \ref{sc:initial}.
\begin{theorem}\label{t53} 
Let $x_t$, $t$ = 1,$\ldots$,$T$, be given by \eqref{genq1}-\eqref{repmainf} and let Assumptions \ref{A2} to \ref{A5} be satisfied. The approximate biases of $\hat{\vartheta}$, $\hat{\vartheta}_{\mu_0}$ and $\hat{\vartheta}_{m}$ are 
\begin{align}
    bias(\hat{\vartheta}) &= \mathcal{S}_T(\vartheta_0) + \mathcal{B}_T(\varphi_0) + o(T^{-1}), \label{eqABcss} \\
    bias(\hat{\vartheta}_{\mu_0}) &= \mathcal{B}_T(\varphi_0) + o(T^{-1}), \label{eqABcssmu0} \\
    bias(\hat{\vartheta}_{m}) &=  \mathcal{B}_T(\varphi_0) + o(T^{-1}). \label{mcss_gen1}
\end{align}
The approximate intrinsic bias is 
\begin{align*}
    T \mathcal{B}_T(\varphi_0) = A^{-1}   \begin{bmatrix}
\iota' \left( A^{-1} \odot \left( G_{1} + F_{1} \right)\right)\iota \\
\vdots    \\
\iota' \left( A^{-1} \odot\left( G_{p+1} + F_{p+1} \right) \right)\iota 
\end{bmatrix}  - \frac{1}{2} A^{-1} \begin{bmatrix}
\iota' \left(\left(A^{-1} C_{0,1} A^{-1} \right) \odot A \right) \iota \\
\vdots  \\
\iota' \left(\left(A^{-1} C_{0,p+1} A^{-1} \right) \odot A \right) \iota
\end{bmatrix}, 
\end{align*}
where $A$ is given in \eqref{genA}, while $C_{0,i}$, $F_i$ and $G_i$ with $i = 1, \ldots, p+1$ are all functions of the lag polynomial $\omega (s, \varphi_0)$ in \eqref{repmainf}, its inverse $\phi (s, \varphi_0) $ in
Assumption \ref{A1} and their derivatives; see \eqref{genC1}-\eqref{genCk} as well as \eqref{genF1}-\eqref{genG2}, respectively. The approximate score bias for $d_0 > 1/2$ is given by
\begin{align*}
    T\mathcal{S}_T(\vartheta_0) =  A^{-1} \frac{\sum_{t = 1}^{\infty} c_{t}(\vartheta_0) D_{\vartheta} c_{t}(\vartheta_0)}{ \sum_{t = 1}^{\infty} c^2_{t}(\vartheta_0) } 
\end{align*}
and for $d_0 < 1/2$ it is
\begin{align*}
    T\mathcal{S}_T(\vartheta_0) = A^{-1}  \begin{bmatrix}
  -\log(T)+\left(\Psi(1-d_0) + (1-2d_0)^{-1}\right) \\
\frac{D_{\varphi_1 } \phi(1;\varphi_0)}{ \phi(1;\varphi_0) } \\
\vdots    \\
\frac{D_{\varphi_p } \phi(1;\varphi_0)}{ \phi(1;\varphi_0) }
\end{bmatrix}
\end{align*}
where $c_{t}(\vartheta)$ is defined in \eqref{convcoef}.
Furthermore, for the approximate intrinsic bias, 
$\mathcal{B}_T(\varphi_0) = O(T^{-1})$, whereas the approximate score bias satisfies $\mathcal{S}_T(\vartheta_0) = O(T^{-1} \log(T))$ when $d_0 < 1/2$ and $\mathcal{S}_T(\vartheta_0) = O(T^{-1})$ when $d_0> 1/2$.
\end{theorem}

The theorem generates five important insights. First, the intrinsic bias can be interpreted as the bias that remains even if the true value of the deterministic component $\mu_0$ were known. Secondly, the intrinsic bias of all
three estimators is the same.  Thirdly, the memory parameter $d$ solely affects the bias through the score bias, not through the intrinsic bias, the latter merely depending on the short-run dynamics $\varphi_0$. Fourthly, the
score bias of the CSS estimator $\hat \vartheta$ dominates its intrinsic bias only in the stationary region because the strength of the constant term is sufficiently strong to impact the stochastic component, see the
discussion following Theorem \ref{t52}.  Fifthly, as \textcite[p.\ 1108]{johansen2016role} note, the key factor to assess the distortion in testing or constructing confidence intervals for $\vartheta_0$ is the relative bias, i.e.\
the ratio of asymptotic bias to asymptotic standard deviation. Taking account of the asymptotic distribution of the three estimators, Theorem \ref{t53} implies that their relative bias is of order $O(T^{-1/2})$ in the
non-stationary region. In the stationary region, the relative bias is of order $O(T^{-1/2} \log(T))$ for the CSS estimator with unknown $\mu_0$, while it is of order $O(T^{-1/2})$ for the CSS estimator with known $\mu_0$ and for
the MCSS estimator. Thus, especially in the stationary region, a $t$-test for the memory parameter or for the short-run dynamics is more accurate when using the MCSS estimator compared to the CSS estimator. Later in the
empirical study, we will exploit this feature to our advantage.

Given the bias expression of $\hat \vartheta_m$ in \eqref{mcss_gen1} the MCSS estimator can be further improved by removing its intrinsic bias. However, $\mathcal{B}_T (\varphi_0)$ depends on the unknown true short-run dynamics
$\varphi_0$, implying that a two-step procedure is required for implementation: In the first step, estimate the model using the MCSS estimator to obtain estimates of the short-run dynamics $\hat{\varphi}_m$. Then, in the second
step, use these estimates to correct for the intrinsic bias. We call this refined estimator the bias-corrected MCSS $(bcm)$ estimator:
\begin{align}
\hat{\vartheta}_{bcm} = \hat{\vartheta}_m -  \mathcal{B}_T(\hat{\varphi}_m). \label{bcmcc}
\end{align}
The following corollary shows the validity of this two-step procedure.  The proof is given in Appendix \ref{val2step}.
\begin{corollary} \label{bcmcorr}
Let $x_t$, $t$ = 1,$\ldots$,$T$, be given by \eqref{genq1}-\eqref{repmainf} and let Assumptions \ref{A2} to \ref{A5} be satisfied. Then, $bias(\hat \vartheta_{bcm}) = o(T^{-1})$.
\end{corollary}
Note that the correction can of course also be based on the estimated exact intrinsic bias $B_T (\hat \varphi_m)$. That is indeed the version we  include in the simulation study in Section \ref{Ssimgen}.

\subsection{Special cases} \label{specialcases}

In this section, we apply the general bias expressions of Theorem \ref{t53} to some special cases of interest. In particular, Section \ref{arfima0d} covers the ARFIMA(0,$d$,0) model and compares the results with those of
\textcite{lieberman2005expansions} and \textcite{johansen2016role}. Section \ref{arfima1d} focuses on the ARFIMA(1,$d$,0) model, comparing it with \textcite{nielsen2005finite}. Lastly, Section \ref{bshortm} looks at short-memory
models, concluding with the biases of the AR(1) model and a comparison with \textcite{tanaka1984asymptotic}.

\subsubsection{ARFIMA(0,\texorpdfstring{\(d\)}{d},0) model} \label{arfima0d}

\textcite{johansen2016role} consider the purely fractional model for $d_0 > 1/2$ when the process involves a finite-number $N_0$ of unobserved pre-sample values. They derive the asymptotic biases for $\hat{d}$ and
$\hat{d}_{\mu_0}$. Their results for $N_0 = 0$ are effectively included in Theorem \ref{t53} when $\omega(L;\varphi) = 1$. For this special setting, the present section adds the asymptotic biases of $\hat{d}$ and $\hat{d}_{\mu_0}$
when $d_0 < 1/2$, as well as the asymptotic bias of the MCSS estimator $\hat{d}_m$. Theorem \ref{t1} provides a summary. Note that no use is made of Assumptions \ref{A1} and \ref{A5}. Note also that $\omega(L;\varphi) = 1$ implies
that $\vartheta = d$.

\begin{theorem}\label{t1} 
  Let $x_t$, $t$ = 1,$\ldots$,$T$, be given by \eqref{genq1} and let $u_t = \epsilon_t$. Let Assumptions \ref{A2} and \ref{A3} be satisfied. The approximate biases of $\hat{d}$, $\hat{d}_{\mu_0}$ and $\hat{d}_{m}$ are as in
  \eqref{eqABcss}, \eqref{eqABcssmu0} and \eqref{mcss_gen1}, respectively. Specifically, the intrinsic bias takes the form $T \mathcal{B}_T = - 3 \zeta_{3} \zeta_{2}^{-2}, $
  while the score bias is now given by 
  \begin{align*}
    T \mathcal{S}_T(d_0) = 
    \begin{cases}
      -\zeta_{2}^{-1}(\Psi(d_0) - \Psi(2d_0-1) ), & \text{for } d_0 > 1/2 \\
      -\zeta_{2}^{-1}\left[\log(T) - ( \Psi(1-d_0)+(1-2d_0)^{-1}) \right], & \text{for } d_0 < 1/2 .
    \end{cases}
  \end{align*}
  $\zeta_{s}$ is Riemann's zeta function $\zeta_{s} = \sum_{j = 1}^{\infty} j^{-s}$, $s>1$, and $\Psi(d) = D\log \Gamma(d)$ denotes the Digamma function.
\end{theorem} 

The expressions for the score and intrinsic biases in Theorem \ref{t53} have simplified considerably. In particular, the intrinsic bias, derived by \cite{johansen2016role} for $d_0>1/2$, is independent of $d_0$ and therefore identical in both regions. The same bias term appears in \textcite{lieberman2004expansions} for the bias of the
estimated memory parameter based on the maximum likelihood estimator in the type-I fractional model with $0 < d_0 < 1/2$ and $\sigma^2$ as well as $\mu$ known, see \textcite[p.\ 1106]{johansen2016role}
for a discussion.  Since $\mathcal{B}_T$ is pivotal, it can now be easily eliminated. The bias-corrected MCSS ($bcm$) estimator defined in Section \ref{S3} is now simply
$\hat{d}_{bcm} = \hat{d}_{m} + T^{-1} 3\zeta_3 \zeta_2^{-2} $, with $bias(\hat{d}_{bcm}) = o(T^{-1})$, as can be directly seen from Corollary \ref{bcmcorr}.

Analysing the theoretical bias terms through numerical comparisons may assist in building up an intuition. Table \ref{table2} therefore presents the theoretical biases, up to $o(T^{-1})$ terms, of the CSS estimator with unknown
and known $\mu_0$ and of the MCSS estimator, for selected values of $d_0$ and $T$. It is evident that the bias of the CSS estimator decreases in both the stationary and non-stationary region as $d_0$ decreases, and decreases
everywhere as $T$ increases. As \textcite[p.\ 1107]{johansen2016role} note, the bias of $\hat d$ is equal to that of $\hat d_{\mu_0}$ when $d_0 = 1$, since in this case, $\mathcal{S}_T(d_0) = 0$. Yet it is curious to see that
bias($\hat{d}$) is actually smaller than bias($\hat{d}_{\mu_0}$) for $d_0 = 1.1$ and $d_0 = 1.2$. In fact, this occurs for all $d_0 > 1$. The reason is that $\mathcal{S}_T(d_0)$ becomes positive for $d_0 > 1$, thereby increasing
the negative intrinsic bias $ \mathcal{B}_T$. The proof is omitted, but the result follows from the fact that $\Psi(d_0) - \Psi(2d_0-1)$ is monotonically decreasing for $d_0 > 1/2$ and from \textcite[eqn.\
6.3.18]{abramowitz1964handbook}, stating that $\Psi(d_0) = \log(d_0) + O(d_0^{-1})$ for $d_0 \rightarrow \infty$. 

\begin{table}[H]
\centering
\resizebox{\textwidth}{!}{%
\begin{tabular}{EEEEEEEEEEEEEEEE}
\hline
&
\multicolumn{1}{c}{bias($\hat{d}$)} &
  \multicolumn{1}{c}{bias($\hat{d}_{\mu_0}$)} &
  \multicolumn{1}{c}{bias($\hat{d}_{m}$)} &
  \multicolumn{1}{l}{} &
\multicolumn{1}{c}{bias($\hat{d}$)} &
  \multicolumn{1}{c}{bias($\hat{d}_{\mu_0}$)} &
  \multicolumn{1}{c}{bias($\hat{d}_{m}$)} &
  \multicolumn{1}{l}{} &
\multicolumn{1}{c}{bias($\hat{d}$)} &
  \multicolumn{1}{c}{bias($\hat{d}_{\mu_0}$)} &
  \multicolumn{1}{c}{bias($\hat{d}_{m}$)} &
  \multicolumn{1}{l}{} &
   \multicolumn{1}{c}{bias($\hat{d}$)} &
  \multicolumn{1}{c}{bias($\hat{d}_{\mu_0}$)} &
  \multicolumn{1}{c}{bias($\hat{d}_{m}$)}  \\ \cline{2-4} \cline{6-8} \cline{10-12}  \cline{14-16}
  \multicolumn{1}{c}{$d_0$ \textbackslash{} $T$} &
 \multicolumn{3}{c}{32} &
  \multicolumn{1}{l}{} &
  \multicolumn{3}{c}{64} &
  \multicolumn{1}{l}{} &
  \multicolumn{3}{c}{128} &
  \multicolumn{1}{l}{} &
  \multicolumn{3}{c}{256} \\ \hline
-0.2 & -9.94 & -4.16 & -4.16 &  & -5.63 & -2.08 & -2.08 &  & -3.14 & -1.04 & -1.04 &  & -1.74 & -0.52 & -0.52 \\
-0.1 & -9.97 & -4.16 & -4.16 & & -5.64 & -2.08 & -2.08 &  & -3.15 & -1.04 & -1.04 &  & -1.74 & -0.52 & -0.52 \\
0.0  & -9.95 & -4.16 & -4.16 & & -5.63 & -2.08 & -2.08 &  & -3.14 & -1.04 & -1.04 &  & -1.74 & -0.52 & -0.52 \\
0.1  & -9.81 & -4.16 & -4.16 & & -5.56 & -2.08 & -2.08 &  & -3.11 & -1.04 & -1.04 &  & -1.72 & -0.52 & -0.52 \\
0.2  & -9.42 & -4.16 & -4.16 & & -5.37 & -2.08 & -2.08 &  & -3.01 & -1.04 & -1.04 &  & -1.67 & -0.52 & -0.52 \\
0.3  & -8.32 & -4.16 & -4.16 & & -4.82 & -2.08 & -2.08 &  & -2.74 & -1.04 & -1.04 &  & -1.53 & -0.52 & -0.52 \\
0.4  & -4.18 & -4.16 & -4.16 & & -2.75 & -2.08 & -2.08 &  & -1.70 & -1.04 & -1.04 &  & -1.02 & -0.52 & -0.52 \\ \hline
0.5   & \multicolumn{1}{c}{-}  & \multicolumn{1}{c}{-}    & \multicolumn{1}{c}{-} &    & \multicolumn{1}{c}{-}      & \multicolumn{1}{c}{-}     & \multicolumn{1}{c}{-}  &    & \multicolumn{1}{c}{-}     & \multicolumn{1}{c}{-}     & \multicolumn{1}{c}{-}  &    & \multicolumn{1}{c}{-} & \multicolumn{1}{c}{-} & \multicolumn{1}{c}{-}      \\ \hline
0.6  & -11.29 & -4.16 & -4.16 &  & -5.64 & -2.08 & -2.08 &  & -2.82 & -1.04 & -1.04 &  & -1.41 & -0.52 & -0.52 \\
0.7  & -6.71 & -4.16 & -4.16 &  & -3.36 & -2.08 & -2.08 &  & -1.68 & -1.04 & -1.04 &  & -0.84 & -0.52 & -0.52 \\
0.8  & -5.26 & -4.16 & -4.16 &  & -2.63 & -2.08 & -2.08 &  & -1.31 & -1.04 & -1.04 &  & -0.66 & -0.52 & -0.52 \\
0.9  & -4.56 & -4.16 & -4.16 &  & -2.28 & -2.08 & -2.08 &  & -1.14 & -1.04 & -1.04 &  & -0.57 & -0.52 & -0.52 \\
1.0  & -4.16 & -4.16 & -4.16 &  & -2.08 & -2.08 & -2.08 &  & -1.04 & -1.04 & -1.04 &  & -0.52 & -0.52 & -0.52 \\
1.1  & -3.91 & -4.16 & -4.16 &  & -1.95 & -2.08 & -2.08 &  & -0.98 & -1.04 & -1.04 &  & -0.49 & -0.52 & -0.52 \\
1.2  & -3.73 & -4.16 & -4.16 &  & -1.87 & -2.08 & -2.08 &  & -0.93 & -1.04 & -1.04 &  & -0.47 & -0.52 & -0.52 \\
\hline
\end{tabular}%
}
\caption{(100 $\times$) Theoretical bias, up to $o(T^{-1})$ terms, of the CSS estimator of $d$ with unknown and known $\mu_0$ and of the MCSS estimator of $d$.}
\label{table2}
\end{table}

As mentioned in Section \ref{model}, \textcite{johansen2016role} generalise the way the process is initialised by allowing for $N_0$ unobserved pre-sample values and by setting $x_t = 0$ only when $t < 1-N_0$, with $N_0$ a fixed
non-negative integer. They show that in the purely fractional model with $d_0 > 1/2$, these unobserved pre-sample values introduce an additional bias term for the CSS estimator. We prove in Theorem \ref{corinitial} in Appendix
\ref{Initschemes} that the MCSS estimator $\hat d_m $ also eliminates the bias associated with the unknown-level parameter $\mu$ in this extended model. It does not, however, address the misspecification bias due to the unobserved
pre-sample values.

\subsubsection{ARFIMA(1,\texorpdfstring{\(d\)}{d},0) model} \label{arfima1d}

In a Monte Carlo simulation performed by \textcite{nielsen2005finite}, the ARFIMA(1,$d$,0) model undergoes thorough analysis. This section provide an explanation of their findings. In the following theorem, we present the
approximate bias expression for this particular model. An extension that does not make use of Assumption \ref{A5} is contained in Theorem \ref{th:ExtARFIMA1d0} in Appendix \ref{Initschemes}.

\begin{theorem}\label{t54} 
  Let $x_t$, $t$ = 1,$\ldots$,$T$, be given by \eqref{genq1} and let $u_t = \varphi u_{t-1} + \epsilon_t$. Let Assumptions \ref{A2} to \ref{A5} be satisfied with $\varphi_0 \neq 0$. The approximate biases of $\hat{\vartheta}$,
  $\hat{\vartheta}_{\mu_0}$ and $\hat{\vartheta}_{m}$ are as in  \eqref{eqABcss}, \eqref{eqABcssmu0} and \eqref{mcss_gen1}, respectively. 
  Specifically, the intrinsic bias takes the form 
  \begin{align}
    T \mathcal{B}_T(\varphi_0) &=   A^{-1}   \begin{bmatrix}
      \iota' \left( A^{-1} \odot \left(G_1 + F_1 \right) \right)\iota \\
      \iota' \left( A^{-1} \odot \left(G_2 + F_2 \right) \right)\iota 
    \end{bmatrix} - \frac{1}{2} A^{-1} \begin{bmatrix}
      \iota' \left(\left(A^{-1} C_{01} A^{-1} \right) \odot A \right) \iota \\
      \iota' \left(\left(A^{-1} C_{02} A^{-1} \right) \odot A \right) \iota
    \end{bmatrix} \label{eqIBarfima1d0}
  \end{align}
  where
  \begin{align*}
    A &= \begin{bmatrix}
      \pi^2/6 &  - \varphi_0^{-1} \log(1-\varphi_0) \\
      - \varphi_0^{-1} \log(1-\varphi_0)  &  (1-\varphi_0^2)^{-1}
    \end{bmatrix}  \\
    C_{01} &= \begin{bmatrix}
      -6 \zeta_3 &  2 \varphi_0^{-1} Li_{2}(-\frac{\varphi_0}{1-\varphi_0}) - \varphi_0^{-1} \log^2(1-\varphi_0) \\
      2 \varphi_0^{-1} Li_{2}(-\frac{\varphi_0}{1-\varphi_0}) - \varphi_0^{-1} \log^2(1-\varphi_0)  & 2 \frac{\log(1 - \varphi_0)}{1-\varphi_0^2}
    \end{bmatrix} \\
    C_{02} &= \begin{bmatrix}
      2 \varphi_0^{-1} Li_{2}(-\frac{\varphi_0}{1-\varphi_0}) - \varphi_0^{-1} \log^2(1-\varphi_0)  & 2 \frac{\log(1 - \varphi_0)}{1-\varphi_0^2} \\
      2 \frac{\log(1 - \varphi_0)}{1-\varphi_0^2} & 0
    \end{bmatrix} \\
    F_{1} &= \begin{bmatrix}
      -2 \zeta_3 & -\varphi_0^{-1}\log^2(1-\varphi_0)   \\
      \varphi_0^{-1} Li_{2}(-\frac{\varphi_0}{1-\varphi_0}) &  \frac{\log(1-\varphi_0)}{1-\varphi_0^2}
    \end{bmatrix} \\
    F_{2} &= \begin{bmatrix}
      \varphi_0^{-1} Li_{2}(-\frac{\varphi_0}{1-\varphi_0})   & \frac{\log(1-\varphi_0)}{1-\varphi_0^2}\\
      0  & 0
    \end{bmatrix} \\
    G_{1} &= \begin{bmatrix}
      -4 \zeta_3 &  2 \varphi_0^{-1} Li_2(-\frac{\varphi_0}{1-\varphi_0}) \\
      - \varphi_0^{-1} \log^2(1- \varphi_0) + \varphi_0^{-1} Li_2(-\frac{\varphi_0}{1-\varphi_0}) & \frac{\log(1-\varphi_0)}{1-\varphi_0^2} - \varphi_0^{-2} \left(\frac{\varphi_0 }{1-\varphi_0 } + \log(1-\varphi_0 )\right) 
    \end{bmatrix}  \\
    G_{2} &= \begin{bmatrix}
      - \varphi_0^{-1} \log^2(1- \varphi_0) + \varphi_0^{-1} Li_2(-\frac{\varphi_0}{1-\varphi_0})  & \frac{\log(1-\varphi_0)}{1-\varphi_0^2} - \varphi_0^{-2} \left(\frac{\varphi_0 }{1-\varphi_0 } + \log(1-\varphi_0 )\right) \\
      2\log(1-\varphi_0) \frac{1}{1-\varphi_0^2}& -2 \frac{\varphi_0}{(1-\varphi_0^2)^2}
    \end{bmatrix}.
  \end{align*}
  The score bias $T \mathcal{S}_T(\vartheta_0)$ for $d_0 > 1/2$ is now equal to
  \begin{align}
      \scriptstyle A^{-1} \left[ (1-\varphi_0)^2 \binom{2d_0-2}{d_0-1} +  \varphi_0  \binom{2d_0}{d_0} \right]^{-1} 
                                    \begin{bmatrix} \scriptstyle (1-\varphi_0)^2  \binom{2d_0-2}{d_0-1} \left( \Psi(2d_0-1)-\Psi(d_0) \right)  + \varphi_0  \binom{2d_0}{d_0} \left( \Psi(2d_0+1)-\Psi(d_0+1) \right) \\
                                      \scriptstyle (\varphi_0- 1)   \binom{2d_0-2}{d_0-1}  +  0.5\binom{2d_0}{d_0}
                                    \end{bmatrix} \label{eqASBnonstat}
  \end{align}
  while for $d_0 < 1/2$ it is
  \begin{align}
    T \mathcal{S}_T(\vartheta_0) = A^{-1} 
    &\begin{bmatrix} -\log(T)+\Psi(1-d_0) + (1-2d_0)^{-1}
      \\
      -\frac{1}{1-\varphi_0}
    \end{bmatrix}. \label{SstatARFI}
  \end{align}
  $\zeta_{s}$ is the Riemann's zeta function $\zeta_{s} = \sum_{j = 1}^{\infty} j^{-s}$, $s>1$, and $\Psi(d) = D\log \Gamma(d)$ denotes the Digamma function, and $Li_{2}(\varphi) = \sum_{i = 1}^{\infty} i^{-2}\varphi^{i}$ is
  the dilogarithm function (Spence's integral). The binomial coefficients are represented using the notation $\binom{\cdot}{\cdot}$.
\end{theorem} 

\begin{figure}[H]
  \centering
  \subfloat[$\mathcal{B}_T(\varphi_0)$ ]{
    \includegraphics[width=0.35\textwidth]{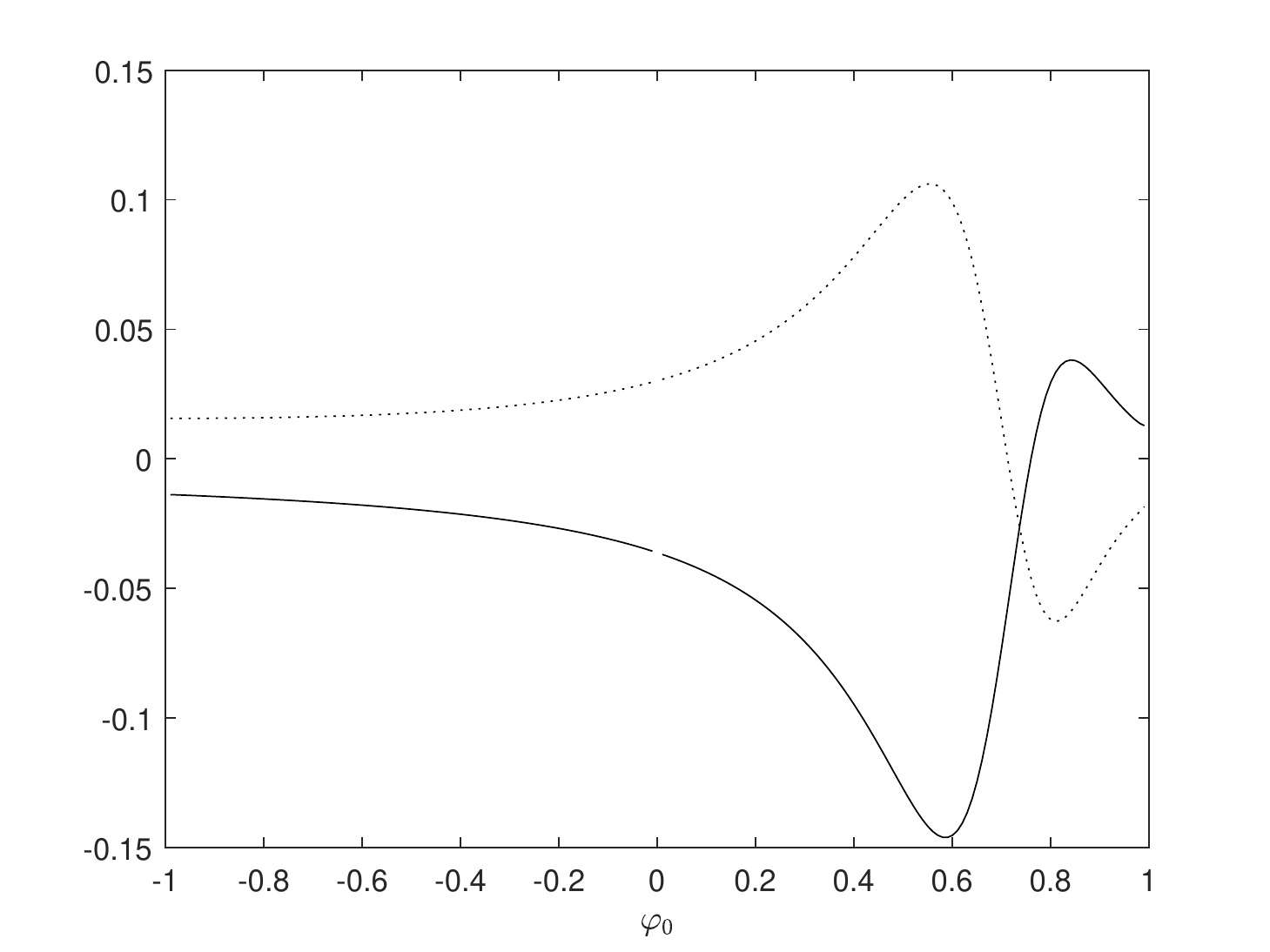}
  }
  \subfloat[$B_T(\varphi_0)$]{
    \includegraphics[width=0.35\textwidth]{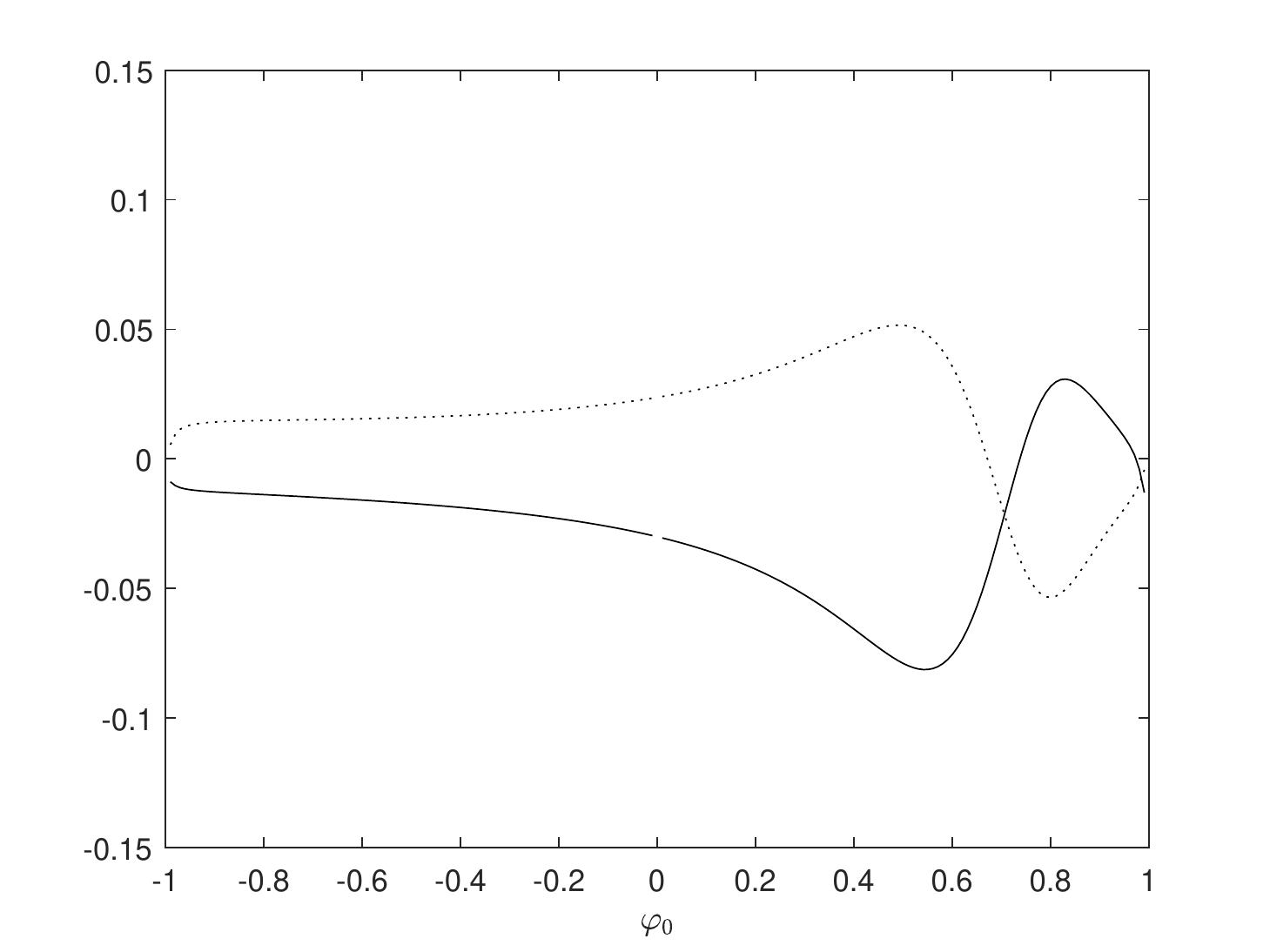}
  }
  \hspace{0mm}
  \subfloat[$\mathcal{S}_T(d_0,\varphi_0)$ for $d_0 = -0.2$ ]
  {
    \includegraphics[width=0.35\textwidth]{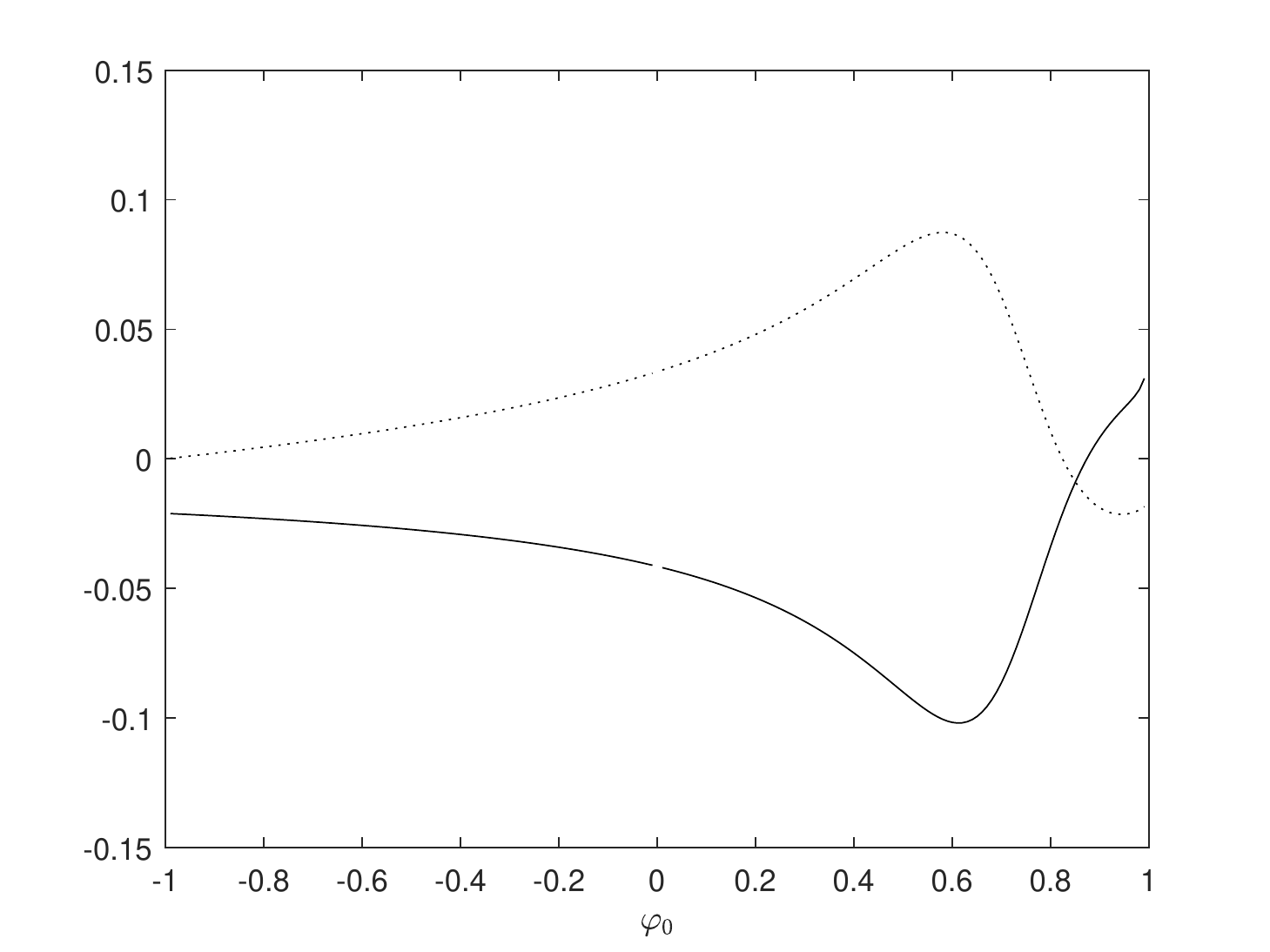}
  }
  \subfloat[$S_T(d_0,\varphi_0)$ for $d_0 = -0.2$]
  {
    \includegraphics[width=0.35\textwidth]{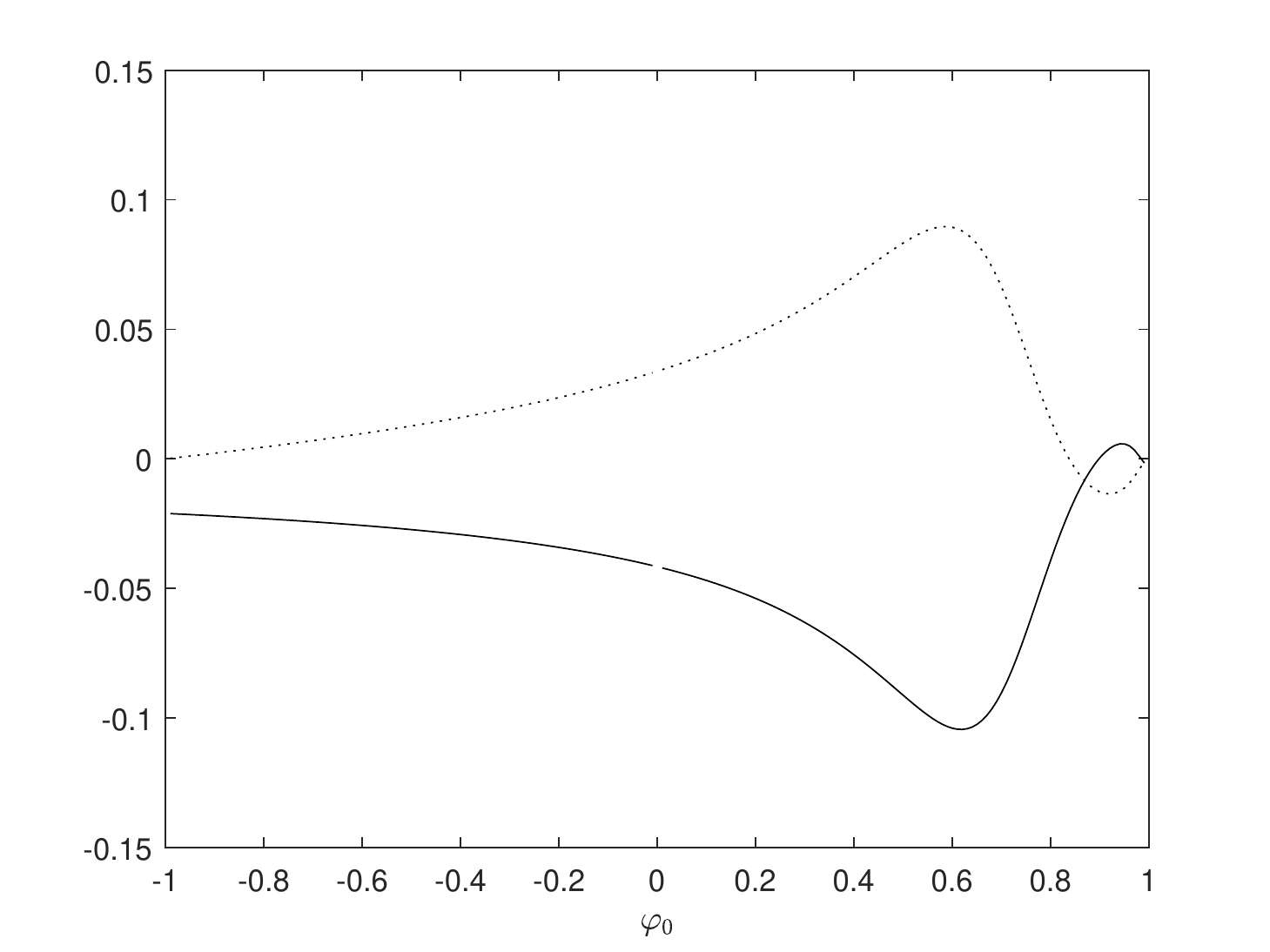}
  }
  \hspace{0mm}
  \subfloat[$\mathcal{S}_T(d_0,\varphi_0)$ for $d_0 = 0.4$ ]
  {
    \includegraphics[width=0.35\textwidth]{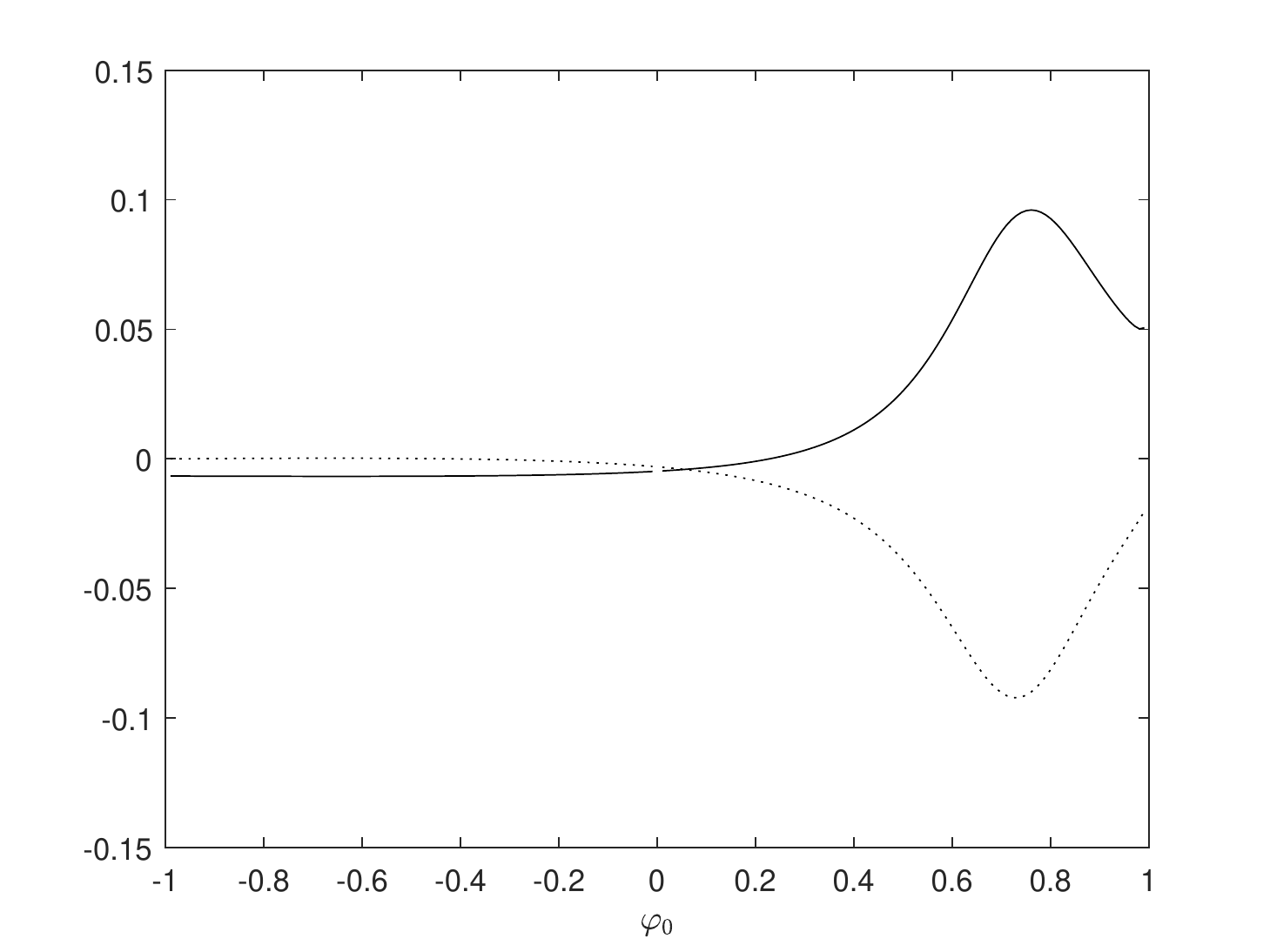}
  }
  \subfloat[$S_T(d_0,\varphi_0)$ for $d_0 = 0.4$]
  {
    \includegraphics[width=0.35\textwidth]{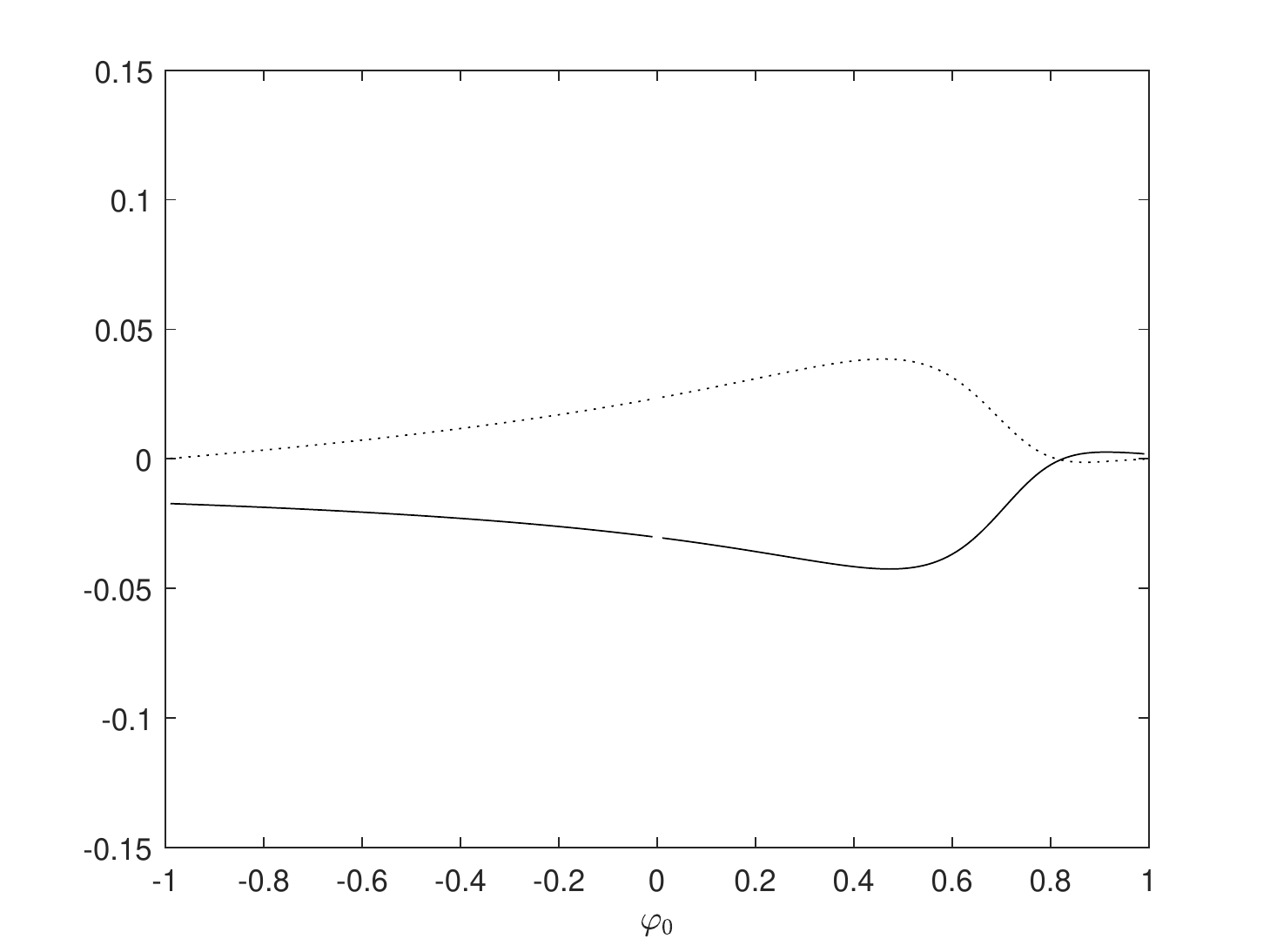}
  }
  \hspace{0mm}
  \subfloat[$\mathcal{S}_T(d_0,\varphi_0)$ for $d_0 = 1$]
  {
    \includegraphics[width=0.35\textwidth]{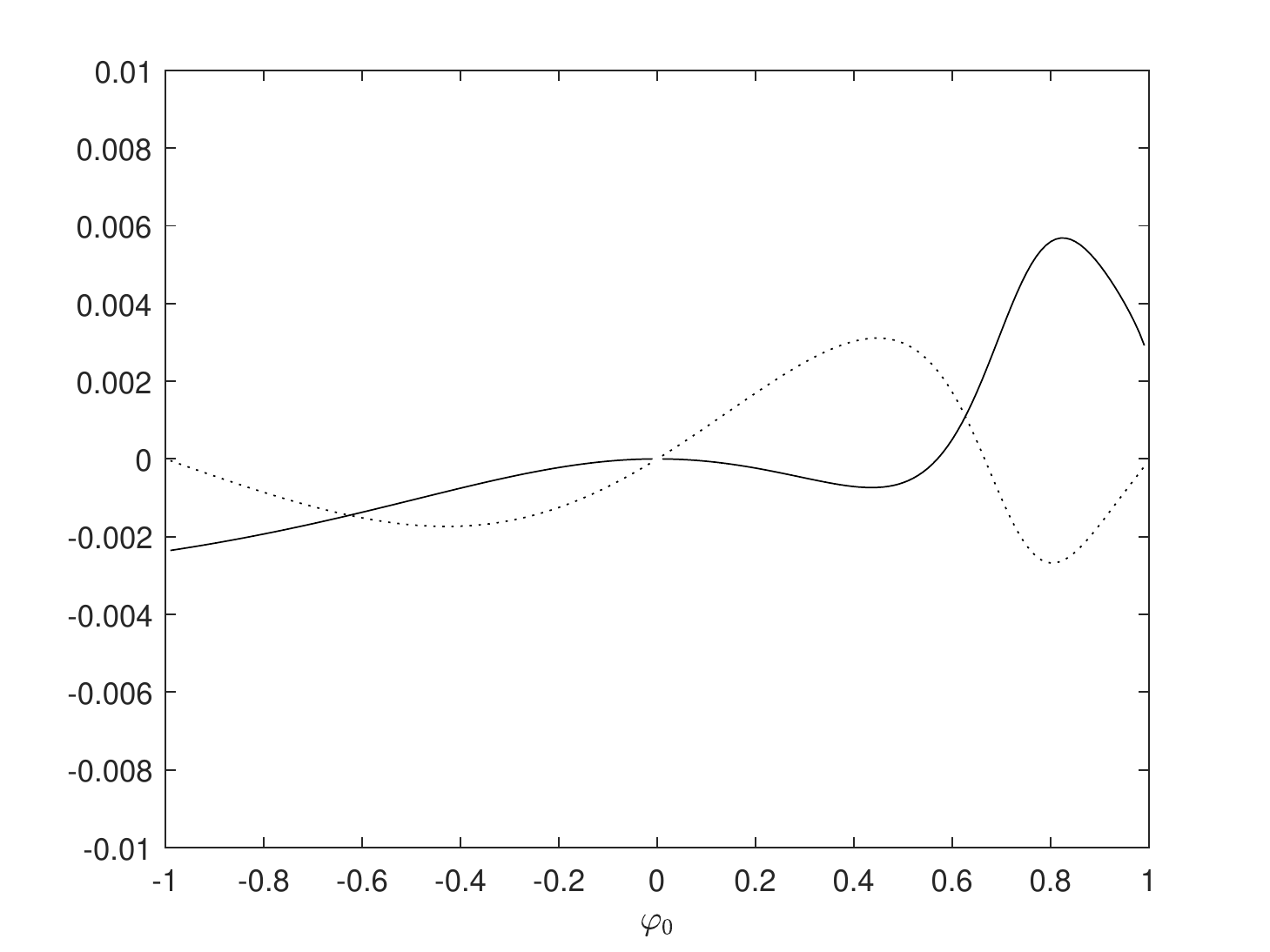}
  }
  \subfloat[$S_T(d_0,\varphi_0)$ for $d_0 = 1$]
  {
    \includegraphics[width=0.35\textwidth]{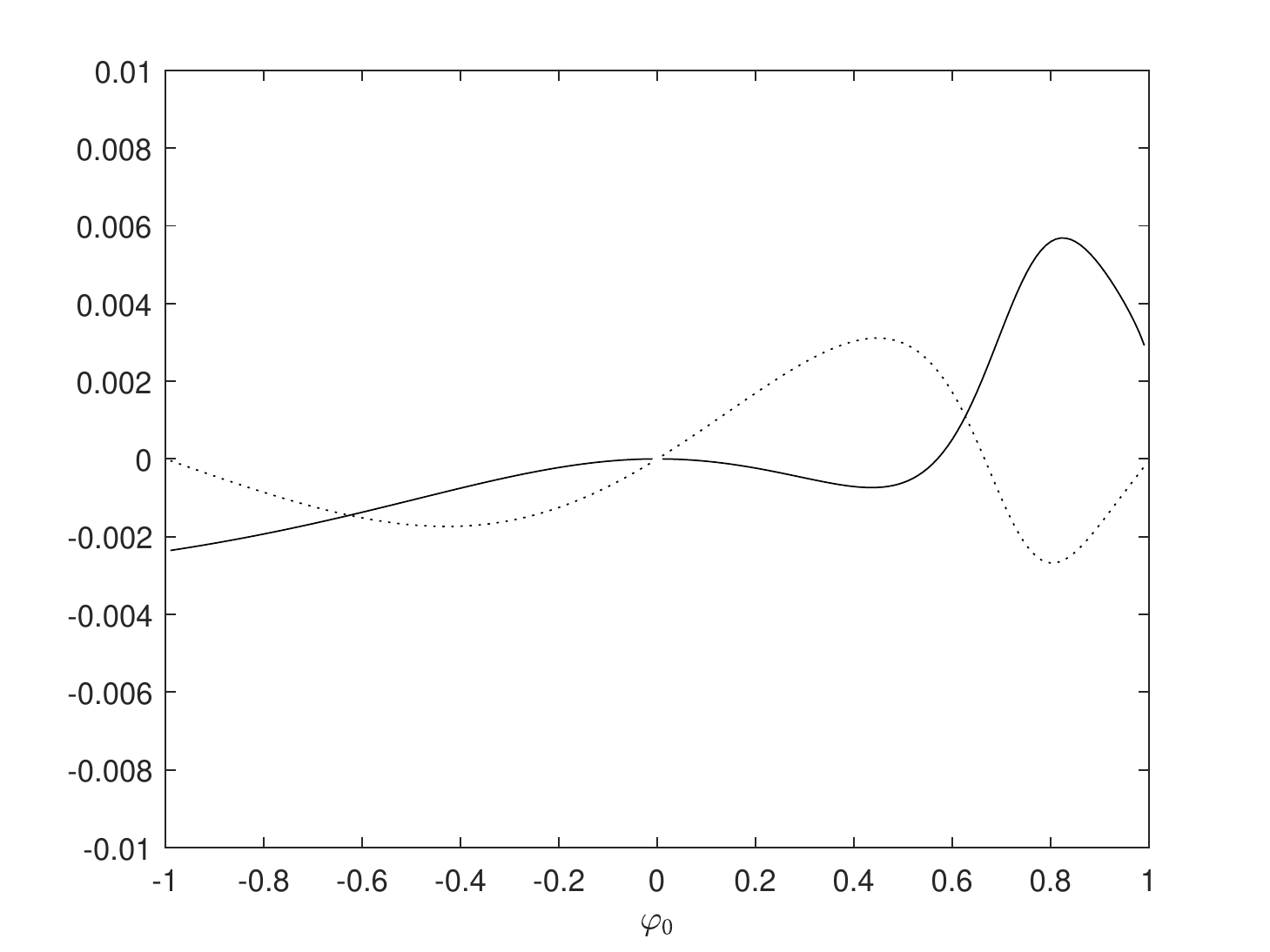}
  }
  \caption{The approximate and exact intrinsic bias for the ARFIMA(1,$d$,0) model are displayed in panel (a) and (b), respectively. The approximate score  bias for the ARFIMA(1,$d$,0) model are displayed in the panels
    (c), (e) and (g), while the exact score biases are in panels (d), (f) and (h). In all cases, $T = 128$.}%
  \label{figic}%
\end{figure}

Figure \ref{figic}(a) plots the approximate intrinsic bias $\mathcal{B}_T (\varphi_0)$ of both the memory parameter and the autoregressive coefficient as given in \eqref{eqIBarfima1d0}. Interestingly, the biases for these two
parameters exhibit near-symmetry. Specifically, the bias for the memory parameter tends to be negative, while the bias for the autoregressive parameter is typically positive. Furthermore, both biases are relatively small and
almost linear when $\varphi_0$ is below 0. However, beyond this value, the absolute value of the biases grows until reaching a peak at around $\varphi_0 = 0.5$. For larger values of $\varphi_0$, the biases decrease rapidly in
absolute value start. This same behaviour was noticed in a Monte Carlo simulation by \textcite{nielsen2005finite}. They noted that the memory parameter's downward bias is particularly pronounced when the AR coefficient is either 0
or 0.4 and that the estimation methods seem robust against stronger positive AR coefficient, such as 0.8. This aligns with the bias expression in Theorem \ref{t54}.

Figure \ref{figic}(b) plots the exact intrinsic bias $B_T (\varphi_0)$ of both the memory parameter and the autoregressive coefficient. $B_T (\varphi_0)$ is based on \eqref{exactBappend} in Appendix \ref{proofthm53}, simplified
appropriately. The patterns of the biases closely resemble those of the approximated intrinsic bias in Figure \ref{figic}(a). However, the specific values differ significantly, particularly in the range between 0 and 0.6. This
discrepancy suggests that the asymptotic approximation can lead to notable distortion, especially within this range. Consequently, we suggest to use the exact intrinsic bias when correcting for it.  As we will observe later, these
biases also align with the findings in our simulation in Section \ref{Ssimgen}: It is suboptimal to use the approximate intrinsic bias when the sample size is small.

Figure \ref{figic} also illustrates the approximate score bias $\mathcal{S}_T (\vartheta_0)$ (see \eqref{eqASBnonstat}-\eqref{SstatARFI}) as well as the exact score bias $S_T (\vartheta_0)$ (based on \eqref{exactSappend}) of the
ARFIMA(1,$d_0$,0) model with the memory parameter $d_0$ taking on values of $-0.2$, 0.4, and 1. Several observations can be made: First, the score bias tends to be more pronounced in the stationary region as compared to the
non-stationary region, which aligns with what we anticipate on the basis of Theorem \ref{t52}. Secondly, there exists a noticeable symmetry between the memory parameter and the autoregressive coefficient. Thirdly, a close match is
observed, on the whole, between the exact and approximate score biases for $d_0 = -0.2$ and $d_0$ = 1. 
However, this correspondence breaks down when $d_0 = 0.4$ and the approximate biases are much larger in absolute value than the exact biases. This distortion arises due to the presence of the term $(1-2d_0)^{-1}$ in
\eqref{SstatARFI}, which diverges as $d_0$ approaches 0.5. It is, therefore, recommendable to employ the exact score biases in practical empirical applications. \textcite{robinson2015efficient} who, in a panel setting, correct for
the score bias of the CSS estimator, also observe that the exact score bias is preferable.  Importantly, this bias is inherently eliminated by the MCSS estimator, obviating the need for additional correction.

\subsubsection{Short-memory models} \label{bshortm}

Theorem \ref{t53} also covers bias expressions in cases where long memory is absent in the model. The resulting expressions are presented in Theorem \ref{t55} below. They are straightforward to derive, so details of the
proof are omitted. It suffices to mention that the matrix $\mathcal{B}_T(\varphi_0)$ is truncated by removing the components related to long memory and by setting $d_0 = 0$ in $\mathcal{S}_T(d_0,\varphi_0)$. It is
important to emphasise that our model is not restricted to ARMA models alone; rather, it encompasses a broader category of short-memory models, with ARMA models being just one particular instance. Indeed, any representation of the
short-run dynamics that conforms to \eqref{repmainf} is admissable, including models like Bloomfield's exponential model.

An extension of the theorem that does not make use of Assumption \ref{A5} is contained in Theorem \ref{th:ExtShortMemory} in Appendix \ref{Initschemes}.

\begin{theorem}\label{t55} 
  Let $x_t$, $t$ = 1,$\ldots$,$T$, be given by \eqref{genq1} with $d_0 = 0$ and let Assumptions \ref{A2} to \ref{A5} be satisfied. Furthermore, when $d$ is set to zero in the respective objective functions, the approximate biases
  of $\hat{\varphi}$, $\hat{\varphi}_{\mu_0}$ and $\hat{\varphi}_{m}$ are as in \eqref{eqABcss}, \eqref{eqABcssmu0} and \eqref{mcss_gen1}, respectively.
  Specifically, the intrinsic bias takes the form
  \begin{align*}
    T  \mathcal{B}_T(\varphi_0) &=   \tilde{A}^{-1}   \begin{bmatrix}
      \iota' \left( \tilde{A}^{-1} \odot \left(\tilde{G}_1 + \tilde{F}_1 \right) \right)\iota \\
      \vdots    \\
      \iota' \left( \tilde{A}^{-1} \odot \left(\tilde{G}_p + \tilde{F}_p \right) \right)\iota 
    \end{bmatrix} - \frac{1}{2} \tilde{A}^{-1} \begin{bmatrix}
      \iota' \left(\left(\tilde{A}^{-1} \tilde{C}_{01} \tilde{A}^{-1} \right) \odot \tilde{A} \right) \iota \\
      \vdots    \\
      \iota' \left(\left(\tilde{A}^{-1} \tilde{C}_{0p} \tilde{A}^{-1} \right) \odot \tilde{A} \right) \iota
    \end{bmatrix} \\
  \end{align*}
  while  the score bias is now given by $ T\mathcal{S}_T(\varphi_0) = \tilde{A}^{-1}  D_{\varphi}\phi(1;\varphi_0) / \phi(1;\varphi_0) $  with 
  \begin{align*}
    \tilde{A} &= \sum_{j = 1}^{\infty} b_{\varphi j}(\varphi_0) b_{\varphi' j}(\varphi_0) \\
    \tilde{F}_m &= \sum_{i = 1}^{\infty}  b_{\varphi \varphi_m i}(\varphi_0) b_{\varphi' i}(\varphi_0) \\
    \tilde{G}_m &= \sum_{k = 1}^{\infty}  \sum_{s = 1}^{\infty} \left(  b_{\varphi_m s}(\varphi_0) b_{\varphi (s+k)}(\varphi_0) + b_{\varphi_m (s+k)}(\varphi_0) b_{\varphi s}(\varphi_0) \right)  b_{\varphi' k}(\varphi_0)\\
    \tilde{C}_{0m} &= \left[ \sum_{i = 1}^{\infty}  b_{\varphi i}(\varphi_0) b_{\varphi' \varphi_m i}(\varphi_0) \right]' + \sum_{i = 1}^{\infty}  b_{\varphi  i}(\varphi_0) b_{\varphi' \varphi_m i}(\varphi_0)  + \sum_{i = 1}^{\infty}  b_{\varphi_m i}(\varphi_0) b_{\varphi \varphi' i}(\varphi_0)
  \end{align*}
  for $m = 1,\ldots,p$. Here, $b_{\cdot i}(\varphi_0)$ is defined in Assumption \ref{A1}.
\end{theorem}

It is instructive to compare these expressions to the corresponding results for stationary and invertible ARFIMA models in Theorem \ref{t53}. In that context, the bias of the short-run dynamics in the stationary region for the CSS
estimator is of order $O(T^{-1} \log(T))$. This, however, is not true for the bias of the short-run dynamics of the CSS estimator for stationary and invertible ARMA models which is of order $O(T^{-1})$, as shown above in
Theorem \ref{t55}. An extension of a stationary ARMA model to an ARFIMA model increases the bias of the short-run dynamics to be of the same order of magnitude as that of the memory parameter. Nevertheless, the MCSS estimator
effectively eliminates this bias by construction, leading to a reduction in the bias order of $\hat{\varphi}$ to $O(T^{-1})$ for general ARFIMA or ARMA models.

Based on Theorem \ref{t55}, we can deduce the analytic bias of an AR(1) model as an illustration. The following corollary presents the expressions describing the analytic biases of the three estimators. The proof is
straightforward and omitted. An generalisation that does not make use of Assumption \ref{A5} is contained in Corollary \ref{cr:ExtShortMemory} in Appendix \ref{Initschemes}.
\begin{corollary}\label{t56} 
  Let $x_t$, $t$ = 1,$\ldots$,$T$, be given by \eqref{genq1} with $d_0 = 0$ and let $u_t = \varphi u_{t-1} + \epsilon_t$. Let Assumptions \ref{A2} to \ref{A5} be satisfied. Furthermore, when $d$ is set to zero in the respective
  objective functions, the approximate biases of $\hat{\varphi}$, $\hat{\varphi}_{\mu_0}$ and $\hat{\varphi}_{m}$ are as in \eqref{eqABcss}, \eqref{eqABcssmu0} and \eqref{mcss_gen1}, respectively.
  Specifically, the intrinsic bias takes the form  $T \mathcal{B}_T(\varphi_0) = -2\varphi_0$ and the score bias is now given by $T\mathcal{S}_T(\varphi_0) = -\varphi_0-1$.
\end{corollary}

$\mathcal{S}_T(\varphi_0)$ and $\mathcal{B}_T (\varphi_0)$ correspond to the bias expressions in the AR(1) model with and without constant term, respectively, derived by \textcite{tanaka1984asymptotic}. Note that Tanaka {\it
  (i)} analyses a well-specified model for an AR(1) process that begins in the infinite past and {\it (ii)} uses exact maximum likelihood to estimate the model, thereby making use of all available information.  We get the same
results because, without unobserved pre-sample values, our model is also well-specified and the MCSS estimator equally makes best use of the available information.

\subsection{The role of unobserved pre-sample values}
\label{sc:initial}

Recall that our model in \eqref{genq1}-\eqref{genq2} has so far been supplemented by Assumption \ref{A5}, i.e.\ by the simplification that not only the observations $x_t$ but also the unobserved error terms $u_t$ and
$\varepsilon_t$ are equal to zero for $t \leq 0$. Assumption \ref{A5} reflects the conditional nature of the CSS estimator. Moreover, the assumption ensures that the exposition of the results in Theorems \ref{t52} and \ref{t53} to
\ref{t55} is not cluttered by bias terms that are due to unobserved pre-sample values.

We now extend that setup and discuss generalisations of our results to different initialisation schemes. It will become plain that our results carry over to these extended settings, with the sole difference that the unobserved
pre-sample values lead to what we call model misspecification bias. As was pointed out before, the modification factor of the MCSS estimator is constructed to eliminate the unknown-level bias. It does not, however, account for
misspecification bias. This is indeed what we showed in Theorem \ref{corinitial} in the context of the ARFIMA(0,$d$,0) model initialised by $N_0$ unobserved pre-sample values, see the remark at the end of Section \ref{arfima0d}
above. A similar result is derived in the present section for the ARFIMA($p_1,d,p_2$) model in \eqref{genq1}-\eqref{genq2}, initialised by an error process $u_t$ that starts in the infinite past. Put differently, Assumption
\ref{A5} is now not imposed. This is the setup used by, for instance, \textcite{hualde2011gaussian} and \textcite{hualde2020truncated}. The following theorem extends our main results in Theorem \ref{t53}, its proof is contained in Appendix
\ref{Initschemes}.

\begin{theorem}\label{th:corinitial}
  Let $x_t$, $t$ = 1,$\ldots$,$T$, be given by \eqref{genq1}-\eqref{genq2} and let Assumptions \ref{A2} to \ref{A1} be satisfied. The biases of $\hat{\vartheta}$, $\hat{\vartheta}_{\mu_0}$ and $\hat{\vartheta}_m$ are given by 
  \begin{align}
      bias(\hat{\vartheta}) &= \mathcal{B}_T(\vartheta_0) + \mathcal{S}_T(\vartheta_0)+ S^{init,\mu}_T(\vartheta_0) + o(T^{-1}) \label{init_1}\\
      bias(\hat{\vartheta}_{\mu_0}) &= \mathcal{B}_T(\vartheta_0) + S^{init,\mu_0}_T(\vartheta_0) + o(T^{-1}) \label{init_2}\\
      bias(\hat{\vartheta}_m) &= \mathcal{B}_T(\vartheta_0)   +S^{init,\mu}_T(\vartheta_0) + o(T^{-1}) \label{init_3},
  \end{align}
  where $\mathcal{B}_T(\varphi_0)$ and  $\mathcal{S}_T(\vartheta_0)$ are defined in Theorem \ref{t53} as before. The new terms $S^{init,\mu}_T(\vartheta_0)$ and $S^{init,\mu_0}_T(\vartheta_0)$ are respectively given by 
  \begin{align}
      T S^{init,\mu}_T(\vartheta_0) &= - A^{-1}  \sum_{r = 0}^{\infty} D_{\vartheta} \left( \frac{1}{2} \left( \sum_{t = 1}^T (g_{t+r}^{(t)}(\vartheta))^2  - \frac{\left( \sum_{t = 1}^T c_t(\vartheta) g_{t+r}^{(t)}(\vartheta)  \right)^2}{\sum_{t = 1}^T c^2_t(\vartheta)} \right) \right) \Bigg|_{\vartheta=\vartheta_0} , \label{eq:init-mu}\\
       T S^{init,\mu_0}_T(\vartheta_0) &=- A^{-1} \sum_{r = 0}^{\infty} D_{\vartheta} \left( \frac{1}{2} \left( \sum_{t = 1}^T (g_{t+r}^{(t)}(\vartheta))^2 \right) \right)\Bigg|_{\vartheta=\vartheta_0},  \label{eq:init-mu0}
  \end{align}
  where $c_t(\vartheta)$ and $g_{t+r}^{(t)}(\vartheta)$ are defined in \eqref{detc} and \eqref{gfunction}.  Furthermore, $S^{init,\mu}_T(\vartheta_0) = O(T^{-1})$ and
  $S^{init,\mu_0}_T(\vartheta_0) = O(T^{-1})$.
\end{theorem}

The components $\mathcal{B}_T(\varphi_0)$ and $\mathcal{S}_T(\vartheta_0)$ are identical to the corresponding quantities in Theorem \ref{t53}. As before, $\mathcal{B}_T(\varphi_0)$ is the intrinsic bias.
$\mathcal{S}_T(\vartheta_0)$ is again the score bias due to estimating the unknown level, yet it is now only one part of the overall score bias: All three bias expressions in Theorem \ref{th:corinitial} now contain a second score
bias term, viz.\ one that is due to the CSS estimator setting all pre-sample values equal to zero. Its form depends on whether or not a constant term is estimated and is denoted by $S_T^{\mathrm{init},\mu}(\vartheta_0)$ or
$S_T^{\mathrm{init},\mu_0}(\vartheta_0)$, respectively\footnote{Note that, in our terminology, we call these misspecification biases exact. The approximate misspecification biases (their asymptotic counterparts) are given in Lemma \ref{apprmb}.}.  The expression in \eqref{eq:init-mu} is part of the bias of the CSS estimator with unknown level and of the MCSS estimator. Specifically, the term involving $c_t(\vartheta)$
shows how initial values feed into the estimation of $\mu$ and thus add bias. The expression in \eqref{eq:init-mu0}, on the other hand, applies when the level is known and the contribution to the bias stems only from the way
initial conditions enter the squared residuals.

As is shown by \textcite{hualde2020truncated}, unobserved pre-sample values do not alter the first-order asymptotic properties of the CSS estimator, yet it is plain from Theorem \ref{th:corinitial} that they do matter for the bias
expressions. Our MCSS estimator removes the unknown-level score bias $\mathcal{S}_T(\vartheta_0)$.  \textcite{johansen2016role} suggest a method to reduce the misspecification score bias due to unobserved initial values: In their model,
the conditional maximum likelihood estimator is biased due to the existence of $N_0$ unobserved pre-sample values; yet the authors show that that bias decreases as the sample size $N$ gets larger, see their Corollary 1. An
application of the results by \textcite{johansen2016role} to our MCSS estimator for the case of $N=0$ is also contained in Appendix \ref{Initschemes}.

\section{Simulation} \label{Ssimgen}

In this section we conduct a simulation study of the finite sample properties of the CSS estimators with unknown and known $\mu_0$ in \eqref{genCSS} and \eqref{genCSSmuknown}, respectively, and of the MCSS estimator in
\eqref{MCSSgen1}, together with the refined bias-corrected version\footnote{Note that we use the exact intrinsic bias $B_T$ as a correction. Simulation results using the approximate intrinsic bias are available upon request.}
thereof in \eqref{bcmcc}. In particular, we take as our DGP the model in \eqref{genq1} with $u_t$ an AR(1) model, i.e.\ $u_t = \varphi_0 u_{t-1} + \epsilon_t$ with $\epsilon_t \sim \textit{NID}(0,1)$. We set $u_0 = 0$ in
accordance with Assumption \ref{A5}.  Without loss of generality, we let $\mu_0 = 0$, since all estimators are invariant to the value of $\mu_0$. In all settings covered by our experiment, we generate $x_t$ for
$T = 32, 64, 128, 256$.  We let the long memory parameter $d_0$ vary. In particular, we set $d_0 = -0.2,-0.1,\ldots, 1.1,1.2$. For the autoregressive parameter we use $\varphi_0 \in \{-0.5,0,0.5 \}$. We compute the estimates using
the optimising intervals $d \in [d_0-5,d_0+5]$ and $\varphi \in [-0.9999,0.9999]$. All results are based on 10,000 replications\footnote{All computations in this paper are done using MATLAB 2019a, see \textcite{MATLAB}. The
  convergence criteria used for numerical optimisation are the default ones. The code for replicating the results in this paper is available on request. }. We use the fractional difference algorithm of
\textcite{jensen2014fast} to generate our models, as well as to compute the objective function of the estimators.

Table \ref{tablear1} and \ref{tablear2} present the Monte Carlo biases (multiplied by 100) of the memory parameter and the autoregressive parameter, respectively. We also report the percentage increase of the bias of the CSS
estimator relative to the bias of the MCSS estimator by $\Delta \% |\text{bias}|$ in the last column of each $T$. In addition, Table \ref{tablear3} and \ref{tablear4} present the Monte Carlo MSE (multiplied by 100) of the memory
parameter and the autoregressive parameter, respectively. We now summarise the main findings.

We first discuss the cases where $\varphi_0 = -0.5$ and $\varphi_0 = 0$. The addition of an autoregressive component to the purely fractional model considerably increases the bias of the CSS estimator of $d$, especially in the stationary region, cf.\
the theoretical biases of the purely fractional model in Table \ref{table2}. The CSS estimator $\hat{d}$ clearly underestimates the true $d_0$, while the $\hat{\varphi}$ overestimates the true $\varphi_0$. The MCSS estimator, however, reduces a large part of this bias. The largest reduction occurs in the stationary region, which is also expected from Theorem \ref{t53}. Importantly, the bias of the MCSS estimator and the bias of the CSS estimator with known $\mu_0$ are close to each other, confirming our
theoretical findings that the leading bias of $\hat{\vartheta}_m$ is the same as that of $\hat{\vartheta}_{\mu_0}$. The bias-corrected MCSS estimator further improves upon the MCSS and CSS estimators, as is expected from Corollary \ref{bcmcorr}. Turning to the MSE, the MCSS estimator consistently outperforms the CSS estimator across nearly all cases. The
largest improvement occurs again in the stationary region, which is where the largest bias reduction takes place. The CSS estimator with known $\mu_0$ performs the best and outperforms the MCSS estimator, while the biases are somewhat similar, the variance of the former estimator is significantly lower because $\mu_0$ is known. This estimator also outperforms the bias-corrected MCSS estimator. Although the bias of the bias-corrected MCSS estimator is lower, its variance appears to be relatively larger.

Additional comments can be made regarding the case $\varphi_0 = 0$. This scenario is both interesting and realistic. Typically, the autoregressive lags of a regression model are unknown, and a lag selection procedure, such as the
method proposed by \textcite{box1990time} or the use of an information criterion as in \textcite{huang2022consistent}, is used to estimate it. It is not unlikely that the lag selection procedure or the information criterion
overestimates the number of lags. The case $\varphi_0 = 0$ resembles a situation in which the lag order in the model is larger than that of the DGP. Then, according to our simulation, the CSS estimator is strongly biased when the
level parameter term is not known, even in comparison to the case with $\varphi_0 = -0.5$. The MCSS estimator, as opposed to the CSS estimator, significantly reduces the bias. Therefore, lag selection procedures when nuisance
parameter are included in the model could be based on the MCSS estimator rather than the CSS estimator, e.g.\ see \textcite{lee2015model}.

Finally, we now discuss the situation where $\varphi_0 = 0.5$. The comments made above are also true for the non-stationary region. In particular, the bias of the MCSS estimator is close to that of the CSS estimator with known
$\mu_0$. In the stationary region, however, the two estimators behave differently. \textcite{baillie2024combining} show that the correlation between the MLE estimates of the short-run dynamics and memory parameter is $-0.94$ for
this setting, which might explain the difficulty for our estimators in distinguishing between the autoregressive and memory component and the possible theoretical mismatch of our estimators. Nevertheless, the differences become
small for $T = 256$. Furthermore, the CSS estimator of performs the worst in terms of the bias, while the other three estimators significantly improve on this estimator. In terms of the MSE, we see that the MSE of the MCSS estimatorof $d$
and that of the CSS estimator with known $\mu_0$ is significantly lower than that of the CSS estimator. As opposed to that, the MSE of the CSS estimator for $\varphi$ is lower than that of the MCSS estimator and also the CSS
estimator with known $\mu_0$ when $T = 32$.

\begin{figure}[H]
  \centering
  
  \subfloat[Density of $\hat d$, $\hat d_{\mu_0}$, $\hat d_m$ for $T$ = 32]
  {
    \includegraphics[width=0.35\textwidth]{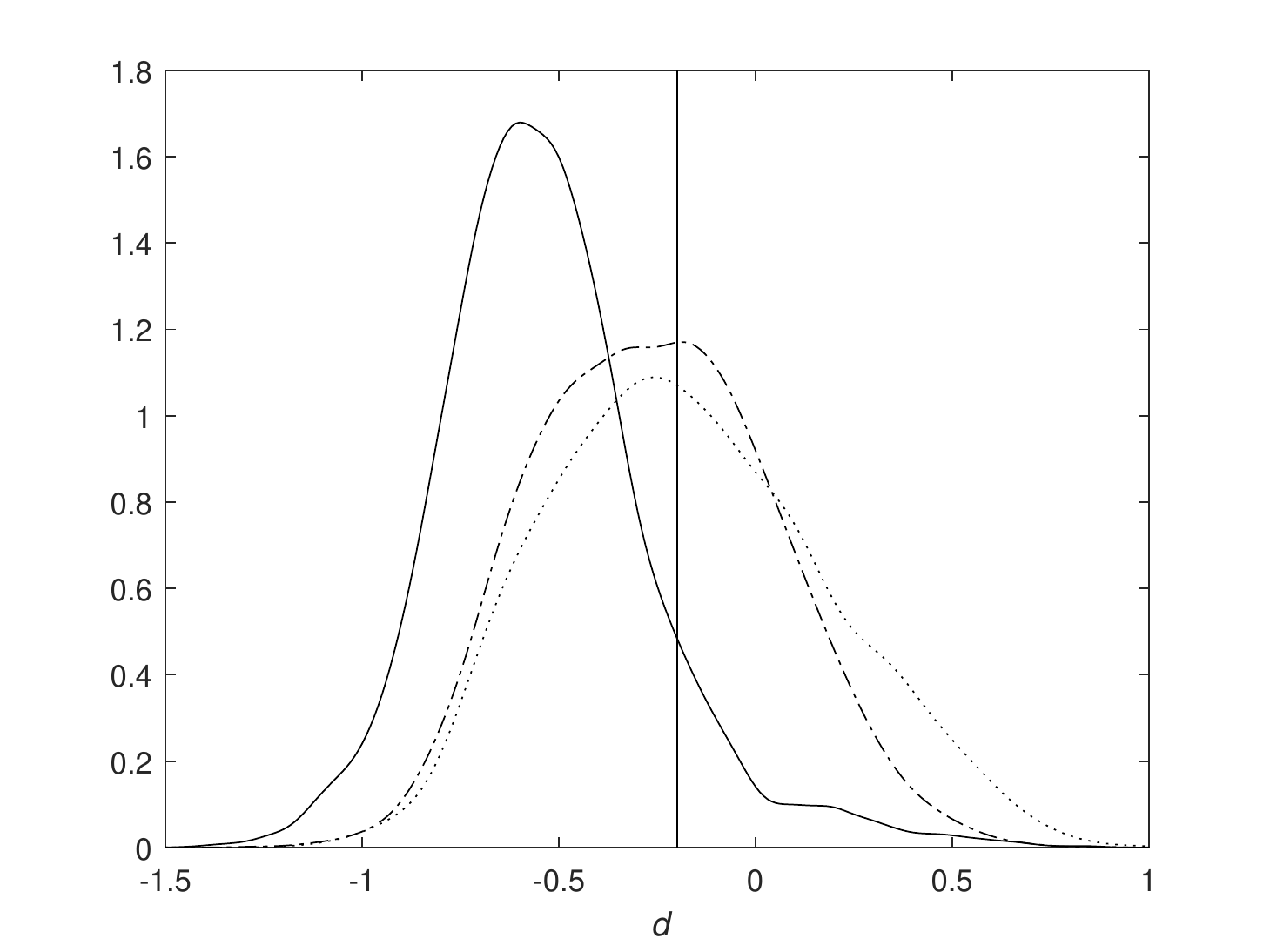}
  }
  \subfloat[Density of $\hat \varphi$, $\hat \varphi_{\mu_0}$, $\hat \varphi_m$ for $T$ = 32]
  {
    \includegraphics[width=0.35\textwidth]{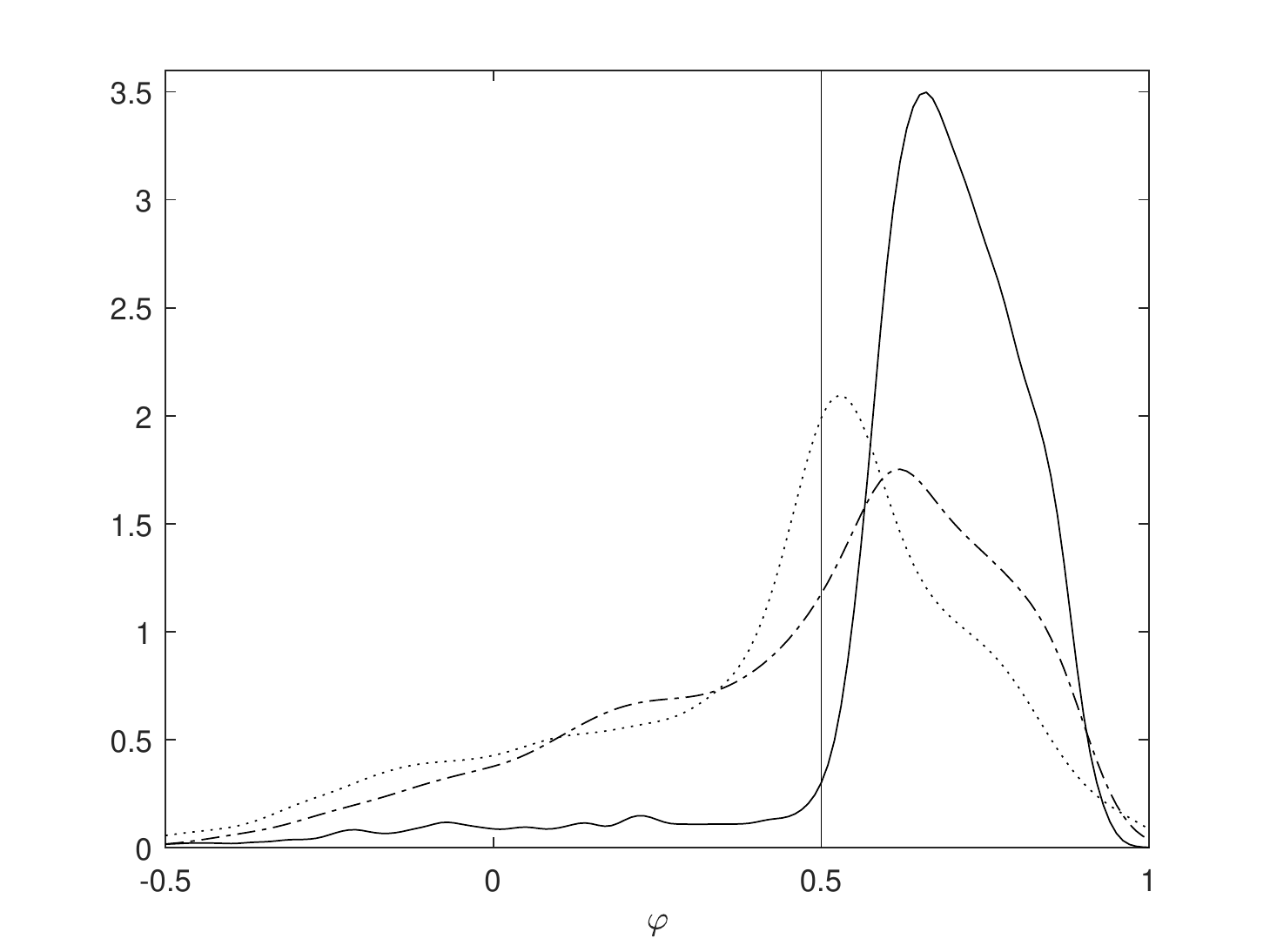}
  }
  \hspace{0mm}
  \subfloat[Density of $\hat d$, $\hat d_{\mu_0}$, $\hat d_m$ for $T$ = 256]
  {
    \includegraphics[width=0.35\textwidth]{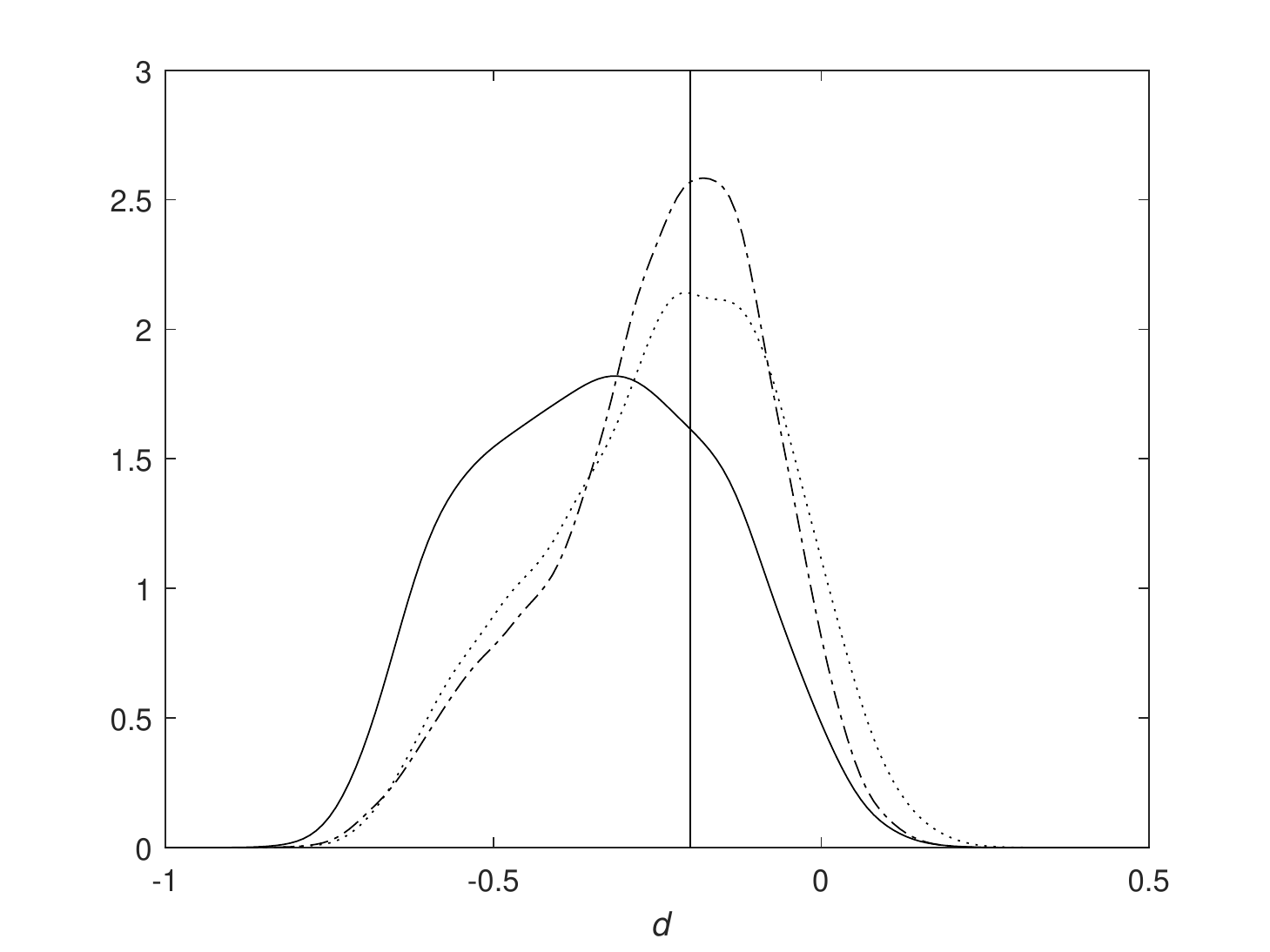}
  }
  \subfloat[Density of $\hat \varphi$, $\hat \varphi_{\mu_0}$, $\hat \varphi_m$ for $T$ = 256]
  {
    \includegraphics[width=0.35\textwidth]{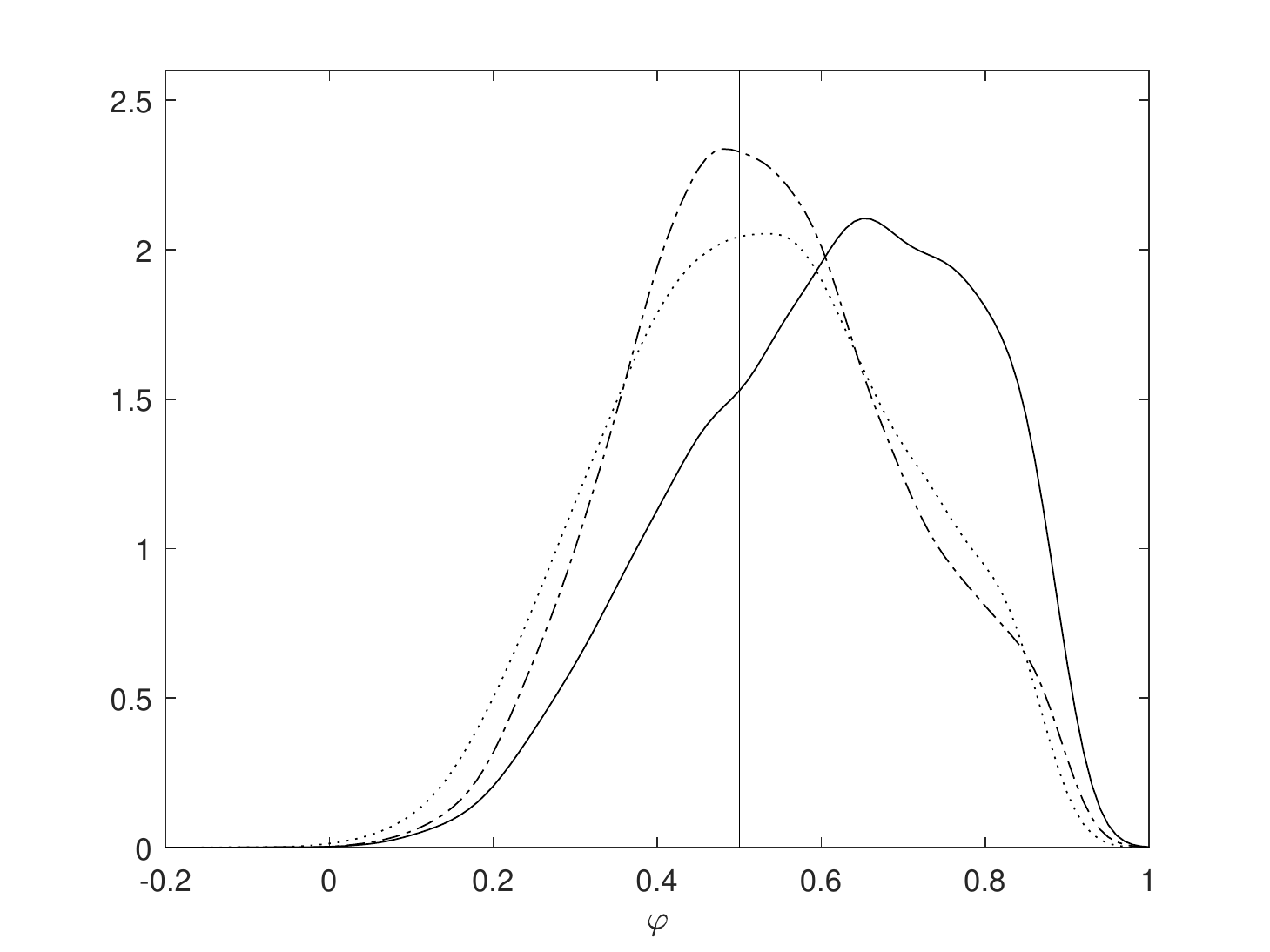}
  }
  \caption{ Density plots of the CSS estimator with unknown level parameter (solid lines), with known level parameter (dashed-dotted lines) as well as of the MCSS estimator (dotted lines) of $d$ (left panels) and $\varphi$ (right panels) in the ARFIMA(1,$d_0$,0) model with $T$ = 32 (upper panels)
    and $T = 256$ (lower panels), where $d_0 = -0.2$ and $\varphi_0 = 0.5$. The density estimates use a normal kernel. }%
  \label{fig711}%
\end{figure}

In order to better understand the differences in the bias and MSE, we have plotted in Figure \ref{fig711} densities of the three estimators of $d$ (left panels) and $\varphi$ (right panels) for $T = 32$ (upper panels) and
$T = 256$ (lower panels), when $d_0 = -0.2$ and $\varphi_0 = 0.5$. It can be seen that the CSS estimators strongly underestimate the true $d_0 = -0.2$ and strongly overestimate the true $\varphi_0 = 0.5$ for $T = 32$. This strong
bias in the CSS estimator contrasts with less variation. The CSS estimator's poor performance extends somewhat to the case $T = 256$. The MCSS and CSS estimators are well-centred, but this centring comes at the cost of an increase
in the variance. This explains the differences in the MSE of the CSS estimator relative to that of the MCSS estimator and the CSS estimator with known $\mu_0$. Also, it appears that for small $T$ the MCSS estimator recentres the
memory parameter relatively more than the autoregressive component, explaining the differences with the bias of the CSS estimator with known $\mu_0$. Nevertheless, these plots show that MCSS density estimates are more similar to
CSS density estimates with known $\mu_0$ than to CSS density estimates with unknown $\mu$. The good finite sample performance of the MCSS estimator is again evident.

\begin{table}[H]
\centering
\resizebox{\textwidth}{!}{%
\begin{tabular}{EEEEEEEEEEEEEEEEEEEEEEEEE}
\hline
\multicolumn{1}{c}{$\varphi_0$} &
 &
  \multicolumn{1}{c}{bias($\hat{d}$)} &
  \multicolumn{1}{c}{bias($\hat{d}_{\mu_0}$)} &
  \multicolumn{1}{c}{bias($\hat{d}_{m}$)} &
   \multicolumn{1}{c}{bias($\hat{d}_{bcm}$)} &
  \multicolumn{1}{c}{$\Delta \% |\text{bias}|$} &
 &
  \multicolumn{1}{c}{bias($\hat{d}$)} &
  \multicolumn{1}{c}{bias($\hat{d}_{\mu_0}$)} &
  \multicolumn{1}{c}{bias($\hat{d}_{m}$)} &
   \multicolumn{1}{c}{bias($\hat{d}_{bcm}$)} &
  \multicolumn{1}{c}{$\Delta \% |\text{bias}|$} &
   &
  \multicolumn{1}{c}{bias($\hat{d}$)} &
  \multicolumn{1}{c}{bias($\hat{d}_{\mu_0}$)} &
  \multicolumn{1}{c}{bias($\hat{d}_{m}$)} &
   \multicolumn{1}{c}{bias($\hat{d}_{bcm}$)} &
  \multicolumn{1}{c}{$\Delta \% |\text{bias}|$} &
   &
  \multicolumn{1}{c}{bias($\hat{d}$)} &
  \multicolumn{1}{c}{bias($\hat{d}_{\mu_0}$)} &
  \multicolumn{1}{c}{bias($\hat{d}_{m}$)} &
  \multicolumn{1}{c}{bias($\hat{d}_{bcm}$)} &
  \multicolumn{1}{c}{$\Delta \% |\text{bias}|$}  \\ \cline{3-7} \cline{9-13} \cline{15-19} \cline{21-25} 
 & \multicolumn{1}{c}{$d_0$ \textbackslash{} $T$} & \multicolumn{5}{c}{32} &  & \multicolumn{5}{c}{64} &  & \multicolumn{5}{c}{128} &  & \multicolumn{5}{c}{256} \\ \hline
  & -0.2 & -37.62 & -9.02  & -12.15 & -6.32  & 209.55  &  & -13.70 & -3.91  & -4.50  & -1.19 & 204.65  &  & -5.91  & -1.89 & -2.00 & -0.24 & 194.61 &  & -2.88  & -0.90 & -0.90 & 0.01  & 219.88 \\
  & -0.1 & -38.59 & -9.02  & -12.20 & -6.36  & 216.22  &  & -13.86 & -3.91  & -4.54  & -1.22 & 205.57  &  & -5.95  & -1.89 & -2.01 & -0.25 & 196.14 &  & -2.89  & -0.90 & -0.90 & 0.01  & 220.69 \\
  & 0.0    & -39.26 & -9.02  & -12.35 & -6.51  & 217.98  &  & -13.86 & -3.91  & -4.58  & -1.26 & 202.82  &  & -5.96  & -1.89 & -2.01 & -0.25 & 195.97 &  & -2.89  & -0.90 & -0.90 & 0.01  & 220.00 \\
  & 0.1  & -39.60 & -9.02  & -12.35 & -6.49  & 220.59  &  & -13.92 & -3.91  & -4.65  & -1.35 & 199.07  &  & -5.94  & -1.89 & -2.02 & -0.26 & 193.49 &  & -2.87  & -0.90 & -0.91 & 0.01  & 216.75 \\
  & 0.2  & -39.80 & -9.02  & -12.49 & -6.64  & 218.60  &  & -13.87 & -3.91  & -4.65  & -1.34 & 198.41  &  & -5.83  & -1.89 & -2.03 & -0.27 & 186.95 &  & -2.82  & -0.90 & -0.91 & 0.00  & 209.18 \\
  & 0.3  & -39.18 & -9.02  & -12.42 & -6.56  & 215.46  &  & -13.54 & -3.91  & -4.67  & -1.36 & 189.92  &  & -5.60  & -1.89 & -2.04 & -0.28 & 173.99 &  & -2.70  & -0.90 & -0.92 & 0.00  & 194.59 \\
  & 0.4  & -38.01 & -9.02  & -12.35 & -6.48  & 207.92  &  & -12.88 & -3.91  & -4.64  & -1.32 & 177.86  &  & -5.24  & -1.89 & -2.05 & -0.29 & 155.29 &  & -2.49  & -0.90 & -0.92 & -0.01 & 169.89 \\ \cline{2-25} 
 \multicolumn{1}{c}{$-0.5$} & 0.5  & -36.47 & -9.02  & -12.17 & -6.30  & 199.53  &  & -11.83 & -3.91  & -4.58  & -1.26 & 158.39  &  & -4.68  & -1.89 & -2.05 & -0.29 & 128.31 &  & -2.18  & -0.90 & -0.93 & -0.02 & 134.69 \\ \cline{2-25} 
  & 0.6  & -33.69 & -9.02  & -12.10 & -6.23  & 178.53  &  & -10.32 & -3.91  & -4.52  & -1.20 & 128.34  &  & -3.98  & -1.89 & -2.03 & -0.27 & 96.41  &  & -1.82  & -0.90 & -0.93 & -0.02 & 94.92  \\
  & 0.7  & -29.72 & -9.02  & -11.68 & -5.82  & 154.40  &  & -8.73  & -3.91  & -4.43  & -1.11 & 97.08   &  & -3.28  & -1.89 & -1.99 & -0.23 & 64.77  &  & -1.49  & -0.90 & -0.93 & -0.02 & 59.79  \\
  & 0.8  & -25.26 & -9.01  & -10.96 & -5.10  & 130.49  &  & -7.25  & -3.91  & -4.29  & -0.97 & 69.05   &  & -2.72  & -1.89 & -1.95 & -0.19 & 39.25  &  & -1.24  & -0.90 & -0.92 & -0.01 & 34.33  \\
  & 0.9  & -20.84 & -9.02  & -10.39 & -4.53  & 100.68  &  & -5.99  & -3.91  & -4.14  & -0.82 & 44.64   &  & -2.35  & -1.89 & -1.93 & -0.17 & 21.77  &  & -1.08  & -0.90 & -0.91 & 0.00  & 17.89  \\
  & 1.0    & -16.97 & -9.02  & -9.82  & -3.95  & 72.78   &  & -5.13  & -3.91  & -4.06  & -0.74 & 26.45   &  & -2.10  & -1.89 & -1.91 & -0.15 & 9.60   &  & -0.98  & -0.90 & -0.91 & 0.01  & 7.59   \\
  & 1.1  & -13.97 & -9.02  & -9.67  & -3.84  & 44.41   &  & -4.46  & -3.91  & -4.00  & -0.68 & 11.50   &  & -1.94  & -1.89 & -1.90 & -0.14 & 2.12   &  & -0.91  & -0.90 & -0.90 & 0.01  & 1.00   \\
  & 1.2  & -11.89 & -9.02  & -9.51  & -3.68  & 25.08   &  & -4.05  & -3.91  & -3.98  & -0.66 & 1.84    &  & -1.85  & -1.89 & -1.90 & -0.14 & -2.73  &  & -0.87  & -0.90 & -0.90 & 0.02  & -3.40   \\ \hline
  &      &        &        &        &        &         &  &        &        &        &       &         &  &        &       &       &       &        &  &        &       &       &       &         \\ \hline
  & -0.2 & -57.19 & -16.48 & -16.99 & -9.58  & 236.65  &  & -36.69 & -10.32 & -12.17 & -6.93 & 201.58  &  & -17.91 & -4.87 & -6.18 & -3.05 & 189.68 &  & -6.90  & -2.09 & -2.36 & -0.66 & 191.75 \\
  & -0.1 & -58.58 & -16.49 & -17.18 & -9.76  & 241.02  &  & -37.44 & -10.31 & -12.22 & -6.98 & 206.35  &  & -18.19 & -4.84 & -6.19 & -3.05 & 194.03 &  & -6.97  & -2.09 & -2.39 & -0.69 & 191.14 \\
  & 0.0    & -59.70 & -16.48 & -17.31 & -9.88  & 244.86  &  & -37.95 & -10.32 & -12.28 & -7.03 & 208.95  &  & -18.48 & -4.86 & -6.24 & -3.11 & 196.00 &  & -7.01  & -2.08 & -2.40 & -0.69 & 192.48 \\
  & 0.1  & -60.38 & -16.50 & -17.42 & -9.97  & 246.56  &  & -37.97 & -10.32 & -12.18 & -6.91 & 211.67  &  & -18.62 & -4.86 & -6.28 & -3.15 & 196.74 &  & -7.01  & -2.09 & -2.39 & -0.68 & 193.12 \\
  & 0.2  & -60.14 & -16.48 & -17.40 & -9.92  & 245.57  &  & -37.64 & -10.33 & -12.17 & -6.89 & 209.18  &  & -18.44 & -4.85 & -6.30 & -3.17 & 192.98 &  & -6.84  & -2.09 & -2.39 & -0.68 & 186.62 \\
  & 0.3  & -58.88 & -16.49 & -17.67 & -10.17 & 233.21  &  & -36.57 & -10.31 & -12.15 & -6.85 & 201.11  &  & -17.77 & -4.85 & -6.21 & -3.07 & 186.04 &  & -6.51  & -2.09 & -2.37 & -0.66 & 174.95 \\
  & 0.4  & -56.06 & -16.49 & -17.81 & -10.30 & 214.74  &  & -34.64 & -10.33 & -12.06 & -6.74 & 187.11  &  & -16.57 & -4.87 & -6.29 & -3.16 & 163.68 &  & -6.00  & -2.09 & -2.36 & -0.65 & 154.47 \\ \cline{2-25} 
 \multicolumn{1}{c}{0} & 0.5  & -51.49 & -16.49 & -18.14 & -10.65 & 183.83  &  & -31.31 & -10.30 & -11.86 & -6.52 & 163.92  &  & -14.55 & -4.83 & -6.03 & -2.89 & 141.40 &  & -5.22  & -2.08 & -2.30 & -0.59 & 126.77 \\ \cline{2-25} 
  & 0.6  & -45.50 & -16.50 & -18.20 & -10.74 & 149.98  &  & -26.72 & -10.32 & -11.57 & -6.24 & 130.98  &  & -12.12 & -4.85 & -5.78 & -2.63 & 109.84 &  & -4.19  & -2.08 & -2.24 & -0.53 & 87.40  \\
  & 0.7  & -38.84 & -16.50 & -18.01 & -10.60 & 115.63  &  & -21.77 & -10.33 & -11.23 & -5.94 & 93.86   &  & -9.37  & -4.85 & -5.41 & -2.25 & 73.28  &  & -3.26  & -2.09 & -2.19 & -0.49 & 48.88  \\
  & 0.8  & -31.76 & -16.49 & -17.34 & -9.92  & 83.18   &  & -17.36 & -10.32 & -10.91 & -5.65 & 59.14   &  & -7.27  & -4.87 & -5.20 & -2.04 & 39.92  &  & -2.65  & -2.09 & -2.14 & -0.44 & 23.78  \\
  & 0.9  & -25.92 & -16.49 & -17.06 & -9.69  & 51.93   &  & -14.10 & -10.32 & -10.76 & -5.53 & 31.10   &  & -5.95  & -4.84 & -5.01 & -1.84 & 18.74  &  & -2.34  & -2.08 & -2.12 & -0.42 & 10.22  \\
  & 1.0    & -21.81 & -16.49 & -16.80 & -9.47  & 29.85   &  & -12.12 & -10.33 & -10.55 & -5.33 & 14.89   &  & -5.32  & -4.87 & -4.95 & -1.77 & 7.49   &  & -2.16  & -2.09 & -2.10 & -0.39 & 3.03   \\
  & 1.1  & -19.14 & -16.51 & -16.75 & -9.44  & 14.28   &  & -11.04 & -10.32 & -10.42 & -5.20 & 5.95    &  & -5.01  & -4.87 & -4.92 & -1.74 & 1.88   &  & -2.07  & -2.08 & -2.06 & -0.35 & 0.44   \\
  & 1.2  & -17.51 & -16.50 & -16.60 & -9.29  & 5.44    &  & -10.52 & -10.32 & -10.35 & -5.12 & 1.60    &  & -4.91  & -4.88 & -4.93 & -1.75 & -0.43  &  & -2.06  & -2.09 & -2.08 & -0.37 & -1.27  \\ \hline
  &      &        &        &        &        &         &  &        &        &        &       &         &  &        &       &       &       &        &  &        &       &       &       &         \\ \hline
  & -0.2 & -33.35 & -5.93  & 2.77   & 9.52   & 1105.96 &  & -24.73 & -5.95  & -0.81  & 5.46  & 2957.56 &  & -19.03 & -5.82 & -3.08 & 1.59  & 517.33 &  & -14.34 & -4.82 & -4.17 & -0.99 & 243.92 \\
  & -0.1 & -33.57 & -5.93  & 2.56   & 9.32   & 1209.35 &  & -24.85 & -5.96  & -0.63  & 5.67  & 3848.19 &  & -19.15 & -5.82 & -3.02 & 1.67  & 534.82 &  & -14.40 & -4.82 & -4.13 & -0.95 & 248.21 \\
  & 0.0    & -32.95 & -5.92  & 2.15   & 8.88   & 1430.67 &  & -24.58 & -5.96  & -0.70  & 5.60  & 3422.44 &  & -19.02 & -5.82 & -3.01 & 1.68  & 531.20 &  & -14.31 & -4.82 & -4.18 & -1.00 & 242.10 \\
  & 0.1  & -31.69 & -5.93  & 1.45   & 8.11   & 2091.57 &  & -23.63 & -5.96  & -1.00  & 5.29  & 2264.08 &  & -18.51 & -5.82 & -3.11 & 1.59  & 495.28 &  & -14.13 & -4.82 & -4.20 & -1.02 & 235.94 \\
  & 0.2  & -28.86 & -5.93  & 0.37   & 6.90   & 7615.84 &  & -21.52 & -5.96  & -1.75  & 4.48  & 1129.85 &  & -17.44 & -5.82 & -3.38 & 1.31  & 415.34 &  & -13.48 & -4.82 & -4.19 & -0.99 & 221.80 \\
  & 0.3  & -24.89 & -5.94  & -1.34  & 4.99   & 1755.66 &  & -18.68 & -5.95  & -2.69  & 3.44  & 594.38  &  & -15.41 & -5.82 & -3.98 & 0.70  & 287.05 &  & -12.08 & -4.82 & -4.34 & -1.13 & 178.18 \\
  & 0.4  & -20.17 & -5.93  & -3.11  & 3.02   & 547.84  &  & -15.00 & -5.96  & -3.83  & 2.17  & 291.96  &  & -12.62 & -5.83 & -4.64 & 0.02  & 171.91 &  & -10.05 & -4.82 & -4.66 & -1.42 & 115.78 \\ \cline{2-25} 
 \multicolumn{1}{c}{0.5} & 0.5  & -15.68 & -5.94  & -4.28  & 1.72   & 266.08  &  & -11.55 & -5.95  & -4.88  & 0.96  & 136.53  &  & -9.64  & -5.82 & -5.11 & -0.46 & 88.55  &  & -7.84  & -4.82 & -4.83 & -1.56 & 62.35  \\ \cline{2-25} 
  & 0.6  & -11.66 & -5.94  & -4.93  & 1.00   & 136.48  &  & -8.97  & -5.95  & -5.53  & 0.18  & 62.16   &  & -7.65  & -5.82 & -5.51 & -0.88 & 39.03  &  & -6.18  & -4.82 & -4.83 & -1.53 & 27.91  \\
  & 0.7  & -8.79  & -5.95  & -5.30  & 0.59   & 65.83   &  & -7.36  & -5.96  & -5.82  & -0.17 & 26.52   &  & -6.52  & -5.82 & -5.72 & -1.12 & 14.03  &  & -5.37  & -4.82 & -4.89 & -1.58 & 9.83   \\
  & 0.8  & -6.97  & -5.93  & -5.48  & 0.40   & 27.06   &  & -6.50  & -5.95  & -5.98  & -0.38 & 8.70    &  & -6.04  & -5.82 & -5.82 & -1.23 & 3.83   &  & -5.03  & -4.82 & -4.88 & -1.57 & 3.10   \\
  & 0.9  & -5.93  & -5.94  & -5.69  & 0.17   & 4.35    &  & -6.02  & -5.95  & -6.03  & -0.44 & -0.14   &  & -5.86  & -5.82 & -5.85 & -1.27 & 0.07   &  & -4.86  & -4.82 & -4.85 & -1.53 & 0.25   \\
  & 1.0    & -5.52  & -5.94  & -5.88  & -0.05  & -6.15   &  & -5.91  & -5.96  & -6.01  & -0.43 & -1.67   &  & -5.77  & -5.83 & -5.85 & -1.27 & -1.35  &  & -4.82  & -4.82 & -4.84 & -1.52 & -0.34  \\
  & 1.1  & -5.32  & -5.95  & -5.96  & -0.13  & -10.73  &  & -5.84  & -5.95  & -6.05  & -0.46 & -3.45   &  & -5.74  & -5.82 & -5.84 & -1.25 & -1.67  &  & -4.81  & -4.82 & -4.81 & -1.49 & -0.09  \\
  & 1.2  & -5.22  & -5.94  & -6.00  & -0.18  & -13.00  &  & -5.83  & -5.96  & -6.06  & -0.47 & -3.76   &  & -5.77  & -5.82 & -5.83 & -1.24 & -0.99  &  & -4.79  & -4.82 & -4.82 & -1.50 & -0.69  \\ \hline
\end{tabular}
}
\caption{(100 $\times$) Monte Carlo bias of the estimated memory parameter for ARFIMA(1,$d_0$,0) of CSS estimator with unknown and known $\mu_0$ and the MCSS estimator, together with the bias-corrected MCSS estimator.}
\label{tablear1}
\end{table}

\begin{table}[H]
\centering
\resizebox{\textwidth}{!}{%
\begin{tabular}{EEEEEEEEEEEEEEEEEEEEEEEEE}
\hline
\multicolumn{1}{c}{$\varphi_0$} &
 &
  \multicolumn{1}{c}{bias($\hat{\varphi}$)} &
  \multicolumn{1}{c}{bias($\hat{\varphi}_{\mu_0}$)} &
  \multicolumn{1}{c}{bias($\hat{\varphi}_{m}$)} &
  \multicolumn{1}{c}{bias($\hat{\varphi}_{bcm}$)} &
  \multicolumn{1}{c}{$\Delta \% |\text{bias}|$} &
 &
 \multicolumn{1}{c}{bias($\hat{\varphi}$)} &
  \multicolumn{1}{c}{bias($\hat{\varphi}_{\mu_0}$) } &
  \multicolumn{1}{c}{bias($\hat{\varphi}_{m}$)} &
  \multicolumn{1}{c}{bias($\hat{\varphi}_{bcm}$)} &
  \multicolumn{1}{c}{$\Delta \% |\text{bias}|$} &
   &
 \multicolumn{1}{c}{bias($\hat{\varphi}$)} &
  \multicolumn{1}{c}{bias($\hat{\varphi}_{\mu_0}$) } &
  \multicolumn{1}{c}{bias($\hat{\varphi}_{m}$)} &
  \multicolumn{1}{c}{bias($\hat{\varphi}_{bcm}$)} &
  \multicolumn{1}{c}{$\Delta \% |\text{bias}|$} &
   &
  \multicolumn{1}{c}{bias($\hat{\varphi}$)} &
  \multicolumn{1}{c}{bias($\hat{\varphi}_{\mu_0}$) } &
  \multicolumn{1}{c}{bias($\hat{\varphi}_{m}$)} &
  \multicolumn{1}{c}{bias($\hat{\varphi}_{bcm}$)} &
  \multicolumn{1}{c}{$\Delta \% |\text{bias}|$}  \\ \cline{3-7} \cline{9-13} \cline{15-19} \cline{21-25} 
 & \multicolumn{1}{c}{$d_0$ \textbackslash{} $T$} & \multicolumn{5}{c}{32} &  & \multicolumn{5}{c}{64} &  & \multicolumn{5}{c}{128} &  & \multicolumn{5}{c}{256} \\ \hline
 & -0.2 & 26.93 & 8.25 & 9.89 & 5.00 & 172.41 &  & 9.43 & 3.67 & 3.96 & 0.94 & 137.96 &  & 3.95 & 1.86 & 1.90 & 0.27 & 107.99 &  & 1.97 & 0.95 & 0.95 & 0.11 & 107.64 \\
 & -0.1 & 27.70 & 8.24 & 9.88 & 4.98 & 180.42 &  & 9.55 & 3.67 & 3.99 & 0.97 & 139.20 &  & 3.97 & 1.86 & 1.90 & 0.28 & 109.35 &  & 1.98 & 0.95 & 0.95 & 0.11 & 108.25 \\
 & 0.0 & 28.22 & 8.25 & 9.94 & 5.06 & 183.75 &  & 9.53 & 3.68 & 4.01 & 0.99 & 137.38 &  & 3.99 & 1.86 & 1.90 & 0.28 & 109.73 &  & 1.98 & 0.95 & 0.95 & 0.11 & 108.10 \\
 & 0.1 & 28.48 & 8.25 & 9.87 & 4.96 & 188.51 &  & 9.59 & 3.67 & 4.07 & 1.06 & 135.59 &  & 3.97 & 1.86 & 1.90 & 0.28 & 108.76 &  & 1.96 & 0.95 & 0.95 & 0.11 & 106.58 \\
 & 0.2 & 28.76 & 8.25 & 9.95 & 5.05 & 189.11 &  & 9.63 & 3.67 & 4.06 & 1.04 & 137.42 &  & 3.91 & 1.86 & 1.91 & 0.28 & 105.22 &  & 1.93 & 0.95 & 0.95 & 0.11 & 102.69 \\
 & 0.3 & 28.40 & 8.24 & 9.82 & 4.91 & 189.15 &  & 9.46 & 3.67 & 4.06 & 1.04 & 132.98 &  & 3.77 & 1.86 & 1.91 & 0.29 & 97.22 &  & 1.86 & 0.95 & 0.95 & 0.12 & 94.89 \\
 & 0.4 & 27.68 & 8.24 & 9.72 & 4.79 & 184.83 &  & 9.07 & 3.67 & 4.01 & 0.99 & 126.00 &  & 3.56 & 1.86 & 1.91 & 0.29 & 85.99 &  & 1.73 & 0.95 & 0.96 & 0.12 & 81.33 \\ \cline{2-25} 
 \multicolumn{1}{c}{$-0.5$} & 0.5 & 26.80 & 8.24 & 9.57 & 4.61 & 180.01 &  & 8.43 & 3.67 & 3.97 & 0.93 & 112.61 &  & 3.23 & 1.86 & 1.91 & 0.29 & 68.64 &  & 1.55 & 0.95 & 0.96 & 0.12 & 61.61 \\ \cline{2-25} 
 & 0.6 & 24.95 & 8.25 & 9.59 & 4.64 & 160.21 &  & 7.43 & 3.67 & 3.94 & 0.90 & 88.78 &  & 2.81 & 1.86 & 1.91 & 0.28 & 47.58 &  & 1.34 & 0.95 & 0.96 & 0.12 & 39.01 \\
 & 0.7 & 22.16 & 8.24 & 9.39 & 4.44 & 135.92 &  & 6.40 & 3.67 & 3.90 & 0.87 & 63.94 &  & 2.40 & 1.86 & 1.89 & 0.27 & 26.62 &  & 1.15 & 0.95 & 0.96 & 0.13 & 19.10 \\
 & 0.8 & 19.05 & 8.24 & 9.01 & 4.01 & 111.49 &  & 5.44 & 3.67 & 3.84 & 0.80 & 41.84 &  & 2.07 & 1.86 & 1.88 & 0.25 & 10.31 &  & 1.01 & 0.95 & 0.96 & 0.12 & 5.13 \\
 & 0.9 & 15.91 & 8.24 & 8.72 & 3.70 & 82.36 &  & 4.62 & 3.67 & 3.76 & 0.72 & 22.86 &  & 1.87 & 1.86 & 1.87 & 0.25 & 0.26 &  & 0.93 & 0.95 & 0.96 & 0.12 & -3.26 \\
 & 1.0 & 13.15 & 8.24 & 8.43 & 3.36 & 56.02 &  & 4.09 & 3.67 & 3.73 & 0.69 & 9.73 &  & 1.74 & 1.86 & 1.86 & 0.24 & -6.41 &  & 0.88 & 0.95 & 0.95 & 0.12 & -7.88 \\
 & 1.1 & 11.07 & 8.24 & 8.47 & 3.43 & 30.58 &  & 3.66 & 3.67 & 3.71 & 0.66 & -1.14 &  & 1.68 & 1.86 & 1.86 & 0.24 & -9.74 &  & 0.85 & 0.95 & 0.95 & 0.11 & -10.29 \\
 & 1.2 & 9.64 & 8.25 & 8.44 & 3.40 & 14.24 &  & 3.43 & 3.67 & 3.70 & 0.66 & -7.41 &  & 1.65 & 1.86 & 1.86 & 0.24 & -11.34 &  & 0.84 & 0.95 & 0.95 & 0.11 & -11.42 \\ \hline
 &  &  &  &  &  &  &  &  &  &  &  &  &  &  &  &  &  &  &  &  &  &  &  &  \\ \hline
 & -0.2 & 41.35 & 12.08 & 10.83 & 8.05 & 281.83 &  & 30.07 & 8.43 & 9.35 & 6.22 & 221.49 &  & 15.59 & 4.18 & 5.18 & 2.86 & 200.96 &  & 6.19 & 1.89 & 2.10 & 0.73 & 195.07 \\
 & -0.1 & 42.40 & 12.09 & 10.93 & 8.16 & 288.02 &  & 30.70 & 8.41 & 9.39 & 6.26 & 227.04 &  & 15.85 & 4.16 & 5.19 & 2.87 & 205.38 &  & 6.26 & 1.89 & 2.13 & 0.76 & 194.01 \\
 & 0.0 & 43.24 & 12.09 & 10.96 & 8.19 & 294.39 &  & 31.12 & 8.43 & 9.43 & 6.30 & 230.15 &  & 16.10 & 4.17 & 5.23 & 2.91 & 208.00 &  & 6.30 & 1.89 & 2.13 & 0.76 & 195.69 \\
 & 0.1 & 43.72 & 12.10 & 10.99 & 8.20 & 297.66 &  & 31.10 & 8.44 & 9.31 & 6.16 & 234.19 &  & 16.22 & 4.17 & 5.24 & 2.94 & 209.35 &  & 6.30 & 1.89 & 2.12 & 0.75 & 196.65 \\
 & 0.2 & 43.49 & 12.08 & 10.95 & 8.12 & 297.07 &  & 30.77 & 8.43 & 9.28 & 6.12 & 231.66 &  & 16.06 & 4.17 & 5.27 & 2.96 & 205.01 &  & 6.14 & 1.90 & 2.12 & 0.75 & 189.58 \\
 & 0.3 & 42.45 & 12.08 & 11.21 & 8.37 & 278.73 &  & 29.82 & 8.42 & 9.25 & 6.07 & 222.42 &  & 15.44 & 4.17 & 5.19 & 2.86 & 197.74 &  & 5.82 & 1.89 & 2.09 & 0.72 & 177.94 \\
 & 0.4 & 40.13 & 12.08 & 11.41 & 8.55 & 251.71 &  & 28.10 & 8.43 & 9.20 & 5.98 & 205.59 &  & 14.34 & 4.18 & 5.26 & 2.95 & 172.47 &  & 5.34 & 1.89 & 2.09 & 0.72 & 155.77 \\ \cline{2-25} 
 \multicolumn{1}{c}{0} & 0.5 & 36.50 & 12.08 & 11.86 & 9.01 & 207.83 &  & 25.23 & 8.42 & 9.09 & 5.83 & 177.41 &  & 12.50 & 4.16 & 5.06 & 2.73 & 147.18 &  & 4.61 & 1.88 & 2.04 & 0.67 & 126.29 \\ \cline{2-25} 
 & 0.6 & 31.89 & 12.09 & 12.16 & 9.30 & 162.37 &  & 21.32 & 8.43 & 8.95 & 5.67 & 138.07 &  & 10.31 & 4.17 & 4.85 & 2.51 & 112.49 &  & 3.67 & 1.88 & 1.99 & 0.62 & 84.58 \\
 & 0.7 & 27.04 & 12.10 & 12.32 & 9.45 & 119.53 &  & 17.23 & 8.43 & 8.81 & 5.51 & 95.50 &  & 7.90 & 4.17 & 4.57 & 2.20 & 72.78 &  & 2.84 & 1.89 & 1.96 & 0.59 & 44.85 \\
 & 0.8 & 21.97 & 12.09 & 12.06 & 9.09 & 82.11 &  & 13.69 & 8.43 & 8.67 & 5.36 & 57.96 &  & 6.09 & 4.18 & 4.41 & 2.02 & 38.20 &  & 2.32 & 1.89 & 1.92 & 0.55 & 20.42 \\
 & 0.9 & 18.00 & 12.10 & 12.11 & 9.11 & 48.60 &  & 11.16 & 8.43 & 8.64 & 5.33 & 29.06 &  & 5.03 & 4.17 & 4.29 & 1.88 & 17.31 &  & 2.07 & 1.89 & 1.92 & 0.55 & 8.32 \\
 & 1.0 & 15.37 & 12.09 & 12.07 & 9.03 & 27.32 &  & 9.71 & 8.43 & 8.52 & 5.18 & 13.97 &  & 4.55 & 4.19 & 4.24 & 1.82 & 7.36 &  & 1.95 & 1.89 & 1.89 & 0.52 & 3.00 \\
 & 1.1 & 13.75 & 12.10 & 12.13 & 9.08 & 13.36 &  & 9.01 & 8.42 & 8.45 & 5.10 & 6.53 &  & 4.36 & 4.19 & 4.23 & 1.80 & 3.20 &  & 1.91 & 1.89 & 1.87 & 0.50 & 2.18 \\
 & 1.2 & 12.88 & 12.09 & 12.08 & 9.00 & 6.65 &  & 8.74 & 8.44 & 8.42 & 5.05 & 3.78 &  & 4.33 & 4.20 & 4.24 & 1.81 & 2.23 &  & 1.92 & 1.90 & 1.89 & 0.51 & 1.97 \\ \hline
 &  &  &  &  &  &  &  &  &  &  &  &  &  &  &  &  &  &  &  &  &  &  &  &  \\ \hline
 & -0.2 & 15.50 & -2.25 & -8.38 & -7.59 & 84.89 &  & 14.41 & 0.66 & -3.60 & -5.26 & 300.20 &  & 12.90 & 2.56 & -0.09 & -2.15 & 13834.33 &  & 10.93 & 3.01 & 2.18 & 0.42 & 400.54 \\
 & -0.1 & 15.65 & -2.24 & -8.21 & -7.42 & 90.58 &  & 14.47 & 0.66 & -3.78 & -5.46 & 283.08 &  & 12.98 & 2.56 & -0.16 & -2.23 & 7966.49 &  & 10.97 & 3.01 & 2.15 & 0.38 & 410.69 \\
 & 0.0 & 15.19 & -2.25 & -7.92 & -7.09 & 91.86 &  & 14.20 & 0.66 & -3.72 & -5.41 & 281.65 &  & 12.83 & 2.55 & -0.17 & -2.25 & 7432.74 &  & 10.88 & 3.01 & 2.19 & 0.43 & 396.56 \\
 & 0.1 & 14.35 & -2.24 & -7.43 & -6.51 & 93.08 &  & 13.40 & 0.66 & -3.47 & -5.14 & 286.12 &  & 12.36 & 2.55 & -0.08 & -2.16 & 14628.76 &  & 10.69 & 3.01 & 2.21 & 0.45 & 383.96 \\
 & 0.2 & 12.35 & -2.24 & -6.74 & -5.64 & 83.05 &  & 11.67 & 0.66 & -2.82 & -4.40 & 313.86 &  & 11.43 & 2.56 & 0.18 & -1.89 & 6182.65 &  & 10.10 & 3.01 & 2.21 & 0.43 & 357.94 \\
 & 0.3 & 9.54 & -2.23 & -5.51 & -4.15 & 73.04 &  & 9.49 & 0.66 & -2.00 & -3.47 & 373.77 &  & 9.73 & 2.55 & 0.75 & -1.30 & 1199.28 &  & 8.87 & 3.01 & 2.38 & 0.58 & 273.09 \\
 & 0.4 & 6.33 & -2.24 & -4.22 & -2.59 & 49.85 &  & 6.79 & 0.66 & -1.05 & -2.36 & 547.99 &  & 7.53 & 2.56 & 1.40 & -0.63 & 438.16 &  & 7.15 & 3.01 & 2.71 & 0.91 & 163.48 \\ \cline{2-25} 
 \multicolumn{1}{c}{0.5} & 0.5 & 3.47 & -2.23 & -3.44 & -1.67 & 0.83 &  & 4.41 & 0.66 & -0.16 & -1.31 & 2683.50 &  & 5.29 & 2.55 & 1.87 & -0.14 & 182.76 &  & 5.36 & 3.01 & 2.93 & 1.08 & 83.22 \\ \cline{2-25} 
 & 0.6 & 1.08 & -2.23 & -3.01 & -1.17 & -64.13 &  & 2.75 & 0.65 & 0.36 & -0.69 & 669.70 &  & 3.89 & 2.55 & 2.25 & 0.26 & 73.17 &  & 4.07 & 3.01 & 2.97 & 1.10 & 36.98 \\ 
 & 0.7 & -0.41 & -2.22 & -2.76 & -0.90 & -85.13 &  & 1.76 & 0.66 & 0.56 & -0.43 & 216.75 &  & 3.12 & 2.55 & 2.44 & 0.48 & 27.78 &  & 3.47 & 3.01 & 3.04 & 1.17 & 14.09 \\
 & 0.8 & -1.37 & -2.24 & -2.69 & -0.85 & -49.04 &  & 1.27 & 0.66 & 0.66 & -0.28 & 92.05 &  & 2.83 & 2.55 & 2.53 & 0.58 & 11.85 &  & 3.24 & 3.01 & 3.04 & 1.17 & 6.45 \\
 & 0.9 & -1.81 & -2.23 & -2.54 & -0.68 & -28.66 &  & 1.02 & 0.66 & 0.70 & -0.23 & 45.54 &  & 2.75 & 2.55 & 2.56 & 0.61 & 7.21 &  & 3.13 & 3.01 & 3.02 & 1.14 & 3.51 \\
 & 1.0 & -1.87 & -2.24 & -2.38 & -0.50 & -21.41 &  & 1.05 & 0.66 & 0.68 & -0.26 & 54.67 &  & 2.73 & 2.56 & 2.57 & 0.61 & 6.19 &  & 3.12 & 3.01 & 3.02 & 1.13 & 3.38 \\
 & 1.1 & -1.84 & -2.23 & -2.31 & -0.42 & -20.49 &  & 1.06 & 0.66 & 0.71 & -0.22 & 49.18 &  & 2.73 & 2.56 & 2.55 & 0.60 & 6.99 &  & 3.12 & 3.01 & 2.99 & 1.10 & 4.26 \\
 & 1.2 & -1.76 & -2.23 & -2.26 & -0.36 & -22.26 &  & 1.11 & 0.67 & 0.73 & -0.21 & 53.33 &  & 2.78 & 2.55 & 2.54 & 0.58 & 9.37 &  & 3.11 & 3.01 & 3.00 & 1.11 & 3.72    \\ \hline
\end{tabular}
}
\caption{(100 $\times$) Monte Carlo bias of the estimated AR coefficient for ARFIMA(1,$d_0$,0) of CSS estimator with unknown and known $\mu_0$ and the MCSS estimator, together with the bias-corrected MCSS estimator.}
\label{tablear2}
\end{table}

\begin{table}[H]
\centering
\resizebox{\textwidth}{!}{%
\begin{tabular}{EEEEEEEEEEEEEEEEEEEEEEEEE}
\hline
\multicolumn{1}{c}{$\varphi_0$} &
 &
 \multicolumn{1}{c}{MSE($\hat{d}$)} &
  \multicolumn{1}{c}{MSE($\hat{d}_{\mu_0}$) } &
  \multicolumn{1}{c}{MSE($\hat{d}_{m}$)} &
  \multicolumn{1}{c}{MSE($\hat{d}_{bcm}$)} &
  \multicolumn{1}{c}{$\Delta \% |\text{MSE}|$} &
 &
 \multicolumn{1}{c}{MSE($\hat{d}$)} &
  \multicolumn{1}{c}{MSE($\hat{d}_{\mu_0}$) } &
  \multicolumn{1}{c}{MSE($\hat{d}_{m}$)} &
  \multicolumn{1}{c}{MSE($\hat{d}_{bcm}$)} &
  \multicolumn{1}{c}{$\Delta \% |\text{MSE}|$} &
   &
 \multicolumn{1}{c}{MSE($\hat{d}$)} &
  \multicolumn{1}{c}{MSE($\hat{d}_{\mu_0}$) } &
  \multicolumn{1}{c}{MSE($\hat{d}_{m}$)} &
  \multicolumn{1}{c}{MSE($\hat{d}_{bcm}$)} &
  \multicolumn{1}{c}{$\Delta \% |\text{MSE}|$} &
   &
  \multicolumn{1}{c}{MSE($\hat{d}$)} &
  \multicolumn{1}{c}{MSE($\hat{d}_{\mu_0}$) } &
  \multicolumn{1}{c}{MSE($\hat{d}_{m}$)} &
  \multicolumn{1}{c}{MSE($\hat{d}_{bcm}$)} &
  \multicolumn{1}{c}{$\Delta \% |\text{MSE}|$}  \\ \cline{3-7} \cline{9-13} \cline{15-19} \cline{21-25} 
 & \multicolumn{1}{c}{$d_0$ \textbackslash{} $T$} & \multicolumn{5}{c}{32} &  & \multicolumn{5}{c}{64} &  & \multicolumn{5}{c}{128} &  & \multicolumn{5}{c}{256} \\ \hline
 & -0.2 & 38.58 & 7.89 & 14.57 & 13.30 & 164.83 &  & 7.04 & 2.21 & 3.37 & 3.13 & 108.65 &  & 1.50 & 0.86 & 1.10 & 1.04 & 36.42 &  & 0.54 & 0.39 & 0.45 & 0.44 & 18.83 \\
 & -0.1 & 40.16 & 7.89 & 14.52 & 13.22 & 176.55 &  & 7.17 & 2.21 & 3.42 & 3.16 & 109.90 &  & 1.52 & 0.86 & 1.10 & 1.04 & 38.22 &  & 0.54 & 0.39 & 0.45 & 0.44 & 19.14 \\
 & 0.0 & 41.46 & 7.89 & 14.69 & 13.38 & 182.33 &  & 7.12 & 2.21 & 3.45 & 3.19 & 106.42 &  & 1.54 & 0.86 & 1.10 & 1.04 & 39.89 &  & 0.54 & 0.39 & 0.45 & 0.44 & 19.45 \\
 & 0.1 & 42.41 & 7.89 & 14.54 & 13.19 & 191.68 &  & 7.36 & 2.21 & 3.57 & 3.32 & 106.19 &  & 1.56 & 0.86 & 1.10 & 1.04 & 42.15 &  & 0.54 & 0.39 & 0.45 & 0.44 & 19.73 \\
 & 0.2 & 43.39 & 7.89 & 14.55 & 13.20 & 198.14 &  & 7.60 & 2.21 & 3.50 & 3.24 & 117.52 &  & 1.55 & 0.86 & 1.09 & 1.03 & 42.69 &  & 0.54 & 0.39 & 0.45 & 0.44 & 19.93 \\
 & 0.3 & 43.27 & 7.89 & 14.08 & 12.74 & 207.44 &  & 7.68 & 2.21 & 3.45 & 3.20 & 122.40 &  & 1.51 & 0.86 & 1.07 & 1.01 & 40.28 &  & 0.53 & 0.39 & 0.45 & 0.43 & 19.95 \\
 & 0.4 & 42.77 & 7.89 & 13.52 & 12.18 & 216.46 &  & 7.60 & 2.21 & 3.29 & 3.03 & 130.76 &  & 1.49 & 0.86 & 1.05 & 0.99 & 42.51 &  & 0.52 & 0.39 & 0.44 & 0.43 & 19.42 \\ \cline{2-25} 
\multicolumn{1}{c}{$-0.5$}  & 0.5 & 42.53 & 7.89 & 12.89 & 11.56 & 230.00 &  & 7.24 & 2.21 & 3.06 & 2.80 & 136.20 &  & 1.43 & 0.86 & 1.01 & 0.95 & 40.59 &  & 0.50 & 0.39 & 0.43 & 0.42 & 17.67 \\ \cline{2-25} 
 & 0.6 & 40.63 & 7.89 & 12.70 & 11.41 & 219.98 &  & 6.45 & 2.21 & 2.90 & 2.63 & 122.43 &  & 1.31 & 0.86 & 0.97 & 0.92 & 34.27 &  & 0.48 & 0.39 & 0.42 & 0.40 & 14.16 \\
 & 0.7 & 36.63 & 7.89 & 11.93 & 10.70 & 207.16 &  & 5.64 & 2.21 & 2.76 & 2.51 & 104.61 &  & 1.16 & 0.86 & 0.94 & 0.88 & 23.98 &  & 0.45 & 0.39 & 0.41 & 0.39 & 9.69 \\
 & 0.8 & 31.57 & 7.89 & 10.76 & 9.64 & 193.36 &  & 4.83 & 2.21 & 2.58 & 2.35 & 86.96 &  & 1.03 & 0.86 & 0.90 & 0.85 & 14.40 &  & 0.42 & 0.39 & 0.40 & 0.39 & 5.81 \\
 & 0.9 & 26.42 & 7.89 & 9.91 & 8.87 & 166.44 &  & 3.98 & 2.21 & 2.41 & 2.17 & 65.47 &  & 0.98 & 0.86 & 0.88 & 0.83 & 10.69 &  & 0.41 & 0.39 & 0.39 & 0.38 & 3.22 \\
 & 1.0 & 21.39 & 7.90 & 8.97 & 7.99 & 138.38 &  & 3.44 & 2.21 & 2.32 & 2.10 & 47.86 &  & 0.91 & 0.86 & 0.87 & 0.82 & 4.69 &  & 0.40 & 0.39 & 0.39 & 0.38 & 1.71 \\
 & 1.1 & 17.14 & 7.89 & 8.81 & 7.90 & 94.52 &  & 2.88 & 2.21 & 2.27 & 2.05 & 26.62 &  & 0.89 & 0.86 & 0.87 & 0.81 & 2.38 &  & 0.39 & 0.39 & 0.39 & 0.38 & 0.86 \\
 & 1.2 & 14.10 & 7.89 & 8.63 & 7.77 & 63.37 &  & 2.56 & 2.21 & 2.27 & 2.06 & 12.70 &  & 0.87 & 0.86 & 0.86 & 0.81 & 1.08 &  & 0.39 & 0.39 & 0.39 & 0.38 & 0.36 \\ \hline
 &  &  &  &  &  &  &  &  &  &  &  &  &  &  &  &  &  &  &  &  &  &  &  &  \\ \hline
 & -0.2 & 54.39 & 16.06 & 21.94 & 20.42 & 147.94 &  & 29.91 & 8.56 & 11.69 & 10.97 & 155.87 &  & 11.74 & 3.23 & 4.87 & 4.64 & 141.09 &  & 2.79 & 1.05 & 1.41 & 1.33 & 97.75 \\
 & -0.1 & 56.32 & 16.06 & 22.01 & 20.47 & 155.85 &  & 30.85 & 8.54 & 11.71 & 10.96 & 163.49 &  & 12.04 & 3.20 & 4.87 & 4.64 & 147.28 &  & 2.84 & 1.06 & 1.43 & 1.35 & 98.77 \\
 & 0.0 & 58.19 & 16.06 & 21.87 & 20.29 & 166.08 &  & 31.56 & 8.55 & 11.75 & 10.98 & 168.62 &  & 12.35 & 3.21 & 4.89 & 4.66 & 152.54 &  & 2.88 & 1.05 & 1.43 & 1.35 & 101.76 \\
 & 0.1 & 59.83 & 16.07 & 21.56 & 19.93 & 177.55 &  & 31.86 & 8.56 & 11.57 & 10.79 & 175.36 &  & 12.62 & 3.22 & 4.89 & 4.67 & 157.84 &  & 2.93 & 1.06 & 1.41 & 1.33 & 107.90 \\
 & 0.2 & 60.51 & 16.07 & 20.92 & 19.27 & 189.20 &  & 31.96 & 8.56 & 11.44 & 10.64 & 179.41 &  & 12.68 & 3.21 & 4.89 & 4.66 & 159.47 &  & 2.88 & 1.06 & 1.40 & 1.32 & 105.33 \\
 & 0.3 & 60.54 & 16.06 & 20.45 & 18.77 & 196.03 &  & 31.58 & 8.55 & 11.22 & 10.42 & 181.60 &  & 12.37 & 3.21 & 4.73 & 4.50 & 161.39 &  & 2.77 & 1.05 & 1.37 & 1.29 & 101.69 \\
 & 0.4 & 59.17 & 16.06 & 19.82 & 18.14 & 198.56 &  & 30.57 & 8.56 & 10.87 & 10.06 & 181.31 &  & 11.80 & 3.22 & 4.75 & 4.53 & 148.55 &  & 2.68 & 1.05 & 1.34 & 1.26 & 99.97 \\ \cline{2-25} 
\multicolumn{1}{c}{0}  & 0.5 & 55.90 & 16.06 & 19.39 & 17.74 & 188.28 &  & 28.37 & 8.55 & 10.40 & 9.61 & 172.73 &  & 10.61 & 3.20 & 4.41 & 4.19 & 140.38 &  & 2.48 & 1.05 & 1.24 & 1.16 & 99.57 \\ \cline{2-25} 
 & 0.6 & 50.68 & 16.07 & 18.75 & 17.18 & 170.28 &  & 24.71 & 8.56 & 9.95 & 9.25 & 148.45 &  & 9.09 & 3.21 & 4.13 & 3.92 & 120.19 &  & 2.09 & 1.05 & 1.16 & 1.08 & 79.60 \\
 & 0.7 & 44.09 & 16.07 & 18.04 & 16.61 & 144.38 &  & 20.53 & 8.56 & 9.50 & 8.91 & 116.02 &  & 7.08 & 3.21 & 3.75 & 3.55 & 88.63 &  & 1.67 & 1.05 & 1.12 & 1.05 & 49.09 \\
 & 0.8 & 36.42 & 16.06 & 17.10 & 15.84 & 112.94 &  & 16.35 & 8.56 & 9.15 & 8.66 & 78.80 &  & 5.37 & 3.23 & 3.52 & 3.32 & 52.39 &  & 1.38 & 1.05 & 1.09 & 1.01 & 26.65 \\
 & 0.9 & 29.58 & 16.07 & 16.80 & 15.72 & 76.14 &  & 13.01 & 8.56 & 9.00 & 8.60 & 44.56 &  & 4.25 & 3.21 & 3.35 & 3.15 & 26.85 &  & 1.24 & 1.05 & 1.08 & 1.01 & 14.15 \\
 & 1.0 & 24.57 & 16.07 & 16.51 & 15.58 & 48.81 &  & 10.90 & 8.56 & 8.79 & 8.43 & 24.01 &  & 3.72 & 3.23 & 3.30 & 3.09 & 12.71 &  & 1.12 & 1.05 & 1.06 & 0.98 & 6.13 \\
 & 1.1 & 20.98 & 16.08 & 16.49 & 15.63 & 27.23 &  & 9.69 & 8.56 & 8.66 & 8.33 & 11.84 &  & 3.45 & 3.23 & 3.28 & 3.07 & 5.28 &  & 1.07 & 1.05 & 1.03 & 0.95 & 3.95 \\
 & 1.2 & 18.70 & 16.07 & 16.36 & 15.54 & 14.28 &  & 9.05 & 8.56 & 8.59 & 8.26 & 5.34 &  & 3.37 & 3.23 & 3.30 & 3.10 & 1.95 &  & 1.06 & 1.06 & 1.05 & 0.97 & 1.20 \\ \hline
 &  &  &  &  &  &  &  &  &  &  &  &  &  &  &  &  &  &  &  &  &  &  &  &  \\ \hline
 & -0.2 & 18.68 & 9.55 & 12.60 & 14.73 & 48.33 &  & 11.81 & 6.57 & 7.58 & 9.06 & 55.91 &  & 8.17 & 4.57 & 5.05 & 5.76 & 61.73 &  & 5.48 & 2.95 & 3.43 & 3.68 & 60.01 \\
 & -0.1 & 19.25 & 9.55 & 12.37 & 14.48 & 55.59 &  & 12.03 & 6.57 & 7.55 & 9.03 & 59.35 &  & 8.28 & 4.57 & 5.02 & 5.72 & 65.00 &  & 5.53 & 2.95 & 3.42 & 3.67 & 61.79 \\
 & 0.0 & 19.73 & 9.55 & 11.91 & 13.95 & 65.71 &  & 12.20 & 6.57 & 7.45 & 8.92 & 63.68 &  & 8.34 & 4.57 & 5.00 & 5.70 & 66.82 &  & 5.53 & 2.95 & 3.43 & 3.68 & 61.53 \\
 & 0.1 & 20.11 & 9.55 & 11.24 & 13.17 & 79.03 &  & 12.28 & 6.57 & 7.23 & 8.67 & 69.89 &  & 8.39 & 4.57 & 4.94 & 5.64 & 69.67 &  & 5.56 & 2.95 & 3.41 & 3.66 & 62.94 \\
 & 0.2 & 19.91 & 9.55 & 10.54 & 12.33 & 88.96 &  & 12.09 & 6.57 & 6.91 & 8.30 & 75.04 &  & 8.29 & 4.57 & 4.84 & 5.54 & 71.17 &  & 5.50 & 2.95 & 3.35 & 3.60 & 64.22 \\
 & 0.3 & 19.18 & 9.55 & 9.92 & 11.53 & 93.28 &  & 11.56 & 6.57 & 6.59 & 7.93 & 75.38 &  & 7.89 & 4.57 & 4.73 & 5.41 & 66.86 &  & 5.22 & 2.95 & 3.27 & 3.51 & 59.64 \\
 & 0.4 & 17.74 & 9.55 & 9.55 & 10.98 & 85.77 &  & 10.49 & 6.57 & 6.41 & 7.68 & 63.58 &  & 7.12 & 4.57 & 4.61 & 5.27 & 54.63 &  & 4.71 & 2.95 & 3.20 & 3.43 & 47.53 \\ \cline{2-25} 
\multicolumn{1}{c}{0.5}  & 0.5 & 15.91 & 9.55 & 9.39 & 10.74 & 69.42 &  & 9.18 & 6.57 & 6.39 & 7.64 & 43.54 &  & 6.10 & 4.57 & 4.53 & 5.19 & 34.67 &  & 4.03 & 2.95 & 3.10 & 3.33 & 30.02 \\ \cline{2-25} 
 & 0.6 & 13.86 & 9.55 & 9.32 & 10.66 & 48.72 &  & 8.04 & 6.57 & 6.48 & 7.74 & 24.16 &  & 5.35 & 4.57 & 4.53 & 5.19 & 17.89 &  & 3.46 & 2.95 & 3.02 & 3.23 & 14.80 \\
 & 0.7 & 12.04 & 9.55 & 9.34 & 10.69 & 28.99 &  & 7.29 & 6.57 & 6.52 & 7.81 & 11.74 &  & 4.88 & 4.57 & 4.57 & 5.23 & 6.87 &  & 3.16 & 2.95 & 3.00 & 3.21 & 5.31 \\
 & 0.8 & 10.89 & 9.55 & 9.42 & 10.79 & 15.57 &  & 6.90 & 6.57 & 6.61 & 7.93 & 4.28 &  & 4.67 & 4.57 & 4.58 & 5.24 & 1.81 &  & 3.03 & 2.95 & 2.98 & 3.20 & 1.60 \\
 & 0.9 & 10.10 & 9.55 & 9.49 & 10.87 & 6.44 &  & 6.67 & 6.57 & 6.61 & 7.93 & 0.87 &  & 4.59 & 4.57 & 4.60 & 5.26 & -0.19 &  & 2.96 & 2.95 & 2.97 & 3.18 & -0.28 \\
 & 1.0 & 9.74 & 9.55 & 9.62 & 11.00 & 1.25 &  & 6.56 & 6.57 & 6.62 & 7.94 & -0.83 &  & 4.54 & 4.57 & 4.59 & 5.25 & -0.99 &  & 2.94 & 2.95 & 2.96 & 3.17 & -0.72 \\
 & 1.1 & 9.56 & 9.55 & 9.61 & 10.99 & -0.51 &  & 6.50 & 6.57 & 6.62 & 7.94 & -1.77 &  & 4.52 & 4.57 & 4.59 & 5.25 & -1.54 &  & 2.93 & 2.95 & 2.95 & 3.16 & -0.66 \\
 & 1.2 & 9.41 & 9.55 & 9.64 & 11.03 & -2.35 &  & 6.46 & 6.57 & 6.60 & 7.92 & -1.99 &  & 4.53 & 4.57 & 4.59 & 5.25 & -1.33 &  & 2.92 & 2.95 & 2.96 & 3.16 & -1.10 \\ \hline
\end{tabular}
}
\caption{(100 $\times$) Empirical MSE of the estimated memory parameter for ARFIMA(1,$d_0$,0) of CSS estimator with unknown and known $\mu_0$ and the MCSS estimator, together with the bias-corrected MCSS estimator.}
\label{tablear3}
\end{table}

\begin{table}[H]
\centering
\resizebox{\textwidth}{!}{%
\begin{tabular}{EEEEEEEEEEEEEEEEEEEEEEEEE}
\hline
\multicolumn{1}{c}{$\varphi_0$} &
&
  \multicolumn{1}{c}{MSE($\hat{\varphi}$)} &
  \multicolumn{1}{c}{MSE($\hat{\varphi}_{\mu_0}$) } &
  \multicolumn{1}{c}{MSE($\hat{\varphi}_{m}$)} &
  \multicolumn{1}{c}{MSE($\hat{\varphi}_{bcm}$)} &
  \multicolumn{1}{c}{$\Delta \% |\text{MSE}|$} &
 &
  \multicolumn{1}{c}{MSE($\hat{\varphi}$)} &
  \multicolumn{1}{c}{MSE($\hat{\varphi}_{\mu_0}$) } &
  \multicolumn{1}{c}{MSE($\hat{\varphi}_{m}$)} &
  \multicolumn{1}{c}{MSE($\hat{\varphi}_{bcm}$)} &
  \multicolumn{1}{c}{$\Delta \% |\text{MSE}|$} &
   &
 \multicolumn{1}{c}{MSE($\hat{\varphi}$)} &
  \multicolumn{1}{c}{MSE($\hat{\varphi}_{\mu_0}$) } &
  \multicolumn{1}{c}{MSE($\hat{\varphi}_{m}$)} &
  \multicolumn{1}{c}{MSE($\hat{\varphi}_{bcm}$)} &
  \multicolumn{1}{c}{$\Delta \% |\text{MSE}|$} &
   &
   \multicolumn{1}{c}{MSE($\hat{\varphi}$)} &
  \multicolumn{1}{c}{MSE($\hat{\varphi}_{\mu_0}$) } &
  \multicolumn{1}{c}{MSE($\hat{\varphi}_{m}$)} &
  \multicolumn{1}{c}{MSE($\hat{\varphi}_{bcm}$)} &
  \multicolumn{1}{c}{$\Delta \% |\text{MSE}|$} \\ \cline{3-7} \cline{9-13} \cline{15-19} \cline{21-25} 
 & \multicolumn{1}{c}{$d_0$ \textbackslash{} $T$} & \multicolumn{5}{c}{32} &  & \multicolumn{5}{c}{64} &  & \multicolumn{5}{c}{128} &  & \multicolumn{5}{c}{256} \\ \hline
 & -0.2 & 28.23 & 7.91 & 10.40 & 10.60 & 171.49 &  & 6.00 & 2.46 & 2.92 & 2.80 & 105.29 &  & 1.32 & 0.98 & 1.05 & 1.00 & 26.59 &  & 0.52 & 0.45 & 0.46 & 0.45 & 11.95 \\
 & -0.1 & 29.35 & 7.91 & 10.35 & 10.53 & 183.70 &  & 6.10 & 2.46 & 2.95 & 2.83 & 106.59 &  & 1.35 & 0.98 & 1.05 & 1.00 & 28.51 &  & 0.52 & 0.45 & 0.46 & 0.45 & 12.10 \\
 & 0.0 & 30.10 & 7.91 & 10.43 & 10.64 & 188.64 &  & 6.00 & 2.46 & 2.98 & 2.85 & 101.49 &  & 1.35 & 0.98 & 1.05 & 1.00 & 29.41 &  & 0.52 & 0.45 & 0.46 & 0.45 & 12.21 \\
 & 0.1 & 30.52 & 7.91 & 10.23 & 10.38 & 198.33 &  & 6.16 & 2.46 & 3.09 & 2.99 & 99.60 &  & 1.37 & 0.98 & 1.05 & 1.00 & 30.75 &  & 0.52 & 0.45 & 0.46 & 0.45 & 12.22 \\
 & 0.2 & 31.28 & 7.91 & 10.32 & 10.50 & 202.98 &  & 6.41 & 2.46 & 3.04 & 2.93 & 110.82 &  & 1.36 & 0.98 & 1.04 & 0.99 & 30.67 &  & 0.52 & 0.45 & 0.46 & 0.45 & 12.07 \\
 & 0.3 & 31.19 & 7.91 & 10.07 & 10.21 & 209.76 &  & 6.46 & 2.46 & 3.03 & 2.91 & 113.64 &  & 1.32 & 0.98 & 1.04 & 0.99 & 26.87 &  & 0.51 & 0.45 & 0.46 & 0.45 & 11.61 \\
 & 0.4 & 30.82 & 7.91 & 9.81 & 9.92 & 214.04 &  & 6.42 & 2.46 & 2.93 & 2.80 & 119.38 &  & 1.32 & 0.98 & 1.03 & 0.98 & 28.24 &  & 0.51 & 0.45 & 0.46 & 0.45 & 10.62 \\ \cline{2-25} 
\multicolumn{1}{c}{$-0.5$}  & 0.5 & 30.61 & 7.91 & 9.48 & 9.55 & 222.79 &  & 6.19 & 2.46 & 2.80 & 2.65 & 120.77 &  & 1.28 & 0.98 & 1.02 & 0.97 & 25.36 &  & 0.50 & 0.45 & 0.46 & 0.44 & 8.84 \\ \cline{2-25} 
 & 0.6 & 29.10 & 7.91 & 9.50 & 9.59 & 206.32 &  & 5.53 & 2.46 & 2.73 & 2.58 & 102.23 &  & 1.21 & 0.98 & 1.01 & 0.96 & 19.70 &  & 0.48 & 0.45 & 0.45 & 0.44 & 6.34 \\
 & 0.7 & 26.22 & 7.90 & 9.26 & 9.34 & 183.24 &  & 4.93 & 2.46 & 2.70 & 2.55 & 82.52 &  & 1.12 & 0.98 & 1.00 & 0.95 & 11.73 &  & 0.47 & 0.45 & 0.45 & 0.44 & 3.75 \\
 & 0.8 & 22.88 & 7.91 & 8.73 & 8.77 & 162.01 &  & 4.35 & 2.46 & 2.62 & 2.47 & 66.02 &  & 1.05 & 0.98 & 0.99 & 0.94 & 5.74 &  & 0.46 & 0.45 & 0.45 & 0.44 & 1.78 \\
 & 0.9 & 19.22 & 7.91 & 8.37 & 8.38 & 129.68 &  & 3.72 & 2.46 & 2.52 & 2.36 & 47.62 &  & 1.03 & 0.98 & 0.99 & 0.94 & 4.31 &  & 0.45 & 0.45 & 0.45 & 0.43 & 0.60 \\
 & 1.0 & 15.81 & 7.90 & 7.98 & 7.92 & 98.09 &  & 3.34 & 2.46 & 2.49 & 2.33 & 34.10 &  & 0.99 & 0.98 & 0.98 & 0.94 & 0.84 &  & 0.45 & 0.45 & 0.45 & 0.43 & -0.01 \\
 & 1.1 & 13.17 & 7.91 & 8.18 & 8.18 & 61.05 &  & 2.89 & 2.46 & 2.46 & 2.30 & 17.31 &  & 0.98 & 0.98 & 0.98 & 0.93 & -0.07 &  & 0.44 & 0.45 & 0.45 & 0.43 & -0.28 \\
 & 1.2 & 11.24 & 7.91 & 8.18 & 8.20 & 37.43 &  & 2.66 & 2.46 & 2.48 & 2.33 & 7.04 &  & 0.98 & 0.98 & 0.98 & 0.93 & -0.44 &  & 0.44 & 0.45 & 0.45 & 0.43 & -0.38 \\ \hline
 &  &  &  &  &  &  &  &  &  &  &  &  &  &  &  &  &  &  &  &  &  &  &  &  \\ \hline
 & -0.2 & 34.38 & 15.17 & 15.24 & 17.91 & 125.60 &  & 24.50 & 9.12 & 10.47 & 11.20 & 134.08 &  & 11.24 & 3.91 & 5.10 & 5.14 & 120.16 &  & 3.09 & 1.44 & 1.74 & 1.68 & 77.98 \\
 & -0.1 & 35.39 & 15.18 & 15.22 & 17.88 & 132.58 &  & 25.13 & 9.11 & 10.48 & 11.21 & 139.84 &  & 11.47 & 3.89 & 5.10 & 5.14 & 124.88 &  & 3.14 & 1.45 & 1.76 & 1.69 & 78.67 \\
 & 0.0 & 36.33 & 15.18 & 15.10 & 17.74 & 140.55 &  & 25.59 & 9.11 & 10.48 & 11.20 & 144.28 &  & 11.74 & 3.90 & 5.12 & 5.16 & 129.15 &  & 3.18 & 1.44 & 1.75 & 1.69 & 81.17 \\
 & 0.1 & 37.13 & 15.19 & 14.97 & 17.56 & 148.08 &  & 25.73 & 9.12 & 10.31 & 11.01 & 149.47 &  & 11.96 & 3.91 & 5.13 & 5.18 & 133.33 &  & 3.22 & 1.45 & 1.73 & 1.67 & 85.65 \\
 & 0.2 & 37.41 & 15.18 & 14.71 & 17.25 & 154.32 &  & 25.73 & 9.12 & 10.23 & 10.90 & 151.63 &  & 12.00 & 3.90 & 5.11 & 5.16 & 134.85 &  & 3.16 & 1.45 & 1.72 & 1.66 & 83.48 \\
 & 0.3 & 37.24 & 15.18 & 14.67 & 17.19 & 153.90 &  & 25.38 & 9.11 & 10.12 & 10.78 & 150.92 &  & 11.71 & 3.90 & 4.98 & 5.01 & 135.36 &  & 3.06 & 1.44 & 1.70 & 1.63 & 80.42 \\
 & 0.4 & 36.08 & 15.18 & 14.55 & 17.05 & 147.91 &  & 24.52 & 9.12 & 9.91 & 10.53 & 147.40 &  & 11.19 & 3.91 & 5.02 & 5.07 & 122.88 &  & 2.96 & 1.44 & 1.67 & 1.60 & 77.75 \\ \cline{2-25} 
\multicolumn{1}{c}{0}  & 0.5 & 33.88 & 15.18 & 14.66 & 17.18 & 131.05 &  & 22.72 & 9.12 & 9.66 & 10.23 & 135.32 &  & 10.12 & 3.89 & 4.77 & 4.79 & 112.32 &  & 2.77 & 1.44 & 1.59 & 1.52 & 74.58 \\ \cline{2-25} 
 & 0.6 & 30.81 & 15.19 & 14.77 & 17.33 & 108.56 &  & 19.88 & 9.12 & 9.45 & 10.03 & 110.36 &  & 8.79 & 3.90 & 4.55 & 4.57 & 93.16 &  & 2.39 & 1.44 & 1.52 & 1.45 & 57.20 \\
 & 0.7 & 27.44 & 15.18 & 14.92 & 17.53 & 83.82 &  & 16.80 & 9.12 & 9.30 & 9.89 & 80.52 &  & 7.03 & 3.90 & 4.27 & 4.25 & 64.81 &  & 2.00 & 1.45 & 1.49 & 1.43 & 33.84 \\
 & 0.8 & 23.61 & 15.18 & 14.83 & 17.40 & 59.27 &  & 13.92 & 9.12 & 9.24 & 9.85 & 50.72 &  & 5.60 & 3.91 & 4.12 & 4.09 & 36.02 &  & 1.73 & 1.44 & 1.47 & 1.40 & 17.87 \\
 & 0.9 & 20.62 & 15.19 & 15.03 & 17.65 & 37.16 &  & 11.79 & 9.12 & 9.27 & 9.91 & 27.08 &  & 4.70 & 3.90 & 4.01 & 3.96 & 17.39 &  & 1.60 & 1.44 & 1.47 & 1.40 & 9.12 \\
 & 1.0 & 18.51 & 15.19 & 15.15 & 17.80 & 22.15 &  & 10.50 & 9.12 & 9.21 & 9.82 & 14.00 &  & 4.31 & 3.92 & 3.98 & 3.91 & 8.28 &  & 1.50 & 1.44 & 1.45 & 1.38 & 3.90 \\
 & 1.1 & 17.14 & 15.19 & 15.28 & 17.95 & 12.13 &  & 9.84 & 9.12 & 9.18 & 9.78 & 7.19 &  & 4.11 & 3.92 & 3.97 & 3.90 & 3.64 &  & 1.47 & 1.44 & 1.43 & 1.36 & 2.87 \\
 & 1.2 & 16.32 & 15.19 & 15.27 & 17.93 & 6.87 &  & 9.52 & 9.13 & 9.15 & 9.74 & 4.01 &  & 4.06 & 3.92 & 3.99 & 3.92 & 1.89 &  & 1.46 & 1.45 & 1.44 & 1.37 & 1.21 \\ \hline
 &  &  &  &  &  &  &  &  &  &  &  &  &  &  &  &  &  &  &  &  &  &  &  &  \\ \hline
 & -0.2 & 6.69 & 9.11 & 10.56 & 13.67 & -36.62 &  & 6.41 & 6.22 & 6.86 & 8.98 & -6.59 &  & 5.52 & 4.30 & 4.68 & 5.80 & 18.11 &  & 4.20 & 2.83 & 3.14 & 3.62 & 33.75 \\
 & -0.1 & 6.87 & 9.11 & 10.43 & 13.54 & -34.14 &  & 6.49 & 6.22 & 6.86 & 8.97 & -5.41 &  & 5.59 & 4.30 & 4.66 & 5.78 & 19.93 &  & 4.23 & 2.83 & 3.13 & 3.61 & 34.89 \\
 & 0.0 & 7.07 & 9.11 & 10.27 & 13.37 & -31.17 &  & 6.59 & 6.22 & 6.80 & 8.91 & -3.15 &  & 5.62 & 4.30 & 4.64 & 5.76 & 21.25 &  & 4.24 & 2.83 & 3.14 & 3.62 & 34.99 \\
 & 0.1 & 7.29 & 9.11 & 10.06 & 13.17 & -27.49 &  & 6.70 & 6.22 & 6.67 & 8.77 & 0.46 &  & 5.67 & 4.30 & 4.59 & 5.71 & 23.50 &  & 4.27 & 2.83 & 3.13 & 3.61 & 36.23 \\
 & 0.2 & 7.64 & 9.11 & 9.80 & 12.95 & -22.05 &  & 6.78 & 6.22 & 6.48 & 8.59 & 4.62 &  & 5.67 & 4.30 & 4.50 & 5.61 & 25.98 &  & 4.24 & 2.83 & 3.08 & 3.55 & 37.67 \\
 & 0.3 & 8.11 & 9.11 & 9.51 & 12.72 & -14.76 &  & 6.79 & 6.22 & 6.32 & 8.45 & 7.43 &  & 5.51 & 4.30 & 4.40 & 5.52 & 25.17 &  & 4.08 & 2.83 & 3.02 & 3.49 & 34.97 \\
 & 0.4 & 8.55 & 9.11 & 9.27 & 12.53 & -7.81 &  & 6.72 & 6.22 & 6.23 & 8.38 & 7.92 &  & 5.22 & 4.30 & 4.32 & 5.43 & 20.74 &  & 3.78 & 2.83 & 2.96 & 3.42 & 27.46 \\ \cline{2-25} 
\multicolumn{1}{c}{0.5}  & 0.5 & 8.89 & 9.10 & 9.19 & 12.51 & -3.27 &  & 6.57 & 6.22 & 6.18 & 8.38 & 6.16 &  & 4.83 & 4.30 & 4.28 & 5.39 & 12.87 &  & 3.40 & 2.83 & 2.91 & 3.35 & 16.87 \\ \cline{2-25} 
 & 0.6 & 9.09 & 9.10 & 9.20 & 12.57 & -1.10 &  & 6.42 & 6.22 & 6.22 & 8.45 & 3.20 &  & 4.56 & 4.30 & 4.29 & 5.41 & 6.20 &  & 3.09 & 2.83 & 2.86 & 3.29 & 8.04 \\
 & 0.7 & 9.15 & 9.10 & 9.21 & 12.63 & -0.65 &  & 6.32 & 6.22 & 6.24 & 8.50 & 1.27 &  & 4.40 & 4.30 & 4.31 & 5.43 & 2.08 &  & 2.94 & 2.83 & 2.86 & 3.30 & 2.63 \\
 & 0.8 & 9.20 & 9.11 & 9.23 & 12.67 & -0.33 &  & 6.26 & 6.22 & 6.29 & 8.57 & -0.34 &  & 4.32 & 4.30 & 4.32 & 5.44 & 0.14 &  & 2.87 & 2.83 & 2.85 & 3.28 & 0.70 \\
 & 0.9 & 9.17 & 9.10 & 9.24 & 12.70 & -0.80 &  & 6.22 & 6.22 & 6.27 & 8.55 & -0.76 &  & 4.30 & 4.30 & 4.33 & 5.45 & -0.67 &  & 2.83 & 2.83 & 2.84 & 3.27 & -0.38 \\
 & 1.0 & 9.10 & 9.11 & 9.25 & 12.72 & -1.65 &  & 6.17 & 6.22 & 6.27 & 8.56 & -1.57 &  & 4.27 & 4.30 & 4.32 & 5.44 & -1.04 &  & 2.82 & 2.83 & 2.84 & 3.27 & -0.62 \\
 & 1.1 & 9.03 & 9.10 & 9.23 & 12.70 & -2.12 &  & 6.14 & 6.22 & 6.27 & 8.55 & -1.98 &  & 4.25 & 4.30 & 4.32 & 5.44 & -1.45 &  & 2.82 & 2.83 & 2.83 & 3.26 & -0.58 \\
 & 1.2 & 8.96 & 9.10 & 9.22 & 12.69 & -2.78 &  & 6.11 & 6.22 & 6.25 & 8.54 & -2.30 &  & 4.26 & 4.30 & 4.32 & 5.44 & -1.34 &  & 2.81 & 2.83 & 2.83 & 3.26 & -0.95  \\ \hline
\end{tabular}
}
\caption{(100 $\times$) Empirical MSE of the estimated AR coefficient for ARFIMA(1,$d_0$,0) of CSS estimator with unknown and known $\mu_0$ and the MCSS estimator, together with the bias-corrected MCSS estimator.}
\label{tablear4}
\end{table}

\section{Empirical examples} \label{illustrations}

As an illustration of the results derived in Section \ref{secgen}, we now present three empirical applications, reconsidering the long-memory modelling of classical datasets.
What all three applications have in common is that the datasets consist of short time series of 79 to 171 observations each, warranting the use of our MCSS estimator to correct the small-sample bias of the
received estimators.

\subsection{Post-second World War real GNP}
\begin{figure}[H]
  \centering
  \subfloat[Post-WW2 real GNP data]{
    \includegraphics[width=0.35\textwidth]{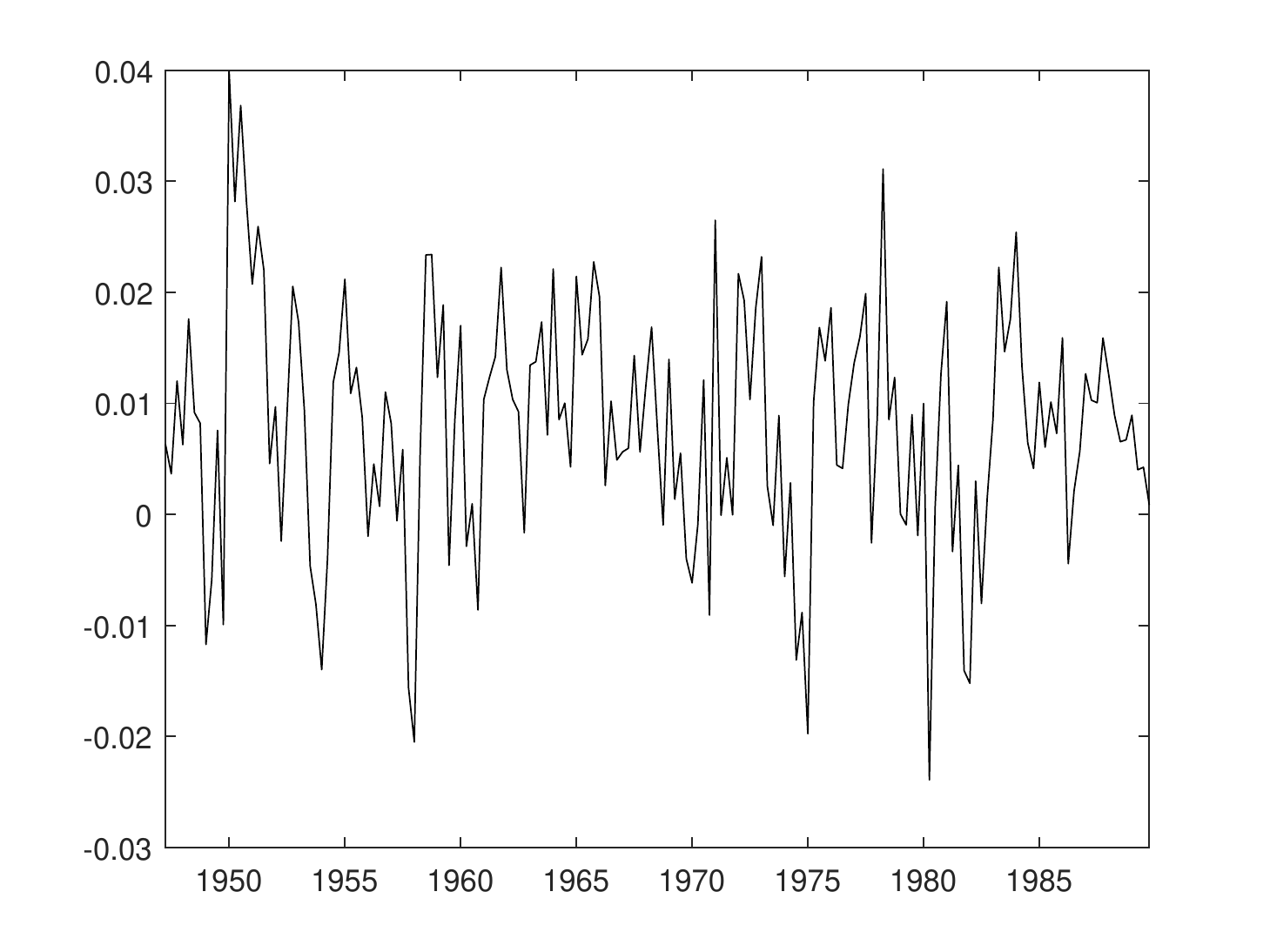}
  }
  \subfloat[Nile data]{
    \includegraphics[width=0.35\textwidth]{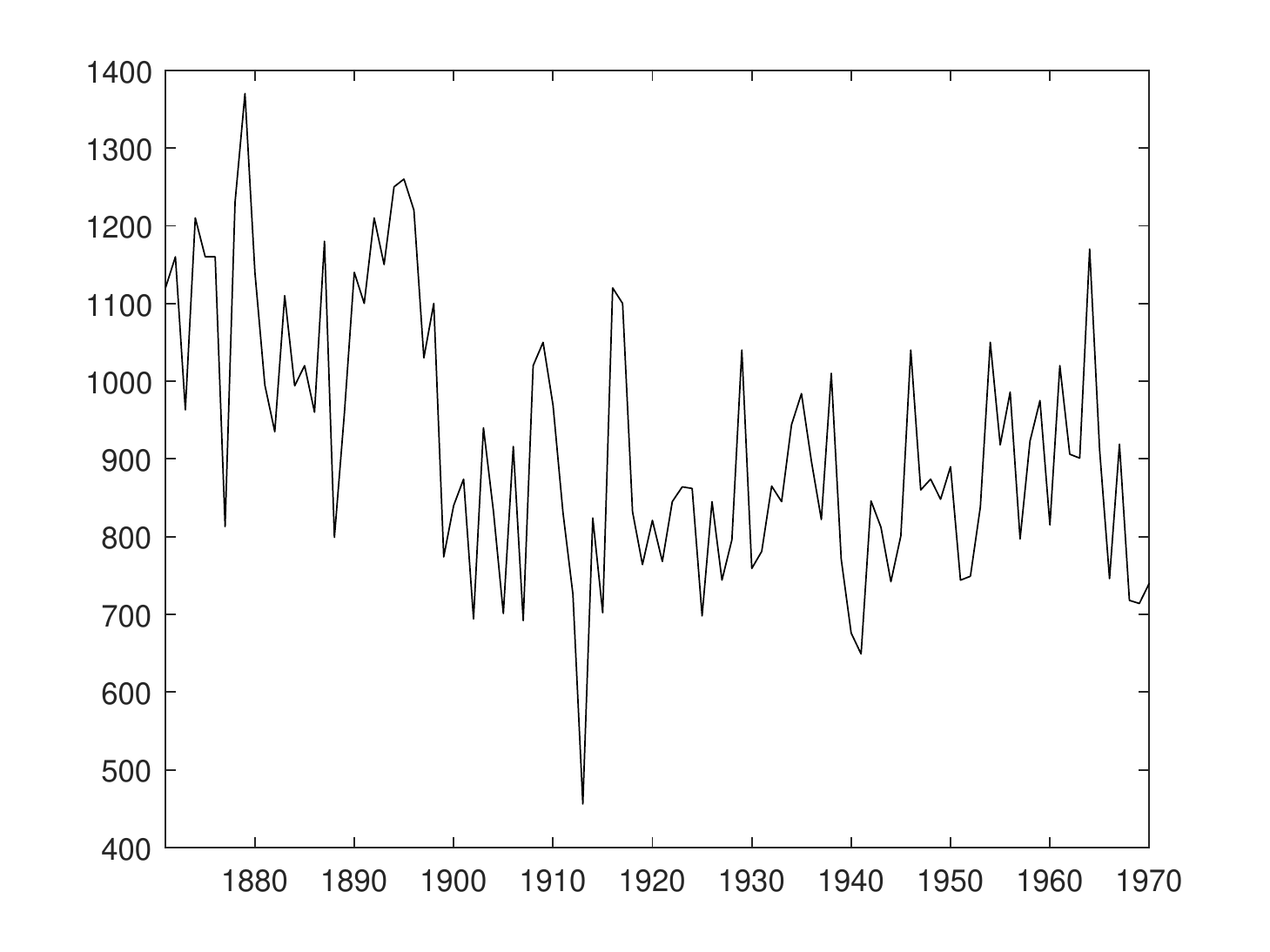}
  }
  \caption{Panel (a) 171 quarterly observations on first differences of log quarterly U.S.\ real GNP for the time period 1947:2 to 1989:4, as in \textcite{sowell1992modeling}. Panel (b) displays 100 annual observations of the
    volume of the Nile for the time period 1871 to 1970.}
  \label{RGNPnile}%
\end{figure}

\textcite{sowell1992modeling} conducted a well-known empirical analysis of the long-memory behaviour of U.S.\ post-Second World War quarterly, seasonally adjusted, log real GNP. The data\footnote{We use the data provided by \textcite{potter1995nonlinear} in the JAE Data Archive who mentions Citibase as his source, as does \textcite{sowell1992modeling}. The dataset can be downloaded from \url{https://journaldata.zbw.eu/dataset/a-nonlinear-approach-to-us-gnp}.} comprise observations from 1947:2 to 1989:4 and are displayed in panel (a) of Figure \ref{RGNPnile}.

\textcite{sowell1992modeling} estimates an ARFIMA(3,$d$,2) type-I model of mean-adjusted first differences using full maximum likelihood (ML), basing the lag order on the Akaike information criterion.  He obtains an estimated
memory parameter of $-0.59$. However, \textcite{smith1997fractional} assert that Sowell's results are substantially biased and especially the memory parameter is strongly underestimated. They propose a simulation-based bias
correction of the profile maximum likelihood (BC-PML) estimator, resulting in $\hat{d} = -0.46$. The BC-PML estimator relies on the assumption that the bias is a linear function in the parameters. However,
\textcite{lieberman2005expansions}\footnote{\textcite{lieberman2005expansions} consider the profile plug-in maximum likelihood estimator instead of the profile maximum likelihood estimator for tractability reasons.}  show this not
to be the case for a simple ARFIMA(0,$d$,0) type-I model. We circumvent this problem by using our MCSS estimator, which does not require the bias to be linear in the parameters. Table \ref{trgnp} presents the CSS and MCSS
estimates of $d$ for the ARFIMA(3,$d$,2) type-II model in \eqref{genq1}, along with the ML estimate of \textcite{sowell1992modeling} and the profile maximum likelihood (PML) as well as BC-PML estimate of
\textcite{smith1997fractional}. It can be noted, first, that the CSS estimate is of similar order of magnitude as the maximum likelihood estimates, compare e.g.\ CSS and (P)ML. Secondly, the bias-correction increases both the CSS
and PML estimates substantially, cf.\ MCSS and BC-PML. In fact, the CSS estimate is increased by a larger margin than the PML estimate. Thirdly, the type-II estimates are less significant than the type-I estimates, and the
significance is reduced by the bias-correction. In conclusion, our results indicate that the long memory parameter is closer to zero than previously thought, even relative to its standard error.

\begin{table}[H]
\centering
\begin{tabular}{l|EEE|EE}\hline
          & \multicolumn{3}{c}{type-I} & \multicolumn{2}{|c}{type-II}                                                                               \\ \cline{2-6}
          & \multicolumn{1}{c}{ML}     & \multicolumn{1}{c}{PML} & \multicolumn{1}{c}{BC-PML} & \multicolumn{1}{|c}{CSS} & \multicolumn{1}{c}{MCSS} \\ \hline
$\hat{d}$ & -0.59                      & -0.61                   & -0.46                      & -0.53                   & -0.26                    \\
SE        & 0.35                       & 0.29                    & 0.29                       & 0.37                    & 0.30                     \\ \hline
\end{tabular}
\caption{The memory parameter estimates for the ARFIMA(3,$d$,2) model and their standard errors. The standard errors are calculated using the inverse of the empirical Hessian matrix.}
\label{trgnp}
\end{table}

\subsection{Extended Nelson-Plosser dataset}

There is a long-standing controversy on whether it is apt to describe the 14 time series in the well-known \textcite{nelson1982trends} dataset, as extended by \textcite{schotman1991bayesian}\footnote{The dataset can be downloaded
  from \url{http://korora.econ.yale.edu/phillips/data/np&enp.dat} and is included in the R package `tseries'.}, by unit root processes. More recently, the literature on long memory processes has broadened the debate by considering
a fractional integration parameter $d$ that can take any value on the real line instead of merely zero or one. Yet the test statistics for the null hypothesis of $d=1$ tend to be close to their critical values, impeding strong
conclusions. Prominent papers are, amongst others, \textcite{crato1994fractional}, \textcite{gil1997testing}, \textcite{shimotsu2010exact} and \textcite{la2019saddlepoint}.

Our enquiry proceeds in two stages: First, we revisit \textcite{crato1994fractional}\footnote{Unfortunately, we did not succeed in replicating the results of \textcite{crato1994fractional}. They use Sowell's Fortran program
  GQSTRFRAC, which is not available to us. Also, \textcite[p.\ 110]{hassler2019time} mentions an error in the autocovariance formula of \textcite[eq.\ (8)]{sowell1992modeling}. We hence exercise caution in interpreting their
  results.} who use profile maximum likelihood (PML) to estimate an ARFIMA type-I model. We compare their PML to the CSS and MCSS estimates of $d$ in our type-II setting. Secondly, we conduct unit root tests and relate them to the results obtained in the frequency-domain setting considered by \textcite{gil1997testing} and \textcite{shimotsu2010exact}. This comparison is of interest
because the MCSS estimator shares one interesting characteristic with frequency-domain estimators, namely that the leading bias of the estimator is not altered by an inclusion of a level parameter, a feature not present in PML or
CSS.

\begin{figure}[H] 
\centering
\includegraphics[scale=0.9]{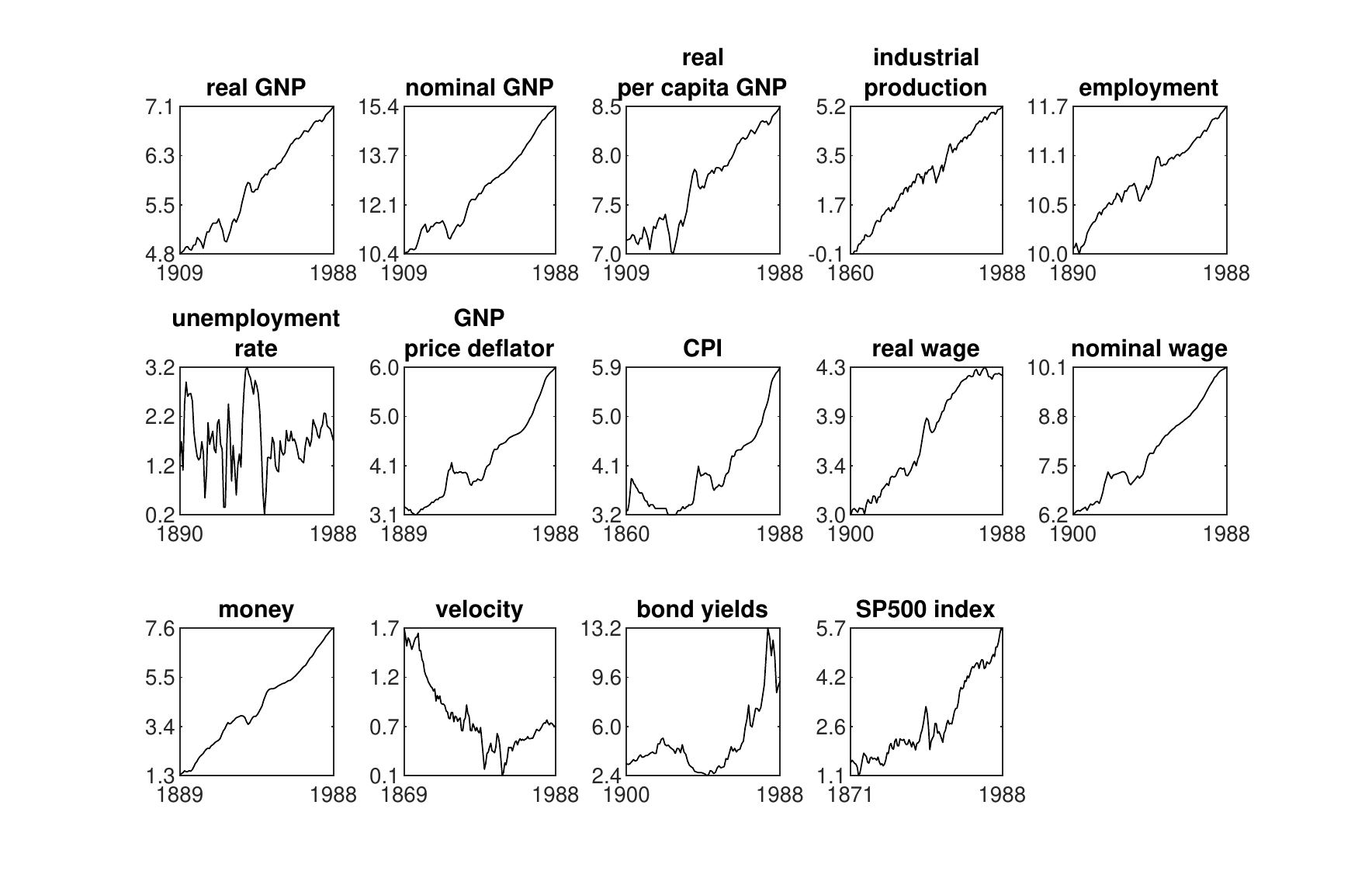}
\caption{ The extended Nelson-Plosser data in levels. All of the series are in logs, except for the bond yield.}
\label{fignp}
\end{figure}

\begin{table}[H]
\centering
\resizebox{\textwidth}{!}{%
\begin{tabular}{l|cc|RRR|RRR|RRR} \hline
                      &     &           & \multicolumn{3}{c|}{PML}      & \multicolumn{3}{c|}{CSS} & \multicolumn{3}{c}{MCSS}                                                                                                                                                                        \\
series                & $T$ & BIC       & \multicolumn{1}{c}{$\hat{d}$} & \multicolumn{1}{c}{SE}   & \multicolumn{1}{c|}{$t$} & \multicolumn{1}{c}{$\hat{d}$} & \multicolumn{1}{c}{SE} & \multicolumn{1}{c|}{$t$} & \multicolumn{1}{c}{$\hat{d}$} & \multicolumn{1}{c}{SE} & \multicolumn{1}{c}{$t$} \\ \hline
real GNP              & 79  & (1,$d$,0) & -0.41                         & 0.21                     & -1.95                    & -0.43                         & 0.21                   & -2.11                    & -0.32                         & 0.23                   & -1.42                   \\
nominal GNP           & 79  & (1,$d$,0) & -0.19                         & 0.24                     & -0.80                    & -0.21                         & 0.25                   & -0.85                    & -0.07                         & 0.26                   & -0.27                   \\
real per capita GNP   & 79  & (1,$d$,0) & -0.43                         & 0.22                     & -1.96                    & -0.44                         & 0.21                   & -2.10                    & -0.33                         & 0.23                   & -1.41                   \\
industrial production & 128 & (1,$d$,0) & -0.64                         & 0.33                     & -1.95                    & -0.59                         & 0.24                   & -2.42                    & -0.46                         & 0.21                   & -2.16                   \\
employment            & 98  & (0,$d$,1) & -0.19                         & 0.12                     & -1.60                    & -0.20                         & 0.12                   & -1.65                    & -0.14                         & 0.13                   & -1.09                   \\
unemployment rate     & 98  & (0,$d$,1) & -0.58                         & 0.11                     & -5.14                    & -0.57                         & 0.11                   & -5.17                    & -0.52                         & 0.11                   & -4.62                   \\
GNP price deflator    & 99  & (1,$d$,0) & -0.39                         & 0.21                     & -1.88                    & -0.40                         & 0.20                   & -1.95                    & 0.22                          & 0.27                  & 0.79                    \\
CPI                   & 128 & (0,$d$,1) & 0.19                          & 0.08                     & 2.24                     & 0.21                          & 0.09                   & 2.40                     & 0.24                          & 0.09                   & 2.60                    \\
real wage             & 88  & (0,$d$,0) & 0.12                          & 0.10                     & 1.16                     & 0.13                          & 0.11                   & 1.19                     & 0.17                          & 0.11                   & 1.59                    \\
nominal wage          & 88  & (1,$d$,0) & -0.21                         & 0.25                     & -0.85                    & -0.23                         & 0.25                   & -0.91                    & -0.07                         & 0.28                   & -0.24                   \\
money                 & 99  & (1,$d$,1) & -0.50                         & 0.22                     & -2.26                    & -0.52                         & 0.21                   & -2.49                    & -0.44                         & 0.26                   & -1.71                   \\
velocity              & 119 & (0,$d$,0) & 0.04                          & 0.08                     & 0.47                     & 0.04                          & 0.08                   & 0.46                     & 0.07                          & 0.08                   & 0.81                    \\
bond yields           & 88  & (0,$d$,1) & -0.19                         & 0.10                     & -1.81                    & -0.20                         & 0.10                   & -1.92                    & -0.15                         & 0.11                   & -1.42                   \\
SP500 index           & 117 & (0,$d$,1) & -0.21                         & 0.10                     & -2.21                    & -0.21                         & 0.09                   & -2.21                    & -0.17                         & 0.10                   & -1.76                   \\ \hline
\end{tabular}
}
\caption{Estimated ARFIMA models of the extended Nelson-Plosser data. The time series are transformed into log-differences, merely bond yields are only in differences. The second column shows the length $T$ of the individual series, the third
  column the model specifications based on the BIC for the profile maximum likelihood (PML) estimator. Subsequent columns then list the estimates of the memory parameter for PML, conditional sum-of-squares (CSS), and modified
      conditional sum-of-squares (MCSS). The empirical Hessian is used to calculate the standard errors, and the $t$-statistics are computed for the unit root null $H_0 \colon d=0$. The PML estimates are computed in R using the ‘arfima’ package, see \textcite{R} and \textcite{arfimapac}.}
\label{fullsamplenp}
\end{table}

The extended Nelson-Plosser dataset consist of 14 annual macroeconomic series, starting between 1860 and 1909 and running to 1988, and are displayed in Figure \ref{fignp}. For the analysis, all of the series are log-differenced\footnote{The ``differencing and adding back'' technique, a
  commonly used method to simplify estimation by removing drift through differencing, has been found to deliver inconsistent CSS estimates in type-II models when the data in levels exhibit a memory parameter of less than 0. As a
  solution to this problem, \textcite{hualde2020truncated} recommend modelling the data in levels instead of first-differences or, alternatively, employing a single dummy variable to capture the initial observation. Implementing this latter approach, our results remain qualitatively the same.}, except for the bond yield, which is merely in differences. Table \ref{fullsamplenp} displays the PML, CSS and MCSS estimates of the memory parameter as well as their respective standard error and the $t$-statistics for testing the
unit root null that $d = 0$. Following \textcite{crato1994fractional}, the model selection is based on the BIC\footnote{\textcite{huang2022consistent} have recently shown the BIC criterion to provide consistent selection of the short-run
  dynamics when based on the CSS estimator in ARFIMA models without constant term.} of the PML estimator. The table reveals that (a) four of the PML $t$-statistics are larger than the 5\% critical values of a two-sided test, with a
further five being borderline cases, (b) the MCSS estimates are consistently larger than the PML and CSS ones, and (c) of the MCSS $t$-statistics, only three lie in or close to the critical region. Another interesting point to note in Table
\ref{fullsamplenp} is that, for the GNP price deflator, PML and CSS provide a long memory estimate of $-0.39$ and $-0.40$, respectively, while the MCSS estimator yields a value of $+0.22$. This disparity may be attributed to the
fact that CSS strongly underestimates the memory parameter when positive AR(1) dynamics are present whereas the MCSS estimator eliminates the bias, as demonstrated in Theorem \ref{t54} and the simulation study presented in Section \ref{Ssimgen}. In summary,
we find greater evidence than in the previous literature in favour of the unit root hypothesis in 11 out of the 14 Nelson-Plosser series.

Let us now turn to the second issue of interest, i.e.\ the comparison of our time-domain estimation to the frequency-domain approaches in \textcite{gil1997testing} and \textcite{shimotsu2010exact}. For the unit root null
hypothesis, \textcite{gil1997testing} employ \citeauthor{robinson1994efficient}'s (\citeyear{robinson1994efficient}) LM-type test based on the Whittle (W) estimator, while \textcite{shimotsu2010exact} uses a $t$-type statistic based on
the extended local Whittle (ELW) objective function. Table \ref{tests} compares the results of the unit root test based on the frequency domain estimators W and ELW with those based on the time domain estimators in Table \ref{fullsamplenp}, a checkmark indicating that $H_0$ is (almost) rejected at the 5\% level. Two important observations can
be made from this table. First, the tests of \textcite{gil1997testing} and \textcite{shimotsu2010exact} give completely different outcomes, confirming the impression that there is presently no consensus in the literature on the unit root
issue. A discussion of the relative merits of the W and ELW estimators is provided in, for instance, \textcite{hualde2011gaussian}.
Secondly, the test decisions of \textcite{gil1997testing} are consistent with the majority of the MCSS tests. They only differ for real GNP, the unemployment rate and CPI, the reason for which could be that \textcite{gil1997testing} capture the short-run
dynamics solely through AR($k$) components, which may be somewhat restrictive considering that the BIC also discovers MA lags.

\begin{table}[H]
\centering
\begin{tabular}{l|ccc|cc}
\hline
                      & \multicolumn{5}{c}{rejection of unit root hypothesis }                                                                                   \\ \cline{2-6} 
                      & \multicolumn{3}{c|}{time domain} & \multicolumn{2}{c}{frequency domain}                                                                  \\
series                & \multicolumn{1}{c}{PML}          & \multicolumn{1}{c}{CSS} & \multicolumn{1}{c|}{MCSS} & \multicolumn{1}{c}{W} & \multicolumn{1}{c}{ELW} \\ \hline
real GNP              & (\checkmark)                     & (\checkmark)              &                           & \checkmark            &                         \\
nominal GNP           &                                  &                         &                           &                       & \checkmark              \\
real per capita GNP   & (\checkmark)                       & (\checkmark)              &                           &    (\checkmark)                     &                         \\
industrial production & (\checkmark)                     & \checkmark              & \checkmark                & \checkmark            &                         \\
employment            &                                  &                         &                           &                       &                         \\
unemployment rate     & \checkmark                       & \checkmark              & \checkmark                &                       &                         \\
GNP price deflator    & (\checkmark)                     &   (\checkmark)                        &                           &                       & \checkmark              \\
CPI                   & \checkmark                       & \checkmark              & \checkmark                &                       & \checkmark              \\
real wage             &                                  &                         &                           &                       &                         \\
nominal wage          &                                  &                         &                           &                       & \checkmark              \\
money                 & \checkmark                       & \checkmark              &                           &                       & \checkmark              \\
velocity              &                                  &                         &                           &                       &                         \\
bond yields           & (\checkmark)                     & (\checkmark)              &                           &                       &                         \\
SP500 index           & \checkmark                       & \checkmark              &                           &                       &                         \\ \hline
\end{tabular}
\caption{Summary of the unit root tests, based on time-domain and frequency-domain estimators.  W denotes the LM-type test of \textcite{gil1997testing} based on the Whittle estimator, while ELW is the LM-type test of
  \textcite{shimotsu2010exact} based on the extended local Whittle estimator.  The presence of a checkmark shows that the null hypothesis of a unit root is rejected at a 5\% significance level against a two-sided fractional
  alternative. A checkmark in parentheses means that the $t$-statistic is just outside the critical region.}
\label{tests}
\end{table}

\subsection{Nile data} \label{nilesec}

We now present an empirical application to the classical dataset\footnote{ The dataset used in this analysis can be obtained from the R package `datasets'.} on the annual water flow volume of the Nile for the years 1871 to 1970. The 100 time-series observations are displayed in panel (b) of  Figure \ref{RGNPnile}. Several
studies have analysed this dataset either in a long memory or short memory framework, with or without the presence of a break in the time series. \textcite{hosking1984modeling} and \textcite{boes1989parameter} focus on long memory without
considering a break. \textcite{macneill1991search}, \textcite{wu2007inference}, \textcite{macneill2020multiple} examine breaks in a short memory time series context. \textcite{atkinson1997detecting} look at breaks in a unit root
model. \textcite{shao2011simple} and \textcite{bet17} address the testing and estimation of a break using a procedure that is robust to long memory although, after identifying a break, they do not proceed to estimating the fractional
parameter. In summary, while there appears to be a consensus on including a break in the model, there is disagreement on whether the dynamics are better described by short or long memory. In particular, the literature currently
does not consider the estimation of the memory parameter that is robust to a break. This is what we aim to achieve.

To that end, we proceed in two steps: First, we extend our model in \eqref{genq1} to incorporate a break, i.e.\ $\mu$ in \eqref{genq1} is replaced by
$    \mu_t(\tau) = \mu + \beta I(t \leq \lfloor \tau T \rfloor), $  
where the break fraction $\tau \in (0,1)$ is assumed unknown. $\mu_t (\tau)$ can be consistently estimated in a type-II fractionally integrated model with $|d_0| < 1/2$, as shown by \textcite{chang2016inference} and
\textcite{iacone2019testing}. It is, however, necessary to generalise our Assumption \ref{A2} such that $q > 1/(1+2d_0)$ moments exist, see \textcite[Theorem 2]{johansen2012necessary}. In a second step, we employ the filtered observations
$\hat{x}_t = x_t - \hat{\mu}_t(\hat{\tau})$ to obtain the CSS estimates $\hat{\vartheta}$ in \eqref{genCSS} and the MCSS estimate $\hat{\vartheta}_m$ in \eqref{MCSSgen1}. The consistency of $\hat{\vartheta}$ in this model follows from similar arguments as in
\textcite[Proposition 1]{robinson2015efficient}, that of $\hat \vartheta_m$ in this model is easily obtained because of its asymptotic equivalence to the CSS estimator, see \eqref{geneq111} in Lemma \ref{gen:lm:md}. The model selection procedure
suggested by \textcite{hualde2011gaussian} is employed, consisting in a preliminary estimator $\tilde{d}$ of $d$ obtained by local Whittle estimation as in \textcite{robinson1995gaussian} before the procedure by \textcite{box1990time} is
applied for selecting the short-run dynamics of $\Delta^{\tilde{d}} \left\{\hat{x}_t\right\}$. The \textcite{lobato1998nonparametric} automatic selection rule of the bandwidth $m$ is used.

In the first step, we find that $\hat \tau = 0.27 $, translating into an estimated break in 1898. This is similar to what most of the aforementioned papers find, and it coincides with the beginning of the
construction of the Lower Aswan Dam in 1899. The estimates of the level and break magnitudes of, respectively, $\hat \mu (\hat \tau) = 849.97$ and $\hat \beta (\hat \tau) = 247.78$ imply that the flow volume was reduced by
22\%. Note that \textcite{hosking1984modeling} implements an alternative adjustment based on the recommendation of \textcite{todini1979hydrological}, namely that the pre-1903 flows are reduced by 8\%.

In the second step, we find a bandwidth of $m = 22$, resulting in a preliminary estimate of $\tilde{d} = -0.05$. The Box-Jenkins procedure indicates that the short-run dynamics are best described by a MA(1) model. The resulting
CSS and MCSS estimates are reported in Table \ref{tableNile}, along with their standard errors and $t$-statistics. The results are unambiguous: the MCSS estimate does not provide evidence of long memory in the Nile data once the
break is incorporated, with the point estimate of the memory parameter being $-0.12$.  The CSS estimate supports this conclusion, with an estimate of $-0.18$. In terms of short-run dynamics, however, CSS and MCSS differ: CSS
estimates the MA coefficient to be $0.30$, an estimate that is statistically significant at the 5\% level. On the other hand, the MCSS estimate of $0.26$ is insignificant. Given the superior finite sample properties of MCSS, our
conclusion is that after incorporating the break, the Nile data is characterised by IID shocks. This finding aligns with that of \textcite{atkinson1997detecting}, supporting their argument that the series can be adequately described
by a white noise process once the break is taken into consideration\footnote{ \textcite{atkinson1997detecting} also identifies an outlier in the year 1913. However, even after removing this outlier, our results remain robust.}.

To corroborate our conclusion regarding the memory parameter, we employ the semi-parametric $t$-type statistic in \textcite{iacone2022semiparametric} to test the null hypothesis $H_0 \colon d_0 = 0$ against the alternative hypothesis
$H_1 \colon d_0 \neq 0$. This test is designed to be robust against breaks and has the advantage that a parametric specification of the shocks is not needed. The test result, omitted to conserve space, is conclusive and supports our
finding: after taking into account the break, there is no evidence that the Nile data exhibits long memory.

\begin{table}[H]
\centering
\begin{tabular}{c|EEE|EEE} \hline
          & \multicolumn{3}{c|}{CSS}      & \multicolumn{3}{c}{MCSS}                                                                                                             \\
Es          & \multicolumn{1}{c}{estimate} & \multicolumn{1}{c}{SE}   & \multicolumn{1}{c|}{$t$} & \multicolumn{1}{c}{estimate} & \multicolumn{1}{c}{SE} & \multicolumn{1}{c}{$t$} \\ \hline
$d$       & -0.18                        & 0.13                     & -1.44                   & -0.12                        & 0.14                   & -0.81                   \\
$\varphi$ & 0.30                         & 0.13                     & 2.36                    & 0.26                         & 0.14                   & 1.88                   \\ \hline
\end{tabular}
\caption{CSS and MCSS estimates of the ARFIMA(0,$d$,1) model for the filtered observations $\hat x_t$ of the Nile data. The MA(1) coefficient is denoted by $\varphi$. The empirical Hessian matrix's inverse is used to calculate the
  standard errors.}
\label{tableNile}
\end{table}

\section{Discussion and outlook}\label{S5}

Practitioners like the CSS estimator due to its simplicity and effectiveness in estimating both stationary and non-stationary ARFIMA models. Recent work by \textcite{hualde2020truncated,hualde2021truncated} provides the asymptotic
justification for using the CSS estimator to estimate models that include deterministic components. However, incorporating a level parameter introduces an additional bias component to the CSS estimator. This bias is due to a
biased score which is particularly pronounced when the data are stationary. To address this issue, we propose modifying the CSS profile objective function to create an unbiased score, resulting in a new estimator which we call the
modified CSS (MCSS) estimator. This new estimator is straightforward to compute and implement, enabling practitioners to obtain more accurate estimates and less distorted tests and confidence intervals. We illustrate the MCSS
estimator by a Monte Carlo simulation and by three classical empirical applications.

Our analysis is for the general ARFIMA($p_1$,$d$,$p_2$) model that includes a constant term and unobserved pre-sample values. Various extensions are conceivable and are of potential interest, yet are beyond the scope of
this paper: First, further deterministic components could be included in the model, e.g.\ a linear time trend: Denoting by $X$ a $T$ $\times$ 2 matrix of a constant and a linear trend it can be shown that the modification term for
the MCSS objective function turns out to be
 \begin{align*}
          m(\vartheta) = \left|(\phi(L;\varphi)\Delta_+^{d}X)'(\phi(L;\varphi)\Delta_+^{d}X)\right|^{\frac{1}{T-2}}. 
\end{align*}  
This modification term is again simple to calculate. Notably, it is equivalent to that in \eqref{genmodificationterm} if $X$ is only a vector of ones and the degrees of freedom in the exponent are replaced by $T-1$.  We expect that
analogous modification terms result for more general deterministic components in $X$. We conjecture that the corresponding MCSS estimators improve on the CSS estimators and that their biases are the same as those of the CSS
estimators with known deterministic components. 

Secondly, while our paper focuses solely on univariate fractional time series, the topic takes on added interest when extended to a panel setting. For instance, \textcite{robinson2015efficient} extend the model presented in
equations \eqref{genq1}-\eqref{genq2} to a panel framework. While the CSS estimator of $\vartheta$ in a panel setting is consistent under large-$T$ asymptotics, its finite sample properties are deficient due to the presence of fixed effects. To
address this issue, the authors propose a bias correction that depends on the true parameters, necessitating the use of estimates to render this correction feasible. However, as the finite sample properties of the CSS estimator is
unsatisfactory, replacing the true values by estimates leads to similarly suboptimal bias corrections. As an alternative to improving the small-sample properties of the CSS estimator a similar modification to the CSS objective
can be made as in Section \ref{S3}. The advantage of this approach is highlighted in the recent work of \textcite{schumann2023role}.

Thirdly, the modification term constructed in this paper is designed to address the unknown-level score bias of the CSS estimator. In principle, our approach can be applied to the elimination of the score bias arising from
other sources. Investigating, for instance, whether the misspecification score bias resulting from unobserved pre-sample values can be eliminated by a modification term derived from our principles is a promising direction
for future research.

Fourthly, the boundary case of $d_0 = 1/2$ is not covered by our theory. Yet our simulations provide evidence that the behaviour of the MCSS estimator for this parameter value is similar to its behaviour for adjacent parameter
ranges. Examining this boundary case analytically is beyond the scope of this paper and therefore left to future work.

\printbibliography

@Manual{R,
  title =        {R: A Language and Environment for Statistical Computing},
  author =       {{R Core Team}},
  organization = {R Foundation for Statistical Computing},
  address =      {Vienna},
  year =         2023,
  version =      {4.4.2},
  url =          {https://www.R-project.org/},
}

@Article{SibbertsenEtAl18,
  author =       {P. Sibbertsen and C. Leschinksi and M. Busch},
  title =        {A multivariate test against spurious long memory},
  journal =      {Journal of Econometrics},
  year =         2018,
  volume =       203,
  pages =        {33--49}
}

@book{abramowitz1964handbook,
  title={Handbook of Mathematical Functions with Formulas, Graphs, and Mathematical Tables},
  author={Abramowitz, Milton and Stegun, Irene A},
  volume={55},
  year={1964},
  publisher={U.S. Government Printing Office}
}

@article{adenstedt1974large,
  title={On large-sample estimation for the mean of a stationary random sequence},
  author={Adenstedt, Rolf K},
  journal={The Annals of Statistics},
  volume={2},
  number={6},
  pages={1095--1107},
  year={1974},
  publisher={JSTOR}
}

@misc{an1993cox,
  title={Cox and Reid’s modification in regression models with correlated errors},
  author={An, S and Bloomfield, P},
  note={Mimeo, North Carolina State University Department of Statistics},
  year={1993}
}

@article{atkinson1997detecting,
  title={Detecting shocks: Outliers and breaks in time series},
  author={Atkinson, Anthony Curtis and Koopman, Siem-Jan and Shephard, Neil},
  journal={Journal of Econometrics},
  volume={80},
  number={2},
  pages={387--422},
  year={1997},
  publisher={Elsevier}
}

@article{barndorff1983formula,
  title={On a formula for the distribution of the maximum likelihood estimator},
  author={O. E. Barndorff-Nielsen},
  journal={Biometrika},
  volume={70},
  number={2},
  pages={343--365},
  year={1983},
  publisher={Oxford University Press}
}

@article{bartolucci2016modified,
  title={Modified profile likelihood for fixed-effects panel data models},
  author={Bartolucci, Francesco and Bellio, Ruggero and Salvan, Alberto and Sartori, Nicola},
  journal={Econometric Reviews},
  volume={35},
  number={7},
  pages={1271--1289},
  year={2016},
  publisher={Taylor \& Francis}
}

@article{beran1995maximum,
  title={Maximum likelihood estimation of the differencing parameter for invertible short and long memory autoregressive integrated moving average models},
  author={Beran, Jan},
  journal={Journal of the Royal Statistical Society: Series B},
  volume={57},
  number={4},
  pages={659--672},
  year={1995},
  publisher={Wiley Online Library}
}

@article{bloomfield1973exponential,
  title={An exponential model for the spectrum of a scalar time series},
  author={Bloomfield, Peter},
  journal={Biometrika},
  volume={60},
  number={2},
  pages={217--226},
  year={1973},
  publisher={Oxford University Press}
}

@article{boes1989parameter,
  title={Parameter estimation in low order fractionally differenced ARMA processes},
  author={Boes, Duane C and Davis, Richard A and Gupta, Sat N},
  journal={Stochastic Hydrology and Hydraulics},
  volume={3},
  pages={97--110},
  year={1989},
  publisher={Springer}
}

@article{hualde2021truncated,
author = {Hualde, Javier and Nielsen, Morten {\O}},
title = {{Truncated sum-of-squares estimation of fractional time series models with generalized power law trend}},
volume = {16},
journal = {Electronic Journal of Statistics},
number = {1},
publisher = {Institute of Mathematical Statistics and Bernoulli Society},
pages = {2884 -- 2946},
keywords = {asymptotic normality, consistency, deterministic trend, Fractional process, generalized polynomial trend, generalized power law trend, noninvertibility, nonstationarity, sum-of-squares estimation},
year = {2022},
%doi = {10.1214/22-EJS2009}
}

@book{box1990time,
  title={Time Series Analysis, Forecasting and Control},
  author={Box, George Edward Pelham and Jenkins, Gwilym},
  year={1990},
  publisher={Holden-Day, Inc.}
}

@article{chang2016inference,
  title={Inference on a structural break in trend with fractionally integrated errors},
  author={Chang, Seong Yeon and Perron, Pierre},
  journal={Journal of Time Series Analysis},
  volume={37},
  number={4},
  pages={555--574},
  year={2016},
  publisher={Wiley Online Library}
}

@article{cheung1994maximum,
  title={On maximum likelihood estimation of the differencing parameter of fractionally-integrated noise with unknown mean},
  author={Cheung, Yin-Wong and Diebold, Francis X},
  journal={Journal of Econometrics},
  volume={62},
  number={2},
  pages={301--316},
  year={1994},
  publisher={Elsevier}
}

@article{chung1993small,
  title={Small sample bias in conditional sum-of-squares estimators of fractionally integrated ARMA models},
  author={Chung, Ching-Fan and Baillie, Richard T},
  journal={Empirical Economics},
  volume={18},
  number={4},
  pages={791--806},
  year={1993},
  publisher={Springer}
}

@incollection{hualde2021frac,
      author = "Javier Hualde and Morten {\O} Nielsen",
      title = "Fractional Integration and Cointegration",
      year = "2023",
      editors={A. Banerjee},
      booktitle={Oxford Research Encyclopedia of Economics and Finance},
      publisher = "Oxford University Press",
      doi = "10.1093/acrefore/9780190625979.013.639",
%      url = "https://oxfordre.com/economics/view/10.1093/acrefore/9780190625979.001.0001/acrefore-9780190625979-e-639"
}

@article{conniffe1987expected,
  title={Expected maximum log likelihood estimation},
  author={Conniffe, Denis},
  journal={Journal of the Royal Statistical Society: Series D},
  volume={36},
  number={4},
  pages={317--329},
  year={1987},
  publisher={Wiley Online Library}
}

@article{cox1987parameter,
  title={Parameter orthogonality and approximate conditional inference},
  author={Cox, David Roxbee and Reid, Nancy},
  journal={Journal of the Royal Statistical Society: Series B},
  volume={49},
  number={1},
  pages={1--18},
  year={1987},
  publisher={Wiley Online Library}
}

@book{cox1994inference,
  title={Inference and Asymptotics},
  author={D. R. Cox and O. E. Barndorff-Nielsen},
  year={1994},
  publisher={CRC Press}
}

@article{crato1994fractional,
  title={Fractional integration analysis of long-run behavior for U.S. macroeconomic time series},
  author={Crato, Nuno and Rothman, Philip},
  journal={Economics Letters},
  volume={45},
  number={3},
  pages={287--291},
  year={1994},
  publisher={Elsevier}
}

@article{dahlhaus1989efficient,
  title={Efficient parameter estimation for self-similar processes},
  author={Dahlhaus, Rainer},
  journal={The Annals of Statistics},
  volume={17},
  number={4},
  pages={1749--1766},
  year={1989},
  publisher={JSTOR}
}

@article{gil1997testing,
  title={Testing of unit root and other nonstationary hypotheses in macroeconomic time series},
  author={Gil-Ala\~na, Luis A and Robinson, Peter M.},
  journal={Journal of Econometrics},
  volume={80},
  number={2},
  pages={241--268},
  year={1997},
  publisher={Elsevier}
}

@book{gradshteyn2014table,
  title={Table of Integrals, Series, and Products},
  author={Gradshteyn, Izrail Solomonovich and Ryzhik, Iosif Moiseevich},
  year={2014},
  publisher={Academic Press}
}

@article{hannan1973asymptotic,
  title={The asymptotic theory of linear time-series models},
  author={Hannan, Edward J},
  journal={Journal of Applied Probability},
  volume={10},
  number={1},
  pages={130--145},
  year={1973},
  publisher={Cambridge University Press}
}

@book{hassler2019time,
  title={Time Series Analysis with Long Memory in View},
  author={Hassler, Uwe},
  year={2019},
  publisher={John Wiley \& Sons}
}

@article{hosking1984modeling,
  title={Modeling persistence in hydrological time series using fractional differencing},
  author={Hosking, J R M},
  journal={Water Resources Research},
  volume={20},
  number={12},
  pages={1898--1908},
  year={1984},
  publisher={Wiley Online Library}
}

@article{hoskingdiff1981,
  title={Fractional differencing},
  author={Hosking, J. R. M.},
  journal={Biometrika},
  volume = {68},
  number = {1},
  pages={165--176},
  year={1981}
}

@article{hualde2011gaussian,
  title={Gaussian pseudo-maximum likelihood estimation of fractional time series models},
  author={Hualde, Javier and Robinson, Peter M.},
  journal={The Annals of Statistics},
  volume={39},
  number={6},
  pages={3152--3181},
  year={2011},
  publisher={Institute of Mathematical Statistics}
}

@article{hualde2020truncated,
  title={Truncated sum of squares estimation of fractional time series models with deterministic trends},
  author={Hualde, Javier and Nielsen, Morten {\O}},
  journal={Econometric Theory},
  volume={36},
  number={4},
  pages={751--772},
  year={2020},
  publisher={Cambridge University Press}
}

@article{huang2022consistent,
  title={Consistent order selection for ARFIMA processes},
  author={Huang, Hsueh-Han and Chan, Ngai Hang and Chen, Kun and Ing, Ching-Kang},
  journal={The Annals of Statistics},
  volume={50},
  number={3},
  pages={1297--1319},
  year={2022},
  publisher={Institute of Mathematical Statistics}
}

@article{iacone2019testing,
  title={Testing the order of fractional integration of a time series in the possible presence of a trend break at an unknown point},
  author={Iacone, Fabrizio and Leybourne, Stephen J and Taylor, AM Robert},
  journal={Econometric Theory},
  volume={35},
  number={6},
  pages={1201--1233},
  year={2019},
  publisher={Cambridge University Press}
}

@article{iacone2022semiparametric,
  title={Semiparametric tests for the order of integration in the possible presence of level breaks},
  author={Iacone, Fabrizio and Nielsen, Morten {\O}  and Taylor, AM Robert},
  journal={Journal of Business \& Economic Statistics},
  volume={40},
  number={2},
  pages={880--896},
  year={2022},
  publisher={Taylor \& Francis}
}

@article{jensen2014fast,
  title={A fast fractional difference algorithm},
  author={Jensen, Andreas Noack and Nielsen, Morten {\O}},
  journal={Journal of Time Series Analysis},
  volume={35},
  number={5},
  pages={428--436},
  year={2014},
  publisher={Wiley Online Library}
}

@article{johansen2008representation,
  title={A representation theory for a class of vector autoregressive models for fractional processes},
  author={Johansen, S{\o}ren},
  journal={Econometric Theory},
  volume={24},
  number={3},
  pages={651--676},
  year={2008},
  publisher={Cambridge University Press}
}

@article{johansen2010likelihood,
  title={Likelihood inference for a nonstationary fractional autoregressive model},
  author={Johansen, S{\o}ren and Nielsen, Morten {\O}},
  journal={Journal of Econometrics},
  volume={158},
  number={1},
  pages={51--66},
  year={2010},
  publisher={Elsevier}
}

@article{johansen2012necessary,
  title={A necessary moment condition for the fractional functional central limit theorem},
  author={Johansen, S{\o}ren and Nielsen, Morten {\O}},
  journal={Econometric Theory},
  volume={28},
  number={3},
  pages={671--679},
  year={2012},
  publisher={Cambridge University Press}
}

@article{johansen2016role,
  title={The role of initial values in conditional sum-of-squares estimation of nonstationary fractional time series models},
  author={Johansen, S{\o}ren and Nielsen, Morten {\O}},
  journal={Econometric Theory},
  volume={32},
  number={5},
  pages={1095--1139},
  year={2016},
  publisher={Cambridge University Press}
}

@article{kalbfleisch1973marginal,
  title={Marginal and conditional likelihoods},
  author={Kalbfleisch, JD and Sprott, DA},
  journal={Sankhy{\=a}: The Indian Journal of Statistics, Series A},
  volume={35},
  number={3},  
  pages={311--328},
  year={1973},
  publisher={JSTOR}
}

@article{la2019saddlepoint,
  title={Saddlepoint approximations for short and long memory time series: A frequency domain approach},
  author={La Vecchia, Davide and Ronchetti, Elvezio},
  journal={Journal of Econometrics},
  volume={213},
  number={2},
  pages={578--592},
  year={2019},
  publisher={Elsevier}
}

@article{laskar1998modified,
  author =       {Laskar, Mizan R. and King, Maxwell L.},
  series =       {Monash University Departement of Econometrics Working Paper},
  title =        {Modified likelihood and related methods for handling nuisance parameters in the linear regression model},
  number =       {5/98},
  doi =          {10.22004/ag.econ.267941},
  year =         {1998}
}

@article{lawley1956general,
  title={A general method for approximating to the distribution of likelihood ratio criteria},
  author={Lawley, Derrick N},
  journal={Biometrika},
  volume={43},
  number={3/4},
  pages={295--303},
  year={1956},
  publisher={JSTOR}
}

@article{lee2015model,
  title={Model selection in the presence of incidental parameters},
  author={Lee, Yoonseok and Phillips, Peter C B},
  journal={Journal of Econometrics},
  volume={188},
  number={2},
  pages={474--489},
  year={2015},
  publisher={Elsevier}
}

@article{li1986fractional,
  title={Fractional time series modelling},
  author={Li, Wai Keung and McLeod, A Ian},
  journal={Biometrika},
  volume={73},
  number={1},
  pages={217--221},
  year={1986},
  publisher={Oxford University Press}
}

@article{liang1987estimating,
  title={Estimating functions and approximate conditional likelihood},
  author={Liang, Kung-Yee},
  journal={Biometrika},
  volume={74},
  number={4},
  pages={695--702},
  year={1987},
  publisher={Oxford University Press}
}

@article{liang1995inference,
  title={Inference based on estimating functions in the presence of nuisance parameters},
  author={Liang, Kung-Yee and Zeger, Scott L},
  journal={Statistical Science},
  volume={10},
  number={2},
  pages={158--173},
  year={1995},
  publisher={Institute of Mathematical Statistics}
}

@article{lieberman2004expansions,
  title={Expansions for the distribution of the maximum likelihood estimator of the fractional difference parameter},
  author={Lieberman, Offer and Phillips, Peter C B},
  journal={Econometric Theory},
  volume={20},
  number={3},
  pages={464--484},
  year={2004},
  publisher={Cambridge University Press}
}

@article{lieberman2005expansions,
  title={Expansions for approximate maximum likelihood estimators of the fractional difference parameter},
  author={Lieberman, Offer and Phillips, Peter C B},
  journal={The Econometrics Journal},
  volume={8},
  number={3},
  pages={367--379},
  year={2005},
  publisher={Oxford University Press Oxford, UK}
}

@article{lobato1998nonparametric,
  title={A nonparametric test for I(0)},
  author={Lobato, Ignacio N and Robinson, Peter M.},
  journal={The Review of Economic Studies},
  volume={65},
  number={3},
  pages={475--495},
  year={1998},
  publisher={Wiley-Blackwell}
}

@article{macneill1991search,
  title={A search for the source of the Nile's change-points},
  author={MacNeill, IB and Tang, SM and Jandhyala, VK},
  journal={Environmetrics},
  volume={2},
  number={3},
  pages={341--375},
  year={1991},
  publisher={Wiley Online Library}
}

@article{macneill2020multiple,
  title={Multiple change-point models for time series},
  author={MacNeill, IB and Jandhyala, VK and Kaul, A and Fotopoulos, SB},
  journal={Environmetrics},
  volume={31},
  number={1},
  pages={e2593},
  year={2020},
  publisher={Wiley Online Library}
}

@article{marinucci1999alternative,
  title={Alternative forms of fractional Brownian motion},
  author={Marinucci, Domenico and Robinson, Peter M.},
  journal={Journal of Statistical Planning and Inference},
  volume={80},
  number={1-2},
  pages={111--122},
  year={1999},
  publisher={Elsevier}
}

@article{marinucci2000weak,
  title={Weak convergence of multivariate fractional processes},
  author={Marinucci, Domenico and Robinson, Peter M.},
  journal={Stochastic Processes and their Applications},
  volume={86},
  number={1},
  pages={103--120},
  year={2000},
  publisher={Elsevier}
}

@article{marinucci2001semiparametric,
  title={Semiparametric fractional cointegration analysis},
  author={Marinucci, Domenico and Robinson, Peter M.},
  journal={Journal of Econometrics},
  volume={105},
  number={1},
  pages={225--247},
  year={2001},
  publisher={Elsevier}
}

@article{martellosio2020adjusted,
  title={Adjusted QMLE for the spatial autoregressive parameter},
  author={Martellosio, Federico and Hillier, Grant},
  journal={Journal of Econometrics},
  volume={219},
  number={2},
  pages={488--506},
  year={2020},
  publisher={Elsevier}
}

@article{mccullagh1990simple,
  title={A simple method for the adjustment of profile likelihoods},
  author={McCullagh, Peter and Tibshirani, Robert},
  journal={Journal of the Royal Statistical Society: Series B},
  volume={52},
  number={2},
  pages={325--344},
  year={1990},
  publisher={Wiley Online Library}
}

@article{nelson1982trends,
  title={Trends and random walks in macroeconmic time series: some evidence and implications},
  author={Nelson, Charles R and Plosser, Charles R},
  journal={Journal of Monetary Economics},
  volume={10},
  number={2},
  pages={139--162},
  year={1982},
  publisher={Elsevier}
}

@article{nielsen2004efficient,
  title={Efficient likelihood inference in nonstationary univariate models},
  author={Nielsen, Morten {\O}},
  journal={Econometric Theory},
  volume={20},
  number={1},
  pages={116--146},
  year={2004},
  publisher={Cambridge University Press}
}

@article{nielsen2005finite,
  title={Finite sample comparison of parametric, semiparametric, and wavelet estimators of fractional integration},
  author={Nielsen, Morten {\O} and Frederiksen, Per Houmann},
  journal={Econometric Reviews},
  volume={24},
  number={4},
  pages={405--443},
  year={2005},
  publisher={Taylor \& Francis}
}

@article{nielsen2015asymptotics,
  title={Asymptotics for the conditional-sum-of-squares estimator in multivariate fractional time-series models},
  author={Nielsen, Morten {\O}},
  journal={Journal of Time Series Analysis},
  volume={36},
  number={2},
  pages={154--188},
  year={2015},
  publisher={Wiley Online Library}
}

@article{potter1995nonlinear,
  title={A nonlinear approach to U.S. GNP},
  author={Potter, Simon M},
  journal={Journal of Applied Econometrics},
  volume={10},
  number={2},
  pages={109--125},
  year={1995},
  publisher={Wiley Online Library}
}

@article{robinson1994efficient,
  title={Efficient tests of nonstationary hypotheses},
  author={Robinson, Peter M.},
  journal={Journal of the American Statistical Association},
  volume={89},
  number={428},
  pages={1420--1437},
  year={1994},
  publisher={Taylor \& Francis}
}

@article{robinson1995gaussian,
  title={Gaussian semiparametric estimation of long range dependence},
  author = {Robinson, Peter M.},
  journal={The Annals of Statistics},
  volume = {23},  
  number = {5},  
  pages={1630--1661},
  year={1995},
  publisher = {Institute of Mathematical Statistics}
}

@article{robinson2003cointegration,
  title={Cointegration in fractional systems with unknown integration orders},
  author={Robinson, Peter M. and Hualde, Javier},
  journal={Econometrica},
  volume={71},
  number={6},
  pages={1727--1766},
  year={2003},
  publisher={Wiley Online Library}
}

@article{robinson2005distance,
  title={The distance between rival nonstationary fractional processes},
  author={Robinson, Peter M.},
  journal={Journal of Econometrics},
  volume={128},
  number={2},
  pages={283--300},
  year={2005},
  publisher={Elsevier}
}

@incollection{robinson2006conditional,
  author    = {Robinson, Peter M.},
  title     = {Conditional-sum-of-squares estimation of models for stationary time series with long memory},
  booktitle = {Time Series and Related Topics: In Memory of Ching-Zong Wei},
  editor    = {Ho, H.-C. and Ing, C.-K. and Lai, T. L.},
  publisher = {Institute of Mathematical Statistics},
  pages     = {130--137},
  year      = {2006}
}

@article{robinson2015efficient,
  title={Efficient inference on fractionally integrated panel data models with fixed effects},
  author={Robinson, Peter M. and Velasco, Carlos},
  journal={Journal of Econometrics},
  volume={185},
  number={2},
  pages={435--452},
  year={2015},
  publisher={Elsevier}
}

@article{robinson2020estimation,
  title={Estimation for dynamic panel data with individual effects},
  author={Robinson, Peter M. and Velasco, Carlos},
  journal={Econometric Theory},
  volume={36},
  number={2},
  pages={185--222},
  year={2020},
  publisher={Cambridge University Press}
}

@article{schotman1991bayesian,
  title={On Bayesian routes to unit roots},
  author={Schotman, Peter C and Van Dijk, Herman K},
  journal={Journal of Applied Econometrics},
  volume={6},
  number={4},
  pages={387--401},
  year={1991},
  publisher={Wiley Online Library}
}

@article{bet17,
author = {Annika Betken},
title = {{Change point estimation based on Wilcoxon tests in the presence of long-range dependence}},
volume = {11},
journal = {Electronic Journal of Statistics},
number = {2},
publisher = {Institute of Mathematical Statistics and Bernoulli Society},
pages = {3633 -- 3672},
keywords = {change point estimation, long-range dependence, self-normalization, Wilcoxon test},
year = {2017},
%doi = {10.1214/17-EJS1323}
}

@article{schumann2023role,
  title={The role of score and information bias in panel data likelihoods},
  author={Schumann, Martin and Severini, Thomas A and Tripathi, Gautam},
  journal={Journal of Econometrics},
  volume={235},
  number={2},
  pages={1215--1238},
  year={2023},
  publisher={Elsevier}
}

@book{severini2000likelihood,
  title={Likelihood Methods in Statistics},
  author={Severini, Thomas A},
  year={2000},
  publisher={Oxford University Press}
}

@article{shao2011simple,
  title={A simple test of changes in mean in the possible presence of long-range dependence},
  author={Shao, Xiaofeng},
  journal={Journal of Time Series Analysis},
  volume={32},
  number={6},
  pages={598--606},
  year={2011},
  publisher={Wiley Online Library}
}

@article{shimotsu2005exact,
  title={Exact local Whittle estimation of fractional integration},
  author={Shimotsu, Katsumi and Phillips, Peter C B},
  journal={The Annals of Statistics},
  volume={33},
  number={4},
  pages={1890--1933},
  year={2005},
  publisher={Institute of Mathematical Statistics}
}

@article{shimotsu2010exact,
  title={Exact local Whittle estimation of fractional integration with unknown mean and time trend},
  author={Shimotsu, Katsumi},
  journal={Econometric Theory},
  volume={26},
  number={2},
  pages={501--540},
  year={2010},
  publisher={Cambridge University Press}
}

@article{smith1997fractional,
  title={Fractional integration with drift: estimation in small samples},
  author={Smith, Anthony A and Sowell, Fallaw and Zin, Stanley E},
  journal={Empirical Economics},
  volume={22},
  number = {1},
  pages={103--116},
  year={1997},
  publisher={Springer}
}

@article{sowell1992modeling,
  title={Modeling long-run behavior with the fractional ARIMA model},
  author={Sowell, Fallaw},
  journal={Journal of Monetary Economics},
  volume={29},
  number={2},
  pages={277--302},
  year={1992},
  publisher={Elsevier}
}

@article{tanaka1984asymptotic,
  title={An asymptotic expansion associated with the maximum likelihood estimators in ARMA models},
  author={Tanaka, Katsuto},
  journal={Journal of the Royal Statistical Society: Series B},
  volume={46},
  number={1},
  pages={58--67},
  year={1984},
  publisher={Wiley Online Library}
}

@article{tanaka1999nonstationary,
  title={The nonstationary fractional unit root},
  author={Tanaka, Katsuto},
  journal={Econometric Theory},
  volume={15},
  number={4},
  pages={549--582},
  year={1999},
  publisher={Cambridge University Press}
}

@book{todini1979hydrological,
  title={Hydrological Simulation of Lake Nasser},
  volume={1},
  editor={Todini, E. and O'Connell, P. E.},
  year={1979},
  publisher={IBM Italia Scientific Centers and Institute of Hydrology},
  address={Italy and Wallingford, Oxfordshire, United Kingdom}
}

@article{velasco2000whittle,
  title={Whittle pseudo-maximum likelihood estimation for nonstationary time series},
  author={Velasco, Carlos and Robinson, Peter M.},
  journal={Journal of the American Statistical Association},
  volume={95},
  number={452},
  pages={1229--1243},
  year={2000},
  publisher={Taylor \& Francis}
}

@article{wu2007inference,
  title={Inference of trends in time series},
  author={Wu, Wei Biao and Zhao, Zhibiao},
  journal={Journal of the Royal Statistical Society: Series B},
  volume={69},
  number={3},
  pages={391--410},
  year={2007},
  publisher={Wiley}
}

@software{MATLAB,
  year =         {2019},
  author =       {{MathWorks Inc.}},
  title =        {MATLAB version: 9.6.0 (R2019a)},
  publisher =    {The MathWorks Inc.},
  address =      {Natick, Massachusetts},
  url =          {https://www.mathworks.com}
}

@book{zygmund1977trigonometric,
  title={Trigonometric Series},
  author={Zygmund, A},
  year={1977},
  publisher={Cambridge University Press}
}

@incollection{Chan2006,
  author = "N. H. Chan and W. Palma",
  title = "Estimation of long-memory time series models: A survey of different likelihood-based methods",
  year = "2006",
  editors = "T. Fomby and D. Terrell",
  booktitle = "Econometric Analysis of Financial and Economic Time Series",
  publisher = "Emerald Group Publishing Limited"
}

@article{baillie2024combining,
  title={Combining long and short memory in time series models: the role of asymptotic correlations of the MLEs},
  author={Baillie, Richard T and Cho, Dooyeon and Rho, Seunghwa},
  journal={Econometrics and Statistics},
  volume={29},
  pages={88--112},
  year={2024},
  publisher={Elsevier}
}

@PhdThesis{arfimapac,
    title = {Persistence and anti-persistence: theory and software},
    author = {Justin Q. Veenstra},
    school = {Western University},
    year = {2012},
  }

@book{McCullagh87,
  author    = {P. McCullagh},
  title     = {Tensor Methods in Statistics},
  publisher = {Chapman and Hall},
  year      = 1987,
}

\appendix
\counterwithin*{equation}{section} 
\renewcommand\theequation{\thesection.\arabic{equation}} 

\section{Appendix} \label{Appendixgeneraldgen}

In this appendix, we provide the proofs of the results in Section \ref{secgen}. The setup is as follows: In Appendix \ref{generalderivgen}, we find expressions for the first three derivatives of $L^*(\vartheta)$, $L_{\mu_0}^*(\vartheta)$ and $L_{m}^*(\vartheta)$, evaluated at $\vartheta = \vartheta_0$. Appendix \ref{genlemmaap} presents some preliminary results that play a central role in approximating
these derivatives. In Appendix \ref{ap3}, we analyse the terms involved in the derivatives and conclude with an asymptotic approximation of the derivatives. In Appendix \ref{generalizationsappendix}, we present the proofs of the main results in Section \ref{secgen}. The appendix concludes in \ref{Initschemes} with an analysis of initial conditions: relaxing Assumption \ref{A5}, we study how unobserved pre-sample values influence the bias of CSS and MCSS under alternative initialisation schemes.

\subsection{Derivatives of the objective functions} \label{generalderivgen}

We first analyse the residuals $\epsilon_t(\vartheta, \mu) = \phi(L;\varphi)\Delta_{+}^{d} x_t- \mu c_t(\vartheta)$  for $t \geq 1$ and introduce some notation. We keep the parameter notation from the main text: let $\vartheta = (d,\varphi')' \in \mathbb{R}^{p+1}$ denote the parameter vector, with true value $\vartheta_0 = (d_0,\varphi_0')'$. 

Clearly, inserting the DGP in \eqref{genq1} into the residuals $\epsilon_t(\vartheta, \mu)$ yields
\begin{align}
    \epsilon_t(\vartheta, \mu) &= \phi(L;\varphi)\Delta_{+}^{d}\{\mu_0 + \Delta_+^{-d_0} u_t\} - \mu c_{t}(d,\varphi) = S^+_t(\vartheta)  - c_{t}(\vartheta) \left( \mu - \mu_0 \right), \label{GenRDEF} 
\end{align}
where the stochastic term $S^+_t(\vartheta)$ is defined as 
\begin{align}
    S^+_t(\vartheta) = \phi(L;\varphi)\Delta_{+}^{d-d_0} u_t \label{Sterm}
\end{align}
and the deterministic term $c_{t}(\vartheta)$, see \eqref{convcoef}, is defined as 
\begin{align}
 c_{t}(\vartheta) = \phi(L;\varphi)\Delta_{+}^{d} I(t \geq 1) = \phi(L;\varphi) \kappa_{0t}(d) = \sum_{j = 0}^{t-1} \phi_j(\varphi) \kappa_{0(t-j)}(d).  \label{detc}
\end{align}
 
The derivative of $\epsilon_t(\vartheta, \mu)$ with respect to $i \in \{\vartheta_k,\vartheta_k \vartheta_j, \vartheta_k \vartheta_j,\vartheta_k \vartheta_j \vartheta_l \}$, for $k, j, l = 1,\ldots, p+1$, evaluated at $\vartheta = \vartheta_0$, is of the form 
\begin{align}
    D_{i}  \epsilon_t(\vartheta_0, \mu) =  S^+_{i t} (\vartheta_0) - c_{ i t}(\vartheta_0)\left( \mu - \mu_0 \right), \label{firstd1}
\end{align}
where 
\begin{align}
     S^+_{i t}(\vartheta_0) = D_{i} S^+_t(\vartheta_0), \label{firststoch}
\end{align}
and
\begin{align}
    c_{ i t}(\vartheta_0) = D_{i}  c_{t}(\vartheta_0).  \label{firstdet} 
\end{align}    

Throughout the appendix, we simplify notation by suppressing the dependence on $\vartheta_0$. For instance, we write $S^+_{i t}$ instead of $S^+_{i t}(\vartheta_0)$. We follow the following convention: the derivative of a function $f(x, y(x))$ with respect to $x$ is written as $D_x f(x, y(x))$, while the partial derivative with respect to $x$ is written as $f_x(x, y(x))$.

The following lemma provide expressions for the derivatives of the stochastic term $S_t^+$ given in \eqref{firststoch}.
\begin{lemma} \label{explicitforms}  
Let Assumption \ref{A5} be satisfied. The stochastic term $S_t^+$ and its derivatives are given by the following expressions. Let $t \geq 1$:
    \begin{align}
        S_t^+ &= \epsilon_t, \label{exp0}\\
        S^+_{m t} &= (-1)^{m^*} \sum_{k = 0}^{t-1} D_m \pi_{k}(0) \epsilon_{t-k},  \label{exp1}\\
        S^+_{z t} &= \sum_{i = 1}^{t-1} b_{zi}(\varphi_0)  \epsilon_{t-i},\label{exp2} \\
        S^+_{ m z  t} &= (-1)^{m^*} \sum_{i = 2}^{t-1} h_{mzi}(\varphi_0)  \epsilon_{t-i},\label{exp5} 
    \end{align}
    where $m \in \{d,dd,ddd \}$ and $m^*$ denotes the number of times $S^+_t(\vartheta)$ is differenced with respect to $d$ and where $z \in \{ \varphi_k, \varphi_k \varphi_j, \varphi_k \varphi_j \varphi_l \}$ for $k,j,l = 1,\ldots,p$ and 
    \begin{align}
     b_{zi}(\varphi_0) &= \sum_{s = 0}^{i-1} \omega_s(\varphi_0) D_{z} \phi_{i-s}(\varphi_0), \label{defb} \\
        h_{mzi}(\varphi_0) &= \sum_{s = 1}^{i-1}  D_m \pi_{i-s}(0) b_{zs}(\varphi_0). \label{defh} 
    \end{align}
    Also,
    \begin{align}
        D_d \pi_j(0) &= j^{-1} I(j \geq 1), \label{dpi1} \\
        D_{dd} \pi_j(0) &= 2j^{-1} a_{j-1} I(j \geq 2), \label{dpi2}
    \end{align}
    where $a_j = I(j \geq 1) \sum_{k = 1}^j k^{-1}$.
\end{lemma}

\begin{proof}[Proof of Lemma \ref{genderivatesLstar}.]
Proof of \eqref{exp0}: Let
\begin{align*}
z_t(\varphi) = \phi(L;\varphi)\{u_t I(t\ge1)\}.
\end{align*}
At $\varphi=\varphi_0$,
\begin{align*}
z_t(\varphi_0) = \phi(L;\varphi_0)u_t = \epsilon_t,
\end{align*}
using $\phi(L;\varphi_0)\omega(L;\varphi_0) = 1$ and Assumption \ref{A5}. By definition $S_t^+(d,\varphi) = \Delta_+^{d-d_0} z_t(\varphi)$.
Thus $S_t^+(d_0,\varphi_0)=z_t(\varphi_0)=\epsilon_t$, proving \eqref{exp0}.

Proof of \eqref{exp1}: Using $\Delta_+^{\alpha}y_t=\sum_{k=0}^{t-1}\pi_k(-\alpha) y_{t-k}$,
\begin{align*}
S^+_{m t}
&= (-1)^{m^*}\sum_{k=0}^{t-1} D_m\pi_k(0) z_{t-k}(\varphi_0)
 = (-1)^{m^*}\sum_{k=0}^{t-1} D_m\pi_k(0) \epsilon_{t-k},
\end{align*}
which is \eqref{exp1} and the identities \eqref{dpi1}-\eqref{dpi2} apply.

Proof of \eqref{exp2}: By
\begin{align*}
S^+_{z t}
&= D_z z_t(\varphi)\big|_{\varphi=\varphi_0}
 = D_z\phi(L;\varphi_0)\{u_tI(t\ge1)\} \\
&= \sum_{i=1}^{t-1} b_{z i}(\varphi_0)\epsilon_{t-i},
\end{align*}
where $b_{z i}(\varphi_0)=\sum_{s=0}^{i-1}\omega_s(\varphi_0)D_z\phi_{i-s}(\varphi_0)$, proving \eqref{exp2}.

Proof of \eqref{exp5}: Since
\begin{align*}
S^+_{m z t}
&= (-1)^{m^*}\sum_{k=1}^{t-1} D_m\pi_k(0) S^+_{zt-k} \\
&= (-1)^{m^*}\sum_{k=1}^{t-1} D_m\pi_k(0)\left(\sum_{s=1}^{t-k-1} b_{z s}(\varphi_0) \epsilon_{t-k-s}\right) \\
&= (-1)^{m^*}\sum_{i=2}^{t-1} \Bigg(\sum_{s=1}^{i-1} D_m\pi_{i-s}(0) b_{z s}(\varphi_0)\Bigg)\epsilon_{t-i} \\
&= (-1)^{m^*}\sum_{i=2}^{t-1} h_{m z i}(\varphi_0)\epsilon_{t-i},
\end{align*}
which is \eqref{exp5} with $h_{m z i}$ as in \eqref{defh}.
\end{proof}

Next, we find expressions for the first three derivatives of $L^*(\vartheta)$, $L^*_{\mu_0}(\vartheta)$ and $L^*_{m}(\vartheta)$ and evaluate them at $\vartheta = \vartheta_0$. We present them in the same order. 

Recall that $L^*(\vartheta)$ in \eqref{genL1} equals $L(\vartheta,\mu(\vartheta))$, where $L(\vartheta,\mu) $ is given in \eqref{genlikmu1} and $\mu(\vartheta) = \hat{\mu}(\vartheta)$ in \eqref{genmu1}. The first derivative of $L^*(\vartheta)$ with respect to $\vartheta_k$ equals
\begin{align*}
    D_{\vartheta_k} L^*(\vartheta) =  L_{\vartheta_k}(\vartheta,\mu(\vartheta)) +  L_{\mu}(\vartheta,\mu(\vartheta)) \mu_{\vartheta_k}(\vartheta). 
\end{align*}
We simplify this expression by noticing that $\hat{\mu}(\vartheta)$ is determined from $L_{\mu}(\vartheta,\mu(\vartheta)) = 0$ such that 
\begin{align}
    D_{\vartheta_k} L^*(\vartheta) =  L_{\vartheta_k}(\vartheta,\mu(\vartheta)).  \label{genproofqw12}
\end{align}
Next, we take the derivative of \eqref{genproofqw12} with respect to $ \vartheta_j$ to get an expression for $D_{\vartheta_k \vartheta_j} L^*(\vartheta)$. Using the chain rule we have that 
\begin{align}
   D_{\vartheta_k \vartheta_j } L^*(\vartheta) &=  L_{\vartheta_k  \vartheta_j}(\vartheta,\mu(\vartheta)) + L_{\vartheta_k \mu}(\vartheta,\mu(\vartheta)) \mu_{\vartheta_j}(\vartheta). \label{genproofqw13}
\end{align}
Taking on both sides the derivative with respect to $\vartheta_j$ of $L_{\mu}(\vartheta,\mu(\vartheta)) = 0$ implies $L_{ \vartheta_j \mu}(\vartheta,\mu(\vartheta)) + L_{\mu \mu}(\vartheta,\mu(\vartheta)) \mu_{\vartheta_j}(\vartheta) = 0$ such that
\begin{align*}
    \mu_{\vartheta_j}(\vartheta) = -\frac{L_{\vartheta_j \mu}(\vartheta,\mu(\vartheta))}{L_{\mu \mu}(\vartheta,\mu(\vartheta)) }. 
\end{align*} 
Lastly, we take the derivative of \eqref{genproofqw13} with respect to $\vartheta_l$ to get an expression for $ D_{\vartheta_k \vartheta_j \vartheta_l } L^*(\vartheta)$. We get that 
\begin{align}
D_{\vartheta_k \vartheta_j \vartheta_l } L^*(\vartheta) = &L_{\vartheta_k  \vartheta_j \vartheta_l}(\vartheta,\mu(\vartheta)) +  L_{\vartheta_k  \vartheta_j \mu}(\vartheta,\mu(\vartheta)) \mu_{\vartheta_l}(\vartheta) +  L_{\vartheta_k  \vartheta_l \mu}(\vartheta,\mu(\vartheta)) \mu_{\vartheta_j}(\vartheta) \nonumber \\
&+   L_{\vartheta_k  \mu \mu}(\vartheta,\mu(\vartheta)) \mu_{\vartheta_l}(\vartheta) \mu_{\vartheta_j}(\vartheta)
+ L_{\vartheta_k  \mu}(\vartheta,\mu(\vartheta)) \mu_{\vartheta_j \vartheta_l}(\vartheta).  \label{genproofqw14}
\end{align}
An expression for $ \mu_{\vartheta_j \vartheta_l}(\vartheta)$ can then be easily found by taking on both sides the derivative with respect to $\vartheta_l$ of $L_{\vartheta_j \mu}(\vartheta,\mu(\vartheta)) + L_{\mu \mu}(\vartheta,\mu(\vartheta)) \mu_{\vartheta_j}(\vartheta) = 0$. We find that
\begin{align*}
  0 = &L_{\vartheta_j \vartheta_l \mu }(\vartheta,\mu(\vartheta)) + L_{ \vartheta_j \mu \mu  }(\vartheta,\mu(\vartheta)) \mu_{\vartheta_l}(\vartheta) \\ &+ 
   \left(  L_{\vartheta_l \mu \mu}(\vartheta,\mu(\vartheta)) +  L_{\mu \mu \mu }(\vartheta,\mu(\vartheta))  \mu_{\vartheta_l}(\vartheta) \right) \mu_{\vartheta_j}(\vartheta) 
   + L_{\mu \mu}(\vartheta,\mu(\vartheta)) \mu_{\vartheta_j \vartheta_l}(\vartheta), 
\end{align*}
and by rewriting 
\begin{align*}
    \mu_{\vartheta_j \vartheta_l}(\vartheta) = &-\frac{L_{ \vartheta_j \vartheta_l \mu}(\vartheta,\mu(\vartheta))}{L_{\mu \mu}(\vartheta,\mu(\vartheta))} -  \mu_{\vartheta_l}(\vartheta)  \frac{  L_{ \vartheta_j  \mu \mu }(\vartheta,\mu(\vartheta))}{L_{\mu \mu}(\vartheta,\mu(\vartheta))}  - \mu_{\vartheta_j}(\vartheta)  \frac{  L_{\vartheta_l \mu \mu }(\vartheta,\mu(\vartheta)) }{L_{\mu \mu}(\vartheta,\mu(\vartheta))} \\
    &-  \mu_{\vartheta_j}(\vartheta) \mu_{\vartheta_l}(\vartheta)  \frac{  L_{\mu \mu \mu }(\vartheta,\mu(\vartheta)) }{L_{\mu \mu}(\vartheta,\mu(\vartheta))}.  
\end{align*}

Next, we find expressions for the first three derivatives of $L^*(\vartheta)$. The derivatives are evaluated at $\vartheta = \vartheta_0$, and recall that we omit the explicit dependence. 
\begin{lemma} \label{genderivatesLstar}
Let the model for the data $x_t$, t = 1,$\ldots$,T, be given by \eqref{genq1} and assume that Assumption \ref{A1} holds. Then the derivatives of $L^*(\vartheta)$, see \eqref{genL1}, evaluated at $\vartheta = \vartheta_0$ are given by
\begin{align}
      D_{\vartheta_k} L^* &=  L_{\vartheta_k}, \label{genqw11}\\
    D_{\vartheta_k \vartheta_j } L^* &=  L_{\vartheta_k  \vartheta_j} + L_{\vartheta_k \mu}\mu_{\vartheta_j} , \label{genqw12}\\
    D_{\vartheta_k \vartheta_j \vartheta_l } L^* &= L_{\vartheta_k  \vartheta_j \vartheta_l} +  L_{\vartheta_k  \vartheta_j \mu} \mu_{\vartheta_l} +  L_{\vartheta_k  \vartheta_l \mu} \mu_{\vartheta_j} +   L_{\vartheta_k  \mu \mu} \mu_{\vartheta_l} \mu_{\vartheta_j}
+ L_{\vartheta_k  \mu} \mu_{\vartheta_j \vartheta_l} , \label{genqw13}
\end{align}
where 
\begin{align*}
     \mu_{\vartheta_j} &= -\frac{L_{ \vartheta_j \mu}}{L_{\mu \mu} }, \\
      \mu_{\vartheta_j \vartheta_l} &= -\frac{L_{\vartheta_j \vartheta_l \mu }}{L_{\mu \mu}} -  \mu_{\vartheta_l} \frac{  L_{ \vartheta_j \mu \mu  }}{L_{\mu \mu}}  - \mu_{\vartheta_j}  \frac{  L_{ \vartheta_l \mu \mu} }{L_{\mu \mu}} 
    -  \mu_{\vartheta_j} \mu_{\vartheta_l}  \frac{  L_{\mu \mu \mu } }{L_{\mu \mu}}.
\end{align*} 
for $k,j,l = 1,\ldots, p+1$.
The partial derivatives of $ L(\vartheta,\mu(\vartheta))$ evaluated at $\vartheta = \vartheta_0$ can be expressed as 
\begin{align*}
    L_{\vartheta_k} 
    &= \sum_{t = 1}^T \left(S_{t} - c_{t}\left(\mu(\vartheta_0)-\mu_0\right)\right) \left( S_{\vartheta_k t} -  c_{\vartheta_k t}\left(\mu(\vartheta_0)-\mu_0\right)\right), \\
     L_{\vartheta_k \vartheta_j } &= \sum_{t = 1}^T \left( S_{\vartheta_j t} -  c_{\vartheta_j t}\left(\mu(\vartheta_0)-\mu_0\right)\right) \left( S_{\vartheta_k t} -  c_{\vartheta_k t}(\vartheta_0)\left(\mu(\vartheta_0)-\mu_0\right)\right) \\
    &\ \ \ + \sum_{t = 1}^T  \left(S_{t} - c_{t}\left(\mu(\vartheta_0)-\mu_0\right)\right) \left( S_{ \vartheta_k  \vartheta_j t} -  c_{\vartheta_k  \vartheta_j t}\left(\mu(\vartheta_0)-\mu_0\right)\right), \\
      L_{\vartheta_k \vartheta_j \vartheta_l} &= \sum_{t = 1}^T \left( S_{\vartheta_j \vartheta_l t} -  c_{\vartheta_j \vartheta_l t}\left(\mu(\vartheta_0)-\mu_0\right)\right) \left( S_{\vartheta_k t} -  c_{\vartheta_k t}\left(\mu(\vartheta_0)-\mu_0\right)\right) \\
    &\ \ \ + \sum_{t = 1}^T  \left(S_{\vartheta_j } - c_{\vartheta_j t}\left(\mu(\vartheta_0)-\mu_0\right)\right) \left( S_{ \vartheta_k \vartheta_l t} -  c_{\vartheta_k \vartheta_l t}\left(\mu(\vartheta_0)-\mu_0\right)\right) \\ 
     &\ \ \ + \sum_{t = 1}^T  \left(S_{\vartheta_l t} - c_{\vartheta_l t}\left(\mu(\vartheta_0)-\mu_0\right)\right) \left( S_{ \vartheta_k \vartheta_j t} -  c_{\vartheta_k \vartheta_j t}\left(\mu(\vartheta_0)-\mu_0\right)\right) \\
      &\ \ \ + \sum_{t = 1}^T  \left(S_{t} - c_{ t}\left(\mu(\vartheta_0)-\mu_0\right)\right) \left( S_{ \vartheta_k \vartheta_j \vartheta_l t} -  c_{\vartheta_k \vartheta_j \vartheta_l t}\left(\mu(\vartheta_0)-\mu_0\right)\right), \\
    L_{\vartheta_k \mu}&= - \sum_{t = 1}^T  c_{ t} \left( S_{\vartheta_k t} -  c_{\vartheta_k t}\left(\mu(\vartheta_0)-\mu_0\right)\right) \\
    &\ \ \ -  \sum_{t = 1}^T  \left( S_{0 t} -  c_{ t} \left(\mu(\vartheta_0)-\mu_0\right)\right) c_{\vartheta_k t},  \\
    L_{\vartheta_k \mu \mu }&=  2 \sum_{t = 1}^T  c_{ t}  c_{\vartheta_k t}, \\
    L_{\vartheta_k \vartheta_j \mu} &= -\sum_{t = 1}^T  c_{\vartheta_j t} \left( S_{\vartheta_k t} -  c_{\vartheta_k t}\left(\mu(\vartheta_0)-\mu_0\right)\right) \\
    &\ \ \ - \sum_{t = 1}^T  c_{t} \left( S_{ \vartheta_k \vartheta_j t} -  c_{\vartheta_k \vartheta_j t}\left(\mu(\vartheta_0)-\mu_0\right)\right) \\ 
     &\ \ \ - \sum_{t = 1}^T  \left(S_{\vartheta_j t} - c_{\vartheta_j t}\left(\mu(\vartheta_0)-\mu_0\right)\right) c_{\vartheta_k  t} \\
      &\ \ \ - \sum_{t = 1}^T  \left(S_{ t} - c_{t}\left(\mu(\vartheta_0)-\mu_0\right)\right)  c_{\vartheta_k \vartheta_j t}, \\
         L_{\mu} &= -\sum_{t = 1}^T \left(S_{t} - c_{t}\left(\mu(\vartheta_0)-\mu_0\right)\right)  c_{t}, \\
            L_{\mu \mu}&= \sum_{t = 1}^T  c^2_{t}, \\
          L_{\mu \mu \mu}&= 0.    
\end{align*}
Here, $\mu(\vartheta_0) = \hat{\mu}(\vartheta_0)$ and the stochastic term $S_{t} = S_{t}^+ + S_{t}^-$ is defined in \eqref{Sterm} and \eqref{Sterm-} and its derivatives in \eqref{firststoch}. The deterministic term $c_{t} $ is defined in \eqref{detc} and its derivatives in \eqref{firstdet}.
\end{lemma}
\begin{proof}[Proof of Lemma \ref{genderivatesLstar}.]
The proof of \eqref{genqw11}, \eqref{genqw12} and \eqref{genqw13} is given in \eqref{genproofqw12},\eqref{genproofqw13} and \eqref{genproofqw14}, respectively. The partial derivatives of  $L(\vartheta,\mu(\vartheta))$ follow from the relationship 
\begin{align*}
    L(\vartheta,\mu) = \frac{1}{2} \sum_{t = 1}^T\epsilon^2_t(d, \varphi, \mu),
\end{align*}
where $\epsilon_t(d, \varphi, \mu)$ is given in \eqref{GenRDEF} and its derivatives are provided in \eqref{firstd1}.
\end{proof}

Next, we find expressions for the first three derivatives of $L_{\mu_0}^*(\vartheta)$, given in \eqref{genlikmu1known}, and evaluate them at $\vartheta = \vartheta_0$.  

\begin{lemma} \label{genderivatesLstarmu}
Let the model for the data $x_t$, t = 1,$\ldots$,T, be given by \eqref{genq1} and assume that Assumption \ref{A1} holds. Then the derivatives of  $L_{\mu_0}^*(\vartheta)$, see \eqref{genlikmu1known}, evaluated at $\vartheta = \vartheta_0$ are given by
\begin{align}
     D_{\vartheta_k} L_{\mu_0}^* &= \sum_{t = 1}^T S_{t} S_{\vartheta_k t} , \label{genmuqw11}\\
     D_{\vartheta_k \vartheta_j} L_{\mu_0}^* &= \sum_{t = 1}^T  S_{\vartheta_j t} S_{\vartheta_k t} + \sum_{t = 1}^T  S_{t}  S_{ \vartheta_k  \vartheta_j t},  \label{genmuqw12} \\ D_{\vartheta_k \vartheta_j \vartheta_l} L_{\mu_0}^* &= \sum_{t = 1}^T  S_{t}  S_{ \vartheta_k \vartheta_j \vartheta_l t}   + \sum_{t = 1}^T  S_{\vartheta_j \vartheta_l t} S_{\vartheta_k t} +  \sum_{t = 1}^T S_{\vartheta_j } S_{ \vartheta_k \vartheta_l t} + \sum_{t = 1}^T  S_{\vartheta_l t} S_{ \vartheta_k \vartheta_j t}, \label{genmuqw13}
\end{align}
for $k,j = 1,\ldots, p+1$.  Here, the stochastic term $S_{t} = S_{t}^+ + S_{t}^-$ is defined in \eqref{Sterm} and \eqref{Sterm-} and its derivatives in \eqref{firststoch}.
\end{lemma}
\begin{proof}[Proof of Lemma \ref{genderivatesLstarmu}.]
Recall that
\begin{align*}
     L_{\mu_0}^*(\vartheta) = \frac{1}{2} \sum_{t = 1}^T \epsilon^2_t(d, \varphi, \mu_0),
\end{align*}
where  $\epsilon_t(d, \varphi, \mu)$ is given in \eqref{GenRDEF}. The second term in \eqref{GenRDEF} becomes zero when $\mu$ is equal to $\mu_0$. This simplifies the proof, which can now be easily derived.
\end{proof}

Finally, we find expressions for the first three derivatives of $L_m^*(\vartheta)$, see \eqref{genmlik}, and evaluate them $\vartheta = \vartheta_0$. 

\begin{lemma} \label{genderivatesLstarMCSS}
Let the model for the data $x_t$, t = 1,$\ldots$,T, be given by \eqref{genq1} and assume that the assumption \ref{A1} holds. Then the derivatives of  $L_m^*(\vartheta)$, given in \eqref{genmlik}, evaluated at $\vartheta = \vartheta_0$ are  given by
\begin{align}
    D_{\vartheta_k} L^* &=  m D_{\vartheta_k} L^*   + m_{\vartheta_k} L^*, \label{genmcssqw11}\\
    D_{\vartheta_k \vartheta_j} L^* &=  m D_{\vartheta_k \vartheta_j} L^* + m_{\vartheta_j} D_{\vartheta_k} L^* + m_{\vartheta_k \vartheta_j} L^* +  m_{\vartheta_k} D_{\vartheta_j}L^* ,  \label{genmcssqw12} \\
    D_{\vartheta_k \vartheta_j \vartheta_l} L^* &=  m D_{\vartheta_k \vartheta_j \vartheta_l} L^* +  m_{\vartheta_l} D_{\vartheta_k \vartheta_j} L^* + m_{\vartheta_j \vartheta_l} D_{\vartheta_k} L^* + m_{\vartheta_j} D_{\vartheta_k \vartheta_l} L^*  \nonumber \\
    &\ \ \ + m_{\vartheta_k \vartheta_j \vartheta_l} L^* +  m_{\vartheta_k \vartheta_j } D_{\vartheta_l}L^* + m_{\vartheta_k \vartheta_l} D_{\vartheta_j}L^* + m_{\vartheta_k} D_{\vartheta_j \vartheta_l}L^*, \label{genmcssqw13}
\end{align}
where the expressions for the derivatives of $L^*(\vartheta_0)$ are given in Lemma \ref{genderivatesLstar}. The modification term $m(\vartheta) = ( \sum_{t = 1}^T c_t^2(\vartheta) )^{1/(T-1)}$, see \eqref{genmodificationterm}, and its derivatives, evaluated at $\vartheta = \vartheta_0$, are given by
\begin{align*}
    m_{\vartheta_k} &= \frac{2}{T-1}  \left( \sum_{t = 1}^T c_t^2  \right)^{-\frac{T-2}{T-1}} \sum_{t = 1}^T  c_t  c_{\vartheta_k t},\\
   m_{\vartheta_k \vartheta_j}  &=  \frac{2}{T-1}  \left( \sum_{t = 1}^T c_t^2  \right)^{-\frac{T-2}{T-1}} \sum_{t = 1}^T \left( c_{\vartheta_j t}  c_{\vartheta_k t} + c_t  c_{\vartheta_k \vartheta_j  t}  \right)  \\ &{\ \ } - 4\frac{T-2}{(T-1)^2}  \left( \sum_{t = 1}^T c_t^2 \right)^{-\frac{2T-3}{T-1}} \sum_{t = 1}^T  c_t  c_{\vartheta_k t} \sum_{t = 1}^T  c_t  c_{\vartheta_j t} , \\
    m_{\vartheta_k \vartheta_j  \vartheta_l }  &=  \frac{2}{T-1}  \left( \sum_{t = 1}^T c_t^2  \right)^{-\frac{T-2}{T-1}} \sum_{t = 1}^T \left( c_{\vartheta_j \vartheta_l t}  c_{\vartheta_k t} +  c_{\vartheta_j t}  c_{\vartheta_k \vartheta_l t} + c_{\vartheta_l t}  c_{\vartheta_k \vartheta_j  t} + c_t  c_{\vartheta_k \vartheta_j \vartheta_l t}  \right)  \\ 
    &\ \ \  - 4\frac{T-2}{(T-1)^2}  \left( \sum_{t = 1}^T c_t^2  \right)^{-\frac{2T-3}{T-1}}  \sum_{t = 1}^T c_t c_{\vartheta_l  t} \sum_{t = 1}^T \left( c_{\vartheta_j t}  c_{\vartheta_k t} + c_t  c_{\vartheta_k \vartheta_j  t}  \right) \\
    &\ \ \ - 4\frac{T-2}{(T-1)^2}  \left( \sum_{t = 1}^T c_t^2 \right)^{-\frac{2T-3}{T-1}} \left(\left( \sum_{t = 1}^T  c_{\vartheta_l t}  c_{\vartheta_k t} + \sum_{t = 1}^T  c_t  c_{\vartheta_k \vartheta_l t} \right) \sum_{t = 1}^T  c_t  c_{\vartheta_j t}  \right.\\ 
    &\left.\ \ \ +  \sum_{t = 1}^T  c_t  c_{\vartheta_k t} \left( \sum_{t = 1}^T  c_{\vartheta_l t}  c_{\vartheta_j t} + \sum_{t = 1}^T  c_t  c_{\vartheta_j \vartheta_l t}  \right) \right) \\
     &\ \ \ + 8\frac{(T-2)(2T-3)}{(T-1)^3}  \left( \sum_{t = 1}^T c_t^2 \right)^{-\frac{3T-4}{T-1}}  \sum_{t = 1}^T c_t c_{\vartheta_l t} \sum_{t = 1}^T  c_t  c_{\vartheta_k t} \sum_{t = 1}^T  c_t  c_{\vartheta_j t}.
\end{align*}
\end{lemma}
\begin{proof}[Proof of Lemma \ref{genderivatesLstarMCSS}.]
Proof is straightforward due to the multiplicative form of the MCSS objective function, see \eqref{genmlik}.
\end{proof}

\subsection{Preliminary results} \label{genlemmaap}

In this section, we present findings that play a central role in the approximation of the derivatives. Appendix \ref{UB} contains useful bounds, while Appendix \ref{UB1} presents results related to fractional coefficients
$\pi_i (a)$ in Section \ref{model}, their derivatives, the weights of the lag polynomial $\omega (s, \varphi)$ in \eqref{repmainf}, its inverse $\phi (s, \varphi)$ in Assumption \ref{A1} along with their derivatives. In Appendix
\ref{UB2}, we investigate the limiting behaviour of the centred product moments, which are particularly relevant in the later expression of the biases. Lastly, Appendix \ref{UB3} focuses on the expectation of the CSS score
function, which is a part of the bias term in the CSS estimator.

\subsubsection{Useful bounds}\label{UB} 
In this section, we provide some general results that are useful for finding the approximation of the derivatives. We sometimes apply them in the remainder without special reference.

\begin{lemma} \label{series}
    For any $d>-1$, as $T\rightarrow \infty$,
     $   1/T^{\left(d+1\right)} \sum_{t = 1}^T t^d \rightarrow 1/(d+1). $ 
\end{lemma}
\begin{proof}[Proof of Lemma \ref{series}] See \textcite[Lemma S.10]{hualde2020truncated}. 
\end{proof}

Next, we present some useful bounds that are frequently used in the remainder of the appendix.

\begin{lemma} \label{genbounds}
For $m \geq 0$ and $c < \infty$, 
\begin{align}
    \sum_{n = 1}^N (1+\log(n))^m n^{\alpha} &\leq c (1+\log(N))^m N^{\alpha + 1} \text{ if } \alpha > -1, \label{lA1} \\
    \sum_{n = N}^{\infty} (1+\log(n))^m n^{\alpha} &\leq c (1+\log(N))^m N^{\alpha + 1} \text{ if } \alpha < -1.\label{lA2}
\end{align}
For $\alpha< 0$, and any $\beta$ it holds that 
\begin{align}
    \sum_{n = 1}^{t-1} n^{\alpha-1} (t-n)^{\beta - 1} \leq c t^{\max(\alpha-1,\beta-1)}. \label{lA3}
\end{align}
For $\alpha\geq 0$, and any $\beta$ it holds that 
\begin{align}
    \sum_{n = 1}^{t-1} n^{\alpha-1} (t-n)^{\beta - 1} \leq c (1 + \log(t)) t^{\max(\alpha+\beta-1,\alpha-1,\beta-1)}. \label{lA4}
\end{align}
For $\alpha+\beta < 1$ and $\beta > 0$ it holds that 
\begin{align}
    \sum_{k = 1}^{\infty} (k+h)^{\alpha-1} k^{\beta-1} (1+\log(k+h))^n \leq c h^{\alpha+\beta-1} (1+\log(h))^n. \label{lA5}
\end{align}

\end{lemma}
\begin{proof}[Proof of Lemma \ref{genbounds}] Proof of \eqref{lA1} and \eqref{lA2}: See \textcite[Lemma A.1]{johansen2016role}. \\
Proof of \eqref{lA3}: The proof follows a similar approach to the proof in \textcite[Lemma 1]{hualde2011gaussian}.
Clearly,
\begin{align*}
    \sum_{n = 1}^{t-1} n^{\alpha-1} (t-n)^{\beta - 1} &\leq  c \sum_{n = 1}^{\lfloor t/2 \rfloor} n^{\alpha-1} (t-n)^{\beta - 1} +  c \sum_{n = \lfloor t/2 \rfloor}^{t-1} n^{\alpha-1} (t-n)^{\beta - 1} \\
    &\leq c t^{\beta - 1} \sum_{n = 1}^{\lfloor t/2 \rfloor} n^{\alpha-1} +  c t^{\alpha-1} \sum_{n = \lfloor t/2 \rfloor}^{t-1} (t-n)^{\beta - 1}, \\
\end{align*}
because $\alpha < 0$ the first summand is $O(1)$ and the second summand is $O(1)$ if $\beta < 0$, $O(\log(t))$ if $\beta = 0$, and $O(t^\beta)$ if $\beta > 0$. \\
Proof of \eqref{lA4} and \eqref{lA5}: See \textcite[Lemma A.5]{johansen2016role}.
\end{proof}

\subsubsection{Bounds for the fractional coefficients and short-run dynamics} \label{UB1}

Next, we present findings concerning the fractional coefficients $\pi_i (a)$ in Section \ref{model}, their derivatives, the weights of the lag polynomial $\omega (s, \varphi)$ in \eqref{repmainf}, its inverse $\phi (s, \varphi)$ in Assumption
\ref{A1} and their derivatives.

\begin{lemma}\label{r11}
    For $m \geq 0$ and $j \geq 1$ it holds that 
    \begin{align}
        |D^m \pi_j(u)| \leq c (1+\log(j))^{m} j^{u-1} \label{r11_1}
    \end{align}
    Under Assumption \ref{A3} and \ref{A1} it follows, as $j \rightarrow \infty$,
    \begin{align}
    \underset{\varphi \in \Phi  }{\sup} \left| \omega_j(\varphi) \right|  &= O(j^{-1-\varsigma}), \label{r11_21}\\
     \underset{\varphi \in \Phi  }{\sup} \left| \phi_j(\varphi) \right|  &= O(j^{-1-\varsigma}), \label{r11_211}\\
     \underset{\varphi \in \Phi  }{\sup} \left| \frac{ \partial \phi_j(\varphi)}{\partial \varphi_i }\right|  &= O(j^{-1-\varsigma}),\label{r11_2} \\
    \underset{\varphi \in \Phi  }{\sup} \left| \frac{ \partial^2 \phi_j(\varphi)}{\partial \varphi_i \partial \varphi_l}\right|  &= O(j^{-1-\varsigma}),\label{r11_3} \\
    \underset{\varphi \in \Phi  }{\sup} \left| \frac{ \partial^3 \phi_j(\varphi)}{\partial \varphi_i \partial \varphi_l \partial \varphi_k}\right|  &= O(j^{-1-\varsigma}),\label{r11_4}
    \end{align}
for $i,l,k = 1,\ldots,p$ and where $1/2 < \varsigma  \leq 1$
\end{lemma}
\begin{proof}[Proof of Lemma \ref{r11}] Proof of \eqref{r11_1}:  See \textcite[Lemma A.3]{johansen2016role} \\
Proof of \eqref{r11_21}-\eqref{r11_4}: See \textcite[page 46]{zygmund1977trigonometric} and \textcite[page 3155 and page 3169]{hualde2011gaussian}.
\end{proof}

Next, we present bounds for the deterministic term in \eqref{detc} and their derivatives and the in-sample terms in \eqref{defb}-\eqref{defh} and the pre-sample terms in \eqref{defc_tr}-\eqref{defh_tr}.
 
\begin{lemma}\label{r12}
For any integer $m^* \geq 0$ and under Assumption \ref{A3} and \ref{A1},
\begin{align}
      \underset{\varphi \in \Phi  }{\sup} \left|\frac{\partial^{m^*} c_t(\vartheta)}{ \partial d^{m^*} } \right| = O(t^{\max(-d,-1-\varsigma)} \log^{m^*}(t)), \label{r12_1}\\
      \underset{\varphi \in \Phi  }{\sup} \left|\frac{\partial^{m^*+1} c_t(\vartheta)}{ \partial d^{m^*} \partial \varphi_i(\varphi)} \right| = O(t^{\max(-d,-1-\varsigma)} \log^{m^*}(t)), \label{r12_2} \\
     \underset{\varphi \in \Phi  }{\sup}\left|\frac{\partial^{m^*+2} c_t(\vartheta)}{ \partial d^{m^*} \partial \varphi_i \partial \varphi_l} \right| = O(t^{\max(-d,-1-\varsigma)} \log^{m^*}(t)), \label{r12_3}\\
      \underset{\varphi \in \Phi  }{\sup}\left|\frac{\partial^{m^*+3} c_t(\vartheta)}{ \partial d^{m^*} \partial \varphi_i \partial \varphi_l \partial \varphi_k } \right| = O(t^{\max(-d,-1-\varsigma)} \log^{m^*}(t)), \label{r12_4} 
\end{align}
for $i,l,k = 1,\ldots,p$ and where $1/2 < \varsigma  \leq 1$. For the in-sample terms $b_{zi}(\varphi)$ and $h_{dzi}(\varphi)$ defined in \eqref{defb} and \eqref{defh} 
\begin{align}
\underset{\varphi \in \Phi  }{\sup}|b_{zj}(\varphi)| &=  O(j^{-1-\varsigma}), \label{bproof} \\  
\underset{\varphi \in \Phi  }{\sup}|h_{dzj}(\varphi)| &= O(j^{-1}), \label{hproof}
\end{align}
where $z \in \{ \varphi_k, \varphi_k \varphi_j, \varphi_k \varphi_j \varphi_l \}$ for $k,j,l = 1,\ldots,p$. 
\end{lemma}
\begin{proof}[Proof of Lemma \ref{r12}] Proof of \eqref{r12_1}-\eqref{r12_4}: We give the proof of \eqref{r12_4} only, as the bounds on derivatives of the weight of the inverse lag polynomials are the same order according to Lemma \ref{r11}, resulting in a similar proof. From $k_{0t}(d) = \pi_{t-1}(1-d)$ and Lemma \ref{r11},
\begin{align*}
    \underset{\varphi\in\Phi}{\sup}  \left|\frac{\partial^{m+3} c_t(\vartheta)}{ \partial d^m \partial \varphi_i \partial \varphi_l \partial \varphi_k } \right|
     &\leq c \sum_{j = 1}^{t-1} \log^m(j) j^{-1-\varsigma} (t-j-1)^{-d}, \\
     &\leq c \log^m(t) \sum_{j = 1}^{t-1} j^{-1-\varsigma} (t-j-1)^{-d},
\end{align*}
because $\varsigma > 1/2$ this summand is $O(t^{\max(-d,-1-\varsigma)})$ from Lemma \ref{genbounds}. \\
Proof of \eqref{hproof} and \eqref{bproof}: From Lemma \ref{r11}, 
\begin{align*}
     \underset{\varphi\in\Phi}{\sup} |b_{zi}(\varphi)| \leq c \sum_{s = 1}^{i-1}  s^{-1-\varsigma} (i-s)^{-1-\varsigma}  = O(i^{-1-\varsigma}).
\end{align*}
The last equality follows from Lemma \ref{genbounds} because $\varsigma > 1/2$. 
From Lemma \ref{genbounds} and the bound above: $    \underset{\varphi\in\Phi}{\sup} |h_{dzi}(\varphi)| = O(\sum_{s = 1}^{i-1} (i-s)^{-1} s^{-1-\zeta} ) = O(i^{-1}).$
\end{proof}

\subsubsection{Limit behaviour of the centred product moments} \label{UB2}

Let $k,j,l = 1,\ldots, p+ 1$, define the centred product moments of the derivative of the stochastic terms in \eqref{firststoch} as
\begin{align}
    M^{+}_{0,\vartheta_k  T} &= \sigma^{-2}_0 T^{-1/2} \sum_{t = 1}^T \left( S^{+}_{t} S^{+}_{\vartheta_k t} - E\left( S^{+}_{t} S^{+}_{\vartheta_k t} \right) \right), \label{genMta} \\
     M^{+}_{0,\vartheta_k \vartheta_j   T} &= \sigma^{-2}_0 T^{-1/2} \sum_{t = 1}^T \left( S^{+}_{t} S^{+}_{\vartheta_k \vartheta_j  t} - E\left( S^{+}_{t} S^{+}_{\vartheta_k \vartheta_j  t} \right) \right), \label{genMtb} \\
     M^{+}_{\vartheta_k,\vartheta_j   T} &= \sigma^{-2}_0 T^{-1/2} \sum_{t = 1}^T \left( S^{+}_{\vartheta_k t} S^{+}_{ \vartheta_j  t} - E\left( S^{+}_{\vartheta_k t} S^{+}_{\vartheta_j  t} \right) \right), \label{genMtc} \\
     M^{+}_{0,\vartheta_k \vartheta_j \vartheta_l  T} &= \sigma^{-2}_0 T^{-1/2} \sum_{t = 1}^T \left( S^{+}_{t} S^{+}_{ \vartheta_k \vartheta_j \vartheta_l  t} - E\left( S^{+}_{t} S^{+}_{\vartheta_k \vartheta_j \vartheta_l  t} \right) \right), \label{genMtd} \\
     M^{+}_{\vartheta_k,\vartheta_j \vartheta_l  T} &= \sigma^{-2}_0 T^{-1/2} \sum_{t = 1}^T \left( S^{+}_{\vartheta_k t} S^{+}_{ \vartheta_j \vartheta_l  t} - E\left( S^{+}_{\vartheta_k t} S^{+}_{\vartheta_j \vartheta_l  t} \right) \right), \label{genMte}
\end{align}
and define some of the corresponding vector forms of the centred product moments as 
\begin{align}
     M^{+}_{0,\vartheta T} &= (M^{+}_{0,\vartheta_1  T}, M^{+}_{0,\vartheta_2  T},\ldots, M^{+}_{0,\vartheta_{p+1}  T})' ,\label{genM1}\\
     M^{+}_{0,\vartheta_k \vartheta T} &= ( M^{+}_{0,\vartheta_k \vartheta_1 T} ,  M^{+}_{0,\vartheta_k \vartheta_2 T} ,\ldots,  M^{+}_{0,\vartheta_k \vartheta_{p+1} T} )' ,\label{genvM2}\\
      M^{+}_{\vartheta_k,\vartheta T} &= ( M^{+}_{\vartheta_k,\vartheta_1 T} ,   M^{+}_{\vartheta_k,\vartheta_2 T} ,\ldots,   M^{+}_{\vartheta_k,\vartheta_{p+1} T} )', \label{genvM3}
\end{align}
and matrix form as 
\begin{align}
 M^{+}_{0,\vartheta \vartheta' T} &=  ( M^{+}_{0,\vartheta_1 \vartheta T}, M^{+}_{0,\vartheta_2 \vartheta T}, \ldots,  M^{+}_{0,\vartheta_{p+1} \vartheta T}  ) ,\label{genM2}\\
       M^{+}_{\vartheta,\vartheta' T} &= (  M^{+}_{\vartheta_1,\vartheta T},  M^{+}_{\vartheta_2,\vartheta T}, \ldots,  M^{+}_{\vartheta_{p+1},\vartheta T}  ) .\label{genM3}
\end{align}

Note that for $S_t^+ = \epsilon_t$, by Lemma \ref{explicitforms}. As a direct consequence, we observe that $E\left( S^+_{t} S^+_{\vartheta_k t} \right) =E\left( S^+_{t} S^+_{\vartheta_k \vartheta_j t} \right) = E\left( S^+_{t} S^+_{\vartheta_k \vartheta_j \vartheta_l  t} \right) = 0$.

We next present a lemma on the limiting behavior of the centred product moments.

\begin{lemma} \label{genlemmma1}
Suppose that Assumptions \ref{A2}-\ref{A5} hold. Then, for $T\rightarrow \infty$, it holds that $M^+_{0,\vartheta T}$ is asymptotic normal with mean zero and the variance of $M^+_{0,\vartheta T}$ is 
\begin{align*}  
     E\left(M^+_{0,\vartheta T} (M^+_{0,\vartheta T})' \right) = A + O(T^{-1} \log(T)), 
\end{align*}
where $A$ is the inverse of the variance-covariance matrix given in \eqref{genA}.
Furthermore, $M^+_{0,\vartheta_k  T} = O_P(1)$, $M^+_{0,\vartheta_k \vartheta_j   T} = O_P(1)$, $ M^+_{\vartheta_k,\vartheta_j   T} = O_P(1)$, $ M^+_{0,\vartheta_k \vartheta_j \vartheta_l  T} = O_P(1)$ and $M^+_{\vartheta_k,\vartheta_j \vartheta_l  T} = O_P(1)$.

\end{lemma}
\begin{proof}[Proof of Lemma \ref{genlemmma1}]
The proof of asymptotic normality of $M^+_{0,\vartheta T}$ and the limiting variance is given in \textcite[(2.54) and (2.55)]{hualde2011gaussian}. The order the rest term comes from the (1,1)-th element of the matrix $E\left(M^+_{0,\vartheta T} (M^+_{0,\vartheta T})' \right) $, 
\begin{align*}
    \sigma_0^{-2} T^{-1}  \sum_{t = 1}^T E\left(S^+_{\vartheta_1t}\right)^2 = T^{-1}  \sum_{t = 1}^T \sum_{k = 1}^{t-1} \frac{1}{k^2} 
    = \sum_{k = 1}^{\infty} \frac{1}{k^2} - T^{-1}  \sum_{t = 1}^T \sum_{k = t}^{\infty} \frac{1}{k^2} 
    = \zeta_2 + O(T^{-1}\log(T)),
\end{align*}
using $\sum_{k = t}^{\infty} k^{-2} = O(t^{-1})$, see \eqref{lA2}. It can be straightforwardly shown that the other elements in this matrix have a rest term of $O(T^{-1})$. We have that $S_t^+ = \epsilon_t$ and therefore the proof of $M^+_{0,\vartheta_k  T} = O_P(1)$, $M^+_{0,\vartheta_k \vartheta_j   T} = O_P(1)$ and  $M^+_{0,\vartheta_k \vartheta_j  \vartheta_l T} = O_P(1)$ are straightforward and can be derived from Lemmata \ref{genbounds}, \ref{r11} and \ref{r12}. The proofs of $ M^+_{\vartheta_k,\vartheta_j   T} = O_P(1)$ and $M^+_{\vartheta_k,\vartheta_j \vartheta_l  T} = O_P(1)$ are analogous, so we will present the proof for $M^+_{\vartheta_k,\vartheta_j \vartheta_l  T} = O_P(1)$. We first show $M^+_{\vartheta_k,\vartheta_j \vartheta_l  T} = O_P(1)$ for the $d$-direction, i.e.\ $k = j = l = 1$. Let $ z_t =S^+_{\vartheta_k t} S^+_{\vartheta_j\vartheta_l t}-E(S^+_{\vartheta_k t} S^+_{\vartheta_j\vartheta_l t})$ and
$M^+_{\vartheta_k,\vartheta_j \vartheta_l  T} = \sigma_0^{-2}T^{-1/2}\sum_{t=1}^T z_t$. From Lemma \ref{explicitforms}
\begin{align*}
    S^+_{\vartheta_1 t} &= \sum_{r = 1}^{t-1} a_r \epsilon_{t-r}, \\
    S^+_{\vartheta_1 \vartheta_1 t} &= \sum_{r = 1}^{t-1} b_r \epsilon_{t-r},  
\end{align*}
with $a_r = - D_d \pi_r(0)$ and $b_r = D_{dd} \pi_r(0)$. By Lemma \ref{r11} we have that
$|a_r| \leq c r^{-1} \log(r)$ and $|b_r| \leq c  r^{-1} \log^2(r)$. 
Hence 
\begin{align}
    \sum_{r = 1}^{\infty} a^2_r < \infty \label{crosspproof1} \\
    \sum_{r = 1}^{\infty} b^2_r < \infty.\label{crosspproof2}
\end{align}

We first show that 
\begin{align}
\sum_{t = 1}^T E(z^2_t) = O(T). \label{toshowcrosspproof}   
\end{align}
Note that $E(z^2_t) = Var( S^+_{\vartheta_k t} S^+_{\vartheta_j\vartheta_l t}) \leq E [ (S^+_{\vartheta_k t})^2 (S^+_{\vartheta_j\vartheta_l t})^2]$ and by Cauchy-Schwarz 
\begin{align}
     E [ (S^+_{\vartheta_k t})^2 (S^+_{\vartheta_j\vartheta_l t})^2] \leq [E(S^+_{\vartheta_k t})^4]^{1/2} [ E(S^+_{\vartheta_j\vartheta_l t})^4]^{1/2}. \label{qwe1b1}
\end{align}
For a linear process $y_t = \sum_{r = 0}^{t-1} e_r \epsilon_{t-r}$ with $\sum_{r = 0}^{t-1} e^2_r < \infty$ and independent and identical distributed errors satisfying $E \epsilon_t = 0$ and $E \epsilon^4_t < \infty$, we have 
\begin{align*}
    E(y^4_t) = \sum_{r = 0}^{t-1} e^4_r E(\epsilon_1^4) + 3  \sum_{r = 0}^{t-1} \sum_{\substack{s=0\\ s\neq r}}^{t-1} e^2_r e^2_s (E(\epsilon^2_1))^2.
\end{align*}
Since $\sum^{t-1}_{r = 0} e_r^4 \leq (\sum_{r = 0}^{\infty} e_r^2)^2$ and $\sum_{r = 0}^{t-1} \sum_{s = 0, s \neq r}^{t-1} e^2_r e^2_s \leq (\sum_{r = 0}^{\infty} e_r^2)^2$, it follows that  
\begin{align}
    E(y^4_t) \leq c \left(\sum_{r = 0}^{\infty} e_r^2\right)^2 \label{boundfourthmoments1}
\end{align}
for $c < \infty$.

Applying \eqref{boundfourthmoments1} with $e_r = a_r$ and $e_r = b_r$ and using \eqref{crosspproof1} and \eqref{crosspproof2}, we obtain  
\begin{align*}
    E(S^+_{\vartheta_k t})^4 &= O(1) \\
     E(S^+_{\vartheta_j\vartheta_l t})^4 &= O(1) .
\end{align*}
Substituting these bound into \eqref{qwe1b1} yields
\begin{align*}
    \sup_t E(z^2_t) <\infty
\end{align*}
and hence \eqref{toshowcrosspproof} follows.

Next, we show that 
\begin{align}
 \sum_{h = 1}^{\infty} \sup_{t \geq 1} |E(z_t z_{t+h})| < \infty.  \label{boundfourthmoments2} 
\end{align}
Fix $h \geq 1$. By Section 3.2, Eq. (3.2) of \textcite{McCullagh87}
\begin{align}
    E(z_t z_{t+h}) &= Cov(S^+_{\vartheta_j\vartheta_l t},S^+_{\vartheta_j\vartheta_l t+h}) Cov(S^+_{\vartheta_k t},S^+_{\vartheta_k t+h}) + Cov(S^+_{\vartheta_j\vartheta_l t},S^+_{\vartheta_k t+h}) Cov(S^+_{\vartheta_k t},S^+_{\vartheta_j\vartheta_l t+h}) \nonumber\\ 
    & \ \ \ + cum(S^+_{\vartheta_j\vartheta_l t},S^+_{\vartheta_j\vartheta_l t+h},S^+_{\vartheta_k t},S^+_{\vartheta_k t+h}).\label{crosspproof4}
\end{align}
Thus it suffices to bound the three terms on the right-hand side.
Using the linear representations 
\begin{align*}
    Cov(S^+_{\vartheta_j\vartheta_l t},S^+_{\vartheta_j\vartheta_l t+h}) &= \sigma_0^2 \sum_{r = 0}^{t-1} a_r a_{r+h} \\
    Cov(S^+_{\vartheta_k t},S^+_{\vartheta_k t+h}) &=  \sigma_0^2 \sum_{r = 0}^{t-1} b_r b_{r+h} \\
    Cov(S^+_{\vartheta_j\vartheta_l t},S^+_{\vartheta_k t+h})&= \sigma_0^2 \sum_{r = 0}^{t-1} a_r b_{r+h} \\
    Cov(S^+_{\vartheta_k t},S^+_{\vartheta_j\vartheta_l t+h}) &= \sigma_0^2 \sum_{r = 0}^{t-1} b_r a_{r+h}.
\end{align*}
Moreover, using $|a_r| \leq c \log(r) r^{-1}$ and $|b_r| \leq c \log^2(r) r^{-1}$, we obtain the bounds
\begin{align*}
    \sup_{t \geq 1} |Cov(S^+_{\vartheta_j\vartheta_l t},S^+_{\vartheta_j\vartheta_l t+h})| &\leq c \log(1+h) h^{-1} \\
    \sup_{t \geq 1} |Cov(S^+_{\vartheta_k t},S^+_{\vartheta_k t+h})|  &\leq c \log^3(1+h) h^{-1} \\ 
    \sup_{t \geq 1} |Cov(S^+_{\vartheta_j\vartheta_l t},S^+_{\vartheta_k t+h})| &\leq c \log^2(1+h) h^{-1} \\ 
    \sup_{t \geq 1} |Cov(S^+_{\vartheta_k t},S^+_{\vartheta_j\vartheta_l t+h})| &\leq c \log^2(1+h) h^{-1}. 
\end{align*}

For the fourth-cumulant term, the only nonzero contributions come when all four innovation indices coincide yielding
\begin{align*}
    cum(S^+_{\vartheta_j\vartheta_l t},S^+_{\vartheta_j\vartheta_l t+h},S^+_{\vartheta_k t},S^+_{\vartheta_k t+h}) = cum(\epsilon_0,\epsilon_0,\epsilon_0,\epsilon_0) \sum_{r = 0}^{t-1} a_r b_r a_{r+h} b_{r+h}.
\end{align*}
Furthermore, 
\begin{align*}
    \sup_{t \geq 1} \sum_{r = 0}^{t-1} |a_r b_r a_{r+h} b_{r+h}| \leq c \log^2(1+h) h^{-2}
\end{align*}
Combining these bounds with \eqref{crosspproof4} gives
\begin{align*}
     \sup_{t \geq 1} |E(z_t z_{t+h})| \leq c \log^4(h) h^{-2}. 
\end{align*}
Since $\sum_{h = 2}^{\infty} \log^4(h) h^{-2} < \infty$,  \eqref{boundfourthmoments2} follows.

Finally, since $E(z_t) = 0$, 
\begin{align*}
    E(\sum_{t = 1}^T z_t)^2 = \sum_{t = 1}^T E(z^2_t) + 2 \sum_{h = 1}^{T-1} \sum_{t= 1}^{T-h} E(z_t z_{t+h}).
\end{align*}
The first term is $O(T)$ by \eqref{boundfourthmoments1} and the second term is $O(T)$ by \eqref{boundfourthmoments2}. Hence  $M^+_{\vartheta_1,\vartheta_1 \vartheta_1  T} = O_P(1)$.

For the other directions, the same arguments applies upon taking the appropriate $a_r$ and $b_r$ and using the corresponding bounds in Lemmata \eqref{r11} or \eqref{r12}.

\end{proof}

\begin{lemma} \label{genlemmma1_1}
Suppose that Assumptions \ref{A2}-\ref{A5} holds. The covariances of  $M^+_{0,\vartheta_k  T}$ and  $M^+_{0,\vartheta_j \vartheta_l}$ are given by 
\begin{align*}
      E\left(  M^+_{0,\vartheta_k  T}  M^+_{0,\vartheta_j \vartheta_l   T} \right) = \sigma_0^{-2} T^{-1}  \sum_{t = 1}^T E \left( S^+_{\vartheta_k t}   S^+_{\vartheta_j \vartheta_l  t} \right), 
\end{align*}
for $k,j,l \in \{1,\ldots,p+1 \}$. 
For $T\rightarrow \infty$, it holds that 
\begin{align}
     \sigma_0^{-2} T^{-1}\sum_{t = 1}^T  E \left( S_{\vartheta_1 t}^+ S_{\vartheta_1 \vartheta_1 t}^+  \right) &= -2 \zeta_3 + O(T^{-1} \log^4(T) ), \label{l1_1a} \\
     \sigma_0^{-2} T^{-1} \sum_{t = 1}^T  E \left( S_{\vartheta_1 t}^+ S_{\vartheta_1 \vartheta_l t}^+  \right) &=    \sum_{i = 2}^{\infty} i^{-1} h_{\vartheta_1 \vartheta_l i}(\varphi_0) + O(T^{-1} \log(T)), \label{l1_1b}\\
     \sigma_0^{-2} T^{-1} \sum_{t = 1}^T  E \left( S_{\vartheta_1 t}^+ S_{\vartheta_j \vartheta_l t}^+  \right) &=  -    \sum_{i = 1}^{\infty}  i^{-1} b_{\vartheta_j \vartheta_l i}(\varphi_0) + O( T^{-1}), \label{l1_1c}\\
    \sigma_0^{-2} T^{-1}\sum_{t = 1}^T  E \left( S_{\vartheta_k t}^+ S_{\vartheta_1 \vartheta_1 t}^+  \right) &=  \sum_{i = 0}^{\infty} D_{dd} \pi_{i}(0)  b_{\vartheta_k i}(\varphi_0)  + O(T^{-1}), \label{l1_1d}\\
     \sigma_0^{-2} T^{-1} \sum_{t = 1}^T  E \left( S_{\vartheta_k t}^+ S_{\vartheta_1 \vartheta_l t}^+  \right) &= -    \sum_{i = 2}^{\infty}   b_{\vartheta_k i}(\varphi_0) h_{\vartheta_1 \vartheta_l i}(\varphi_0) + O( T^{-1}), \label{l1_1e} \\
     \sigma_0^{-2} T^{-1} \sum_{t = 1}^T  E \left( S_{\vartheta_k t}^+ S_{\vartheta_j \vartheta_l t}^+  \right) &=     \sum_{i = 1}^{\infty} b_{\vartheta_k i}(\varphi_0) b_{\vartheta_j \vartheta_l i}(\varphi_0) + O( T^{-1}), \label{l1_1f}
\end{align}
for $k,j,l \in \{2,\ldots,p+1 \}$.

\end{lemma}
\begin{proof}[Proof of Lemma \ref{genlemmma1_1}]
From Lemma \ref{explicitforms} we have that  $S^+_{t} = \epsilon_t$ and $ S^+_{\vartheta_k t}$, $S^+_{\vartheta_j \vartheta_l t}$ are weighted sums of $\epsilon_{1},\ldots,\epsilon_{t-1}$, so that  
\begin{align}
      E\left(  M^+_{0,\vartheta_k  T}  M^+_{0,\vartheta_j \vartheta_l   T} \right) &= 
      \sigma_0^{-4} T^{-1} E \left( \sum_{t = 1}^T  S^+_{t} S^+_{\vartheta_k t} \sum_{s = 1}^T S^+_{s} S^+_{\vartheta_j \vartheta_l  s} \right) \nonumber\\ 
      &=  \sigma_0^{-4} T^{-1}  \sum_{t = 1}^T E \left( \left(S^+_{t} \right)^2  S^+_{\vartheta_k t} S^+_{\vartheta_j \vartheta_l  t} \right) \nonumber \\
      &= \sigma_0^{-2} T^{-1}  \sum_{t = 1}^T E \left( S^+_{\vartheta_k t}   S^+_{\vartheta_j \vartheta_l  t} \right), \label{EXP1} 
\end{align}
where the last inequality uses the independence of $\left(S^+_{t} \right)^2$ and $S^+_{\vartheta_k t} S^+_{\vartheta_j \vartheta_l  t}$. 

Proof of \eqref{l1_1a}: Consider the case $k=j=l = 1$ for \eqref{EXP1}. We have 
\begin{align*}
      \sigma_0^{-2} T^{-1} \sum_{t = 1}^T  E \left( S_{\vartheta_1 t}^+ S_{\vartheta_1 \vartheta_1 t}^+  \right) &=  -  \sigma_0^{-2} T^{-1} \sum_{t = 1}^T  E \left( \left( \sum_{k = 0}^{t-1} D_d \pi_{k}(0) \epsilon_{t-k} \right) \left( \sum_{k = 0}^{t-1} D_{dd} \pi_{k}(0) \epsilon_{t-k} \right) \right) \\
     &= - T^{-1} \sum_{t = 1}^T \sum_{k = 0}^{t-1} D_{dd} \pi_{k}(0)  D_d \pi_{k}(0) \\
     &= -  \sum_{k = 0}^{\infty} D_{dd} \pi_{k}(0)  D_d \pi_{k}(0) + T^{-1} \sum_{t = 1}^T \sum_{k = t}^{\infty} D_{dd} \pi_{k}(0)  D_d \pi_{k}(0).
\end{align*}
Then from \eqref{dpi1} and \eqref{dpi2}
  $  - \sum_{k = 0}^{\infty} D_{dd} \pi_{k}(0)  D_d \pi_{k}(0) = -2 \sum_{ k = 2}^{\infty} k^{-2} \sum_{j = 1}^{k-1} j^{-1} = -2 \zeta_3, $
where the last equality follows from \textcite[Lemma B.2]{johansen2016role}
and from Lemmata \ref{genbounds}  and \ref{r11}  
\begin{align*}
     T^{-1} \sum_{t = 1}^T \sum_{k = t}^{\infty} D_{dd} \pi_{k}(0)  D_d \pi_{k}(0) =  O(T^{-1} \sum_{t = 1}^T \sum_{k = t}^{\infty} (1+\log(k))^3 k^{-2} )  =  O(T^{-1} \log^4(T) ).
\end{align*}

Proof of \eqref{l1_1b}: Consider the case $k=j =1$ and $l \geq 2$ for \eqref{EXP1}. Then 
\begin{align*}
    \sigma_0^{-2} T^{-1} \sum_{t = 1}^T  E \left( S_{\vartheta_1 t}^+ S_{\vartheta_1 \vartheta_l t}^+  \right) &=   \sigma_0^{-2} T^{-1} \sum_{t = 1}^T  E \left( \left( \sum_{k = 0}^{t-1} D_d \pi_{k}(0) \epsilon_{t-k} \right) \left( \sum_{i = 2}^{t-1} h_{\vartheta_1 \vartheta_l i}(\varphi_0)  \epsilon_{t-i} \right) \right) \\
    &=    T^{-1} \sum_{t = 1}^T  \sum_{i = 2}^{t-1}  i^{-1}  h_{\vartheta_1 \vartheta_l i}(\varphi_0) \\
    &=     \sum_{i = 2}^{\infty} i^{-1} h_{\vartheta_1 \vartheta_l i}(\varphi_0) - T^{-1} \sum_{t = 1}^T  \sum_{i = t}^{\infty}  i^{-1} h_{\vartheta_1 \vartheta_l i}(\varphi_0),
\end{align*}
where
\begin{align*}
   T^{-1} \sum_{t = 1}^T  \sum_{i = t}^{\infty} i^{-1} h_{\vartheta_1 \vartheta_l i}(\varphi_0) = O( T^{-1} \sum_{t = 1}^T  \sum_{i = t}^{\infty}  i^{-2} ) =   O( T^{-1} \log(T)) .
\end{align*}

Proof of \eqref{l1_1c}: Consider the case $k= 1$ and $j,l \geq 2$ for \eqref{EXP1}. Then
\begin{align*}
    \sigma_0^{-2} T^{-1} \sum_{t = 1}^T  E \left( S_{\vartheta_1 t}^+ S_{\vartheta_j \vartheta_l t}^+  \right) &=  -  \sigma_0^{-2} T^{-1} \sum_{t = 1}^T  E \left( \left( \sum_{k = 0}^{t-1} D_d \pi_{k}(0) \epsilon_{t-k} \right) \left(\sum_{i = 1}^{t-1} b_{\vartheta_j \vartheta_l i}(\varphi_0)  \epsilon_{t-i} \right) \right) \\
    &=  -  T^{-1} \sum_{t = 1}^T  \sum_{i = 1}^{t-1}  i^{-1}  b_{\vartheta_k \vartheta_l i}(\varphi_0) \\
    &= -    \sum_{i = 1}^{\infty}  i^{-1} b_{\vartheta_j \vartheta_l i}(\varphi_0) +  T^{-1} \sum_{t = 1}^T  \sum_{i = t}^{\infty}   i^{-1} b_{\vartheta_k \vartheta_l i}(\varphi_0) ,
\end{align*}
where 
\begin{align*}
 T^{-1} \sum_{t = 1}^T  \sum_{i = t}^{\infty}   i^{-1} b_{\vartheta_j \vartheta_l i}(\varphi_0) = O( T^{-1} \sum_{t = 1}^T  \sum_{i = t}^{\infty}  i^{-2-\varsigma} )  = O( T^{-1}) .
\end{align*}

Proof of \eqref{l1_1d}: Consider the case $k \geq 2$ and $j=l=1$ for \eqref{EXP1}. Then 
\begin{align*}
      \sigma_0^{-2} T^{-1} \sum_{t = 1}^T  E \left( S_{\vartheta_k t}^+ S_{\vartheta_1 \vartheta_1 t}^+  \right) &=    \sigma_0^{-2} T^{-1} \sum_{t = 1}^T  E \left( \left( \sum_{i = 1}^{t-1} b_{\vartheta_k i}(\varphi_0)  \epsilon_{t-i} \right) \left( \sum_{k = 0}^{t-1} D_{dd} \pi_{k}(0) \epsilon_{t-k} \right) \right) \\
     &=  T^{-1} \sum_{t = 1}^T \sum_{i = 0}^{t-1} D_{dd} \pi_{i}(0)  b_{\vartheta_k i}(\varphi_0)  \\
     &= \sum_{i = 0}^{\infty} D_{dd} \pi_{i}(0)  b_{\vartheta_k i}(\varphi_0) - T^{-1} \sum_{t = 1}^T \sum_{i = t}^{\infty} D_{dd} \pi_{i}(0)  b_{\vartheta_k i}(\varphi_0), 
\end{align*}
where
\begin{align*}
 T^{-1} \sum_{t = 1}^T \sum_{i = t}^{\infty} D_{dd} \pi_{i}(0)  b_{\vartheta_k i}(\varphi_0) = O(T^{-1} \sum_{t = 1}^T \sum_{i = t}^{\infty} (1+\log(k))^2 i^{-2-\varsigma} ) = O(T^{-1} ).
\end{align*}

Proof of \eqref{l1_1e}: Consider the case $k,l \geq 2$ and $j=1$ for \eqref{EXP1}. Then 
\begin{align*}
    \sigma_0^{-2} T^{-1} \sum_{t = 1}^T  E \left( S_{\vartheta_k t}^+ S_{\vartheta_1 \vartheta_l t}^+  \right) &=    -\sigma_0^{-2} T^{-1} \sum_{t = 1}^T  E \left( \left( \sum_{i = 1}^{t-1} b_{\vartheta_k i}(\varphi_0)  \epsilon_{t-i}  \right) \left( \sum_{i = 2}^{t-1} h_{\vartheta_1 \vartheta_l i}(\varphi_0)  \epsilon_{t-i} \right) \right) \\
    &=  -  T^{-1} \sum_{t = 1}^T  \sum_{i = 2}^{t-1}   b_{\vartheta_k i}(\varphi_0) h_{\vartheta_1 \vartheta_l i}(\varphi_0)  \\
    &=  -    \sum_{i = 2}^{\infty}   b_{\vartheta_k i}(\varphi_0) h_{\vartheta_1 \vartheta_l i}(\varphi_0) +  T^{-1} \sum_{t = 1}^T  \sum_{i = t}^{\infty}    b_{\vartheta_k i}(\varphi_0) h_{\vartheta_1 \vartheta_l i}(\varphi_0),  
\end{align*}
where 
\begin{align*}
 T^{-1} \sum_{t = 1}^T  \sum_{i = t}^{\infty}   b_{\vartheta_k i}(\varphi_0) h_{\vartheta_1 \vartheta_l i}(\varphi_0)   = O( T^{-1} \sum_{t = 1}^T  \sum_{i = t}^{\infty}  i^{-2-  \varsigma} ) =  O( T^{-1}). 
\end{align*}

Proof of \eqref{l1_1f}: Consider the case $k,j,l \geq 2$ for \eqref{EXP1}. Then
\begin{align*}
    \sigma_0^{-2} T^{-1} \sum_{t = 1}^T  E \left( S_{\vartheta_k t}^+ S_{\vartheta_j \vartheta_l t}^+  \right) &=   \sigma_0^{-2} T^{-1} \sum_{t = 1}^T  E \left( \left( \sum_{i = 1}^{t-1} b_{\vartheta_k i}(\varphi_0)  \epsilon_{t-i}  \right) \left(\sum_{i = 1}^{t-1} b_{\vartheta_j \vartheta_l i}(\varphi_0)  \epsilon_{t-i} \right) \right) \\
    &=   T^{-1} \sum_{t = 1}^T  \sum_{i = 1}^{t-1}  b_{\vartheta_k i}(\varphi_0)   b_{\vartheta_j \vartheta_l i}(\varphi_0) \\
    &= \sum_{i = 1}^{\infty}   b_{\vartheta_k i}(\varphi_0)   b_{\vartheta_j \vartheta_l i}(\varphi_0) -  T^{-1} \sum_{t = 1}^T  \sum_{i = t}^{\infty}   b_{\vartheta_k i}(\varphi_0)   b_{\vartheta_j \vartheta_l i}(\varphi_0), 
\end{align*}
where 
\begin{align*}
 T^{-1} \sum_{t = 1}^T  \sum_{i = t}^{\infty}   b_{\vartheta_k i}(\varphi_0)   b_{\vartheta_j \vartheta_l i}(\varphi_0) = O( T^{-1} \sum_{t = 1}^T  \sum_{i = t}^{\infty}  i^{-2-2\varsigma} ) =   O( T^{-1}) .
\end{align*}
\end{proof}

\begin{lemma} \label{genlemma891}
    Suppose that Assumptions \ref{A2}-\ref{A5} holds. The covariances of  $M^+_{0,\vartheta_1  T}$ and $M^+_{\vartheta_1 ,\vartheta_1}$ are given by 
\begin{align}
    E\left(  M^+_{0,\vartheta_1  T}  M^+_{\vartheta_1 ,\vartheta_1   T} \right) &= -4\zeta_3 + O(T^{-1} \log^2(T)), \label{ab11} \\
    E\left(  M^+_{0,\vartheta_1  T}  M^+_{\vartheta_1 ,\vartheta_l   T} \right) &= \sum_{k = 1}^{\infty} k^{-1} \sum_{s = 1}^{\infty} \left(  s^{-1} b_{\vartheta_l (s+k)}(\varphi_0) + (s+k)^{-1} b_{\vartheta_l s}(\varphi_0) \right) \nonumber \\
    &\ \ \ + O(T^{-1}\log(T)), \label{ab21}\\ 
    E\left(  M^+_{0,\vartheta_1  T}  M^+_{\vartheta_n ,\vartheta_l   T} \right) &= -\sum_{k = 1}^{\infty} k^{-1} \sum_{s = 1}^{\infty} \left(  b_{\vartheta_n s}(\varphi_0) b_{\vartheta_l (s+k)}(\varphi_0) + b_{\vartheta_n (s+k)}(\varphi_0) b_{\vartheta_l s}(\varphi_0) \right) \nonumber \\  
    &\ \ \ + O(T^{-1}\log(T)), \label{ab31}\\
    E\left(  M^+_{0,\vartheta_l  T}  M^+_{\vartheta_1 ,\vartheta_1   T} \right) &= 2 \sum^{\infty}_{k = 1} b_{\vartheta_l k}(\varphi_0) \sum_{s = 1}^{\infty} s^{-1} (s+k)^{-1} + O(T^{-1}\log(T)),\label{ab41}\\
    E\left(  M^+_{0,\vartheta_m  T}  M^+_{\vartheta_1 ,\vartheta_l   T} \right) &= -\sum_{k = 1}^{\infty} b_{\vartheta_m k}(\varphi_0)  \sum_{s = 1}^{\infty} \left(  s^{-1} b_{\vartheta_l (s+k)}(\varphi_0) + (s+k)^{-1} b_{\vartheta_l s}(\varphi_0) \right) \nonumber \\ 
    &\ \ \ + O(T^{-1}\log(T)),\label{ab51}\\ 
    E\left(  M^+_{0,\vartheta_m  T}  M^+_{\vartheta_n ,\vartheta_l   T} \right) &= \sum_{k = 1}^{\infty} b_{\vartheta_m k}(\varphi_0)  \sum_{s = 1}^{\infty} \left(  b_{\vartheta_n s}(\varphi_0) b_{\vartheta_l (s+k)}(\varphi_0) + b_{\vartheta_n (s+k)}(\varphi_0) b_{\vartheta_l s}(\varphi_0) \right) \nonumber \\
    &\ \ \ + O(T^{-1}\log(T)), \label{ab61}
\end{align}
for $k,j,l \in \{2,\ldots,p+1 \}$.

\end{lemma}

\begin{proof}[Proof of Lemma \ref{genlemma891}] We have that 
\begin{align*}
      E\left(  M^+_{0,\vartheta_k  T}  M^+_{\vartheta_j ,\vartheta_l   T} \right) =  \sigma^{-4}_0 T^{-1}  E\left(  \sum_{t = 1}^T  S^+_{t} S^+_{\vartheta_k t}  \sum_{s = 1}^T  S^+_{\vartheta_j s} S^+_{ \vartheta_l  s}  \right). 
\end{align*}
The expectation of $S^+_{t} S^+_{\vartheta_k t}  S^+_{\vartheta_j s} S^+_{ \vartheta_l}$ equals zero for $s \leq t$ so that what only matters is 
\begin{align}
\sigma^{-4}_0 T^{-1}  E\left(  \sum_{t = 1}^T  S^+_{t} S^+_{\vartheta_k t}  \sum_{s = t+1}^T  S^+_{\vartheta_j s} S^+_{ \vartheta_l  s}  \right). \label{abgen}
\end{align}
Now we consider the different cases. 

Proof of \eqref{ab11}: Consider the case $k,j,l = 1$ for \eqref{abgen}.  From Lemma \ref{explicitforms}
\begin{align*}
-\sigma^{-4}_0 T^{-1}  E\left(  \sum_{t = 1}^T  \epsilon_t  \sum_{k = 0}^{t-1} D_{d} \pi_{k}(0) \epsilon_{t-k}   \sum_{s = t+1}^T \sum_{n = 0}^{s-1} D_{d} \pi_{n}(0) \epsilon_{s-n} \sum_{a = 0}^{s-1} D_{d} \pi_{a}(0) \epsilon_{s-a}  \right).  
\end{align*}
Only the contributions of the form $\epsilon^2_t \epsilon^2_{t-k}$ are non-zero such that what only matter is if $s-n = t$ and $s-a = t-k$ or if $s-a = t$ and $s-n = t-k$ and since both contributions are equal we get 
\begin{align*}
-2\sigma^{-4}_0 T^{-1}   \sum_{t = 1}^T    \sum_{k = 0}^{t-1}  \sum_{s = t+1}^T  D_{d} \pi_{k}(0) D_{d} \pi_{s-t}(0) D_{d} \pi_{s-t+k}(0)  E \left( \epsilon^2_t \epsilon^2_{t-k}   \right). 
\end{align*}
Plugging in $D_{d} \pi_{k}(0) D_{d} = k^{-1} I(k \geq 1)$ and $D_{dd} \pi_j(0) = 2j^{-1} a_{j-1} I(j \geq 2)$, with $a_j = I(j \geq 1) \sum_{k = 1}^j k^{-1}$, see \eqref{dpi1} and \eqref{dpi2}, yields 
\begin{align*}
-2 T^{-1}   \sum_{t = 1}^T    \sum_{k = 1}^{t-1}   \sum_{s = t+1}^T  k^{-1}  (s-t)^{-1} (s-t+k)^{-1},  
\end{align*}
or, equivalently,
\begin{align*}
    -2 T^{-1}   \sum_{t = 1}^T   
\sum_{s = t+1}^T  \sum_{k = 1}^{t-1}   (t-k)^{-1}  (s-t)^{-1} (s-k)^{-1} ,
\end{align*}
which can be written as 
\begin{align*}
    &-2 T^{-1}   \sum_{t = 1}^T   
\sum_{s = t+1}^T  \sum_{k = -\infty}^{t-1}   (t-k)^{-1}  (s-t)^{-1} (s-k)^{-1} \\ &+ 2T^{-1}   \sum_{t = 1}^T   
\sum_{s = t+1}^T (s-t)^{-1}  \sum_{k = -\infty}^{0}  (t-k)^{-1}  (s-k)^{-1}. 
\end{align*}
For the first term, we have 
\begin{align*}
     -2 T^{-1}   \sum_{t = 1}^T   
\sum_{s = t+1}^T  \sum_{k = -\infty}^{t-1}   (t-k)^{-1}  (s-t)^{-1} (s-k)^{-1}  = -4\zeta_3,
\end{align*}
see \textcite[Lemma B.2]{johansen2016role}. For the second term, we have 
\begin{align*}
    O(T^{-1}   \sum_{t = 1}^T   
\sum_{s = t+1}^T (s-t)^{-1}  \sum_{k = -\infty}^{0}  (t-k)^{-1}  (s-k)^{-1} ) &=  O(T^{-1}   \sum_{t = 1}^T   
\sum_{s = t+1}^T (s-t)^{-1}  \sum_{k = -\infty}^{0}  (s-k)^{-2}  ) \\
&= O(T^{-1}  \log^2(T) ). 
\end{align*}

Proof of \eqref{ab21}: Consider the case $k,j = 1$ and $l > 1$ for \eqref{abgen}. From Lemma \ref{explicitforms} 
\begin{align*}
\sigma^{-4}_0 T^{-1}  E\left(  \sum_{t = 1}^T  \epsilon_t  \sum_{k = 0}^{t-1} D_{d} \pi_{k}(0) \epsilon_{t-k}   \sum_{s = t+1}^T \sum_{n = 0}^{s-1} D_{d} \pi_{n}(0) \epsilon_{s-n}\sum_{a = 1}^{s-1} b_{\vartheta_l a}(\varphi_0)  \epsilon_{s-a}  \right) . 
\end{align*}
Only the contributions of the form $\epsilon^2_t \epsilon^2_{t-k}$ are non-zero such that
\begin{align*}
\sigma^{-4}_0 T^{-1}  E\left(  \sum_{t = 1}^T  \epsilon_t  \sum_{k = 0}^{t-1} D_{d} \pi_{k}(0) \epsilon_{t-k}   \sum_{s = t+1}^T \left(  \pi_{s-t}(0) \epsilon_{t} b_{\vartheta_l (s-t+k)}(\varphi_0)  \epsilon_{t-k} + \pi_{s-t+k}(0) \epsilon_{t-k} b_{\vartheta_l (s-t)}(\varphi_0)  \epsilon_{t}, 
\right)   \right) 
\end{align*}
and plugging in the definition $D_{d} \pi_{k}(0)$, see Lemma \ref{explicitforms}, gives 
\begin{align}
T^{-1}  \sum_{t = 1}^T  \sum_{k = 1}^{t-1} k^{-1} \sum_{s = t+1}^T \left(  (s-t)^{-1} b_{\vartheta_l (s-t+k)}(\varphi_0)  + (s-t+k)^{-1} b_{\vartheta_l (s-t)}(\varphi_0) \label{abc2} 
\right) . 
\end{align}
The first term in \eqref{abc2} is
\begin{align*}
   T^{-1}  \sum_{t = 1}^T  \sum_{k = 1}^{t-1} k^{-1} \sum_{s = t+1}^T (s-t)^{-1} b_{\vartheta_l (s-t+k)}(\varphi_0) &=  T^{-1}  \sum_{t = 1}^T  \sum_{k = 1}^{t-1} k^{-1} \sum_{s = 1}^{T-t} s^{-1} b_{\vartheta_l (s+k)}(\varphi_0) \\
   &= \sum_{k = 1}^{\infty} k^{-1} \sum_{s = 1}^{\infty} s^{-1} b_{\vartheta_l (s+k)}(\varphi_0) \\&\ \ \ - T^{-1}  \sum_{t = 1}^T  \sum_{k = 1}^{t-1} k^{-1} \sum_{s = T-t+1}^{\infty} s^{-1} b_{\vartheta_l (s+k)}(\varphi_0) \\
   & \ \ \ - T^{-1}  \sum_{t = 1}^T  \sum_{k = t}^{\infty} k^{-1} \sum_{s = 1}^{\infty} s^{-1} b_{\vartheta_l (s+k)}(\varphi_0), 
\end{align*}
where
\begin{align*}
    T^{-1}  \sum_{t = 1}^T  \sum_{k = 1}^{t-1} k^{-1} \sum_{s = T-t+1}^{\infty} s^{-1} b_{\vartheta_l (s+k)}(\varphi_0) &= O(T^{-1}  \sum_{t = 1}^T  \sum_{k = 1}^{t-1} k^{-1} \sum_{s = T-t+1}^{\infty} s^{-1} (s+k)^{-1-\varsigma}) 
     = O(T^{-1} \log(T)  ), 
\end{align*}
and 
\begin{align*}
     T^{-1}  \sum_{t = 1}^T  \sum_{k = t}^{\infty} k^{-1} \sum_{s = 1}^{\infty} s^{-1} b_{\vartheta_l (s+k)}(\varphi_0) &= O(  T^{-1}  \sum_{t = 1}^T  \sum_{k = t}^{\infty} k^{-1} \sum_{s = 1}^{\infty} s^{-1} (s+k)^{-1-\varsigma}) = O(  T^{-1}  ).
\end{align*}
The second term in \eqref{abc2} is
\begin{align*}
T^{-1}  \sum_{t = 1}^T  \sum_{k = 1}^{t-1} k^{-1} \sum_{s = t+1}^T (s-t+k)^{-1} b_{\vartheta_l (s-t)}(\varphi_0)    &= T^{-1}  \sum_{t = 1}^T  \sum_{k = 1}^{t-1} k^{-1} \sum_{s = 1}^{T-t} (s+k)^{-1} b_{\vartheta_l s}(\varphi_0) \\
&= \sum_{k = 1}^{\infty} k^{-1} \sum_{s = 1}^{\infty} (s+k)^{-1} b_{\vartheta_l s}(\varphi_0) \\ 
& \ \ \ - T^{-1}  \sum_{t = 1}^T  \sum_{k = 1}^{t-1} k^{-1} \sum_{s = T-t+1}^{\infty} (s+k)^{-1} b_{\vartheta_l s}(\varphi_0) \\
& \ \ \ - T^{-1}  \sum_{t = 1}^T  \sum_{k = t}^{\infty} k^{-1} \sum_{s = 1}^{\infty} (s+k)^{-1} b_{\vartheta_l s}(\varphi_0),
\end{align*}
where 
\begin{align*}
    T^{-1}  \sum_{t = 1}^T  \sum_{k = 1}^{t-1} k^{-1} \sum_{s = T-t+1}^{\infty} (s+k)^{-1} b_{\vartheta_l s}(\varphi_0) &= O( T^{-1}  \sum_{t = 1}^T  \sum_{k = 1}^{t-1} k^{-1} \sum_{s = T-t+1}^{\infty} (s+k)^{-1} s^{-1-\varsigma} ) \\
    &= O( T^{-1} \log(T)),
\end{align*}
and 
\begin{align*}
    T^{-1}  \sum_{t = 1}^T  \sum_{k = t}^{\infty} k^{-1} \sum_{s = 1}^{\infty} (s+k)^{-1} b_{\vartheta_l s}(\varphi_0)&= O(   T^{-1}  \sum_{t = 1}^T  \sum_{k = t}^{\infty} k^{-1} \sum_{s = 1}^{\infty} (s+k)^{-1} s^{-1-\varsigma}) 
     &= O(   T^{-1}  \log(T) ). 
\end{align*}
Proof of \eqref{ab31}-\eqref{ab61}: The proof is omitted, as it follows a similar step as in the proof of \eqref{ab21}. 
\end{proof}

\subsubsection{Expectation of the score function}\label{UB3}

The following lemma will be used to calculate 
the expectation of the score function of $L^*(\vartheta)$. 

\begin{lemma} \label{genlemma88} Suppose that Assumptions \ref{A2}-\ref{A5} holds. Then 
 $  E\left( \sum_{s = 1}^T  c_{s} S_{s}^+ \sum_{t = 1}^T c_{t} S_{\vartheta_l t}^+ \right) = \sigma_0^2 \sum_{t = 1}^T  c_{t} c_{\vartheta_l t}, $
for $l \in \{1,\ldots,p+1 \}$. 
\end{lemma}
\begin{proof}[Proof of Lemma \ref{genlemma88}]
We first show the proof for $l = 1$, i.e.\ $\vartheta_1 = d$. From $S_{s}^+= \epsilon_s $ and $S_{d t}^+ =  -\sum_{k = 0}^{t-1} k^{-1}  \epsilon_{t-k} $, see Lemma \ref{explicitforms}, we find 
\begin{align*}
    \sum_{t = 1}^T  c_{t} S_{d t}^+  &= - \sum_{t = 1}^{T-1} \epsilon_t \sum_{k = t + 1}^T  c_{k} \frac{1}{k-t} = - \sum_{t = 1}^{T-1} \epsilon_t \sum_{k = 1}^T c_{k} D\pi_{k-t}(u)|_{u = 0} , 
\end{align*}
where  $D\pi_{k-t}(u)|_{u = 0} = (k-t)^{-1} I(k-t \geq 1)$ and hence 
\begin{align*}
E(\sum_{s = 1}^T  c_{s} S_{s}^+ \sum_{t = 1}^T  c_{t} S_{dt}^+) &= - \sigma^2_0 \sum_{t = 1}^{T-1} c_{t}\sum_{k = 1}^T c_{k} D\pi_{k-t}(u)|_{u = 0} =  - \sigma^2_0 \sum_{k = 1}^T c_{k}  \sum_{t = 1}^{T-1}  c_{t} D\pi_{k-t}(u)|_{u = 0}, 
\end{align*}
Next, we show that $\sum_{t = 1}^{T-1}  c_{t} D\pi_{k-t}(u)|_{u = 0} = c_{dk}$. From $c_{t}= \sum_{j = 0}^{t-1} \phi_j(\varphi) \kappa_{0(t-j)}(d)$ we find that 
\begin{align*}
     \sum_{t = 1}^{T-1} c_{t} D\pi_{k-t}(u)|_{u = 0} &= \sum_{t = 1}^{T-1} D\pi_{k-t}(u)|_{u = 0} \sum_{j = 0}^{t-1} \phi_j(\varphi) \kappa_{0(t-j)}(d) \\
     &= \sum_{j = 0}^{t-1} \phi_j(\varphi)  \sum_{t = 1}^{T-1} D\pi_{k-t}(u)|_{u = 0}\kappa_{0(t-j)}(d) \\
    &= \sum_{j = 0}^{t-1} \phi_j(\varphi)  \sum_{t = 1}^{T-1} D\pi_{(k-j)-(t-j)}(u)|_{u = 0}\kappa_{0(t-j)}(d) \\
    &= \sum_{j = 0}^{t-1} \phi_j(\varphi)  \sum_{m = 1-j}^{T-1} D\pi_{(k-j)-m}(u)|_{u = 0}\kappa_{0m}(d) \\
    &= \sum_{j = 0}^{t-1} \phi_j(\varphi)  \sum_{m = 1}^{k-j} D\pi_{(k-j)-m}(u)|_{u = 0}\kappa_{0m}(d) \\     
     &= \sum_{j = 0}^{k-1} \phi_j(\varphi)  \kappa_{1(k-j)}(d) \\    
    &=  c_{d k},
\end{align*}
where the second last equality follows from \textcite[Lemma A.4]{johansen2016role}. We next give a proof for $l \in \{2,\ldots,p+1 \}$, i.e.\ $\varphi_{n}$ for $n \in \{1,\ldots,p \}$. From Lemma \ref{explicitforms} it follows
\begin{align*}
   \sum_{t = 1}^T c_{t} S_{\varphi_n t}^+  &=  \sum_{t = 1}^{T} \epsilon_t \sum_{s = t}^{T} c_{s} b_{\varphi_n(s-t)},  
\end{align*}
so that
\begin{align*}
E(  \sum_{t = 1}^T c_{t} S_{\varphi_n t}^+\sum_{s = 1}^T  c_{s} S_{s}^+  ) &=    \sum_{t = 1}^T  c_{t} E( S_{\varphi_n t}^+\sum_{s = 1}^T c_{s} S_{s}^+ ) =  \sigma^2_0 \sum_{t = 1}^{T} c_{t} \sum_{k = 1}^{t-1} c_k b_{n(t-k)}.  
\end{align*}
We need to show that $\sum_{k = 1}^{t-1} c_k b_{n(t-k)} = c_{\varphi_{n} t} $. We find that 
\begin{align*}
    \sum_{k = 1}^{t-1} c_k b_{n(t-k)}  &= \sum_{k = 1}^{t-1} \sum_{j = 1}^{k} \phi_{k-j}(\varphi) \kappa_{0j}(d) \sum_{i = 1}^{t-k} \omega_{t-k-i}(\varphi) D_{\varphi_{n}} \phi_{i}(\varphi) \\
    &=  \sum_{j = 1}^{t-1}  \kappa_{0j}(d) \sum_{i = 1}^{t-1}  D_{\varphi_{n}} \phi_{i}(\varphi) \sum_{k = 1}^{t-1}  \phi_{k-j}(\varphi)  \omega_{t-k-i}(\varphi)\\ 
     &=  \sum_{j = 1}^{t-1}  \kappa_{0j}(d) \sum_{i = 1}^{t-1}  D_{\varphi_{n}} \phi_{i}(\varphi) \sum_{k = j}^{t-i}  \phi_{k-j}(\varphi)  \omega_{t-k-i}(\varphi)\\ 
      &=  \sum_{j = 1}^{t-1}  \kappa_{0j}(d) \sum_{i = 0}^{t-j}  D_{\varphi_{n}} \phi_{i}(\varphi) \sum_{k = 0}^{t-j-i}  \phi_{k}(\varphi)  \omega_{(t-j-i)-k}(\varphi)\\ 
      &= \sum_{j = 1}^{t-1}  \kappa_{0j}(d)  D_{\varphi_{n}} \phi_{t-j}(\varphi),
\end{align*}
since $\sum_{k = 0}^{t-j-i}  \phi_{k}(\varphi)  \omega_{(t-j-i)-k}(\varphi) = 1$ and follows from the identity $\phi(L;\varphi)  \omega(L;\varphi) = 1$.
\end{proof}

The following lemma finds the expectation of $D L^*(\vartheta_0) )$  and $L^*(\vartheta_0)$.

\begin{lemma} \label{genlemmaexpectations}
Let the model for the data $x_t$, t = 1,$\ldots$,T, be given by \eqref{genq1} and let Assumptions \ref{A2}-\ref{A5} be satisfied. Then 
\begin{align}
  E\left( D_{\vartheta_k} L^*(\vartheta_0)\right) &= -\sigma^2_0 \frac{\sum_{t = 1}^T c_{t}(\vartheta_0) c_{\vartheta_k t}(\vartheta_0)}{\sum_{t = 1}^T c^2_{t}(\vartheta_0)}, \label{genEDLmu} \\
   E\left( L^*(\vartheta_0) \right) &=  \sigma^2_0 \frac{T-1}{2},  \label{genELmu}
\end{align}
for $k \in \{1,\ldots,p+1 \}$. 
\end{lemma}
\begin{proof}[Proof of Lemma \ref{genlemmaexpectations}.] The proofs are omitted since it follows straightforwardly from Lemmata \ref{genderivatesLstar} and \ref{genlemma88}.
\end{proof}

\subsection{Approximation of the derivatives} \label{ap3}

In this section, we provide approximations for the first three derivatives of $L^*(\vartheta)$, $L^*_{\mu_0}(\vartheta)$ and $L^*_{m}(\vartheta)$ evaluated at $\vartheta = \vartheta_0$. Before that, we present results that analyse the terms involved in these derivatives. Specifically, we examine the order of magnitude of functions that incorporate the derivatives of the deterministic term $c_{t}(\vartheta)$ and the derivatives of the stochastic term $S^+_t(\vartheta)$, as well as the product moments that contain these terms. This analysis is divided into two parts. In Section \ref{gennon1}, we focus on the non-stationary region, where $d_0 > 1/2$. Then, in Section \ref{genstat1}, we explore the stationary region, where $d_0 < 1/2$. The reason for conducting separate analyses is that the order of magnitude varies depending on the region. Each section concludes with an approximation of the derivatives.

\subsubsection{Non-stationary region}\label{gennon1}

In Lemmata \ref{genlemmaaaa2n} and \ref{genlemmaa99n}, we investigate the order of magnitude of functions involving the deterministic term $c_{t}(\vartheta)$ and its derivatives and the stochastic term $S^+_{t}$ and its derivatives and the product moments containing these. In Lemma \ref{genlemmaa1}, we investigate the order of magnitude involving the modification term $m(\vartheta)$ and derivatives of these. These lemmata are then used to find asymptotic results for the first three derivatives of $L^*$, $L^*_{\mu_0}$ and $L^*_{m}$ in Lemmata \ref{asyappgennon1}, \ref{asyappgennon2}, and \ref{asyappgennon3}, respectively.

\begin{lemma} \label{genlemmaaaa2n} Suppose that Assumptions \ref{A3}-\ref{A5} holds. Let $d > 1/2$, then we have that:
\begin{align}
     \sum_{t = 1}^T c^2_{t}(\vartheta) &= \sum_{t = 1}^{\infty} c^2_{t}(\vartheta) + O(T^{\max(1-2d,-1-2\varsigma) }), \label{2na} \\
    \sum_{t = 1}^T c_{t}(\vartheta) c_{\vartheta_k t}(\vartheta) &= \sum_{t = 1}^{\infty} c_{t}(\vartheta) c_{\vartheta_k t}(\vartheta) \nonumber \\  
    &\ \ \ + O(T^{\max(1-2d,-1-2\varsigma) }  \log(T) I( k = 1) + T^{\max(1-2d,-1-2\varsigma) }   I( k > 1)),\label{2nb} \\
    \sum_{t = 1}^T c_{s t}(\vartheta) c_{i t}(\vartheta) &= O(1), \label{2nc}
\end{align} 
where $s \in \{0,\vartheta_{\tilde{k}},\vartheta_{\tilde{k}} \vartheta_{\tilde{k}}, \vartheta_{\tilde{k}} \vartheta_j,\vartheta_{\tilde{k}} \vartheta_{\tilde{k}} \vartheta_{\tilde{l}} \}$, $i \in \{0,\vartheta_k,\vartheta_k \vartheta_j, \vartheta_k \vartheta_j,\vartheta_k \vartheta_j \vartheta_l \}$ and $\tilde{k},\tilde{j},\tilde{l},k,j,l = 1,\ldots, p+1$. Here, $c_{0t}(\vartheta)$ refers to $c_{t}(\vartheta)$. 

\end{lemma}
\begin{proof}[Proof of Lemma \ref{genlemmaaaa2n}]
Proof of \eqref{2na}: Given that 
  $   \sum_{t = 1}^T c^2_{t}(\vartheta) =  \sum_{t = 1}^{\infty} c^2_{t}(\vartheta) - \sum_{t = T+1}^{\infty} c^2_{t}(\vartheta), $
and using $c_t(\vartheta)  = O(t^{\max(-d,-1-\varsigma)})$, see \eqref{r12_1} in Lemma \ref{r12}, we can deduce that 
\begin{align*}
    \sum_{t = T+1}^{\infty} c^2_{t}(\vartheta) &= O\left(\sum_{t = T+1}^{\infty} t^{\max(-2d,-2-2\varsigma)} \right) =  O\left(T^{\max(1-2d,-1-2\varsigma)}\right),
\end{align*}
where the last equality follows from \eqref{lA2} in Lemma \ref{genbounds}. \\
Proof of \eqref{2nb}: Given that
 $   \sum_{t = 1}^T c_{t}(\vartheta) c_{\vartheta_k t}(\vartheta)  = \sum_{t = 1}^{\infty} c_{t}(\vartheta) c_{\vartheta_k t}(\vartheta) - \sum_{t = T+1}^{\infty} c_{t}(\vartheta) c_{\vartheta_k t}(\vartheta).  $
For $\vartheta_1$, using $c_{\vartheta_1 t}(\vartheta)  = O(\log(T) t^{\max(-d,-1-\varsigma)})$, see \eqref{r12_1} in Lemma \ref{r12}, we can deduce that
\begin{align*}
    \sum_{t = T+1}^{\infty} c_{t}(\vartheta) c_{\vartheta_1 t}(\vartheta) &= O\left( \sum_{t = T+1}^{\infty} \log(t) t^{\max(-2d,-2-2\varsigma)} \right) = O\left( \log(T) T^{\max(1-2d,-1-2\varsigma)} \right),
\end{align*}
where the last equality follows from \eqref{lA2} in Lemma \ref{genbounds}.
Regarding $\vartheta_s$, $s \geq 2$, it can be shown that $c_{\vartheta_s t}(\vartheta)  = O(t^{\max(-d,-1-\varsigma)})$, see \eqref{r12_2} in Lemma \ref{r12}. The proof follows similarly as in the proof of \eqref{2na}.\\
Proof of \eqref{2nc}: We observe that we can establish an upper bound for $|c_{st}(\vartheta)|$ as $c \log^3(t) t^{\max(-d,-1-\varsigma)}$, see Lemma \ref{r12}, where $c$ is a generic
arbitrarily large positive constant. Consequently, we proceed to evaluate the summation 
\begin{align*}
     \sum_{t = 1}^T c_{s t}(\vartheta) c_{i t}(\vartheta) &\leq c \sum_{t = 1}^T \log^6(t) t^{\max(-2d,-2-2\varsigma)} \\
     &\leq c \log^6(T) T^{\max(1-2d,-1-2\varsigma)},  
\end{align*}
where the last inequality follows from \eqref{lA2} in Lemma \ref{genbounds}. Since $d$ and $\varsigma$ are both greater than 1/2, this term is $O(1)$.
\end{proof}

\begin{lemma} \label{genlemmaa1} Suppose that Assumptions \ref{A3}-\ref{A5} holds. Let $d > 1/2$, then we have that:
\begin{align}
    m(\vartheta) &= 1 + O(T^{-1}), \label{genab1} \\
    m_{i}(\vartheta) &=   O(T^{-1}), \label{genab2} 
\end{align}
\end{lemma}
where $i \in \{\vartheta_k,\vartheta_k \vartheta_j, \vartheta_k \vartheta_j,\vartheta_k \vartheta_j \vartheta_l \}$ and $k,j,l = 1,\ldots, p+1$.
\begin{proof}[Proof of Lemma \ref{genlemmaa1}] Proof of \eqref{genab1}: The expression for $m(\vartheta)$, as provided in \eqref{genmodificationterm}, can be represented as 
  $  m(\vartheta) =  e^{\frac{1}{T-1} \log \left( \sum_{t = 1}^T c^2_t(\vartheta)  \right)}.$
By employing the expansion $e^{b} = \sum_{k = 0}^{\infty} \frac{b^k}{k !}$ and considering \eqref{2nc} in Lemma \ref{genlemmaaaa2n}, we have that
 $    m(\vartheta) = 1 + O\left(T^{-1}\right).$
Proof of \eqref{genab2}: The derivatives of $m(\vartheta)$ are given in Lemma \ref{genderivatesLstarMCSS}. Proof follows directly from \eqref{2nc} in Lemma \ref{genlemmaaaa2n}. 
\end{proof}

\begin{lemma} \label{genlemmaa99n}
Suppose that Assumptions \ref{A2}-\ref{A5} hold. Let $d_0 > \frac{1}{2}$. Then
\begin{align} 
    \sum_{t = 1}^T S_{st}^+ c_{it} = O_{P}(1) \label{qwn1} 
\end{align}
where $s \in \{0,\vartheta_{\tilde{k}},\vartheta_{\tilde{k}} \vartheta_{\tilde{k}}, \vartheta_{\tilde{k}} \vartheta_j,\vartheta_{\tilde{k}} \vartheta_{\tilde{k}} \vartheta_{\tilde{l}} \}$, $i \in \{0,\vartheta_k,\vartheta_k \vartheta_j, \vartheta_k \vartheta_j,\vartheta_k \vartheta_j \vartheta_l \}$ and $\tilde{k},\tilde{j},\tilde{l},k,j,l = 1,\ldots, p+1$. Here, $c_{0t}(\vartheta)$ refers to $c_{t}(\vartheta)$ and $S_{0t}^+$ to $S_{t}^+$. 
\end{lemma}

\begin{proof}[Proof of Lemma \ref{genlemmaa99n}] Proof of \eqref{qwn1}: Note that $S_{t}^+(\vartheta_0) = \epsilon_t $, and as a consequence, the results for $s = 0$ directly follow from \eqref{2nc} in Lemma \ref{genlemmaaaa2n}.
Next, we provide a general proof. To begin with, we observe that from Lemma \ref{explicitforms}, $S_{st}^+$ can be expressed as
 $  S_{st}^+ = \sum_{k = 1}^{t-1} v_{st} \epsilon_{t-k}, $
where the weights $v_{st}$ depend on $s$. From \eqref{r11_1} in Lemma \ref{r11} and \eqref{bproof} in Lemma \ref{r12}, it follows that $|v_{st}| \leq c \log^3(t) t^{-1}$. Also, from the proof of \eqref{2nc}, we have established a bound for $|c_{st}(\vartheta)|$ as $c \log^3(t) t^{\max(-d,-1-\varsigma)}$. 

Firstly, we note that 
 $   \sum_{t = T+1}^{\infty} S_{s t}^+ c_{it}=  \sum_{k = 1}^{\infty} \epsilon_k \sum_{t = \max(T,k)+1}^{\infty}  c_{it} v_{s(t-k)} . $
For small $\delta>0$, we bound $\log^3(t) \leq c t^{\delta}$ and use the bounds $|c_{st}(\vartheta)| \leq c \log^3(t) t^{\max(-d,-1-\varsigma)} \leq c t^{\max(-d,-1)+\delta}$ and $|v_{s k}| \leq c \log^3(k) k^{-1} \leq c k^{-1+\delta}$, $t^{\max(-d,-1)+\delta} \leq (t-k)^{-2\delta} k^{\max(-d,-1)+2\delta}$. Then, we obtain
\begin{align*}
    Var\left( \sum_{t = T+1}^{\infty} S_{s t}^+ c_{it} \right) &\leq  c \sum_{k = 1}^{\infty}  \left( \sum_{t = \max(T,k)+1}^{\infty}  (t-k)^{-\delta-1} k^{\max(-d,-1)+2\delta}  \right)^2 \\
    &\leq c \sum_{k = 1}^{\infty} k^{\max(-2d,-2)+4\delta}  \left( \sum_{t = \max(T,k)+1}^{\infty}  (t-k)^{-\delta-1}   \right)^2. 
\end{align*}
Since $\sum_{t = \max(T,k)+1}^{\infty} (t-k)^{-\delta-1} \rightarrow 0$ as $T \rightarrow \infty$ and because $\sum_{k = 1}^{\infty} k^{\max(-2d,-2)+4\delta} < \infty$, we conclude, by the dominated convergence theorem, that this variance converges to zero.
\end{proof}

\begin{lemma} \label{asyappgennon1}
Let the model for the data $x_t$, t = 1,$\ldots$,T, be given by \eqref{genq1} and let Assumptions \ref{A2}-\ref{A5} be satisfied with $d_0 > 1/2$. Then the normalised derivatives of the likelihood function $L^*$, see \eqref{genL1}, satisfy
\begin{align}
    \sigma_0^{-2} T^{-1/2} D_{\vartheta} L^*(\vartheta_0)  &= A_{0} + T^{-1/2} A_{1}, \label{17a} \\
    \sigma_0^{-2} T^{-1} D_{\vartheta \vartheta'} L^*(\vartheta_0)  &= B_{0} + T^{-1/2}  B_{1} + O_P(T^{-1} \log(T) ), \label{17b}\\
    \sigma_0^{-2} T^{-1} D_{\vartheta_i \vartheta \vartheta'} L^*(\vartheta_0)  &= C_{0i} + O_P(T^{-1/2}), \label{17c}
\end{align}
for $i = 1,\ldots,p+1$ and where
\begin{align*}
A_{0} &= M_{0\vartheta}^{+}, \ \ \ \ E(A_{1}) = E(\sigma^{-2}_0 D_{\vartheta} L^*(\vartheta_0) ) = O(1), \\
B_{0} &= A, \ \ \ \ B_1 = M_{\vartheta,\vartheta' T}^{+} + M_{0,\vartheta \vartheta' T}^{+}, 
\end{align*}
Here,
$M_{0\vartheta}^{+}$, $M_{0,\vartheta \vartheta' T}^{+}$ and $M_{\vartheta,\vartheta' T}^{+}$ are given in \eqref{genM1}, \eqref{genM2}, \eqref{genM3} respectively,
and $A$ is the inverse of the variance-covariance matrix given in \eqref{genA}. The expression for $C_{0i}$, $i = 1,\ldots,p+1$, is given in \eqref{genC1} and \eqref{genCk}. 
\end{lemma}

\begin{proof}[Proof of Lemma \ref{asyappgennon1}]  Proof of \eqref{17a}:
From Lemma \ref{genderivatesLstar}, we have that 
\begin{align*}
\sigma_0^{-2} T^{-1/2} D_{\vartheta_k} L^*   &= \sigma_0^{-2} T^{-1/2} \sum_{t = 1}^T S_{t}  S_{\vartheta_k t} - \sigma_0^{-2} T^{-1/2} \left(\mu(\vartheta_0)-\mu_0\right) \sum_{t = 1}^T S_{t}   c_{\vartheta_k t} \\
     &\ \ \ - \sigma_0^{-2} T^{-1/2} \left(\mu(\vartheta_0)-\mu_0\right)  \sum_{t = 1}^T  S_{\vartheta_k t}   c_{t} + \sigma_0^{-2} T^{-1/2} \left(\mu(\vartheta_0)-\mu_0\right)^2  \sum_{t = 1}^T c_{t}  c_{\vartheta_k t} \\
     &=  M_{0\vartheta_k}^{+} + T^{-1/2} A_{1}(k), 
\end{align*}
with $A_{1}(k)$ given by
\begin{align*}
    A_{1}(k) &=  - \sigma_0^{-2}  \left(\mu(\vartheta_0)-\mu_0\right) \sum_{t = 1}^T S_{t}   c_{\vartheta_k t} \\
    &\ \ \ - \sigma_0^{-2} \left(\mu(\vartheta_0)-\mu_0\right)  \sum_{t = 1}^T  S_{\vartheta_k t}   c_{0t} + \sigma_0^{-2}\left(\mu(\vartheta_0)-\mu_0\right)^2  \sum_{t = 1}^T c_{t}  c_{\vartheta_k t}, 
\end{align*}
since $E( M_{0\vartheta_k}^{+}) = 0$ it follows that $E\left(A_{1}(k)\right) =   E\left(\sigma_0^{-2}  D_{\vartheta_k} L^*\right)$  and from Lemmata \ref{genlemmaexpectations} and \ref{genlemmaaaa2n} we find that $E\left(\sigma_0^{-2}  D_{\vartheta_k} L^*\right) = O(1)$.

Proof of \eqref{17b}: From Lemma \ref{genderivatesLstar} we have that 
\begin{align*}
   \sigma_0^{-2} T^{-1} D_{\vartheta_k \vartheta_j } L^* &= \sigma_0^{-2} T^{-1} L_{\vartheta_k  \vartheta_j} -  \sigma_0^{-2} T^{-1} \frac{L_{\mu \vartheta_j} L_{\mu \vartheta_k}}{L_{\mu \mu} },
\end{align*}
where  $\sigma_0^{-2} T^{-1} L_{\vartheta_k \mu}\mu_{\vartheta_j}/L_{\mu \mu} = O_P(T^{-1})$ from Lemmata \ref{genlemmaaaa2n} and \ref{genlemmaa99n}. Thus we get 
\begin{align*}
     \sigma_0^{-2} T^{-1} D_{\vartheta_k \vartheta_j } L^* &=  \sigma_0^{-2} T^{-1} \sum_{t = 1}^T \left( S_{\vartheta_j t}^+ -  c_{\vartheta_j t}(\vartheta_0)\left(\mu(\vartheta_0)-\mu_0\right)\right) \left( S_{\vartheta_k t}^+ -  c_{\vartheta_k t}(\vartheta_0)\left(\mu(\vartheta_0)-\mu_0\right)\right) \\
    &\ \ \ +  \sigma_0^{-2} T^{-1}  \sum_{t = 1}^T  \left(S_{t}^+ - c_{t}(\vartheta_0)\left(\mu(\vartheta_0)-\mu_0\right)\right) \left( S_{ \vartheta_k  \vartheta_j t}^+ -  c_{\vartheta_k  \vartheta_j t}(\vartheta_0)\left(\mu(\vartheta_0)-\mu_0\right)\right) \\
    &\ \ \ +  O_P(T^{-1} ), 
\end{align*}
ignoring terms that are of order $T^{-1}$ we get
\begin{align*}
 \sigma_0^{-2} T^{-1} D_{\vartheta_k \vartheta_j } L^* &= \sigma_0^{-2} T^{-1} \sum_{t = 1}^T S_{\vartheta_j t}^+ S_{\vartheta_k t}^+ + \sigma_0^{-2} T^{-1} \sum_{t = 1}^T S_{t}^+ S_{ \vartheta_k  \vartheta_j t}^+ +   O_P(T^{-1}) \\
 &= \sigma_0^{-2} T^{-1} \sum_{t = 1}^T E S_{\vartheta_j t}^+ S_{\vartheta_k t}^+ + T^{-1/2} \left( M^+_{\vartheta_j,\vartheta_k T} + M^+_{0,\vartheta_j \vartheta_k T}\right) +  O_P(T^{-1} ).
\end{align*}

We notice that $\sigma_0^{-2} T^{-1} \sum_{t = 1}^T E  \left( S_{\vartheta_j t}^+ S_{\vartheta_k t}^+ \right)  = E \left( M_{0,\vartheta_j} M_{0,\vartheta_k} \right)$ and is already covered in Lemma \ref{genlemmma1}.

Proof of \eqref{17c}: For the third derivative it can be shown from Lemmata \ref{genlemmaaaa2n} and \ref{genlemmaa99n} that the extra terms involving derivatives $\mu_{\vartheta_k}$ and $\mu_{\vartheta_k \vartheta_k}$, see Lemma \ref{genderivatesLstar}, can be ignored and we find
\begin{align*}
    \sigma_0^{-2} T^{-1} D_{\vartheta_k \vartheta_j \vartheta_l} L^*  &= \sigma_0^{-2} T^{-1}  \sum_{t = 1}^T \left( S_{\vartheta_j \vartheta_l t}^+ -  c_{\vartheta_j \vartheta_l t}\left(\mu(\vartheta_0)-\mu_0\right)\right) \left( S_{\vartheta_k t}^+ -  c_{\vartheta_k t}\left(\mu(\vartheta_0)-\mu_0\right)\right) \\
    & \ \ \ + \sigma_0^{-2} T^{-1} \sum_{t = 1}^T  \left(S_{\vartheta_j }^+ - c_{\vartheta_j t}\left(\mu(\vartheta_0)-\mu_0\right)\right) \left( S_{ \vartheta_k \vartheta_l t}^+ -  c_{\vartheta_k \vartheta_l t}\left(\mu(\vartheta_0)-\mu_0\right)\right) \\ 
     &\ \ \ + \sigma_0^{-2} T^{-1} \sum_{t = 1}^T  \left(S_{\vartheta_l t}^+ - c_{\vartheta_l t}\left(\mu(\vartheta_0)-\mu_0\right)\right) \left( S_{ \vartheta_k \vartheta_j t}^+ -  c_{\vartheta_k \vartheta_j t}\left(\mu(\vartheta_0)-\mu_0\right)\right)\\
     &\ \ \ + O_P(T^{-1}) \\ 
     &=  \sigma_0^{-2} T^{-1} \sum_{t = 1}^T E\left( S_{\vartheta_j \vartheta_l t}^+ S_{\vartheta_k t}^+ \right) +  \sigma_0^{-2} T^{-1} \sum_{t = 1}^T  E\left( S_{\vartheta_j }^+ S_{ \vartheta_k \vartheta_l t}^+ \right) \\ 
     &\ \ \ + \sigma_0^{-2} T^{-1} \sum_{t = 1}^T   E\left( S_{\vartheta_l t}^+ S_{ \vartheta_k \vartheta_j t}^+ \right) \\
     &\ \ \ + T^{-1/2} \left(  M^+_{0,\vartheta_k \vartheta_j \vartheta_l  T}  + M^+_{\vartheta_k ,\vartheta_{j}\vartheta_{l}   T} + M^+_{\vartheta_j ,\vartheta_{l}\vartheta_{k}   T} + M^+_{\vartheta_l ,\vartheta_{j}\vartheta_{k}   T} \right) + O_P(T^{-1}) \\
     &=  \sigma_0^{-2} T^{-1} \sum_{t = 1}^T E\left( S_{\vartheta_j \vartheta_l t}^+ S_{\vartheta_k t}^+ \right) +  \sigma_0^{-2} T^{-1} \sum_{t = 1}^T  E\left( S_{\vartheta_j }^+ S_{ \vartheta_k \vartheta_l t}^+ \right) \\ 
     &\ \ \ + \sigma_0^{-2} T^{-1} \sum_{t = 1}^T   E\left( S_{\vartheta_l t}^+ S_{ \vartheta_k \vartheta_j t}^+ \right) + O_P(T^{-1/2}),
\end{align*}
where the second-to-last equality uses Lemmata \ref{genlemmaaaa2n} and \ref{genlemmaa99n}  and the last equality uses Lemma \ref{genlemmma1}. The terms in this expression are given in Lemma \ref{genlemmma1_1}. In matrix notation, we can therefore  define $C_{0i}$ in \eqref{17c} as follows 
\begin{align}
C_{01} = \begin{pmatrix}
 C_{01}(1,1)  & C_{01}(1,2) \\
C_{01}(2,1) & C_{01}(2,2)
\end{pmatrix},  \label{genC1}
\end{align}
where the elements are given by
\begin{align*}
     C_{01}(1,1) &= -6 \zeta_3, \\
     C_{01}(1,2) &= 2 \sum_{i = 2}^{\infty} i^{-1} h_{d \varphi' i}(\varphi_0) + \sum_{i = 0}^{\infty} D_{dd} \pi_i(0) b_{\varphi'i}(\varphi_0),\\
     C_{01}(2,1) &= 2 \sum_{i = 2}^{\infty} i^{-1} h_{d \varphi i}(\varphi_0) + \sum_{i = 0}^{\infty} D_{dd} \pi_i(0) b_{\varphi i}(\varphi_0),\\
     C_{01}(2,2) &=  -\sum_{i = 1}^{\infty} i^{-1} b_{\varphi \varphi'i}(\varphi_0) -\sum_{i = 2}^{\infty}   b_{\vartheta i}(\varphi_0) h_{d \vartheta' i}(\varphi_0) - \left(\sum_{i = 2}^{\infty}   b_{\vartheta i}(\varphi_0) h_{d \vartheta' i}(\varphi_0)\right)',
\end{align*}
and for $k = 1,\dots, p$ we have that
\begin{align}
C_{0(k+1)} = \begin{pmatrix}
 C_{0(k+1)}(1,1)  & C_{0(k+1)}(1,2) \\
C_{0(k+1)}(2,1) & C_{0(k+1)}(2,2),
\end{pmatrix}  \label{genCk}
\end{align}
where the elements are given by
\begin{align*}
     C_{0(k+1)}(1,1) &= 2 \sum_{i = 2}^{\infty} i^{-1} h_{d \varphi_k i}(\varphi_0)  + \sum_{i = 0}^{\infty} D_{dd} \pi_i(0) b_{\varphi_k i}(\varphi_0),   \\
     C_{0(k+1)}(1,2) &= -\sum_{i = 1}^{\infty} i^{-1} b_{\varphi' \varphi_k i}(\varphi_0) -    \sum_{i = 2}^{\infty}   b_{\varphi_k i}(\varphi_0) h_{d \varphi'  i}(\varphi_0) -  \sum_{i = 2}^{\infty}   b_{\varphi'  i}(\varphi_0) h_{d \varphi_k  i}(\varphi_0),   \\
     C_{0(k+1)}(2,1) &= -\sum_{i = 1}^{\infty} i^{-1} b_{\varphi \varphi_k i}(\varphi_0) -    \sum_{i = 2}^{\infty}   b_{\varphi_k i}(\varphi_0) h_{d \varphi  i}(\varphi_0) -  \sum_{i = 2}^{\infty}   b_{\varphi  i}(\varphi_0) h_{d \varphi_k  i}(\varphi_0),  \\
    C_{0(k+1)}(2,2) &= \left( \sum_{i = 1}^{\infty}  b_{\varphi i}(\varphi_0) b_{\varphi' \varphi_k i}(\varphi_0) \right)' + \sum_{i = 1}^{\infty}  b_{\varphi  i}(\varphi_0) b_{\varphi' \varphi_k i}(\varphi_0)  + \sum_{i = 1}^{\infty}  b_{\varphi_k i}(\varphi_0) b_{\varphi \varphi' i}(\varphi_0).
\end{align*}
\end{proof}

\begin{lemma} \label{asyappgennon2}
Let the model for the data $x_t$, t = 1,$\ldots$,$T$, be given by \eqref{genq1} and let Assumptions \ref{A2}-\ref{A1} be satisfied with $d_0 > 1/2$. Then the normalised derivatives of the likelihood function $L_{\mu_0}^*$, see \eqref{genlikmu1known}, satisfy
\begin{align}
    \sigma_0^{-2} T^{-1/2} D_{\vartheta} L_{\mu_0}^*(\vartheta_0)  &= A_{0} + T^{-1/2}A_1, \\
    \sigma_0^{-2} T^{-1} D_{\vartheta \vartheta'} L_{\mu_0}^* (\vartheta_0)  &= B_{0} + T^{-1/2}  B_{1} + O_P(T^{-1} \log(T) ), \\
    \sigma_0^{-2} T^{-1} D_{\vartheta_i \vartheta \vartheta'} L^*(\vartheta_0)  &= C_{0i} + O_P(T^{-1/2}),
\end{align}
for $i = 1,\ldots,p+1$ and where
\begin{align*}
A_{0} &= M_{0,\vartheta T}, \ \ \ \ E(A_{1}) = E(\sigma^{-2}_0 D_{\vartheta} L^*(\vartheta_0) ) = O(1), \\
B_{0} &= A, \ \ \ \ B_1 = M_{\vartheta,\vartheta' T}^{+} + M_{0,\vartheta \vartheta' T}^{+}, 
\end{align*}
Here,
$M_{0,\vartheta T}$, $M_{0,\vartheta \vartheta' T}^{+}$ and $M_{\vartheta,\vartheta' T}^{+}$ are given in \eqref{genM1}, \eqref{genM2} and \eqref{genM3}, respectively,
and $A$ is the inverse of the variance-covariance matrix given in \eqref{genA}. The expression for $C_{0i}$, $i = 1,\ldots,p+1$, is given in \eqref{genC1} and \eqref{genCk}. 
\end{lemma}

\begin{proof}[Proof of Lemma \ref{asyappgennon2}]
The proof is omitted and follows from the same approach as in the proof of Lemma  \ref{asyappgennon1} but is much easier since the constant term is known. 
\end{proof}

\begin{lemma} \label{asyappgennon3}
Let the model for the data $x_t$, t = 1,$\ldots$,T, be given by \eqref{genq1} and let Assumptions \ref{A2}-\ref{A5} be satisfied with $d_0 > 1/2$. Then the normalised derivatives of the likelihood function $L_m^*$, see \eqref{genmlik}, satisfy
\begin{align}
    \sigma_0^{-2} T^{-1/2} D_{\vartheta} L_m^*(\vartheta_0)  &= A_{0} + T^{-1/2} A_{1} +O(T^{-1}), \\
    \sigma_0^{-2} T^{-1} D_{\vartheta \vartheta'} L_m^*(\vartheta_0)  &= B_{0} + T^{-1/2}  B_{1} + O_P(T^{-1} \log(T) ), \\
     \sigma_0^{-2} T^{-1} D_{\vartheta_i \vartheta \vartheta'} L^*(\vartheta_0)  &= C_{0i} + O_P(T^{-1/2}), 
\end{align}
for $i = 1,\ldots,p+1$ and where
\begin{align*}
A_{0} &= M_{0\vartheta}^{+}, \ \ \ \ E(A_{1}) = E(\sigma^{-2}_0 D_{\vartheta} L^*(\vartheta_0) ) = 0, \\
B_{0} &= A, \ \ \ \ B_1 = M_{\vartheta,\vartheta' T}^{+} + M_{0,\vartheta \vartheta' T}^{+}, 
\end{align*}
Here,
$M_{0\vartheta}^{+}$, $M_{0,\vartheta \vartheta' T}^{+}$ and $M_{\vartheta,\vartheta' T}^{+}$ are given in \eqref{genM1}, \eqref{genM2} and \eqref{genM3}, respectively,
and $A$ is the inverse of the variance-covariance matrix given in \eqref{genA}. The expression for $C_{0i}$, $i = 1,\ldots,p+1$, is given in \eqref{genC1} and \eqref{genCk}.
\end{lemma}

\begin{proof}[Proof of Lemma \ref{asyappgennon3}]
The proof is omitted and follows from Lemma \ref{asyappgennon1} and the asymptotic behaviour of the modification term and its derivatives in Lemma \ref{genlemmaa1}. 
\end{proof}

\subsubsection{Stationary region}\label{genstat1}

In Lemmata \ref{genlemmaaaa2s} and \ref{genlemmaa99s}, we investigate the order of magnitude of functions involving the deterministic term $c_{t}(\vartheta)$ and its derivatives and the stochastic term $S^+_{t}$ and its derivatives and the product moments containing these. In Lemma \ref{genlemmaa1stat}, we investigate the order of magnitude involving the modification term $m(\vartheta)$ and derivatives of these. These lemmata are then used to find asymptotic results for the first three derivatives of $L^*$, $L^*_{\mu_0}$ and $L^*_{m}$ in Lemmata \ref{asyappgenstat1}, \ref{asyappgenstat2}, and \ref{asyappgenstat3}, respectively.

\begin{lemma} \label{genlemmaaaa2s} Suppose that Assumptions \ref{A3}-\ref{A5} holds. Let $d < 1/2$, then we have that:
\begin{align}
      \frac{1}{T^{1-2d} }\sum_{t = 1}^T c^2_{t}(\vartheta) &= \phi^2(1;\varphi)       \frac{1}{T^{1-2d} } \sum_{t = 1}^T \kappa^2_{0t}(d) + o(1), \label{2sa}\\
    \frac{1}{T^{1-2d} }\sum_{t = 1}^T c_{t}(\vartheta) c_{d t}(\vartheta) &= \phi^2(1;\varphi) \frac{1}{T^{1-2d} } \sum_{t = 1}^T \kappa_{0t}(d) \kappa_{1t}(d)   + o(1),  \label{2sb}\\
    \frac{1}{T^{1-2d} }\sum_{t = 1}^T c_{t}(\vartheta) c_{\varphi_{k} t}(\vartheta) &= \phi(1;\varphi) D_{\varphi_{k}}\phi(1;\varphi)  \frac{1}{T^{1-2d}}  \sum_{t = 1}^T \kappa^2_{0t}(d) + o(1), \label{2sc}\\
        \frac{1}{T^{1-2d} } \sum_{t = 1}^T \kappa_{0t}(d) \kappa_{1t}(d)  &=  - \left( \log(T) - \Psi(1-d)\right) \frac{1}{T^{1-2d} }  \sum_{t = 1}^T \kappa^2_{0t}(d) \nonumber \\ 
    &\ \ \ + \frac{1}{\Gamma(1-d)^2 (1-2d)^2}  + o(1),   \label{2sc1}\\
     \frac{1}{T^{1-2d}}\sum_{t = 1}^T k^2_{0t}(d) &\rightarrow \frac{1}{\Gamma(1-d)^2 (1-2d)}, \label{2sc2} \\
     \sum_{t = 1}^T c_{t}(\vartheta) c_{\vartheta_k t}(\vartheta) &= O(T^{1-2d} \log(T)), \label{2sd1} \\
     \sum_{t = 1}^T c_{t}(\vartheta) c_{\vartheta_k \vartheta_j t}(\vartheta) &= O(T^{1-2d} \log^2(T)), \label{2sd2} \\
     \sum_{t = 1}^T c_{t}(\vartheta) c_{\vartheta_k \vartheta_j \vartheta_l t}(\vartheta) &= O(T^{1-2d} \log^3(T)), \label{2sd3} \\
      \sum_{t = 1}^T c_{\vartheta_k t}(\vartheta) c_{\vartheta_j t}(\vartheta) &= O(T^{1-2d} \log^2(T)), \label{2sd4} \\
      \sum_{t = 1}^T c_{\vartheta_k t}(\vartheta) c_{\vartheta_j \vartheta_l  t}(\vartheta) &= O(T^{1-2d} \log^3(T)), \label{2sd5}  
\end{align} 
for $k,j,l = 1,\ldots, p+1$.
\end{lemma}
\begin{proof}[Proof of Lemma \ref{genlemmaaaa2s}]
Proof of \eqref{2sa}: See \textcite[Lemma S.15]{hualde2020truncated}.

Proof of \eqref{2sb}: 
By summation by parts 
\begin{align*}
    c_t(\vartheta) &=  \sum_{j = 0}^{t-1} \phi_j(\varphi) \kappa_{0(t-j)}(d) = \kappa_{0t}(d) \sum_{j = 0}^{t-1} \phi_j(\varphi) - \sum_{j = 0}^{t-2} \left(\kappa_{0(t-j)}(d) - \kappa_{0(t-j-1)}(d) \right) \sum_{k = j+1}^{t-1}  \phi_k(\varphi),
\end{align*}
From $\kappa_{0(t-j)}(d) - \kappa_{0(t-j-1)} = \pi_{t-j-1}(1-d) - \pi_{t-j-2}(1-d) = \pi_{t-j-1}(-d)$, see \textcite[Lemma A.4]{johansen2016role}, we have
\begin{align*}
     c_t(\vartheta) &=  \kappa_{0t}(d) \sum_{j = 0}^{\infty} \phi_j(\varphi) -  \kappa_{0t}(d) \sum_{j = t}^{\infty} \phi_j(\varphi) - \sum_{j = 0}^{t-2} \pi_{t-j-1}(-d) 
    \sum_{k = j+1}^{t-1}  \phi_k(\varphi),
\end{align*}
Notice that 
\begin{align*}
     \sum_{j = 0}^{t-2} \pi_{t-j-1}(-d) 
    \sum_{k = j+1}^{t-1}  \phi_k(\varphi) = \sum_{j = 1}^{t-1} \pi_j(-d) \sum_{k = 1}^j  \phi_{t-k}(\varphi),
\end{align*}
therefore
\begin{align}
     c_t(\vartheta) &= \kappa_{0t}(d) \sum_{j = 0}^{\infty} \phi_j(\varphi) -  \kappa_{0t}(d) \sum_{j = t}^{\infty} \phi_j(\varphi) - \sum_{j = 1}^{t-1} \pi_j(-d) \sum_{k = 1}^j  \phi_{t-k}(\varphi), \label{2sbcc}
\end{align}
Taking the derivative of $c_t(\vartheta)$ with respect to $d$ gives
\begin{align*}
    D_{d}   c_t(\vartheta) &= \kappa_{1t}(d) \sum_{j = 0}^{\infty} \phi_j(\varphi) -  \kappa_{1t}(d) \sum_{j = t}^{\infty} \phi_j(\varphi) - \sum_{j = 1}^{t-1} D_d \pi_j(-d) \sum_{k = 1}^j  \phi_{t-k}(\varphi).
\end{align*}

The first term of $c_t(\vartheta)$ is bounded by $O(t^{-d})$ from \eqref{r11_1}. The second term is bounded 
by $O(t^{-d-\varsigma})$ from \eqref{r11_1} and \eqref{r11_21} and from the same arguments the third term is bounded by $O\left(\sum_{j = 1}^{t-1}j^{-d-1} \sum_{k = 1}^j  (t-k)^{-1-\varsigma}  \right) = O(t^{-d-\varsigma})$ which also involves employing Lemma \ref{series}.

The first term of $D_{d} c_t(\vartheta)$ is bounded by $O(\log(t) t^{-d})$ from \eqref{r11_1}. The second term is bounded by $O(\log(t)t^{-d-\varsigma})$ from \eqref{r11_1} and \eqref{r11_21} and from the same arguments the third term is bounded by $O\left(\sum_{j = 1}^{t-1} \log(j) j^{-d-1} \sum_{k = 1}^j  (t-k)^{-1-\varsigma}  \right) = O(\log(t) t^{-d-\varsigma})$ which also involves employing Lemma \ref{series}. 

The leading term of $\sum_{t = 1}^T c_{t}(\vartheta) c_{d t}(\vartheta)$ involves only the first term of $c_t(\vartheta)$ and $D_{d} c_t(\vartheta)$ and the remainder term is bounded by 
\begin{align*}
    O(\sum_{t=1}^T \log(t) t^{-2d-\varsigma} ) = O(\log(T)\sum_{t=1}^T t^{-2d-\varsigma} ) 
\end{align*}
This term is $O(\log(T))$ when $-2d-\varsigma < -1$, $O(\log^2(T))$ when  $-2d-\varsigma = -1$, and $O(\log(T)  T^{1-2d-\varsigma})$ when $-2d-\varsigma > -1$. The proof is now completed. 

Proof of \eqref{2sc}: 
Taking the derivative of $c_t(\vartheta)$ in \eqref{2sbcc} with respect to $\varphi_k$ gives
\begin{align*}
    D_{\varphi_k}   c_t(\vartheta) &= \kappa_{0t}(d) \sum_{j = 0}^{\infty}  D_{\varphi_k}  \phi_j(\varphi) -  \kappa_{0t}(d)  \sum_{j = t}^{\infty} D_{\varphi_k}  \phi_j(\varphi) \\
      &\ \ \ - \sum_{j = 1}^{t-1} \pi_j(-d) \sum_{k = 1}^j  D_{\varphi_k}  \phi_{t-k}(\varphi) .
\end{align*}
The first term of $ D_{\varphi_k}  c_t(\vartheta)$ is bounded by $O( t^{-d})$ from \eqref{r11_1} and \eqref{r11_2}. The second term is bounded by $O(t^{-d-\varsigma})$ from \eqref{r11_1} and \eqref{r11_2} and from the same arguments the third term is bounded by $O\left(\sum_{j = 1}^{t-1}  j^{-d-1} \sum_{k = 1}^j  (t-k)^{-1-\varsigma}  \right) = O( t^{-d-\varsigma})$ which also involves employing Lemma \ref{series}. 
The leading term of $\sum_{t = 1}^T c_{t}(\vartheta) c_{\varphi_k t}(\vartheta)$ involves only the first term of $c_t(\vartheta)$ and $D_{\varphi_k}   c_t(\vartheta)$ and the remainder term is bounded by 
  $  O(\sum_{t=1}^T  t^{-2d-\varsigma} ). $
This term is $O(1)$ when $-2d-\varsigma < -1$, $O(\log(T))$ when  $-2d-\varsigma = -1$, and $O(T^{1-2d-\varsigma})$ when $-2d-\varsigma > -1$. The proof is now completed. 

Proof of \eqref{2sc1}: We note that 
 $   \kappa_{1t}(d) = - \kappa_{0t}(d) \left( \Psi(t-d) - \Psi(1-d) \right), $
then 
\begin{align}
    \frac{1}{T^{1-2d} } \sum_{t = 1}^T \kappa_{0t}(d) \kappa_{1t}(d)  =  - \frac{1}{T^{1-2d} } \sum_{t = 1}^T \kappa^2_{0t}(d) \Psi(t-d) +  \Psi(1-d)\frac{1}{T^{1-2d} } \sum_{t = 1}^T \kappa^2_{0t}(d), \label{ert1}
\end{align}
We evaluate the first term in \eqref{ert1}. We have that 
\begin{align}
   \frac{1}{T^{1-2d} } \sum_{t = 1}^T \kappa^2_{0t}(d) \Psi(t-d) &= \Psi(T-d)   \frac{1}{T^{1-2d} }  \sum_{t = 1}^T \kappa^2_{0t}(d) \nonumber \\
   &\ \ \ +   \frac{1}{T^{1-2d} }  \sum_{t = 1}^T k^2_{0t}(d) \left(\Psi(t-d)-\Psi(T-d)\right). \label{ert2}
\end{align}
For a fixed $d$, 
 $   \Psi(t + d) = \log(t) + O(t^{-1}), $
see \textcite[eqn.\ 6.3.18]{abramowitz1964handbook}, hence the second term in \eqref{ert2} is 
\begin{align*}
\frac{1}{T^{1-2d} } \sum_{t = 1}^{T} \kappa^2_{0t}(d) \left(  \Psi(t-d) - \Psi(T-d) \right) &= \frac{1}{T^{1-2d} }\sum_{t = 1}^{T} \kappa^2_{0t}(d) \log(t/T) + o(1). 
\end{align*}
The first term in \eqref{ert2} is  
\begin{align*}
     \Psi(T-d) \frac{1}{T^{1-2d} }  \sum_{t = 1}^T \kappa^2_{0t}(d) = \log(T) \frac{1}{T^{1-2d} }  \sum_{t = 1}^T \kappa^2_{0t}(d)   + o(1).
\end{align*}
Thus we find for \eqref{ert2} that 
\begin{align}
   \frac{1}{T^{1-2d} } \sum_{t = 1}^T \kappa^2_{0t}(d) \Psi(t-d) &=  \log(T) \frac{1}{T^{1-2d} }  \sum_{t = 1}^T \kappa^2_{0t}(d) \nonumber \\
   &\ \ \ +   \frac{1}{T^{1-2d} }\sum_{t = 1}^{T} \kappa^2_{0t}(d) \log(t/T) + o(1).\label{ert3}
\end{align}
The second term in \eqref{ert3} is 
\begin{align*}
      \frac{1}{T^{1-2d}}\sum_{t = 1}^{T} \kappa^2_{0t}(d) \log(t/T) &= \frac{1}{\Gamma\left(1-d\right)^2} \frac{1}{T^{1-2d}} \sum_{t = 1}^T \log(t/T) t^{-2d} + o(1) \\
     &\rightarrow - \frac{1}{\Gamma(1-d)^2 (1-2d)^2},
\end{align*}
from Stirling's approximation, see \textcite[page 257 6.1.47]{abramowitz1964handbook},
\begin{align}
    \pi_t(d) \sim \frac{1}{\Gamma(d)} t^{d-1} + O(t^{d-2}), \label{stirl}
\end{align}
and the last line follows from \textcite[Lemma S.10]{hualde2020truncated}. Thus \eqref{ert3} equals
\begin{align}
    \frac{1}{T^{1-2d} } \sum_{t = 1}^T \kappa^2_{0t}(d) \Psi(t-d) &=  \log(T) \frac{1}{T^{1-2d} }  \sum_{t = 1}^T \kappa^2_{0t}(d) - \frac{1}{\Gamma(1-d)^2 (1-2d)^2} + o(1). \label{ert4}
\end{align}
Plugging in \eqref{ert4} to \eqref{ert1} gives 
\begin{align*}
    \frac{1}{T^{1-2d} } \sum_{t = 1}^T \kappa_{0t}(d) \kappa_{1t}(d)  &=  - \left( \log(T) - \Psi(1-d)\right) \frac{1}{T^{1-2d} }  \sum_{t = 1}^T \kappa^2_{0t}(d) \nonumber \\ 
    &\ \ \ + \frac{1}{\Gamma(1-d)^2 (1-2d)^2}  + o(1), 
\end{align*}
completing the proof.

Proof of \eqref{2sc2}: From Stirling's approximation \eqref{stirl}
\begin{align*}
     \frac{1}{T^{1-2d}}\sum_{t = 1}^T k^2_{0t}(d) = \frac{1}{\Gamma\left(1-d\right)^2} \frac{1}{T^{1-2d}} \sum_{t = 1}^T t^{-2d} + o(1) \rightarrow \frac{1}{\Gamma(1-d)^2 (1-2d)},
\end{align*}
and the last line follows from Lemma \ref{series}.\\

Proof of \eqref{2sd1}-\eqref{2sd5}: The proofs can be straightforwardly deduced from the provided bounds in Lemma \ref{r12}, together with the application of Lemma \ref{series}.
\end{proof}

\begin{lemma} \label{genlemmaa1stat} Suppose that Assumptions \ref{A3}-\ref{A5} holds. Let $d < 1/2$, then we have that:
\begin{align}
    m(\vartheta) &= 1 + O(T^{-1} \log(T)), \label{genab1s} \\
    m_{\vartheta_k}(\vartheta) &=  O(T^{-1} \log(T)) \label{genab2s}, \\
     m_{\vartheta_k \vartheta_j}(\vartheta) &=  O(T^{-1}\log^2(T)), \label{genab3s} \\
     m_{\vartheta_k \vartheta_j \vartheta_l}(\vartheta) &=  O(T^{-1}\log^3(T)), \label{genab4s} 
\end{align}
for $k,j,l = 1,\ldots, p+1$.
\end{lemma}

\begin{proof}[Proof of Lemma \ref{genlemmaa1stat}]
Proof of \eqref{genab1s}: 
The expression for $m(\vartheta)$, as provided in \eqref{genmodificationterm}, can be represented as 
 $   m(\vartheta) = \left( \frac{1}{T^{1-2d}} \sum_{t = 1}^T c^2_t(\vartheta)  \right)^{\frac{1}{T-1}} \left(T^{1-2d}  \right)^{\frac{1}{T-1}}. $
By employing the expansion $e^{b} = \sum_{k = 0}^{\infty} \frac{b^k}{k !}$ and considering \eqref{2sa} in Lemma \ref{genlemmaaaa2s}, we have that
\begin{align*}
    \left( \frac{1}{T^{1-2d}} \sum_{t = 1}^T c^2_t(\vartheta)  \right)^{\frac{1}{T-1}} &= e^{ (T-1)^{-1} \log \left( T^{-1+2d} \sum_{t = 1}^T  c^2_t(\vartheta)  \right) } =   1 + O(T^{-1}).
\end{align*}
From the same expansion we have that 
 $   \left(T^{1-2d}  \right)^{\frac{1}{T-1}} = e^{ (T-1)^{-1} (1-2d)  \log(T) } = 1+O(T^{-1}\log(T)).                            $
We conclude that: 
 $   m(\vartheta) = (1+O(T^{-1}))(1+O(T^{-1}\log(T)))  = 1+O(T^{-1}\log(T)). $

Proof of \eqref{genab2s}-\eqref{genab4s}:  The derivatives of $m(\vartheta)$ are given in Lemma \ref{genderivatesLstarMCSS}. Proof follows directly from \eqref{2sd1}-\eqref{2sd5} in Lemma \ref{genlemmaaaa2s}. 
\end{proof}

\begin{lemma} \label{genlemmaa99s}
Suppose that Assumptions \ref{A2}-\ref{A5} hold. Let $d_0 < \frac{1}{2}$. Then
\begin{align} 
    &\sum_{t = 1}^T S_{t}^+ c_t = O_{P}(T^{1/2-d_0}), \ \ \ 
    \sum_{t = 1}^T S_{t}^+ c_{\vartheta_k t} = O_{P}(T^{1/2-d_0} \log(T)), \nonumber\\
    &\sum_{t = 1}^T S_{t}^+(\vartheta) c_{\vartheta_k \vartheta_j t} = O_{P}(T^{1/2-d_0} \log^2(T)), \label{qws1} \\
    & \sum_{t = 1}^T S_{\vartheta_l t}^+ c_t = O_{P}(T^{1/2-d_0} \log(T)),
    \sum_{t = 1}^T S_{\vartheta_l t}^+ c_{\vartheta_k t} = O_{P}(T^{1/2-d_0} \log^{2}(T)), \nonumber\\
    &\sum_{t = 1}^T S_{\vartheta_l t}^+ c_{\vartheta_k \vartheta_j t}= O_{P}(T^{1/2-d_0} \log^{3}(T)), \label{qws2}\\
     &\sum_{t = 1}^T S_{\vartheta_l \vartheta_n t}^+ c_t = O_{P}(T^{1/2-d_0} \log^{2}(T)),
    \sum_{t = 1}^T S_{\vartheta_l \vartheta_n t}^+ c_{\vartheta_k t} = O_{P}(T^{1/2-d_0} \log^{3}(T)), \nonumber \\
    &\sum_{t = 1}^T S_{\vartheta_l \vartheta_n t}^+ c_{\vartheta_k \vartheta_j t} = O_{P}(T^{1/2-d_0} \log^{4}(T)). \label{qws3}
\end{align}
for $l,n,k,j = 0,\ldots,p+1$.
\end{lemma}

\begin{proof}[Proof of Lemma \ref{genlemmaa99s}] Proof of \eqref{qws1}: Note that $S_{t}^+ = \epsilon_t $ such that the results follow from Lemma \ref{genlemmaaaa2s}.\\
Proof of \eqref{qws2}: Due to their similarity and relative simplicity, we exclusively show $\sum_{t = 1}^T S_{\vartheta_l t}^+ c_{\vartheta_z \vartheta_j t} = O_{P}(T^{1/2-d_0} \log^3(T))$, omitting the proofs for $\sum_{t = 1}^T S_{\vartheta_l t}^+ c_t  = O_{P}(T^{1/2-d_0} \log(T))$ and $\sum_{t = 1}^T S_{\vartheta_l t}^+ c_{\vartheta_z t} = O_{P}(T^{1/2-d_0} \log^{2}(T))$. First, consider $l = 1$. By Lemma \ref{explicitforms}, 
$S^{+}_{\vartheta_1 t} = - \sum_{k = 0}^{t-1} D\pi_k(0) \epsilon_{t-k}, $
resulting in
 $   \sum_{t = 1}^{T} S_{\vartheta_1 t}^+(\vartheta)  c_{\vartheta_z \vartheta_j k} = -\sum_{t = 1}^{T-1} \epsilon_t \sum_{k = t+1}^T c_{\vartheta_z \vartheta_j k}  D\pi_{k-t}(0). $

Now, we analyse three scenarios: the first case involves $z = 1$ and $j = 1$; the second case involves $z = 1$ and $j>1$; and the third case encompasses $z>1$ and $j > 1$.

\textit{Case I: z = 1 and j = 1.} First let $0 < d_0 < 1/2$. We use the following bounds $|c_{\vartheta_1 \vartheta_1 t}| \leq c t^{-d_0} \log^2(t)$ and $|D\pi_{t}(0)| \leq ct^{-1} I(t \geq 1)$. Then 
\begin{align*}
    Var(\sum_{t = 1}^{T} S_{\vartheta_1 t}^+ c_{\vartheta_1 \vartheta_1 t}) &\leq c \sum_{t = 1}^{T-1} \left( \sum_{k =t+1}^T \log^2(k)k^{-d_0}  (k-t)^{-1} \right)^2 \\  
    &\leq c \sum_{t = 1}^{T-1} \left( \sum_{k = 1}^{T-t} \log^2(t+k) (t+k)^{-d_0}  k^{-1} \right)^2 \\  
    &\leq c \sum_{t = 1}^{T} t^{-2d_0} \log^4(T) \left( \sum_{k = 1}^{T}   k^{-1} \right)^2 \\ 
    &\leq c T^{1-2d_0} \log^6(T),
\end{align*}
because $\sum_{k = 1}^{T}   k^{-2d} = O(T^{1-2d})$ for $d< 1/2$. 

Second, let $d_0 \leq 0$. Then 
\begin{align*}
    Var(\sum_{t = 1}^{T} S_{\vartheta_1 t}^+ c_{\vartheta_1 \vartheta_1 t}) &\leq c \sum_{t = 1}^{T-1} \left( \sum_{k =t+1}^T \log^2(k)k^{-d_0}  (k-t)^{-1} \right)^2 \\  
    &\leq c \sum_{t = 1}^{T-1} \left( \sum_{k = 1}^{T-t} \log^2(t+k) (t+k)^{-d_0}  k^{-1} \right)^2 \\  
    &\leq c  T^{-2d_0}   \log^4(T) \sum_{t = 1}^{T} \left( \sum_{k = 1}^{T-t}   k^{-1} \right)^2 \\ 
    &\leq c  T^{-2d_0}   \log^4(T) \sum_{t = 1}^{T} \log^2(T-t+1) \\
     &\leq c  T^{1-2d_0}   \log^6(T),
\end{align*}
because 
\begin{align*}
     \sum_{t = 1}^{T} \log^2(T-t+1) &= \sum_{t = 1}^{T} \log^2(t) \\
                                &\leq  T\log^2(T)\\
\end{align*}
This shows that $\sum_{t = 1}^{T} S_{\vartheta_1 t}^+(\vartheta)  c_{\vartheta_1 \vartheta_1 k} = O_P(T^{1/2-d_0} \log^3(T)) $.

\textit{Case II: z = 1 and j > 1.} The proof of this case follows in a similar way to that in \textit{Case I} but now with the bound $|c_{\vartheta_1 \vartheta_j t}| \leq c t^{-d_0} \log(t)$. To conclude, $\sum_{t = 1}^{T} S_{\vartheta_1 t}^+(\vartheta)  c_{\vartheta_1 \vartheta_j k} = O_P(T^{1/2-d_0} \log^2(T)) $.

\textit{Case III: z > 1 and j > 1.} The proof of this case follows in a similar way to that in \textit{Case I} but now with the bound $|c_{\vartheta_z \vartheta_j t}| \leq c t^{-d_0}$. To conclude, $\sum_{t = 1}^{T} S_{\vartheta_1 t}^+(\vartheta)  c_{\vartheta_z \vartheta_j k} = O_P(T^{1/2-d_0} \log(T)) $.

Finally, consider $l > 1$. By Lemma \ref{explicitforms},
$S^+_{\vartheta_l t} = \sum_{i = 1}^{t-1} b_{\vartheta_li} \epsilon_{t-i}, $
resulting in
 $   \sum_{t = 1}^{T} S_{\vartheta_l t}^+  c_{\vartheta_z \vartheta_j  t} = \sum_{t = 1}^{T-1} \epsilon_t \sum_{k = t+1}^T c_{\vartheta_z \vartheta_j k}  b_{\vartheta_l(k-t)}. $

Now, we analyse the same three scenarios again: the first case involves $z = 1$ and $j = 1$; the second case involves $z = 1$ and $j>1$; and the third case encompasses $z>1$ and $j > 1$.

\textit{Case I: z = 1 and j = 1.} Let $0< d_0 < 1/2$. Given a small $\delta > 0$ to be chosen later, we use the following bounds $|c_{\vartheta_1 \vartheta_1 t}| \leq c t^{-d_0} \log^2(t)$ and $|b_{\vartheta_j t}| \leq c t^{-1-\varsigma+\delta}$. Then  
\begin{align*}
      Var(\sum_{t = 1}^{T} S_{\vartheta_l t}^+(\vartheta) c_{\vartheta_1 \vartheta_1 t}(\vartheta)) &\leq c \sum_{t = 1}^{T-1} \left( \sum_{k =t+1}^T \log^2(k) k^{-d_0}  (k-t)^{-1-\varsigma+\delta} \right)^2 \\  
    &\leq c \sum_{t = 1}^{T-1} \left( \sum_{k = 1}^{T-t} \log^2(t+k)(t+k)^{-d_0}  k^{-1-\varsigma+\delta} \right)^2 \\  
    &\leq c \sum_{t = 1}^{T} t^{-2d_0} \log^4(T) \left( \sum_{k = 1}^{T}   k^{-1-\varsigma+\delta} \right)^2 \\ 
    &\leq c  T^{1-2d_0} \log^4(T),
\end{align*}
because $\sum_{k = 1}^{T}   k^{-1-\varsigma+\delta} = O(1)$ since $-\varsigma+\delta < 0$ from choosing $\delta$ to satisfy $0< \delta < \varsigma < 1/2$. This show that $\sum_{t = 1}^{T} S_{\vartheta_l t}^+ c_{\vartheta_1 \vartheta_1 t} = O_P( T^{1/2-d_0} \log^2(T))$.

For $d_0 \leq 0$, replace the third inequality by using $(t+k)^{-d_0} \leq T^{-d_0}$ to get the same conclusion. 

\textit{Case II: z = 1 and j > 1.} The proof of this case follows in a similar way to that in \textit{Case I} but now with the bound $|c_{\vartheta_1 \vartheta_j t}| \leq c t^{-d_0} \log(t)$. To conclude that $\sum_{t = 1}^{T} S_{\vartheta_l t}^+(\vartheta)  c_{\vartheta_1 \vartheta_j k} = O_P(T^{1/2-d_0} \log(T)) $.

\textit{Case III: z > 1 and j > 1.} The proof of this case follows in a similar way to that in \textit{Case I} but now with the bound $|c_{\vartheta_z \vartheta_j t}| \leq c t^{-d_0}$. To conclude that $\sum_{t = 1}^{T} S_{\vartheta_l t}^+  c_{\vartheta_z \vartheta_j k} = O_P(T^{1/2-d_0}) $.

Proof of \eqref{qws3}: The proofs follow from similar arguments as in the proof of \eqref{qws2} and are therefore omitted.
\end{proof}

\begin{lemma} \label{asyappgenstat1}
Let the model for the data $x_t$, t = 1,$\ldots$,T, be given by \eqref{genq1} and let Assumptions \ref{A2}-\ref{A5} be satisfied with $d_0 < 1/2$. Then the normalised derivatives of the likelihood function $L^*$, see \eqref{genL1}, satisfy
\begin{align}
    \sigma_0^{-2} T^{-1/2} D_{\vartheta} L^*(\vartheta_0)  &= A_{0} + T^{-1/2} A_{1}, \label{23a}\\
    \sigma_0^{-2} T^{-1} D_{\vartheta \vartheta'} L^*(\vartheta_0)  &= B_{0} + T^{-1/2}  B_{1} + O_P(T^{-1} \log^2(T) ), \label{23b}\\
     \sigma_0^{-2} T^{-1} D_{\vartheta_i \vartheta \vartheta'} L^*(\vartheta_0)  &= C_{0i} + O_P(T^{-1/2}), \label{23c} 
\end{align}
for $i = 1,\ldots,p+1$ and where
\begin{align*}
A_{0} &= M_{0\vartheta}^{+}, \ \ \ \ E(A_{1}) = E(\sigma^{-2}_0 D_{\vartheta} L^*(\vartheta_0) ) = O(\log(T)), \\
B_{0} &= A, \ \ \ \ B_1 = M_{\vartheta,\vartheta' T}^{+} + M_{0,\vartheta \vartheta' T}^{+}, 
\end{align*}
Here,
$M_{0\vartheta}^{+}$, $M_{0,\vartheta \vartheta' T}^{+}$ and $M_{\vartheta,\vartheta' T}^{+}$ are given in \eqref{genM1}, \eqref{genM2} and \eqref{genM3}, respectively,
and $A$ is the inverse of the variance-covariance matrix given in \eqref{genA}.The expression for $C_{0i}$, $i = 1,\ldots,p+1$, is given in \eqref{genC1} and \eqref{genCk}. 
\end{lemma}

\begin{proof}[Proof of Lemma \ref{asyappgenstat1}] Proof of \eqref{23a}:
From Lemma \ref{genderivatesLstar}, we have that 
\begin{align*}
\sigma_0^{-2} T^{-1/2} D_{\vartheta_k} L^*   &= \sigma_0^{-2} T^{-1/2} \sum_{t = 1}^T S_{t}^+  S_{\vartheta_k t}^+ - \sigma_0^{-2} T^{-1/2} \left(\mu(\vartheta_0)-\mu_0\right) \sum_{t = 1}^T S_{t}^+   c_{\vartheta_k t} \\
     & \ \ \ - \sigma_0^{-2} T^{-1/2} \left(\mu(\vartheta_0)-\mu_0\right)  \sum_{t = 1}^T  S_{\vartheta_k t}^+   c_{t} + \sigma_0^{-2} T^{-1/2} \left(\mu(\vartheta_0)-\mu_0\right)^2  \sum_{t = 1}^T c_{t}  c_{\vartheta_k t} \\
     &=  M_{0\vartheta_k}^{+} + T^{-1/2} A_1,
\end{align*}
with elements of $A_1$ given by
\begin{align*}
    A_{1}(k) &=  - \sigma_0^{-2}  \left(\mu(\vartheta_0)-\mu_0\right) \sum_{t = 1}^T S_{t}^+   c_{\vartheta_k t} \\
    &\ \ \ - \sigma_0^{-2} \left(\mu(\vartheta_0)-\mu_0\right)  \sum_{t = 1}^T  S_{\vartheta_k t}^+   c_{0t} + \sigma_0^{-2}\left(\mu(\vartheta_0)-\mu_0\right)^2  \sum_{t = 1}^T c_{t}  c_{\vartheta_k t}, 
\end{align*}
since $E( M_{0\vartheta_k}^{+}) = 0$ it follows that $E\left(A_{1}(k)\right) =   E\left(\sigma_0^{-2}  D_{\vartheta_k} L^*\right)$ and from Lemmata \ref{genlemmaexpectations} and \ref{genlemmaaaa2s} we find that $E(A) = O(\log(T))$.

Proof of \eqref{23b}: From Lemma \ref{genderivatesLstar} we have that 
\begin{align*}
   \sigma_0^{-2} T^{-1} D_{\vartheta_k \vartheta_j } L^* &= \sigma_0^{-2} T^{-1} L_{\vartheta_k  \vartheta_j} -  \sigma_0^{-2} T^{-1} \frac{L_{\mu \vartheta_j} L_{\mu \vartheta_k}}{L_{\mu \mu} },
\end{align*}
where  $\sigma_0^{-2} T^{-1} L_{\vartheta_k \mu}\mu_{\vartheta_j}/L_{\mu \mu} = O_P(T^{-1}\log^2(T)$ from Lemmata \ref{genlemmaaaa2s} and \ref{genlemmaa99s}. Thus we get 
\begin{align*}
     \sigma_0^{-2} T^{-1} D_{\vartheta_k \vartheta_j } L^* &=  \sigma_0^{-2} T^{-1} \sum_{t = 1}^T \left( S_{\vartheta_j t}^+ -  c_{\vartheta_j t}(\vartheta_0)\left(\mu(\vartheta_0)-\mu_0\right)\right) \left( S_{\vartheta_k t}^+ -  c_{\vartheta_k t}(\vartheta_0)\left(\mu(\vartheta_0)-\mu_0\right)\right) \\
    &\ \ \ +  \sigma_0^{-2} T^{-1}  \sum_{t = 1}^T  \left(S_{t}^+ - c_{t}(\vartheta_0)\left(\mu(\vartheta_0)-\mu_0\right)\right) \left( S_{ \vartheta_k  \vartheta_j t}^+ -  c_{\vartheta_k  \vartheta_j t}(\vartheta_0)\left(\mu(\vartheta_0)-\mu_0\right)\right) \\
    &\ \ \ +  O_P(T^{-1} \log^2(T) ), \\
\end{align*}
ignoring terms that are of order $T^{-1} \log^2(T)$ we get
\begin{align*}
 \sigma_0^{-2} T^{-1} D_{\vartheta_k \vartheta_j } L^* &= \sigma_0^{-2} T^{-1} \sum_{t = 1}^T S_{\vartheta_j t}^+ S_{\vartheta_k t}^+ + \sigma_0^{-2} T^{-1} \sum_{t = 1}^T S_{t}^+ S_{ \vartheta_k  \vartheta_j t}^+ +   O_P(T^{-1}) \\
 &= \sigma_0^{-2} T^{-1} \sum_{t = 1}^T E S_{\vartheta_j t}^+ S_{\vartheta_k t}^+ + T^{-1/2} \left( M^+_{\vartheta_j,\vartheta_k T} + M^+_{0,\vartheta_j \vartheta_k T}\right) +  O_P(T^{-1} \log^2(T)).
\end{align*}

We notice that $\sigma_0^{-2} T^{-1} \sum_{t = 1}^T E  \left( S_{\vartheta_j t}^+ S_{\vartheta_k t}^+ \right)  = E \left( M^+_{0,\vartheta_j} M^+_{0,\vartheta_k} \right)$ and is already covered in Lemma \ref{genlemmma1}.

Proof of \eqref{23c}: The proof strategy closely resembles that used in the derivation of \eqref{17c} in Lemma \ref{asyappgennon1} and is thus omitted. It is worth noting that, for this approximation, the terms provided in Lemmata \ref{genlemmaaaa2s} and \ref{genlemmaa99s} can be considered negligible.
\end{proof}

\begin{lemma} \label{asyappgenstat2}
Let the model for the data $x_t$, t = 1,$\ldots$,T, be given by \eqref{genq1} and let Assumptions \ref{A2}-\ref{A5} be satisfied with $d_0 < 1/2$. Then the normalised derivatives of the likelihood function $L_{\mu_0}^*$, see \eqref{genlikmu1known}, satisfy
\begin{align}
    \sigma_0^{-2} T^{-1/2} D_{\vartheta} L_{\mu_0}^*(\vartheta_0)  &= A_{0}, \\
    \sigma_0^{-2} T^{-1} D_{\vartheta \vartheta'} L_{\mu_0}^* (\vartheta_0)  &= B_{0} + T^{-1/2}  B_{1} + O_P(T^{-1} \log(T) ), \\
     \sigma_0^{-2} T^{-1} D_{\vartheta_i \vartheta \vartheta'} L^*(\vartheta_0)  &= C_{0i} + O_P(T^{-1/2}), 
\end{align}
for $i = 1,\ldots,p+1$ and where
\begin{align*}
A_{0} &= M_{0\vartheta}^{+},\\
B_{0} &= A, \ \ \ \ B_1 = M_{\vartheta,\vartheta' T}^{+} + M_{0,\vartheta \vartheta' T}^{+}, 
\end{align*}
Here,
$M_{0\vartheta}^{+}$, $M_{0,\vartheta \vartheta' T}^{+}$ and $M_{\vartheta,\vartheta' T}^{+}$ are given in \eqref{genM1}, \eqref{genM2} and \eqref{genM3}, respectively,
and $A$ is the inverse of the variance-covariance matrix given in \eqref{genA}. The expression for $C_{0i}$, $i = 1,\ldots,p+1$, is given in \eqref{genC1} and \eqref{genCk}. 
\end{lemma}

\begin{proof}[Proof of Lemma \ref{asyappgenstat2}]
The proof is omitted and follows from the same approach as in the proof of Lemma  \ref{asyappgenstat1} but is much easier since the constant term is known. 
\end{proof}

\begin{lemma} \label{asyappgenstat3}
Let the model for the data $x_t$, t = 1,$\ldots$,T, be given by \eqref{genq1} and let Assumptions \ref{A2}-\ref{A5} be satisfied with $d_0 < 1/2$. Then the normalised derivatives of the likelihood function $L_m^*$, see \eqref{genmlik}, satisfy
\begin{align}
    \sigma_0^{-2} T^{-1/2} D_{\vartheta} L_m^*(\vartheta_0)  &= A_{0} + T^{-1/2} A_{1} +O_P(T^{-1} \log(T)), \\
    \sigma_0^{-2} T^{-1} D_{\vartheta \vartheta'} L_m^*(\vartheta_0)  &= B_{0} + T^{-1/2}  B_{1} + O_P(T^{-1} \log^2(T) ), \\
     \sigma_0^{-2} T^{-1} D_{\vartheta_i \vartheta \vartheta'} L^*(\vartheta_0)  &= C_{0i} + O_P(T^{-1/2}), 
\end{align}
for $i = 1,\ldots,p+1$ and where
\begin{align*}
A_{0} &= M_{0\vartheta}^{+}, \ \ \ \ E(A_{1}) = E(\sigma^{-2}_0 D_{\vartheta} L_m^*(\vartheta_0) ) = 0, \\
B_{0} &= A, \ \ \ \ B_1 = M_{\vartheta,\vartheta' T}^{+} + M_{0,\vartheta \vartheta' T}^{+}, 
\end{align*}
Here,
$M_{0\vartheta}^{+}$, $M_{0,\vartheta \vartheta' T}^{+}$ and $M_{\vartheta,\vartheta' T}^{+}$ are given in \eqref{genM1}, \eqref{genM2} and \eqref{genM3}, respectively,
and $A$ is the inverse of the variance-covariance matrix given in \eqref{genA}. The expression for $C_{0i}$, $i = 1,\ldots,p+1$, is given in \eqref{genC1} and \eqref{genCk}. 
\end{lemma}

\begin{proof}[Proof of Lemma \ref{asyappgenstat3}]
The proof is omitted and follows from Lemma  \ref{asyappgenstat1} and the asymptotic behaviour of the modification term and its derivatives in Lemma \ref{genlemmaa1stat}. 
\end{proof}

\subsection{Proof of the main results} \label{generalizationsappendix}

In this section, we provide the proofs for the main results presented in Section \ref{secgen}.

\subsubsection{Proof of Theorem \ref{t52}} \label{proofthm52}

The proof for the CSS score follows directly from Lemmata \ref{genlemmaexpectations}, \ref{genlemmaaaa2n} and \ref{genlemmaaaa2s}. The proof for the CSS score with known $\mu_0$ follows directly from Lemma \ref{genderivatesLstar} and from $E( S_{t}^+ S_{\vartheta_k t}^+) = 0$.

\subsubsection{Proof of Lemma \ref{gen:lm:md}} \label{lemma31}

The first property is readily established since 
 $   \sum_{t = 1}^T c^2_t(\vartheta) \geq c^2_1 = \phi_0(\varphi) \kappa_{01}(d) = 1, $
from $\kappa_{01}(d)  = 1$ and $\phi_0(\varphi) = 1$.

In a special case without the short-run dynamics  $ \omega(L;\varphi) =\phi(L;\varphi) = 1$, $c_t(\vartheta) = \kappa_{0t}(d)$. Then
 $   \sum_{t = 1}^T  \kappa^2_{0t}(d) \leq \kappa^2_{01}(d) + \kappa^2_{02}(d) = 1 + (1-d)^2,  $
because  $\kappa_{01}(d) = 1$ for all $d$ and $\kappa_{02}(d) = \pi_1(1-d) = 1-d$ for all $d$, see \textcite[Lemma A.4]{johansen2016role}. Thus $\kappa_{02}(d) = 0$ only if $d = 1$ and from the recursive relationship $\pi_j(a) = \frac{j-1+a}{j} \pi_{j-1}(a)$ for $j \geq 1$ and for all $a$, see for instance p.96 in \textcite{hassler2019time}, it follows that $\kappa_{0n}(d) = 0$ for all $n \geq 2$ when $d = 1$. Thus $m(\vartheta) = 1$ if $d = 1$ and $\omega(L;\varphi) = 1$.

The proof of the second property is given in Lemmata \ref{genlemmaa1} and \ref{genlemmaa1stat} for $d > 1/2$ and $d < 1/2$, respectively. For $d$ = 1/2, we notice that $c_t(\vartheta)  = O(t^{-1/2})$, see \eqref{r12_1} in Lemma \ref{r12}, and therefore 
 $   \sum_{t = 1}^T c^2_t(\vartheta) = O( \sum_{t = 1}^T t^{-1}) = O(\log(T)). $
Then 
\begin{align*}
    m(\vartheta) &= \left( \frac{1}{\log(T)} \sum_{t = 1}^T c^2_t(\vartheta)  \right)^{\frac{1}{T-1}} \left(\log(T)  \right)^{\frac{1}{T-1}}.
\end{align*}
By employing the expansion $e^{b} = \sum_{k = 0}^{\infty} \frac{b^k}{k !}$
\begin{align*}
    \left( \frac{1}{\log(T)} \sum_{t = 1}^T c^2_t(\vartheta)  \right)^{\frac{1}{T-1}} &= e^{ (T-1)^{-1} \log \left( \log(T)^{-1} \sum_{t = 1}^T  c^2_t(\vartheta)  \right) } =   1 + O(T^{-1}).
\end{align*}
From the same expansion 
 $   \left(\log(T)  \right)^{\frac{1}{T-1}} = e^{ (T-1)^{-1} \log( \log(T)) } = 1+O(T^{-1}\log(\log(T))).                 $  
We conclude that 
 $   m(\vartheta) = (1+O(T^{-1}))(1+O(T^{-1}\log(\log(T))))  = 1+O(T^{-1} \log(\log(T))). $

\subsubsection{Proof of Theorem \ref{t51}} \label{proofthm51}

The MCSS estimator is 
\begin{align*}
    \hat{\vartheta}_{m} &= \operatorname*{argmin}_{\vartheta \in \Theta}  L^*_{m}(\vartheta) = \operatorname*{argmin}_{\vartheta \in \Theta}  \log \left(m(\vartheta)\frac{2}{T}L^*(\vartheta)\right).
\end{align*}
Define the new objective function as $\tilde{L}(\vartheta) = \log \left(m(\vartheta)\frac{2}{T}L^*(\vartheta)\right) = \log(m(\vartheta)) + \log\left(\frac{2}{T}L^*(\vartheta)\right)$. We note that $R(\vartheta) = \frac{2}{T}L^*(\vartheta) $ is the same objective function as in \textcite{hualde2020truncated}. 
Fix $\epsilon > 0$ and let $M_{\epsilon} = \{\vartheta \in \Theta: |\vartheta-\vartheta_0| < \epsilon \}$
and $\bar{M}_{\epsilon} = \{\vartheta \in \Theta: |\vartheta-\vartheta_0| \geq \epsilon \}$. Then 
\begin{align*}
    \text{Pr}\left(\hat{\vartheta}_{m} \in \bar{M} \right) &= \text{Pr}\left(\inf_{\vartheta \in \bar{M}_{\epsilon}}\tilde{L}(\vartheta) \leq \inf_{\vartheta \in M_{\epsilon}} \tilde{L}(\vartheta) \right), \\ 
    &\leq \text{Pr}\left(\inf_{\vartheta \in \bar{M}_{\epsilon}}\tilde{L}(\vartheta) \leq  \tilde{L}(\vartheta_0) \right), \\
    &\leq \text{Pr}\left(\inf_{\vartheta \in \bar{M}_{\epsilon}} \log(R(\vartheta))-\log(R(\vartheta_0)) \leq  \log(m(\vartheta_0))- \inf_{\vartheta \in \Theta} \log(m(\vartheta)) \right),.
\end{align*}
From \textcite{hualde2020truncated}, as $T \rightarrow \infty$, 
 $   \text{Pr}\left(\inf_{\vartheta \in \bar{M}_{\epsilon}} \log(R(\vartheta))-\log(R(\vartheta_0)) \leq 0 \right) \rightarrow 0. $
To prove consistency, it remains to show that  
 $   \log(m(\vartheta_0))- \inf_{\vartheta \in \Theta} \log(m(\vartheta)) \rightarrow 0, $
which is already established in Lemmata \ref{genlemmaaaa2n} and \ref{genlemmaaaa2s}.

To show the asymptotic normality of $\hat{\vartheta}_{m}$, we proceed with a usual Taylor expansion of the score function, 
 $   0 =  D_{\vartheta}L_{m}^*(\hat{\vartheta}_{m}) =  D_{\vartheta} L_{m}^*(\vartheta_0) + \left(\hat{\vartheta}_{m}-\vartheta_0\right)  D_{\vartheta \vartheta'} L_{m}^*(\vartheta^*), $
where $\vartheta^*$ is an intermediate value satisfying $|\vartheta^*-\vartheta_0| \leq |\hat{\vartheta}_{m} - \vartheta_0| \overset{P}{\rightarrow} 0$. The product moments within $D_{\vartheta \vartheta'}L^*(\vartheta)$ have been demonstrated in \textcite[Lemma C.4]{johansen2010likelihood} and \textcite[ Lemma A.8(i)]{johansen2012necessary} to exhibit tightness or equicontinuity in a neighbourhood of $\vartheta_0$. This allows us to apply \textcite[Lemma A.3]{johansen2010likelihood} and conclude that $D_{\vartheta \vartheta'} L_{m}^*(\vartheta^*) = D_{\vartheta \vartheta'}L_{m}^*(\vartheta_0) + o_P(1)$. Consequently, we proceed to analyse $D_{\vartheta} L_{m}^*(\vartheta_0)$ and $D_{\vartheta \vartheta'}L_{m}^*(\vartheta_0)$. According to Lemmata \ref{asyappgennon3} and \ref{asyappgenstat3}, we find that $ \sigma^2_0 T^{-1/2}D_{\vartheta} L_{m}^*(\vartheta_0) =  M_{0\vartheta}^{+} + O_P(T^{-1/2}\log(T))$ and $ \sigma^2_0 T^{-1}D_{\vartheta \vartheta'}L_{m}^*(\vartheta_0) = A + O_P(T^{-1/2})$ so that the final result follows from Lemma \ref{genlemmma1}.

\subsubsection{Proof of Theorem \ref{t53}} \label{proofthm53}

First, we consider the bias of $\hat{\vartheta}$. A Taylor series expansion of $D_{\vartheta}L^*(\hat{\vartheta}) = 0$ around $\vartheta_0$ gives 
\begin{align}
0 = D_{\vartheta}L^*(\hat{\vartheta}) = D_{\vartheta}L^*(\vartheta_0)   + D_{\vartheta \vartheta'}L^*(\vartheta_0) (\hat{\vartheta}-\vartheta_0) + \frac{1}{2} \begin{bmatrix}
(\hat{\vartheta}-\vartheta_0)' D_{\vartheta_1 \vartheta \vartheta'}L(\vartheta^*) (\hat{\vartheta}-\vartheta_0) \\
\vdots  \\
(\hat{\vartheta}-\vartheta_0)' D_{\vartheta_{p+1} \vartheta \vartheta'}L(\vartheta^*) (\hat{\vartheta}-\vartheta_0) 
\end{bmatrix}, \label{gentaylorexp}
\end{align}
where $\vartheta^*$ is an intermediate value which is allowed to vary across the different rows of $D_{\vartheta_i \vartheta \vartheta'}L(\vartheta^*) $ for $i = 1,\ldots,p+1$ and satisfies $|\vartheta^* - \vartheta_0| \leq |\hat{\vartheta} - \vartheta_0| \xrightarrow[]{P} 0$. We then insert $\hat{\vartheta}-\vartheta_0 = T^{-1/2} G_{1T} + T^{-1} G_{2T} + O_p(T^{-3/2}) $ and find
 {\small
\begin{align*}
    G_{1T} &= - T^{1/2} (D_{\vartheta \vartheta'}L^*(\vartheta_0))^{-1}  D_{\vartheta}L^*(\vartheta_0), \\
    G_{2T} &= - \frac{1}{2} T (D_{\vartheta \vartheta'}L^*(\vartheta_0))^{-1} \begin{bmatrix}
D_{\vartheta}L^*(\vartheta_0)' (D_{\vartheta \vartheta'}L^*(\vartheta_0))^{-1} D_{\vartheta_1 \vartheta \vartheta'}L(\vartheta^*) (D_{\vartheta \vartheta'}L^*(\vartheta_0))^{-1} D_{\vartheta}L^*(\vartheta_0) \\
\vdots  \\
D_{\vartheta}L^*(\vartheta_0)' (D_{\vartheta \vartheta'}L^*(\vartheta_0))^{-1} D_{\vartheta_{p+1} \vartheta \vartheta'}L(\vartheta^*) (D_{\vartheta \vartheta'}L^*(\vartheta_0))^{-1} D_{\vartheta}L^*(\vartheta_0)
\end{bmatrix} ,
\end{align*}
}
which we write as 
{\small
\begin{align}
    T^{1/2} (\hat{\vartheta}-\vartheta_0) &= - (T^{-1} D_{\vartheta \vartheta'}L^*(\vartheta_0))^{-1} T^{-1/2}D_{\vartheta}L^*(\vartheta_0) - \frac{1}{2} T^{-1/2} (T^{-1} D_{\vartheta \vartheta'}L^*(\vartheta_0))^{-1}  \nonumber \\
    &\begin{bmatrix}
T^{-1/2}D_{\vartheta}L^*(\vartheta_0)' ( T^{-1} D_{\vartheta \vartheta'}L^*(\vartheta_0))^{-1} T^{-1} D_{\vartheta_1 \vartheta \vartheta'}L(\vartheta^*) (T^{-1} D_{\vartheta \vartheta'}L^*(\vartheta_0))^{-1} T^{-1/2} D_{\vartheta}L^*(\vartheta_0) \\
\vdots  \\
T^{-1/2} D_{\vartheta}L^*(\vartheta_0)' (T^{-1} D_{\vartheta \vartheta'}L^*(\vartheta_0))^{-1} T^{-1} D_{\vartheta_{p+1} \vartheta \vartheta'}L(\vartheta^*) (T^{-1} D_{\vartheta \vartheta'}L^*(\vartheta_0))^{-1} T^{-1/2}D_{\vartheta}L^*(\vartheta_0)
\end{bmatrix} \nonumber\\
&\ \ \ + o_P(T^{-1/2}).  \label{genexpension}
\end{align}
}

First we note that, as in Appendix \ref{proofthm51}, we can apply \textcite[Lemma A.3]{johansen2010likelihood} to conclude that $ D_{\vartheta_i \vartheta \vartheta'}L(\vartheta^*) =  D_{\vartheta_i \vartheta \vartheta'}L(\vartheta_0) + o_P(1)$ for $i = 1,\ldots, p+1$. Consequently, we plug in the derivatives in Lemma \ref{asyappgennon1} and \ref{asyappgenstat1} into the expansion \eqref{genexpension} and find
 {\small
\begin{align}
    T^{1/2} (\hat{\vartheta}-\vartheta_0) &= - (B_{0} + T^{-1/2}  B_{1})^{-1} \left(A_{0} + T^{-1/2} A_{1}\right) - \frac{1}{2} T^{-1/2} (B_{0} + T^{-1/2}  B_{1})^{-1} \nonumber \\
    &\begin{bmatrix}
\left( A_{0} + T^{-1/2} A_{1} \right)' ( B_{0} + T^{-1/2}  B_{1} )^{-1}  C_{01}  (B_{0} + T^{-1/2}  B_{1})^{-1} \left( A_{0} + T^{-1/2} A_{1} \right) \nonumber \\
\vdots  \\
\left( A_{0} + T^{-1/2} A_{1} \right)'  (B_{0} + T^{-1/2}  B_{1})^{-1}  C_{0(p+1))} (B_{0} + T^{-1/2}  B_{1})^{-1} \left( A_{0} + T^{-1/2} A_{1} \right)
\end{bmatrix} \label{genexp}\\
&\ \ \ + o_P(T^{-1/2}).
\end{align}
}
Using the Woodbury matrix identity
\begin{align*}
    (B_0 + T^{-1/2} B_1)^{-1} &= B_0^{-1} - T^{-1/2} B_0^{-1} (I+T^{-1/2} B_1 B_0^{-1})^{-1} B_1 B_0^{-1} \\
    &= B_0^{-1} - T^{-1/2} B_0^{-1} B_1 B_0^{-1} + O_P(T^{-1}),
\end{align*}
and hence \eqref{genexp} reduces to 
 {\small
\begin{align*}
    T^{1/2} (\hat{\vartheta}-\vartheta_0) &= -B_0^{-1} A_0 - T^{-1/2} \left(B_0^{-1}A_1 - B_0^{-1}B_1B_0^{-1} A_0 + \frac{1}{2} B_0^{-1} \begin{bmatrix}
A_0'B_0^{-1} C_{0,1} B_0^{-1} A_0 \\
\vdots  \\
A_0'B_0^{-1} C_{0,p+1} B_0^{-1} A_0
\end{bmatrix} \right) \\
&\ \ \ +  o_P(T^{-1/2}).
\end{align*}
}
We find that $E(A_{0}) = E( M_{0\vartheta}^{+}) =  0$ so that 
 {\small
\begin{align}
    T E(\hat{\vartheta}-\vartheta_0) &= -\left(B_0^{-1} E(A_1) - B_0^{-1} E(B_1B_0^{-1} A_0) + \frac{1}{2} B_0^{-1} \begin{bmatrix}
E(A_0'B_0^{-1} C_{0,1} B_0^{-1} A_0) \\
\vdots  \\
E(A_0'B_0^{-1} C_{0,p+1} B_0^{-1} A_0)
\end{bmatrix} \right) \nonumber \\
&\ \ \ +  o(1). \label{exactbias1}
\end{align}
}
We rewrite
 $   E(A_0'B_0^{-1} C_{0,i} B_0^{-1} A_0) = \iota' \left(\left(B_0^{-1} C_{0,i} B_0^{-1} \right) \odot E(A_0 A_0') \right) \iota, $
and from Lemma \ref{genlemmma1}  we have that 
 $   E(A_0 A_0') =  E\left(M^+_{0,\vartheta T} (M^+_{0,\vartheta T})' \right) = A + o(1) .$
We also rewrite 
\begin{align*}
  E \left(B_1 B_0^{-1} A_0 \right)  &= 
  E \left(\left( M_{\vartheta,\vartheta' T}^{+} + M_{0,\vartheta \vartheta' T}^{+} \right )A^{-1} M^+_{0,\vartheta T} \right) \\
  &= E \left( M_{\vartheta,\vartheta' T}^{+} A^{-1} M^+_{0,\vartheta T} \right) + E \left( M_{0,\vartheta \vartheta' T}^{+} A^{-1} M^+_{0,\vartheta T} \right)\\
  &=
  \begin{bmatrix}
\iota' \left( A^{-1} \odot E\left(M_{\vartheta_1,\vartheta T}^{+} \left(M^+_{0,\vartheta T}\right)'  \right) \right)\iota \\
\vdots    \\
\iota' \left( A^{-1} \odot E\left(M_{\vartheta_{p+1},\vartheta T}^{+} \left(M^+_{0,\vartheta T}\right)'  \right) \right)\iota 
\end{bmatrix} \\
&\ \ \ +  \begin{bmatrix}
\iota' \left( A^{-1} \odot E\left(M_{0,\vartheta_1\vartheta T}^{+} \left(M^+_{0,\vartheta T}\right)'  \right) \right)\iota \\
\vdots    \\
\iota' \left( A^{-1} \odot E\left(M_{0,\vartheta_{p+1}\vartheta T}^{+} \left(M^+_{0,\vartheta T}\right)'  \right) \right)\iota 
\end{bmatrix}.
\end{align*}

From Lemma \ref{genlemmma1_1} we have that
 $    E\left(M_{0,\vartheta_k\vartheta T}^{+} \left(M^+_{0,\vartheta T}\right)'  \right) = F_{k} + o(1),  $
for $k = 1,\ldots,p+1$, with 
\begin{align}
    F_{1} = \begin{pmatrix}
-2 \zeta_3 &  \sum_{i = 0}^{\infty} D_{dd} \pi_i(0) b_{\varphi'i}(\varphi_0)  \\
 \sum_{i = 2}^{\infty} i^{-1} h_{d \varphi i}(\varphi_0)  &-    \sum_{i = 2}^{\infty}    h_{d \varphi i}(\varphi_0) b_{\varphi' i}(\varphi_0) 
\end{pmatrix},  \label{genF1}
\end{align}
and for $m = 1,\dots, p$ it follows that
\begin{align}
    F_{m+1} = \begin{pmatrix}
  \sum_{i = 2}^{\infty} i^{-1} h_{d \varphi_m i}(\varphi_0)    & - \sum_{i = 2}^{\infty}    h_{d \varphi_{m} i}(\varphi_0) b_{\varphi' i}(\varphi_0)  \\
 -\sum_{i = 1}^{\infty} i^{-1} b_{\varphi \varphi_m i}(\varphi_0)  & \sum_{i = 1}^{\infty}  b_{\varphi \varphi_m i}(\varphi_0) b_{\varphi' i}(\varphi_0) 
\end{pmatrix},  \label{genF2}
\end{align}
From Lemma \ref{genlemma891} we have that
 $    E\left(M_{\vartheta_k,\vartheta T}^{+} \left(M^+_{0,\vartheta T}\right)'  \right)  = G_{k} + o(1)  $
for $k = 1,\ldots,p+1$, with
\begin{align}
   G_{1} = \begin{pmatrix}
G_{1}(1,1) &  G_{1}(1,2) \\
G_{1}(2,1) & G_{1}(2,2) 
\end{pmatrix},  \label{genG1}
\end{align}
where the elements are given by
\begin{align*}
G_{1}(1,1) &=  -4 \zeta_3, \\
G_{1}(1,2) &=  2 \sum^{\infty}_{k = 1} b_{\varphi' k}(\varphi_0) \sum_{s = 1}^{\infty} s^{-1} (s+k)^{-1}, \\
G_{1}(2,1) &=  \sum_{k = 1}^{\infty} k^{-1} \sum_{s = 1}^{\infty} \left(  s^{-1} b_{\varphi (s+k)}(\varphi_0) + (s+k)^{-1} b_{\varphi s}(\varphi_0) \right), \\
G_{1}(2,2) &= -\sum_{k = 1}^{\infty} \sum_{s = 1}^{\infty} \left(  s^{-1} b_{\varphi (s+k)}(\varphi_0) + (s+k)^{-1} b_{\varphi s}(\varphi_0) \right)  b_{\varphi' k}(\varphi_0),
\end{align*}
and for $m = 1,\dots, p$ it follows that 
\begin{align}
    G_{m+1} = \begin{pmatrix}
 G_{m+1}(1,1)  & G_{m+1}(1,2) \\
G_{m+1}(2,1) & G_{m+1}(2,2)
\end{pmatrix}  \label{genG2}
\end{align}
where the elements are given by
\begin{align*}
 G_{m+1}(1,1) &= \sum_{k = 1}^{\infty} k^{-1} \sum_{s = 1}^{\infty} \left(  s^{-1} b_{\varphi_m (s+k)}(\varphi_0) + (s+k)^{-1} b_{\varphi_m s}(\varphi_0) \right), \\
 G_{m+1}(1,2) &= -\sum_{k = 1}^{\infty} b_{\varphi' k}(\varphi_0)  \sum_{s = 1}^{\infty} \left(  s^{-1} b_{\varphi_m (s+k)}(\varphi_0) + (s+k)^{-1} b_{\varphi_m s}(\varphi_0) \right), \\
 G_{m+1}(2,1) &=  -\sum_{k = 1}^{\infty} k^{-1} \sum_{s = 1}^{\infty} \left(  b_{\varphi_m s}(\varphi_0) b_{\varphi (s+k)}(\varphi_0) + b_{\varphi_m (s+k)}(\varphi_0) b_{\varphi s}(\varphi_0) \right), \\
 G_{m+1}(2,2) &= \sum_{k = 1}^{\infty}  \sum_{s = 1}^{\infty} \left(  b_{\varphi_m s}(\varphi_0) b_{\varphi (s+k)}(\varphi_0) + b_{\varphi_m (s+k)}(\varphi_0) b_{\varphi s}(\varphi_0) \right)  b_{\varphi' k}(\varphi_0).
\end{align*}

The expressions for $C_{01}$ and $C_{0m+1}$ for $m = 1,\dots, p$  are given in \eqref{genC1} and \eqref{genCk}, respectively.

Then
\begin{align*}
    T E(\hat{\vartheta}-\vartheta_0) &= -A^{-1} E(\sigma^{-2}_0 D_{\vartheta} L^*(\vartheta_0) ) + A^{-1}   \begin{bmatrix}
\iota' \left( A^{-1} \odot G_1 \right)\iota \\
\vdots    \\
\iota' \left( A^{-1} \odot G_{p+1} \right)\iota 
\end{bmatrix}
+  A^{-1}   \begin{bmatrix}
\iota' \left( A^{-1} \odot F_1 \right)\iota \\
\vdots    \\
\iota' \left( A^{-1} \odot F_{p+1} \right)\iota 
\end{bmatrix} \\
&\ \ \ - \frac{1}{2} A^{-1} \begin{bmatrix}
\iota' \left(\left(A^{-1} C_{0,1} A^{-1} \right) \odot A \right) \iota \\
\vdots  \\
\iota' \left(\left(A^{-1} C_{0,p+1} A^{-1} \right) \odot A \right) \iota,
\end{bmatrix} +  o(1)
\end{align*}

Hence we can write 
\begin{align}
    E\left( \hat{\vartheta} - \vartheta_0 \right) = S_T(d_0,\varphi_0) + \mathcal{B}_T(\varphi_0) + o(T^{-1}), \label{trian11}
\end{align}
where 
\begin{align*}
    T S_T(d_0,\varphi_0) &= - A^{-1} \left[ \sigma^{-2}_0 E \left( \DLvt^*(\vartheta_0) \right)  \right] = A^{-1}  \frac{\sum_{t = 1}^T c_{t}(\vartheta_0) c_{\vartheta t}(\vartheta_0)}{\sum_{t = 1}^T c^2_{t}(\vartheta_0)},
\end{align*}
from Lemma \ref{genlemmaexpectations} and 
\begin{align}
    T \mathcal{B}_T(\varphi_0) = A^{-1}   \begin{bmatrix}
\iota' \left( A^{-1} \odot \left( G_{1} + F_{1} \right)\right)\iota \\
\vdots    \\
\iota' \left( A^{-1} \odot\left( G_{p+1} + F_{p+1} \right) \right)\iota 
\end{bmatrix}  - \frac{1}{2} A^{-1} \begin{bmatrix}
\iota' \left(\left(A^{-1} C_{0,1} A^{-1} \right) \odot A_T \right) \iota \\
\vdots  \\
\iota' \left(\left(A^{-1} C_{0,p+1} A^{-1} \right) \odot A_T \right) \iota
\end{bmatrix}. \label{BtINT}
\end{align}

For $d_0 > 1/2$, it follows from Lemma \ref{genlemmaaaa2n}, and for $d_0 < 1/2$ it follows from Lemma \ref{genlemmaaaa2s} that: 
\begin{align*}
    T S_T(d_0,\varphi_0) = T\mathcal{S}_T(d_0,\varphi_0) + o(1)
\end{align*}
where 
\begin{align}
T\mathcal{S}_T(d_0,\varphi_0) = 
\begin{cases} 
A^{-1} \frac{\sum_{t = 1}^{\infty} c_{t}(\vartheta_0) c_{\vartheta t}(\vartheta_0)}{ \sum_{t = 1}^{\infty} c^2_{t}(\vartheta_0)}, & \text{if } d_0 > 1/2, \\[10pt]
A^{-1}  
\begin{bmatrix}
  -\log(T)+\left(\Psi(1-d_0) + (1-2d_0)^{-1}\right) \\
\frac{D_{\varphi_1 } \phi(1;\varphi_0)}{ \phi(1;\varphi_0) } \\
\vdots    \\
\frac{D_{\varphi_p } \phi(1;\varphi_0)}{ \phi(1;\varphi_0) }
\end{bmatrix}, & \text{if } d_0 < 1/2,
\end{cases}
\label{niceS_T}
\end{align}
completing the proof for the bias of $\hat{\vartheta}$. It follows from Lemmata \ref{asyappgennon2}, \ref{asyappgennon3}, \ref{asyappgenstat2} and \ref{asyappgenstat3} that analogues of \eqref{trian11} also hold for $\hat{\vartheta}_{\mu_0}$ and $\hat{\vartheta}_m$. Replacing $\DLvt^*(\vartheta_0)$ by $\DLvt_{\mu_0}^*(\vartheta_0)$, it is clear that the expected score term $\mathcal{S}_T(d_0,\varphi_0)$ gets eliminated. Similarly,  $E (\DLvt_{m}^*(\vartheta_0)) = 0$ by construction, and the proof is completed.  

Observe that we additionally write $S_T(d_0,\varphi_0)$ and $B_T(\varphi_0)$, where we use the exact expectations of the expressions in \eqref{exactbias1}, referred to as $F_{T,i}$, $G_{T,i}$, $A_T$ and $C_{T,0i}$, instead of the asymptotic expectations of $F_{i}$, $G_{i}$, $A$ and $C_{0i}$. Then, the ``exact'' bias is given by  
\begin{align*}
    E\left( \hat{\vartheta} - \vartheta_0 \right) = S_T(d_0,\varphi_0) + B_T(\varphi_0) + o(T^{-1}), 
\end{align*}
where
\begin{align}
    T S_T(d_0,\varphi_0) &= - A^{-1} \left[ \sigma^{-2}_0 E \left( \DLvt^*(\vartheta_0) \right)  \right] = A^{-1}  \frac{\sum_{t = 1}^T c_{t}(\vartheta_0) c_{\vartheta t}(\vartheta_0)}{\sum_{t = 1}^T c^2_{t}(\vartheta_0)}, \label{exactSappend}
\end{align}
and 
\begin{align}
    T B_T(\varphi_0) = A^{-1}   \begin{bmatrix}
\iota' \left( A^{-1} \odot \left( G_{T,1} + F_{T,1} \right)\right)\iota \\
\vdots    \\
\iota' \left( A^{-1} \odot\left( G_{T,p+1} + F_{T,p+1} \right) \right)\iota 
\end{bmatrix}  - \frac{1}{2} A^{-1} \begin{bmatrix}
\iota' \left(\left(A^{-1} C_{T,0,1} A^{-1} \right) \odot A_T \right) \iota \\
\vdots  \\
\iota' \left(\left(A^{-1} C_{T,0,p+1} A^{-1} \right) \odot A_T \right) \iota
\end{bmatrix} . \label{exactBappend}
\end{align}
With $F_{T,k} = E\left(M_{0,\vartheta_k\vartheta T}^{+} \left(M^+_{0,\vartheta T}\right)'  \right)$ such that  
\begin{align*}
    F_{T,1} = \begin{pmatrix}
-T^{-1} \sum_{t = 1}^T \sum_{i = 0}^{t-1}  D_{dd} \pi_i(0)  D_{d} \pi_i(0) &  T^{-1} \sum_{t = 1}^T \sum_{i = 0}^{t-1} D_{dd} \pi_i(0) b_{\varphi'i}(\varphi_0)  \\
 T^{-1} \sum_{t = 1}^T \sum_{i = 2}^{t-1} i^{-1} h_{d \varphi i}(\varphi_0)  & - T^{-1} \sum_{t = 1}^T \sum_{i = 2}^{t-1}    h_{d \varphi i}(\varphi_0) b_{\varphi' i}(\varphi_0) 
\end{pmatrix}, 
\end{align*}
and for $m = 1,\dots, p$ we have
\begin{align*}
    F_{T,m+1} = \begin{pmatrix}
  T^{-1} \sum_{t = 1}^T \sum_{i = 2}^{t-1}i^{-1} h_{d \varphi_m i}(\varphi_0)    & - T^{-1}\sum_{t = 1}^T \sum_{i = 2}^{t-1}    h_{d \varphi_{m} i}(\varphi_0) b_{\varphi' i}(\varphi_0)  \\
 -T^{-1} \sum_{t = 1}^T \sum_{i = 1}^{t-1} i^{-1} b_{\varphi \varphi_m i}(\varphi_0)  & T^{-1} \sum_{t = 1}^T \sum_{i = 1}^{t-1}  b_{\varphi \varphi_m i}(\varphi_0) b_{\varphi' i}(\varphi_0) 
\end{pmatrix},  
\end{align*}

With $G_{T,k}  = E\left(M_{\vartheta_k,\vartheta T}^{+} \left(M^+_{0,\vartheta T}\right)'  \right)$ such that 
\begin{align*}
   G_{T,1} = \begin{pmatrix}
G_{T,1}(1,1) &  G_{T,1}(1,2) \\
G_{T,1}(2,1) & G_{T,1}(2,2) 
\end{pmatrix}, 
\end{align*}
where the elements are given by
\begin{align*}
G_{T,1}(1,1) &= -2T^{-1} \sum_{t = 1}^T \sum_{k = 1}^{t-1} \sum_{s = t+1}^T k^{-1}(s-t)^{-1}(s-t+k)^{-1}, \\
G_{T,1}(1,2) &=  2T^{-1} \sum_{t = 1}^T \sum_{k = 1}^{t-1} \sum_{s = t+1}^T b_{\varphi' k}(\varphi_0) (s-t)^{-1} (s-t+k)^{-1}, \\
G_{T,1}(2,1) &=  T^{-1} \sum_{t = 1}^T \sum_{k = 1}^{t-1} \sum_{s = t+1}^T k^{-1}\left(  (s-t)^{-1} b_{\varphi (s-t+k)}(\varphi_0) + (s-t+k)^{-1} b_{\varphi (s-t)}(\varphi_0) \right), \\
G_{T,1}(2,2) &= - T^{-1} \sum_{t = 1}^T \sum_{k = 1}^{t-1} \sum_{s = t+1}^T \left(  (s-t)^{-1} b_{\varphi (s-t+k)}(\varphi_0) + (s-t+k)^{-1} b_{\varphi (s-t)}(\varphi_0) \right)  b_{\varphi' k}(\varphi_0),
\end{align*}
and for $m = 1,\dots, p$ we have
\begin{align*}
    G_{m+1,T} = \begin{pmatrix}
 G_{T,m+1}(1,1)  & G_{T,m+1}(1,2) \\
G_{T,m+1,T}(2,1) & G_{T,m+1}(2,2)
\end{pmatrix},  
\end{align*}
where the elements are given by
\begin{align*}
 G_{T,m+1}(1,1) &=T^{-1} \sum_{t = 1}^T \sum_{k = 1}^{t-1} k^{-1}\sum_{s = t+1}^T \left(  (s-t)^{-1} b_{\varphi_m (s-t+k)}(\varphi_0) + (s-t+k)^{-1} b_{\varphi_m (s-t)}(\varphi_0) \right), \\
 G_{T,m+1}(1,2) &= - T^{-1} \sum_{t = 1}^T \sum_{k = 1}^{t-1}b_{\varphi' k}(\varphi_0) \sum_{s = t+1}^T \left(  (s-t)^{-1} b_{\varphi_m (s-t+k)}(\varphi_0) + (s-t+k)^{-1} b_{\varphi_m (s-t)}(\varphi_0) \right), \\
 G_{T,m+1}(2,1) &=  - T^{-1} \sum_{t = 1}^T \sum_{k = 1}^{t-1}k^{-1} \sum_{s = t+1}^T \left(  b_{\varphi_m (s-t)}(\varphi_0) b_{\varphi (s-t+k)}(\varphi_0) + b_{\varphi_m (s-t+k)}(\varphi_0) b_{\varphi (s-t)}(\varphi_0) \right), \\
 G_{T,m+1}(2,2) &= T^{-1} \sum_{t = 1}^T \sum_{k = 1}^{t-1} \sum_{s = t+1}^T  \left(  b_{\varphi_m (s-t)}(\varphi_0) b_{\varphi (s-t+k)}(\varphi_0) + b_{\varphi_m (s-t+k)}(\varphi_0) b_{\varphi (s-t)}(\varphi_0) \right)  b_{\varphi' k}(\varphi_0),
\end{align*}

With $ A_T = E\left(M^+_{0,\vartheta T} (M^+_{0,\vartheta T})' \right)$  such that 
\begin{align*}
    A_T = \begin{pmatrix}
 T^{-1}  \sum_{t = 1}^T \sum_{j = 1}^{t-1} \frac{1}{j^2}  & - T^{-1}  \sum_{t = 1}^T \sum_{j = 1}^{t-1} b_{\varphi' j}(\varphi_0)/j \\
- T^{-1}  \sum_{t = 1}^T \sum_{j = 1}^{t-1} b_{\varphi j}(\varphi_0)/j  & T^{-1}  \sum_{t = 1}^T \sum_{j = 1}^{t-1} b_{\varphi j}(\varphi_0) b_{\varphi' j}(\varphi_0).
\end{pmatrix} 
\end{align*}

Lastly, $C_{T,0i}$ is defined as follows 
\begin{align*}
C_{T,01} = \begin{pmatrix}
 C_{T,01}(1,1)  & C_{T,01}(1,2) \\
C_{T,01}(2,1) & C_{T,01}(2,2)
\end{pmatrix}, 
\end{align*}
where the elements are given by
\begin{align*}
     C_{T,01}(1,1) &= -3 T^{-1} \sum_{t = 1}^T \sum_{i = 0}^{t-1}  D_{dd} \pi_i(0)  D_{d} \pi_i(0), \\
     C_{T,01}(1,2) &= 2 T^{-1} \sum_{t = 1}^T \sum_{i = 2}^{t-1}  i^{-1} h_{d \varphi' i}(\varphi_0) + T^{-1} \sum_{t = 1}^T \sum_{i = 0}^{t-1}  D_{dd} \pi_i(0) b_{\varphi'i}(\varphi_0),\\
     C_{T,01}(2,1) &= 2 T^{-1} \sum_{t = 1}^T \sum_{i = 2}^{t-1}  i^{-1} h_{d \varphi i}(\varphi_0) + T^{-1} \sum_{t = 1}^T \sum_{i = 0}^{t-1}  D_{dd} \pi_i(0) b_{\varphi i}(\varphi_0),\\
     C_{T,01}(2,2) &=  -T^{-1} \sum_{t = 1}^T \sum_{i = 1}^{t-1}  i^{-1} b_{\varphi \varphi'i}(\varphi_0) -T^{-1} \sum_{t = 1}^T \sum_{i = 2}^{t-1}    b_{\vartheta i}(\varphi_0) h_{d \vartheta' i}(\varphi_0) \\
     &\ \ \ - \left(T^{-1} \sum_{t = 1}^T \sum_{i = 2}^{t-1}    b_{\vartheta i}(\varphi_0) h_{d \vartheta' i}(\varphi_0)\right)',
\end{align*}
and for $k = 1,\dots, p$ we have that
\begin{align*}
C_{T,0(k+1)} = \begin{pmatrix}
 C_{T,0(k+1)}(1,1)  & C_{T,0(k+1)}(1,2) \\
C_{T,0(k+1)}(2,1) & C_{T,0(k+1)}(2,2),
\end{pmatrix}  
\end{align*}
where the elements are given by
\begin{align*}
     C_{T,0(k+1)}(1,1) &= 2 T^{-1} \sum_{t = 1}^T \sum_{i = 2}^{t-1}  i^{-1} h_{d \varphi_k i}(\varphi_0)  + T^{-1} \sum_{t = 1}^T \sum_{i = 0}^{t-1}  D_{dd} \pi_i(0) b_{\varphi_k i}(\varphi_0),   \\
     C_{T,0(k+1)}(1,2) &= -T^{-1} \sum_{t = 1}^T \sum_{i = 1}^{t-1}  i^{-1} b_{\varphi' \varphi_k i}(\varphi_0) -    T^{-1} \sum_{t = 1}^T \sum_{i = 2}^{t-1}    b_{\varphi_k i}(\varphi_0) h_{d \varphi'  i}(\varphi_0) \\ 
     &\ \ \ -  T^{-1} \sum_{t = 1}^T \sum_{i = 2}^{t-1}   b_{\varphi'  i}(\varphi_0) h_{d \varphi_k  i}(\varphi_0),   \\
     C_{T,0(k+1)}(2,1) &= -T^{-1} \sum_{t = 1}^T \sum_{i = 1}^{t-1}  i^{-1} b_{\varphi \varphi_k i}(\varphi_0) -    T^{-1} \sum_{t = 1}^T \sum_{i = 2}^{t-1}    b_{\varphi_k i}(\varphi_0) h_{d \varphi  i}(\varphi_0) \\
     & \ \ \ -  T^{-1} \sum_{t = 1}^T \sum_{i = 2}^{t-1}    b_{\varphi  i}(\varphi_0) h_{d \varphi_k  i}(\varphi_0),  \\
    C_{T,0(k+1)}(2,2) &= \left( T^{-1} \sum_{t = 1}^T \sum_{i = 1}^{t-1}   b_{\varphi i}(\varphi_0) b_{\varphi' \varphi_k i}(\varphi_0) \right)' + T^{-1} \sum_{t = 1}^T \sum_{i = 1}^{t-1}   b_{\varphi  i}(\varphi_0) b_{\varphi' \varphi_k i}(\varphi_0)  \\
    &\ \ \ + T^{-1} \sum_{t = 1}^T \sum_{i = 1}^{t-1}   b_{\varphi_k i}(\varphi_0) b_{\varphi \varphi' i}(\varphi_0).
\end{align*}

\subsubsection{Proof of Corollary \ref{bcmcorr}} \label{val2step}

We need to show that 
 $   T \left(\mathcal{B}_T(\hat{\varphi}_m) - \mathcal{B}_T(\varphi_0) \right) = o_P(1). $
Inspecting the expression $\mathcal{B}_T(\varphi_0)$ in \eqref{BtINT} it follows that it is sufficient to show that 
\begin{align}
    b_{\varphi_k i}(\hat{\varphi}_m)-b_{\varphi_k i}(\varphi_0) = O_P(T^{-1/2} i^{-1-\varsigma}), \label{prb1} \\
    b_{\varphi_k \varphi_j i}(\hat{\varphi}_m)-b_{\varphi_k \varphi_j i}(\varphi_0) = O_P(T^{-1/2} i^{-1-\varsigma}), \label{prb2}
\end{align}
for $k,j = 1,\ldots p+1$, where $b_{ \cdot i}$ is defined in \eqref{defb}. The proof strategy of \eqref{prb1} and \eqref{prb2} are similar and closely related to the proof in \textcite[Lemma S.7]{hualde2020truncated}. Therefore, we only give the proof of \eqref{prb1}.

Note, 
\begin{align}
      b_{\varphi_k i}(\hat{\varphi}_m)-b_{\varphi_k i}(\varphi_0) &= \sum_{s = 0}^{i-1} \omega_s(\varphi_0) \left( \frac{\partial \phi_{i-s}(\hat{\varphi}_m)}{\partial \varphi_k} - \frac{\partial \phi_{i-s}(\varphi_0)}{\partial \varphi_k} \right) \nonumber \\
      &\ \ \ + \sum_{s = 0}^{i-1} \left(\omega_s(\hat{\varphi}_m)-\omega_s(\varphi_0) \right) \frac{\partial \phi_{i-s}(\varphi_0)}{\partial \varphi_k} \nonumber  \\
      &\ \ \ + \sum_{s = 0}^{i-1} \left(\omega_s(\hat{\varphi}_m)-\omega_s(\varphi_0) \right) \left( \frac{\partial \phi_{i-s}(\hat{\varphi}_m)}{\partial \varphi_k} - \frac{\partial \phi_{i-s}(\varphi_0)}{\partial \varphi_k} \right) \label{bproof1}
\end{align}
Fix $\varepsilon < 1/2$. Then 
\begin{align}
    \frac{\partial \phi_{i}(\hat{\varphi}_m)}{\partial \varphi_k} - \frac{\partial \phi_{i}(\varphi_0)}{\partial \varphi_k} = \left(   \frac{\partial \phi_{i}(\hat{\varphi}_m)}{\partial \varphi_k} - \frac{\partial \phi_{i}(\varphi_0)}{\partial \varphi_k}  \right) \left( I(\left\| \hat{\varphi}_m - \varphi_0 \right\| < \varepsilon) + I(\left\| \hat{\varphi}_m - \varphi_0 \right\| \geq \varepsilon ) \right), \label{val21}
\end{align}
so by the mean value theorem the left-hand side of \eqref{val21} is bounded by 
\begin{align}
      \underset{\left\| \varphi - \varphi_0 \right\| < \varepsilon }{\sup} \left\| \frac{\partial^2 \phi_{i}(\varphi)}{\partial \varphi_k \partial \varphi} \right\| \left\| \hat{\varphi}_m - \varphi_0 \right\| + c  \underset{\varphi \in  }{\sup} \left|  \frac{\partial \phi_{i}(\varphi)}{\partial \varphi_k} \right| \frac{ \left\| \hat{\varphi}_m - \varphi_0 \right\|^N }{\varepsilon^N}, \label{val22}
\end{align}
for any arbitrarily large fixed number $N$. Then by \eqref{r11_2} and the $\sqrt{T}$-consistency of $\hat{\varphi}_m$, the second term in \eqref{val22} is of smaller order, whereas by \eqref{r11_3}, the first one is $O_P(T^{-1/2}i^{-1-\varsigma})$. This implies that the first term on the right-hand side of \eqref{bproof1} is $O_P(T^{-1/2} i^{-1-\varsigma})$ by \eqref{lA3} of Lemma \ref{genbounds}. Next, a similar proof concludes that 
\begin{align*}
    \omega_s(\hat{\varphi}_m)-\omega_s(\varphi_0) = O_P(T^{-1/2} s^{-1-\varsigma})
\end{align*}
so that the second term of \eqref{bproof1} is $O_P(T^{-1/2} i^{-1-\varsigma})$. Finally, combining the arguments for the first two terms, the third term on the right-hand
side of \eqref{bproof1} is of smaller order, to conclude the proof of \eqref{prb1}.

\subsubsection{Proof of Theorem \ref{t54}} \label{proofthm54}

The lag polynomial for AR(1) specification is given by
$\omega(L;\varphi) = (1-\varphi L )^{-1} = \sum_{j = 0}^{\infty} \omega_j(\varphi) L^j$, where $\omega_j(\varphi) = \varphi^j$, $j\geq 0$. The inverse lag polynomial is given by 
$\phi(L;\varphi) = \omega^{-1}(L;\varphi) = (1-\varphi L ) = \sum_{j = 0}^{\infty} \phi_j(\varphi) L^j,$ where $\phi_0(\varphi) = 1$ and $\phi_1(\varphi) = -\varphi$, $\phi_s(\varphi) = 0$, $s \geq 2$. Taking the derivatives of $\phi_s(\varphi)$ with respect to $\varphi$ yields: $\partial \phi_0(\varphi)/\partial \varphi = 0$, $\partial \phi_1(\varphi)/\partial \varphi = -1$, and $\partial \phi_s(\varphi)/\partial \varphi = 0$, $s \geq 2$. Then $b_{\varphi j}(\varphi_0) = \sum_{k = 0}^{j-1} \omega_k(\varphi_0) \partial \phi_{j-k}(\varphi_0)/\partial \varphi = -\varphi_0^{j-1}$, for $j \geq 1$. Since $\partial^2 \phi_j(\varphi)/\partial \varphi^2 = 0$  for $j \geq 1$ we get $b_{\varphi \varphi j}(\varphi_0) = \sum_{k = 0}^{j-1} \omega_k(\varphi_0) \partial^2 \phi_{j-k}(\varphi_0)/\partial \varphi^2 = 0$. 

We need the following expansions in deriving the bias expressions 
\begin{align}
    \sum_{k = 0}^{\infty} x^{k} &= \frac{1}{1-x},  \label{expans1} \\
    \sum_{k = 1}^{\infty} k^{-1} x^k &= - \log(1-x), \label{expans2}\\
    \sum_{k = 1}^{\infty} (k+1)^{-1} x^{k+1}\sum_{n = 1}^k n^{-1} &= \frac{1}{2} \log^2(1-x) , \label{expans3} \\
    \sum_{k = 1}^{\infty} x^k \sum_{s = 1}^{\infty} s^{-1} (s+k)^{-1} &= -Li_{2} (-\frac{x}{1-x}) , \label{expans4}
\end{align}
where $|x|<1$.
The first three expansions are well known and can be found in \textcite{gradshteyn2014table} on pages 7 (0.231-1), 44 (1.513-4), and 45 (1.516-1) respectively. The last expansion makes use of a couple of results. First, we have the expression:
\begin{align*}
    \sum_{s = 1}^{\infty} s^{-1}(s+k)^{-1} &= k^{-1} \left( \Psi(k+1) + \gamma \right), \\
     \Psi(k+1) + \gamma  &= \int_{0}^{1} (1-t)^{-1}(1-t^k) dt, 
\end{align*}
These results can be found in \textcite{abramowitz1964handbook} on page 259, 6.3.16 and 6.3.22 respectively. Using these results and \eqref{expans2}, we get
\begin{align*}
     \sum_{k = 1}^{\infty} x^k \sum_{s = 1}^{\infty} s^{-1} (s+k)^{-1} &=  \sum_{k = 1}^{\infty} x^k   k^{-1}  \int_{0}^{1} (1-t)^{-1}(1-t^k) dt \\
     &= \int_{0}^{1} (1-t)^{-1}\sum_{k = 1}^{\infty} x^k   k^{-1}  (1-t^k) dt \\
     &= \int_{0}^{1} (1-t)^{-1}\log\left( \frac{1-xt}{1-x} \right) dt, 
\end{align*}
Then a change of variable of integration to get 
\begin{align*}
    \int_{0}^{1} (1-t)^{-1}\log\left( \frac{1-xt}{1-x} \right) dt &= - \int_{1}^{(1-x)^{-1}} (1-u)^{-1} \log(u) du  = - Li_2(-\frac{x}{1-x}),
\end{align*}
where $Li_{2}(\varphi) = \sum_{i = 1}^{\infty} i^{-2}\varphi^{i}$, or alternatively $Li_2(1-v) = \int_{1}^v (1-t)^{-1}\log(t) dt $, is the dilogarithm function (Spence's integral), see \textcite[page 1004, 27.7.1]{abramowitz1964handbook} for the integral representation where a slight different definition of the dilogarithm function is used, namely $f(x) = Li_{2}(1-x)$.

First, we find the expressions for the intrinsic bias $\mathcal{B}_T(\vartheta_0)$ in \eqref{BtINT}. The expressions for $A$ in \eqref{genA}, and its inverse  $A^{-1}$ follows from using \eqref{expans1} and \eqref{expans2}: 
\begin{align*}
    A = \begin{pmatrix}
\pi^2/6 &  - \varphi_0^{-1} \log(1-\varphi_0) \\
 - \varphi_0^{-1} \log(1-\varphi_0)  &  (1-\varphi_0^2)^{-1}
\end{pmatrix} ,
\end{align*}
and 
\begin{align*}
    A^{-1} = \frac{\varphi_0}{\pi^2 \varphi_0^2 - 6 (1-\varphi_0^2)
    \log^2(1-\varphi_0)} \begin{pmatrix}
6 \varphi_0&  6 \log(1-\varphi_0) (1-\varphi_0^2) \\
6 \log(1-\varphi_0) (1-\varphi_0^2)   &  \pi^2 \varphi_0(1-\varphi_0^2)
\end{pmatrix}.  
\end{align*}

Next, we find the expressions for $C_{01}$ and $C_{02}$ in \eqref{genC1} and \eqref{genCk}, respectively. Using \eqref{expans2} and \eqref{expans3} we find 
\begin{align*}
C_{01} = \begin{pmatrix}
  -6 \zeta_3 &  2 \varphi_0^{-1} Li_{2}(-\frac{\varphi_0}{1-\varphi_0}) - \varphi_0^{-1} \log^2(1-\varphi_0) \\
 2 \varphi_0^{-1} Li_{2}(-\frac{\varphi_0}{1-\varphi_0}) - \varphi_0^{-1} \log^2(1-\varphi_0)  & 2 \frac{\log(1 - \varphi_0)}{1-\varphi_0^2}
\end{pmatrix}, 
\end{align*}
and  
\begin{align*}
C_{02} = \begin{pmatrix}
 2 \varphi_0^{-1} Li_{2}(-\frac{\varphi_0}{1-\varphi_0}) - \varphi_0^{-1} \log^2(1-\varphi_0)  & 2 \frac{\log(1 - \varphi_0)}{1-\varphi_0^2} \\
 2 \frac{\log(1 - \varphi_0)}{1-\varphi_0^2} & 0
\end{pmatrix}.
\end{align*}

Next, we find expressions for $F_1$ and $F_2$ in \eqref{genF1} and \eqref{genF2}, respectively,. Using \eqref{expans2} and \eqref{expans3} we find 
\begin{align*}
    F_{1} = \begin{pmatrix}
-2 \zeta_3 & -\varphi_0^{-1}\log^2(1-\varphi_0)   \\
  \varphi_0^{-1} Li_{2}(-\frac{\varphi_0}{1-\varphi_0}) &  \frac{\log(1-\varphi_0)}{1-\varphi_0^2}
\end{pmatrix},  
\end{align*}
and 
\begin{align*}
    F_{2} = \begin{pmatrix}
  \varphi_0^{-1} Li_{2}(-\frac{\varphi_0}{1-\varphi_0})   & \frac{\log(1-\varphi_0)}{1-\varphi_0^2}\\
 0  & 0
\end{pmatrix}.
\end{align*}

Next, we find expressions for $G_1$ and $G_2$ in \eqref{genG1} and \eqref{genG2}, respectively. Using \eqref{expans1},\eqref{expans2} and \eqref{expans4}  we find 
\begin{align*}
   G_{1} = \begin{pmatrix}
-4 \zeta_3 &  2 \varphi_0^{-1} Li_2(-\frac{\varphi_0}{1-\varphi_0}) \\
 - \varphi_0^{-1} \log^2(1- \varphi_0) + \varphi_0^{-1} Li_2(-\frac{\varphi_0}{1-\varphi_0}) & \frac{\log(1-\varphi_0)}{1-\varphi_0^2} - \varphi_0^{-2} \left(\frac{\varphi_0 }{1-\varphi_0 } + \log(1-\varphi_0 )\right) 
\end{pmatrix},  
\end{align*}
and
\begin{align*}
    G_{2} = \begin{pmatrix}
 - \varphi_0^{-1} \log^2(1- \varphi_0) + \varphi_0^{-1} Li_2(-\frac{\varphi_0}{1-\varphi_0})  & \frac{\log(1-\varphi_0)}{1-\varphi_0^2} - \varphi_0^{-2} \left(\frac{\varphi_0 }{1-\varphi_0 } + \log(1-\varphi_0 )\right) \\
2\log(1-\varphi_0) \frac{1}{1-\varphi_0^2}& -2 \frac{\varphi_0}{(1-\varphi_0^2)^2}
\end{pmatrix}.
\end{align*}

Second, we find the expressions for the intrinsic bias $\mathcal{S}_T(\vartheta_0)$ in \eqref{niceS_T}. First we consider the non-stationary region, i.e.\ $d_0 > 1/2$. We have that
 $   c_t(\vartheta) = \sum_{j = 0}^{t-1} \phi_j(\varphi) \kappa_{0(t-j)}(d) =  \kappa_{0t}(d) - \varphi \kappa_{0(t-1)}(d) I(t \geq 2),   $
and, therefore, 
\begin{align}
    \sum_{t = 1}^{\infty} c^2_t(d,\varphi)
        &=  (1+\varphi^2)\sum_{t = 1}^{\infty} \kappa^2_{0t}(d) - 2\varphi \sum_{t = 1}^{\infty} \kappa_{0t}(d) \kappa_{0(t+1)}(d) \nonumber  \\
        &=  (1+\varphi^2)\sum_{t = 1}^{\infty} \kappa^2_{0t}(d) + 2\varphi \sum_{t = 1}^{\infty} \left(\kappa_{0(t+1)}(d)- \kappa_{0t}(d) \right) \kappa_{0(t+1)}(d) - 2  \varphi  \sum_{t = 1}^{\infty} \kappa^2_{0(t+1)}(d) \nonumber \\
        &=  (1+\varphi^2- 2\varphi)\sum_{t = 1}^{\infty} \kappa^2_{0t}(d) + 2\varphi \sum_{t = 1}^{\infty} \kappa_{0t}(d)\kappa_{0t}(1+d), \label{ctarfi}
\end{align}
where we used the properties $\pi_{0}(u) = 1$ and  $\pi_{t}(u) - \pi_{t-1}(u) = \pi_t(u-1)$ for any $u$, see \textcite[Lemma A.4]{johansen2016role}. The first summand in this expression is given in \textcite[Lemma B.1]{johansen2016role}, i.e.\ , 
\begin{align*}
    \sum_{t = 1}^{\infty} \kappa^2_{0t}(d) =  \binom{2d-2}{d-1},
\end{align*}
where $d > 1/2$. The second term can be derived using a similar proof strategy.
We have that 
\begin{align*}
   \sum_{t = 1}^{\infty} \kappa_{0t}(d)\kappa_{0t}(1+d) &= \frac{1}{\Gamma(1-d) \Gamma(-d)} \sum_{t = 0}^{\infty} \frac{\Gamma(1-d+t)\Gamma(-d+t)}{\Gamma(t) t!} = 0.5  \binom{2d}{d},
\end{align*}
where the last equality follows from \textcite[p. 556, eqn. 15.1.20)]{abramowitz1964handbook}.
We conclude that 
\begin{align*}
    \sum_{t = 1}^{\infty} c^2_t = (1-\varphi)^2 \binom{2d-2}{d-1} +  \varphi  \binom{2d}{d}. 
\end{align*}

Next, we find an expression for $\sum_{t = 1}^{\infty} c_t(d,\varphi) D_{d} c_t(d,\varphi)$. Taking the derivative of the left and right-hand side of \eqref{ctarfi} with respect to $d$ gives 
\begin{align*}
     2 \sum_{t = 1}^{\infty} c_t(d,\varphi) D_{d} c_t(d,\varphi) =  2(1-\varphi)^2   \sum_{t = 1}^{\infty} \kappa_{1t}(d)  \kappa_{0t}(d)   +2 \varphi  \sum_{t = 1}^{\infty} \left( \kappa_{1t}(d) \kappa_{0t}(1+d) + \kappa_{0t}(d) \kappa_{1t}(1+d) \right) . 
\end{align*}
The first summand in this expression is given in \textcite[Lemma B.1]{johansen2016role}, i.e.\ , 
\begin{align*}
      \sum_{t = 1}^{\infty} \kappa_{0t}(d)\kappa_{1t}(d) = \binom{2d-2}{d-1} \left( \Psi(2d-1)-\Psi(d) \right),
\end{align*}
where $d > 1/2$. The second term can be obtained using a similar approach as in \textcite[Lemma B.1]{johansen2016role}, and is given by 
\begin{align*}
    D_d \sum_{t = 1}^{\infty} \kappa_{0t}(d)\kappa_{0t}(1+d)   = \binom{2d}{d} \left( \Psi(2d+1)-\Psi(d+1) \right).
\end{align*}
We conclude that 
\begin{align*}
     \sum_{t = 1}^{\infty} c_t(d,\varphi) D_{d} c_t(d,\varphi) &= (1-\varphi)^2  \binom{2d-2}{d-1} \left( \Psi(2d-1)-\Psi(d) \right)  \\
     &\ \ \ + \varphi  \binom{2d}{d} \left( \Psi(2d+1)-\Psi(d+1) \right).
\end{align*}

Next, we find an expression for $\sum_{t = 1}^{\infty} c_t(d,\varphi) D_{\varphi} c_t(d,\varphi)$. Taking the derivative of the left-hand side and right-hand side of \eqref{ctarfi} with respect to $\varphi$ gives 
\begin{align*}
     2 \sum_{t = 1}^{\infty} c_t(d,\varphi) D_{\varphi} c_t(d,\varphi) = (2\varphi- 2)\sum_{t = 1}^{\infty} \kappa^2_{0t}(d) + 2\sum_{t = 1}^{\infty} \kappa_{0t}(d)\kappa_{0t}(1+d).
\end{align*}
Note that the expressions for the two summands are given above. We conclude that 
\begin{align*}
    \sum_{t = 1}^{\infty} c_t(d,\varphi) D_{\varphi} c_t(d,\varphi) = (\varphi- 1)   \binom{2d-2}{d-1}  + 0.5 \binom{2d}{d}.
\end{align*}
Therefore the score bias $ \mathcal{S}_T(d_0,\varphi_0)$ for $d_0 > 1/2$ is given by
\begin{align*}
     T \mathcal{S}_T(d_0,\varphi_0)  &= A^{-1} \left[ (1-\varphi_0)^2 \binom{2d_0-2}{d_0-1} +  \varphi_0  \binom{2d_0}{d_0} \right]^{-1} \times
     \\ 
     &\begin{bmatrix} (1-\varphi_0)^2  \binom{2d_0-2}{d_0-1} 
     \left( \Psi(2d_0-1)-\Psi(d_0) \right)  + \varphi_0  \binom{2d_0}{d_0} \left( \Psi(2d_0+1)-\Psi(d_0+1) \right)
  \\
(\varphi_0- 1)   \binom{2d_0-2}{d_0-1}  +  0.5\binom{2d_0}{d_0}
\end{bmatrix},
\end{align*}
The score bias $\mathcal{S}_T(d_0,\varphi_0)$ for $d_0 < 1/2$ is given by
\begin{align*}
     T \mathcal{S}_T(d_0,\varphi_0) = A^{-1} 
     &\begin{bmatrix} -\log(T)+\Psi(1-d_0) + (1-2d_0)^{-1}
  \\
-\frac{1}{1-\varphi_0}
\end{bmatrix}, 
\end{align*}
because $\phi(1;\varphi) = 1-\varphi$.

\subsection{The impact of unobserved pre-sample values} \label{Initschemes}

So far our analysis has been conducted under the simplifying assumption that all pre-sample values of the observables $x_t$ as well as of the error terms $u_t$ and $\varepsilon_t$ are zero, cf.\ Assumption \ref{A5}. In this subsection, we relax this assumption
and investigate how alternative initialisation schemes for the unobserved
history of the process affect the bias of the CSS and MCSS estimators.

Our first initialisation scheme in Section \ref{proofcorrcorinitial} involves Johansen and Nielsen’s (2016) nonstationary type-II process with a finite number of unobserved pre-sample observations. We derive explicit bias
expressions for our MCSS estimator for that data-generating process.

For our second initialisation scheme, see Section \ref{proofthcorinitial1}, we consider both stationary and nonstationary type-II processes with an error term $u_t$ that started in the infinite past. This is our model in \eqref{genq1}-\eqref{genq2} supplemented by Assumptions \ref{A2}-\ref{A1} {\it but not} Assumption \ref{A5}. It is also the model considered  by Hualde and Nielsen (2020). 
We extend the main results of the
paper, including Theorems \ref{t52} and \ref{t53} to \ref{t55} as well as Corollary \ref{t56}, to this setting.

\subsubsection{Initialisation scheme I: \texorpdfstring{$N_0$}{N0} unobserved pre-sample values} \label{proofcorrcorinitial}

Consider the model analysed by \textcite{johansen2016role}, i.e.\ set $\omega(L;\varphi) = 1$  in \eqref{genq1}-\eqref{genq2}, assume that $d_0 > 1/2$, and introduce $N_0$ unobserved pre-sample values by replacing the truncation operator $ \Delta^{d}_+$ by $ \Delta_{-N_0}^{d}$, see Section \ref{model}. As
a result, the model of interest becomes
\begin{align}
    \Delta_{-N_0}^{d} \left(x_t - \mu \right)= \epsilon_t. \label{modelintial}
\end{align}
The following theorem is referred to at the end of Section \ref{arfima0d}. The bias expressions for the CSS estimator with $\mu$ known and estimated are given in \textcite[Theorem 2]{johansen2016role} and mentioned here for completeness.

\begin{theorem}\label{corinitial}
  Let $x_t$, $t$ = 1,$\ldots$,$T$, be given by \eqref{modelintial} and let Assumptions \ref{A2} and \ref{A3} be satisfied. All expectations and biases below are taken \emph{conditional on the pre-sample observations} $\{x_t:t<1\}$
  or, equivalently, on the pre-sample path $x_{-N_0+1},\ldots,x_0$. For the non-stationary region, i.e.\ when $d_0 > 1/2$, the conditional bias of the estimators are as follows:
  \begin{align*}
  bias(\hat{d}) &= \mathcal{B}_T(d_0) + \mathcal{S}_T(d_0)+ \mathcal{S}^{init,\mu}_T(d_0) + o(T^{-1}) \\
  bias(\hat{d}_{\mu_0}) &= \mathcal{B}_T(d_0) + \mathcal{S}_T(d_0)+ \mathcal{S}^{init,\mu_0}_T(d_0) + o(T^{-1}) \\
  bias(\hat{d}_m) &= \mathcal{B}_T(d_0) + \mathcal{S}^{init,\mu}_T(d_0) + o(T^{-1}).
  \end{align*}
  The intrinsic bias $\mathcal{B}_T(d_0)$ and the unknown-level score bias $\mathcal{S}_T (d_0)$ are given in Theorem \ref{t1}. The score bias due to pre-sample values depends on whether or not a constant term is estimated:
  \begin{align}
      T\,\mathcal{S}_T^{\mathrm{init},\mu}(d_0)
&= - \lim_{T\to\infty} A^{-1}
 D\left[
\frac{1}{2}\left(
\sum_{t=1}^{T} (\eta_{0t}(d))^2
-\frac{\Big(\sum_{t=1}^{T}\kappa_{0t}(d)\,\eta_{0t}(d)\Big)^2}
{\sum_{t=1}^{T}\kappa^2_{0t}(d)}
\right)
\right]\Bigg|_{d=d_0},\\
T\,\mathcal{S}_T^{\mathrm{init},\mu_0}(d_0)
&= - \lim_{T\to\infty} A^{-1} D
\left[
\frac{1}{2}\sum_{t=1}^{T} (\eta_{0t}(d))^2
\right]\Bigg|_{d=d_0}.
  \end{align}
  where $\kappa_{0t}(d_0)$, $\eta_{0t}(d_0)$ and $D_d \eta_{0t}(d_0)$ are given in \eqref{detc}, \eqref{eta_jn} and \eqref{Deta_jn}, respectively. 
\end{theorem}
To prove the bias of the MCSS estimator, we note first that it is analogous to the bias expression in \eqref{trian}, i.e.\
\begin{align}
    E\left( \hat{d}_m - d_0 \right) = S_T(d_0) + B_T(d_0) + o(T^{-1}), \label{trian1}
\end{align}
where $S_T(d_0) = - A^{-1} T^{-1} [ \sigma^{-2}_0 E \left( D L_m^*(d_0) \right) ]$ and $B_T(d_0)$ is given in \eqref{exactBappend}.
To this end, we see from the first three derivatives of the CSS objective function in \textcite[Lemma B.4]{johansen2016role} that the unobserved pre-sample values enter only the score bias, while the remaining terms are unaffected, regardless of whether the model includes unobserved pre-sample values. Using their expressions of the derivatives and the simple multiplicative structure of the MCSS objective function, it can be shown that the derivatives of the MCSS objective are the same to those in Lemma \ref{asyappgennon3} or \textcite[Lemma B.4]{johansen2016role}, except for the expectation of the score, which differs. This implies that we can use \eqref{trian1} to find the bias of the MCSS estimator. The exact and approximate intrinsic biases remain unaffected by the unobserved pre-sample values, and the latter is given in Theorem \ref{t1}, i.e. $\mathcal{B}_T (d_0) = - 3 \zeta_{3} \zeta_{2}^{-2}/T$.
Therefore, to complete our proof of Theorem \ref{corinitial} the task reduces to finding the score bias $- A^{-1} T^{-1} [ \sigma^{-2}_0 E \left( D L_m^*(d_0) \right) ]$ for model \eqref{modelintial}. The next lemma provides $E(D L_m^*(d_0))$.
\begin{lemma}\label{jnmcss} 
  Let $x_t$, $t$ = 1,$\ldots$,$T$, be given by \eqref{genq1} and let $u_t = \epsilon_t$. Let Assumptions \ref{A2} and \ref{A3} be satisfied. All expectations and biases below are taken \emph{conditional on the pre-sample observations} $\{x_t:t<1\}$ or, equivalently, on the pre-sample path $x_{-N_0+1},\ldots,x_0$. Then the conditional expectation of $D L_m^*(d_0)$ is
  \begin{align*}
      E(D L_m^*(d_0)) = \sigma^2_0 D\left[
\frac{1}{2}\left(
\sum_{t=1}^{T} (\eta_{0t}(d))^2
-\frac{\Big(\sum_{t=1}^{T}\kappa_{0t}(d)\,\eta_{0t}(d)\Big)^2}
{\sum_{t=1}^{T}\kappa^2_{0t}(d)}
\right)
\right]\Bigg|_{d=d_0} + O(T^{-1}),
  \end{align*}
  where $\kappa_{0t}(d_0)$, $\eta_{0t}(d_0)$ and $D_d \eta_{0t}(d_0)$ are given in \eqref{detc}, \eqref{eta_jn} and \eqref{Deta_jn}, respectively. 
\end{lemma}
\begin{proof}
We first introduce expressions similar to \textcite[page 1104 and page 1105]{johansen2016role} and let
\begin{align}
    \xi_{N_0,T}(d_0) &= D\left[ \frac{1}{2}\left( \sum_{t=1}^{T} (\eta_{0t}(d))^2 -\frac{\Big(\sum_{t=1}^{T}\kappa_{0t}(d) \eta_{0t}(d)\Big)^2} {\sum_{t=1}^{T}\kappa^2_{0t}(d)} \right) \right]\Bigg|_{d=d_0},\\
 \eta_{0t}(d_0) &= - \sum_{n=-N_0+1}^{0} \pi_{t-n}(-d_0)(x_n - \mu_0)/\sigma_0, \label{eta_jn}\\
 D_d\eta_{0t}(d_0) &=  \eta_{1t}(d_0) = \sum_{k=1}^{t-1} k^{-1} \sum_{n=-N_0+1}^{0} \pi_{t-n-k}(-d_0)(x_n - \mu_0)/\sigma_0, \label{Deta_jn}
\end{align}
where $\kappa_{0t}(d) = \pi_{t-1}(1-d)$ and $\kappa_{1t}(d) = D_d \kappa_{0t}(d) = -D \pi_{t-1}(1-d)$.  

The MCSS objective function is defined as $L^*_m(d) = m(d) L^*(d)$. Taking its first derivative gives $DL^*_m(d) = Dm(d) L^*(d) + m(d) DL^*(d)$. Hence, $EDL^*_m(d_0) = Dm(d_0) E L^*(d_0) + m(d_0) E DL^*(d_0)$. We therefore need to
find $E L^*(d_0)$ and $EDL^*(d_0)$. To improve readability, we suppress the explicit dependence of $\eta$, $\kappa$ and $\mu$ on $d_0$.

Note that $L^*(d_0) = \frac{1}{2} \sum_{t=1}^{T} \epsilon_t^2(d_0)$, where $\epsilon_t(d_0)$ is given by $\epsilon_t(d_0) = \epsilon_t + \sigma_0 \eta_{0t} - \kappa_{0t} (\hat{\mu} - \mu_0)$, see \textcite[equation (B.1)]{johansen2016role}, with $\hat{\mu} - \mu_0 = (\sum_{t = 1}^T \epsilon_t \kappa_{0t} + \sigma_0  \sum_{t = 1}^T \eta_{0t} \kappa_{0t})/(\sum_{t = 1}^T \kappa^2_{0t})$.
\begin{align*}
    E L^*(d_0) = \frac{1}{2} \sum_{t=1}^{T} E( \epsilon_t^2 + \sigma^2_0 \eta_{0t}^2 + \kappa_{0t}^2 (\hat{\mu} - \mu_0)^2 + 2 \epsilon_t \sigma_0 \eta_{0t} - 2 \epsilon_t \kappa_{0t} (\hat{\mu} - \mu_0) - 2 \eta_{0t} \sigma_0 \kappa_{0t} (\hat{\mu} - \mu_0) ).
\end{align*}
Since $\epsilon_t$ is \textit{IID}(0,$\sigma_0^2$) we can simplify the terms inside this expectation. To this end, $E\epsilon_t^2 = \sigma_0^2$, and $\kappa_{0t}^2 E (\hat{\mu} - \mu_0)^2 = \kappa_{0t}^2   (\sigma_0^2\sum_{t = 1}^T \kappa^2_{0t} +  \sigma_0^2(\sum_{t = 1}^T \eta_{0t} \kappa_{0t})^2)/(\sum_{t = 1}^T \kappa^2_{0t})^2$, and $E\epsilon_t \eta_{0t} = 0$, and $E \epsilon_t \kappa_{0t} (\hat{\mu} - \mu_0) = (\sigma^2_0 \kappa^2_{0t})/ \sum_{t = 1}^T \kappa^2_{0t}$, and $E \eta_{0t} \kappa_{0t} (\hat{\mu} - \mu_0) = \eta_{0t} \kappa_{0t} \sigma_0 \sum_{t = 1}^T \eta_{0t} \kappa_{0t} / \sum_{t = 1}^T \kappa^2_{0t}$. Thus,
\begin{align*}
E L^*(d_0) =  \frac{1}{2}\sigma^2_0(T-1)  + \frac{1}{2} \sigma^2_0 \sum_{t=1}^{T} \eta^2_{0t} -  \frac{1}{2} \sigma^2_0\frac{ \left( \sum_{t = 1}^T \eta_{0t} \kappa_{0t} \right)^2}{\sum_{t = 1}^T \kappa^2_{0t}}.
\end{align*}

\textcite[Lemma B.4]{johansen2016role} show that 
\begin{align*}
    E[\sigma^{-2}_0 T^{-1/2}D L^*(d_0)] = T^{-1/2} ( \xi_{N_0,T}(d_0) + \tau_T),
\end{align*}
with $\tau_T = - \sum_{t = 1}^T \kappa_{0t}\kappa_{1t}/\sum_{t = 1}^T \kappa^2_{0t}$. Note that the first term on right-hand side of this expression is due to unobserved pre-sample values and the second term due to the presence of the nuisance parameter $\mu$. Indeed, the second term is the same as in Lemma \ref{genlemmaexpectations}

Then using both results and $EDL^*_m(d_0) =  Dm(d_0) E L^*(d_0) + m(d_0) E DL^*(d_0)$ gives the expectation of $\sigma^{-2}_0 T^{-1/2}D L_m^*(d_0)$: 
\begin{align*}
E(\sigma^{-2}_0 T^{-1/2}D L_m^*(d_0)) 
&=  \sigma^{-2}_0 T^{-1/2} Dm(d_0) \frac{1}{2}\sigma^2_0(T-1) +  m(d_0)  T^{-1/2} \tau_T \\
&\ \ \ + \sigma^{-2}_0 T^{-1/2} Dm(d_0) \frac{1}{2} \sum_{t=1}^{T} \eta^2_{0t} - \frac{1}{2}\frac{ \left( \sum_{t = 1}^T \eta_{0t} \kappa_{0t} \right)^2}{\sum_{t = 1}^T \kappa^2_{0t}} \sigma^{-2}_0 T^{-1/2} Dm(d_0) \\
&\ \ \ +  m(d_0)  T^{-1/2}  \xi_{N_0,T}(d_0).
\end{align*}
From $m(d) = ( \sum_{t = 1}^T \kappa_{0t}^2(d) )^{1/(T-1)}$ it follows
\begin{align*}
    \sigma^{-2}_0 T^{-1/2} Dm(d_0) \frac{1}{2}\sigma^2_0(T-1) +  m(d_0)  T^{-1/2} \tau_T = 0.
\end{align*}
Therefore,
\begin{align*}
E(\sigma^{-2}_0 T^{-1/2}D L_m^*(d_0)) &= \sigma^{-2}_0 T^{-1/2} Dm(d_0) \frac{1}{2} \sum_{t=1}^{T} \eta^2_{0t} - \frac{1}{2}\frac{ \left( \sum_{t = 1}^T \eta_{0t} \kappa_{0t} \right)^2}{\sum_{t = 1}^T \kappa^2_{0t}} \sigma^{-2}_0 T^{-1/2} Dm(d_0) \\
&\ \ \ +  m(d_0)  T^{-1/2}  \xi_{N_0,T}(d_0).
\end{align*}
The summations in this expression are all bounded in $T$, see \textcite[Lemma B.4]{johansen2016role} and from $m(d) = 1 + O(T^{-1})$ (Lemma \ref{gen:lm:md}), $D m(d) = O(T^{-1})$ (Lemma \ref{genlemmaa1}) we conclude that
$ E(\sigma^{-2}_0 T^{-1/2}D L_m^*(d_0)) = T^{-1/2}   \xi_{N_0,T}(d_0) + O(T^{-3/2}) 
$. 
\end{proof}

\subsubsection{Initialisation scheme II: Infinite pre-sample values} \label{proofthcorinitial1}

This appendix contains a proof of Theorem \ref{th:corinitial}, i.e.\ of the extension of Theorem \ref{t53} to the model in \eqref{genq1}-\eqref{genq2} without Assumption \ref{A5}. The model now allows for unobserved pre-sample
values entering $x_t$ through the infinite representation of the lag polynomial $\omega$ in \eqref{repmainf}. It corresponds to the model in, for instance, Hualde \& Nielsen (2020). After the proof of Theorem \ref{th:corinitial},
this appendix also presents analogous extensions of Theorems \ref{t54} and \ref{t55} and Corollary \ref{t56}.

The outline of the proof of Theorem \ref{th:corinitial} is as follows: Recall that the bias of the CSS and MCSS estimators admits the decomposition in \eqref{trian} into an intrinsic component $B_T$ and a score-induced
component $S_T$. We analyse the terms in the decomposition in turn: {\it (i)} It will follow from Lemma \ref{totalcntrmoments} that $B_T$ (and hence $\mathcal{B}_T$) is unchanged, and {\it (ii)} it turns out that the only effect of
relaxing Assumption \ref{A5} is on the expectation of the score, analysed in Theorem \ref{genlemmaexpectations-}. Theorem \ref{th:corinitial} then follows by combining these two results.  

The intrinsic bias $B_T$ depends on correlations among centred product moments of the derivative of the stochastic term $\phi(L;\varphi)\Delta_{+}^{d-d_0} u_t$ in \eqref{Sterm} which now decomposes into an in-sample and a pre-sample part:
\begin{align}
    S^+_t(\vartheta) = \phi(L;\varphi)\Delta_{+}^{d-d_0} u^+_t, \nonumber \\
    S^-_t(\vartheta) = \phi(L;\varphi)\Delta_{+}^{d-d_0} u^-_t, \label{Sterm-}
\end{align}
where $u_t = u_t^+ + u_t^-$ and
\begin{align*}
     u_t^+ &= \sum_{j = 0}^{t-1} \omega_j(\varphi_0) \epsilon_{t-j}, \\
     u_t^- &= \sum_{j = t}^{\infty} \omega_j(\varphi_0) \epsilon_{t-j}.
\end{align*}
Lemma \ref{explicitforms} provides the derivatives of $S_t^{+}(\vartheta)$. The next lemma provides expressions of the derivatives of $S_t^{-}(\vartheta)$ in \eqref{Sterm-}.

\begin{lemma}\label{explicitforms_minus}
For $t \geq 1$ and $r \geq t$, define
\begin{align}
    g_r^{(t)}(\vartheta) = \sum_{i = 0}^{t-1} \pi_i(d_0-d) \sum_{j = 0}^{t-i-1} \phi_{j}(\varphi) \omega_{r-i-j}(\varphi_0). \label{gfunction}
\end{align}
The pre-sample stochastic term $S_t^-$ and its derivatives are given by the following expressions. Let $t \ge 1$:
\begin{align}
S_t^- &= \sum_{r=t}^{\infty}  g_r^{(t)}(\vartheta_0) \epsilon_{t-r}, \label{exp0m}\\
S^-_{m t} &= \sum_{r=t}^{\infty} D_m g_r^{(t)}(\vartheta_0) \epsilon_{t-r}, \label{exp1m}\\
S^-_{z t} &= \sum_{r=t}^{\infty} D_z g_r^{(t)}(\vartheta_0) \epsilon_{t-r}, \label{exp2m}\\
S^-_{m z t} &= \sum_{r=t}^{\infty} D_m D_z g_r^{(t)}(\vartheta_0) \epsilon_{t-r}, \label{exp5m}
\end{align}
   where $m \in \{d,dd,ddd \}$ and $z \in \{ \varphi_k, \varphi_k \varphi_j, \varphi_k \varphi_j \varphi_l \}$ for $k,j,l = 1,\ldots,p$ and 
\begin{align}
g_r^{(t)}(\vartheta_0) &= \sum_{i=0}^{t-1}\phi_i(\varphi_0) \omega_{r-i}(\varphi_0), \qquad r\ge t, \label{defc_tr}\\
D_m g_r^{(t)}(\vartheta_0) &= (-1)^{m^*} \sum_{i=1}^{t-1} D_m \pi_i(0)g_r^{(t-i)}(\vartheta_0) , \qquad r\ge t, \label{defl_tr}\\
D_z g_r^{(t)}(\vartheta_0) &= \sum_{i=0}^{t-1}  \omega_{r-i}(\varphi_0) D_z\phi_i(\varphi_0) , \qquad r\ge t, \label{defb_tr}\\
D_m D_z g_r^{(t)}(\vartheta_0) &= (-1)^{m^*} \sum_{i=1}^{t-1} D_m \pi_i(0) D_z g_r^{(t-i)}(\vartheta_0), \qquad r\ge t. \label{defh_tr}
\end{align}
where $m^*$ denotes the number of times $S^-_t(\vartheta)$ is differenced with respect to $d$.
The identities \eqref{dpi1}-\eqref{dpi2} hold as before.    
\end{lemma}

\begin{proof}[Proof of Lemma \ref{explicitforms_minus}]
Proof of \eqref{exp0m}: By definition $
S_t^-(d,\varphi)=\phi(L;\varphi) \Delta^{d-d_0}\{u_t^- I(t\ge1)\}$. At $(d,\varphi)=(d_0,\varphi_0)$, $\Delta^{0}$ is the identity, so
\begin{align*}
S_t^-(\vartheta_0)=\phi(L;\varphi_0)\{u_t^- I(t\ge1)\}
=\sum_{i=0}^{t-1}\phi_i(\varphi_0)\,u_{t-i}^-
=\sum_{r=t}^{\infty}\Big(\sum_{i=0}^{t-1}\phi_i(\varphi_0) \omega_{r-i}(\varphi_0)\Big)\epsilon_{t-r},
\end{align*}
which is \eqref{exp0m} with $g_r^{(t)}(\vartheta_0)$ as in \eqref{defc_tr}. 

Proof of \eqref{exp1m}: Writing $S_t^-(d,\varphi)=  \Delta^{d-d_0} \phi(L;\varphi) \{u_t^- I(t\ge1)\}$ and using the derivative of $\Delta_+^{\alpha} y_t=\sum_{k=0}^{t-1}\pi_k(-\alpha) y_{t-k}$ at $\alpha=0$ yields
\begin{align*}
S^-_{m t}
=(-1)^{m^*}\sum_{k=1}^{t-1} D_m\pi_k(0)\, \phi(L;\varphi_0)\{u_{t-k}^- I(t-k\ge1)\}.
\end{align*}
Apply $S_t^-(\vartheta_0)$ with $t$ replaced by $t-k$ and reindex to obtain \eqref{exp1m}. The weights $D_d\pi_j(0)$ and $D_{dd}\pi_j(0)$ are given by \eqref{dpi1}-\eqref{dpi2}.

Proof of \eqref{exp2m}: Only $\phi(L;\varphi)$ depends on $\varphi$, hence
\begin{align*}
S^-_{z t}= D_z\phi(L;\varphi_0)\{u_t^- I(t\ge1)\}
=\sum_{i=0}^{t-1} D_z\phi_i(\varphi_0) u_{t-i}^-,    
\end{align*}
and substituting $u_{t-i}^-=\sum_{j=t-i}^{\infty}\omega_j(\varphi_0)\epsilon_{t-i-j}$ yields \eqref{exp2m} with $D_z g_r^{(t)}(\vartheta_0)$ as in \eqref{defb_tr}.

Proof of \eqref{exp5m}: Differentiating $S_t^-(d,\varphi)=  \Delta^{d-d_0} \phi(L;\varphi) \{u_t^- I(t\ge1)\}$ $m^*$ times w.r.t.\ $d$ via $\Delta_+$ and differentiate it w.r.t.\ $z$ via $\phi(L;\varphi)$ evaluate at $\vartheta = \vartheta_0$ yields
\begin{align*}
S^-_{m z t}= \sum_{k=1}^{t-1} D_m\pi_k(0) S^-_{zt-k}
=  (-1)^{m^*} \sum_{r=t}^{\infty}\Big(\sum_{k=1}^{t-1} D_m \pi_k(0)  b^{(t-k)}_{z r}(\varphi_0)\Big)\epsilon_{t-r},
\end{align*}
which is \eqref{exp5m} with $D_m D_z g_r^{(t)}(\vartheta_0)$ as in \eqref{defh_tr}.
\end{proof}

The next lemma provides bounds for the pre-sample term $g_r^{(t)}(\vartheta_0)$ and its derivatives in \eqref{defc_tr}-\eqref{defh_tr}, see Lemma \ref{explicitforms_minus}. The corresponding expression in the context of $S^+_t$ where presented in Lemma A.8
\begin{lemma}\label{r12-}
Let Assumptions \ref{A3} and \ref{A1} hold. For the pre-sample terms $g_r^{(t)}(\vartheta_0)$,
$D_m g_r^{(t)}(\vartheta_0)$,
$D_z g_r^{(t)}(\vartheta_0)$, 
$D_m D_z g_r^{(t)}(\vartheta_0)$,
defined in \eqref{defc_tr}-\eqref{defh_tr} uniformly in $t\ge 1$ 
\begin{align}
\underset{\varphi_0 \in\Phi}{\sup} |g_r^{(t)}(\vartheta_0)|
&= O(r^{-1-\varsigma}), \label{krtproof} \\
\underset{\varphi_0 \in\Phi}{\sup}|D_m g_r^{(t)}(\vartheta_0)|
&= O(r^{-1-\varsigma} \log^{m^*+ 1} (r)), \label{lrtproof} \\
\underset{\varphi_0\in\Phi}{\sup} |D_z g_r^{(t)}(\vartheta_0)|
&= O(r^{-1-\varsigma}), \label{brtproof} \\ 
\underset{\varphi_0\in\Phi}{\sup} |D_m D_z g_r^{(t)}(\vartheta_0)|
&= O(r^{-1-\varsigma} \log^{m^*+ 1} (r)) \label{hrtproof},
\end{align}
where $m^*$ denotes the order of $m$.
\end{lemma}
\begin{proof}[Proof of Lemma \ref{r12-}] 
Proof of \eqref{krtproof}: 
Using the identity
$\sum_{i=0}^{r}\phi_i(\varphi)\omega_{r-i}(\varphi)=0$ for $r\ge1$:
\begin{align*}
g_r^{(t)}(\vartheta_0)
&= -\sum_{i=t}^{r}\phi_i(\varphi_0) \omega_{r-i}(\varphi_0) \\
&\le c \sum_{i=t}^{r-1} i^{-1-\varsigma}(r-i)^{-1-\varsigma}
 \le c \sum_{i=1}^{r-1} i^{-1-\varsigma}(r-i)^{-1-\varsigma}
= O(r^{-1-\varsigma}),
\end{align*}
uniformly in $t\ge1$ and $\varphi_0 \in\Phi$.

Proof of \eqref{lrtproof}: 
Using \eqref{r11_1} and \eqref{krtproof}:
\begin{align*}
|D_m g_r^{(t)}(\vartheta_0)|
\le \sum_{i=1}^{t-1} |D_{m^*}\pi_i(0)| |g^{(t-i)}_{r}(\varphi_0)|
 \le c r^{-1-\varsigma}\sum_{i=1}^{r-1} \log^{m^*} (i) i^{-1} = O(\log^{m^* + 1}(r) r^{-1-\varsigma})
\end{align*}
uniformly in $\varphi_0\in\Phi$.

Proof of \eqref{brtproof}: 
Since only $D_z\phi(\varphi)$ depends on $z$:
\begin{align*}
|D_m D_z g_r^{(t)}(\vartheta_0)|
&\le \sum_{i=0}^{t-1} |D_z\phi_i(\varphi_0)| |\omega_{r-i}(\varphi_0)|
 &\le c \sum_{i=1}^{t-1} i^{-1-\varsigma}(r-i)^{-1-\varsigma} 
 \le\ c\sum_{i=1}^{r-1} i^{-1-\varsigma}(r-i)^{-1-\varsigma}\\
&= O(r^{-1-\varsigma}),
\end{align*}
uniformly in $\varphi_0\in\Phi$.

Proof of \eqref{hrtproof}:
Using \eqref{r11_1} and \eqref{brtproof}:
\begin{align*}
|D_m D_z g_r^{(t)}(\vartheta_0)|
\le \sum_{i=1}^{t-1} |D_m\pi_i(0)| |D_z g_r^{(t-i)}(\vartheta_0)|
\le c r^{-1-\varsigma}\sum_{i=1}^{r-1} \log^{m^*} (i) i^{-1} = O(\log^{m^* + 1}(r) r^{-1-\varsigma})
\end{align*}
uniformly in $\varphi_0\in\Phi$.
\end{proof}

Define the pre-sample centred product moments for $S_t^{-}$ by replacing $S_t^{+}$ by $S_t^{-}$ in \eqref{genMta}-\eqref{genMte} and denote the resulting quantities by $M^{-}_{\cdot,\cdot\,T}$. The following lemma provides bounds for $M^{-}_{\cdot,\cdot\,T}$ and will be used in Lemma \ref{totalcntrmoments}.
\begin{lemma} \label{genlemmma1-}
Suppose that Assumptions \ref{A2}-\ref{A1} hold. Then, for $T\rightarrow \infty$, $M^-_{0,\vartheta_k  T} = O_P(T^{-1/2})$, $M^-_{0,\vartheta_k \vartheta_j   T} = O_P(T^{-1/2})$, $ M^-_{\vartheta_k,\vartheta_j   T} = O_P(T^{-1/2})$, $ M^-_{0,\vartheta_k \vartheta_j \vartheta_l  T} = O_P(T^{-1/2})$ and $M^-_{\vartheta_k,\vartheta_j \vartheta_l  T} = O_P(T^{-1/2})$ for $k,j,l = 1,\ldots,p+1$. 
\end{lemma}
\begin{proof}[Proof of Lemma \ref{genlemmma1-}]
We only cover $M^-_{0,\vartheta_k T}=O_p(T^{-1/2})$, the other cases follow from identical arguments. 
We first show $M^-_{0,\vartheta_k T}=O_p(T^{-1/2})$ for the $d$-direction, i.e.\ $k = 1$. Let $ z_t = S_t^- S^-_{\vartheta_k t}-E(S_t^-S^-_{\vartheta_k t})$ and
$M^-_{0,\vartheta_k T} = \sigma_0^{-2}T^{-1/2}\sum_{t=1}^T z_t$. By Minkowski's inequality 
\begin{align}
    \left( E \left(\sum_{t=1}^T z_t \right)^2 \right)^{1/2} \leq  \sum_{t=1}^T (E(z^2_t))^{1/2}.\label{qwe1a}
\end{align}
Moreover, $E(z^2_t) = Var( S_t^- S^-_{\vartheta_k t}) \leq E [ (S_t^-)^2 (S^-_{\vartheta_k t})^2]$ and by Cauchy-Schwarz 
\begin{align}
    E [ (S_t^-)^2 (S^-_{\vartheta_k t})^2] \leq [E(S_t^-)^4]^{1/2} [ E(S^-_{\vartheta_k t})^4]^{1/2}. \label{qwe1b}
\end{align}
For a linear process $y_t = \sum_{r = t}^{\infty} a_r \epsilon_{t-r}$ with $\sum_{r = t}^{\infty} a^2_r < \infty$ and independent and identical distributed errors satisfying $E \epsilon_t = 0$ and $E \epsilon^4_t < \infty$, we have 
\begin{align*}
    E(y^4_t) = \sum_{r = t}^{\infty} a^4_r E(\epsilon_1^4) + 3  \sum_{r = t}^{\infty} \sum_{\substack{s=t\\ s\neq r}}^{\infty} a^2_r a^2_s (E(\epsilon^2_1))^2,
\end{align*}
Since $\sum^{\infty}_{r = t} a_r^4 \leq (\sum_{r = t}^{\infty} a_r^2)^2$ and $\sum_{r = t}^{\infty} \sum_{s = t, s \neq r}^{\infty} a^2_r a^2_s \leq (\sum_{r = t}^{\infty} a_r^2)^2$, it follows that  
\begin{align}
    E(y^4_t) \leq c \left(\sum_{r = t}^{\infty} a_r^2\right)^2 \label{boundfourthmoments}
\end{align}
for $c < \infty$.

Applying \eqref{boundfourthmoments} with $a_r = g^{(t)}_{r}$ and $a_r = D_d g^{(t)}_{r}$ and using
$\sum_{r =t}^{\infty} (g^{(t)}_{r})^2=O(t^{-1-2\varsigma})$ and
$\sum_{r =t}^{\infty} (D_z g_r^{(t)})^2=O(t^{-1-2\varsigma})$ from \eqref{krtproof} and \eqref{lrtproof}, we obtain  
\begin{align*}
    E(S_t^-)^4 &= O(t^{-2-4\varsigma}) \\
     E(S^-_{\vartheta_k t})^4 &= O(t^{-2-4\varsigma}) .
\end{align*}
Substituting these bound into \eqref{qwe1b} gives
\begin{align*}
    E(z^2_t) = O(t^{-2-4\varsigma}).
\end{align*}
Hence by \eqref{qwe1a}
\begin{align*}
\left( E \left(\sum_{t=1}^T z_t \right)^2 \right)^{1/2} 
\leq c \sum_{t = 1}^T   t^{-1-2\varsigma}.
\end{align*}
Since $\varsigma > 1/2$, $ \sum_{t = 1}^T   t^{-1-2\varsigma} = O(1)$ and therefore $M^-_{0,\vartheta_k T}=O_p(T^{-1/2})$ for $k = 1$. For the $\varphi$-direction, i.e.\ any $k > 2$, the same arguments applies upon taking $a_r = D_{\varphi_k} g^{(t)}_{r}$ in \eqref{boundfourthmoments} and using the corresponding bound in \eqref{hrtproof}. 
\end{proof}

Write $S_t(\vartheta) = S^-_t(\vartheta) + S^+_t(\vartheta)$. Define the centred product moments for $S_t$ by replacing $S_t^{+}$ by $S_t$ in \eqref{genMta}-\eqref{genMte} and denote the resulting quantities by $M_{\cdot,\cdot\,T}$. The following lemma provides bounds and is the analogue of Lemma \ref{genlemmma1}. 
\begin{lemma} \label{totalcntrmoments}
Suppose that Assumptions \ref{A2}-\ref{A1} hold. Then, for $T\rightarrow \infty$,
\begin{align*}
 M_{0,\vartheta_k  T} &= M^+_{0,\vartheta_k  T} + O_P(T^{-1/2})\\
 M_{0,\vartheta_k \vartheta_j   T} &= M^+_{0,\vartheta_k \vartheta_j   T} + O_P(T^{-1/2}) \\
M_{\vartheta_k,\vartheta_j   T} &= M^+_{\vartheta_k,\vartheta_j   T} + O_P(T^{-1/2}) \\
 M_{0,\vartheta_k \vartheta_j \vartheta_l  T} &= M^+_{0,\vartheta_k \vartheta_j \vartheta_l  T} + O_P(T^{-1/2}) \\
 M_{\vartheta_k,\vartheta_j \vartheta_l  T} &= M^+_{\vartheta_k,\vartheta_j \vartheta_l  T} + O_P(T^{-1/2}),
\end{align*}
for $k,j,l = 1,\ldots,p+1$.
\end{lemma}
\begin{proof}
    We present the argument for $M_{0,\vartheta_k  T}$. The other cases follow from identical arguments. Using $S_t = S^+_t + S^-_t$ and $S_{\vartheta_k t} = S^+_{\vartheta_k t} + S^-_{\vartheta_k t}$
    \begin{align*}
        S_t S_{\vartheta_k t} = S^+_t S^+_{\vartheta_k t} + S^+_t S^-_{\vartheta_k t} +  S^-_tS^+_{\vartheta_k t} + S^-_t S^-_{\vartheta_k t}.
    \end{align*}
 Therefore, writing $M_{0,\vartheta_k  T} - M^+_{0,\vartheta_k  T}$ as the sum of the centred product moments related to  $S^+_t S^-_{\vartheta_k t}$, $S^-_tS^+_{\vartheta_k t}$ and $S^-_t S^-_{\vartheta_k t}$ yields
\begin{align*}
    M_{0,\vartheta_k  T} - M^+_{0,\vartheta_k  T} = M^{+,-}_{0,\vartheta_k  T} + M^{-,+}_{0,\vartheta_k  T} + M^{-}_{0,\vartheta_k  T},
\end{align*}
where $M^{-}_{0,\vartheta_k  T}$ is defined in Lemma \ref{genlemmma1-} and $M^{+,-}_{0,\vartheta_k  T}$ (resp. $M^{-,+}_{0,\vartheta_k  T}$) denote the centred product moments obtained by replacing $ S_t S_{\vartheta_k t}$ with $S^+_t S^-_{\vartheta_k t}$ (resp. $S^-_t S^-_{\vartheta_k t}$).
By the same argument as in the proof of Lemma \ref{genlemmma1-}(Minkowski's inequality and Cauchy-Scharz) and using the fourth moments bounds for $S_t^+$ and $S_t^-$ derived in the proof of Lemmata \ref{genlemmma1} and \ref{genlemmma1-}, it follows that 
\begin{align*}
    M^{+,-}_{0,\vartheta_k  T} &= O_P(T^{-1/2}), \\
    M^{-,+}_{0,\vartheta_k  T} &= O_P(T^{-1/2}). 
\end{align*}
 Lemma \ref{genlemmma1-} yields $M^{-}_{0,\vartheta_k  T} = O_P(T^{-1/2})$. Hence $M_{0,\vartheta_k  T} = M^{+}_{0,\vartheta_k  T} + O_P(T^{-1/2})$. 
\end{proof}
The pre-samples contribution to the centred product moments is $O_P(T^{-1/2})$ and therefore negligible relative to the in-sample centred moments in Lemma \ref{genlemmma1}, so $B_T$ and thus $\mathcal{B}_T$ are unchanged.

The following theorem establishes the expected values of the scores when Assumption \ref{A5} is discarded, that is, when $u_t$ is initialised in the infinite past. It is the extension of Theorem \ref{t52} and it forms the
key step in generalising Theorem \ref{t53} to Theorem \ref{th:corinitial}.
\begin{theorem} \label{genlemmaexpectations-}
Let the model for the data $x_t$, t = 1,$\ldots$,T, be given by \eqref{genq1} and let Assumptions \ref{A2}-\ref{A1} be satisfied. Then, for $k \in \{1,\ldots,p+1 \}$,
\begin{align*}
  E\left( D_{\vartheta_k} L^*(\vartheta_0)\right) &= -\sigma^2_0 \frac{\sum_{t = 1}^T c_{t}(\vartheta_0) c_{\vartheta_k t}(\vartheta_0)}{\sum_{t = 1}^T c^2_{t}(\vartheta_0)} \\ & \ \ \ + \sigma_0^2 \sum_{r = 0}^{\infty} D_{\vartheta_k} \left( \frac{1}{2} \left( \sum_{t = 1}^T (g_{t+r}^{(t)} (\vartheta_0) )^2 - \frac{\left( \sum_{t = 1}^T c_t(\vartheta_0) g_{t+r}^{(t)}(\vartheta_0)  \right)^2}{\sum_{t = 1}^T c^2_t(\vartheta_0)} \right) \right) ,\\
  E\left( D_{\vartheta_k} L_{\mu_0}^*(\vartheta_0)\right) &= \sigma_0^2 \sum_{r = 0}^{\infty} D_{\vartheta_k} \left( \frac{1}{2} \left( \sum_{t = 1}^T (g_{t+r}^{(t)}(\vartheta_0))^2  \right) \right) ,\\
  E\left( D_{\vartheta_k} L_m^*(\vartheta_0)\right) &= \sigma_0^2 \sum_{r = 0}^{\infty} D_{\vartheta_k} \left( \frac{1}{2} \left( \sum_{t = 1}^T (g_{t+r}^{(t)}(\vartheta_0))^2 - \frac{\left( \sum_{t = 1}^T c_t(\vartheta_0) g_{t+r}^{(t)}(\vartheta_0)  \right)^2}{\sum_{t = 1}^T c^2_t(\vartheta_0)} \right) \right) + o(1),
\end{align*}
where $c_t(\vartheta)$ and $g_{t+r}^{(t)}(\vartheta)$ are defined in \eqref{detc} and \eqref{gfunction}, respectively. Furthermore, for $L^*$, we have
\begin{align*}
E(D_d L^*(\vartheta_0)) 
  &= O\left( \log(T) I(d_0 < 1/2) + I(d_0 > 1/2) \right), \\
E(D_\varphi L^*(\vartheta_0)) 
  &= O(1),
\end{align*}
and for $L_{\mu_0}^*$ and $L_m^*$, we have
\begin{align*}
E(D_d L_\bullet^*(\vartheta_0)) 
  &= O(1), \\
E(D_\varphi L_\bullet^*(\vartheta_0)) 
  &= O(1),
\end{align*}
where $L_\bullet^* \in \{L_{\mu_0}, L_m^*\}$.
\end{theorem}
\begin{proof}[Proof of Theorem \ref{genlemmaexpectations-}.] Start from the decomposition
\begin{align*}
    E D_{\vartheta_k} L^* &= \sum_{t = 1}^T E(S^-_t + S^+_t - c_t(\hat{\mu} - \mu_0))(S^-_{\vartheta_kt} + S^+_{\vartheta_kt} - c_{\vartheta_kt}(\hat{\mu} - \mu_0)) \\
    &= \sum_{t = 1}^T E\left(S^+_t  - c_t \frac{\sum_{t = 1}^T c_t S^+_t}{\sum_{t = 1}^T c^2_t}\right)\left( S^+_{\vartheta_kt}  - c_{\vartheta_kt} \frac{\sum_{t = 1}^T c_t S^+_t}{\sum_{t = 1}^Tc^2_t}\right) \\
    &\ \ \ + \sum_{t = 1}^T E\left(S^-_t - c_t \frac{\sum_{t = 1}^T c_t S^-_t}{\sum_{t = 1}^T c^2_t}\right)\left(S^-_{\vartheta_kt}  - c_{\vartheta_kt} \frac{\sum_{t = 1}^T c_t S^-_t}{\sum_{t = 1}^T c^2_t}\right) 
\end{align*}
By Lemma \ref{genlemmaexpectations}, the first sum equals $-\sigma^2_0 \frac{\sum_{t = 1}^T c_{t} c_{\vartheta_k t}}{\sum_{t = 1}^T c^2_{t}}$.
Using simple algebra, the second sum equals $\sigma^2_0$ times 
\begin{align}
    \sum_{t = 1}^T \sum_{r = t}^{\infty} g_r^{(t)} D_{\vartheta_k} g_r^{(t)} -\frac{1}{\sum_{t = 1}^T c^2_t} \sum_{t = 1}^T  c_{\vartheta_k t} \sum_{s = 1}^T c_s \sum_{r = 0}^{\infty} g_{s+r}^{(s)} g_{t+r}^{(t)} \nonumber \\ 
    -  \frac{1}{\sum_{t = 1}^T c^2_t}\sum_{t = 1}^T  c_{t} \sum_{s = 1}^T c_s \sum_{r = 0}^{\infty} g_{s+r}^{(s)}  D_{\vartheta_k} g_{t+r}^{(t)} + \frac{1}{(\sum_{t = 1}^T c^2_t)^2}       \sum_{r = 0}^{\infty} (\sum_{t = 1}^T c_t g_{t+r}^{(t)})^2       \sum_{t = 1}^T c_{t}c_{\vartheta_k t} \label{initbiasterm}
\end{align}
or, equivalently, in the more compact form 
\begin{align*}
    \sigma_0^2 \sum_{r = 0}^{\infty} D_{\vartheta_k} \left( \frac{1}{2} \left( \sum_{t = 1}^T (g_{t+r}^{(t)})^2 - \frac{\left( \sum_{t = 1}^T c_t g_{t+r}^{(t)}  \right)^2}{\sum_{t = 1}^T c^2_t} \right) \right)
\end{align*}
which gives the desired result.

For $L^*_{\mu_0}$, the term with $c_t$ drops out, and the same calculation gives only the $\sum_{t = 1}^T (g_{t+r}^{(t)})^2$ term.

For $L_m^*$, we use
\begin{align*}
     E\left( D_{\vartheta_k} L_m^* \right) = m E (D_{\vartheta_k} L^* )  + D_{\vartheta_k} m E ( L^* ),
\end{align*}
Decompose $E(L^*)$ as
\begin{align*}
     E ( L^* ) &= \frac{1}{2} \sum_{t = 1}^T E(S_t^{+}  + S_t^{-} - c_t (\hat{\mu} - \mu_0) )^2 \\
     &= \frac{1}{2} \sum_{t = 1}^T E\left(S_t^{+}  - c_t \frac{\sum_{t = 1}^T c_t S^+_t}{\sum_{t = 1}^T c^2_t} \right)^2 \\
     &\ \ \ + \frac{1}{2} \sum_{t = 1}^T E\left(S_t^{-}  - c_t \frac{\sum_{t = 1}^T c_t S^-_t}{\sum_{t = 1}^T c^2_t} \right)^2.
\end{align*}
By the definition of $m$, the following equality holds:
\begin{align*}
    -m \sigma^2_0 \frac{\sum_{t = 1}^T c_{t} c_{\vartheta_k t}}{\sum_{t = 1}^T c^2_{t}} + D_{\vartheta_k} m \frac{1}{2} \sum_{t = 1}^T E\left(S_t^{+}  - c_t \frac{\sum_{t = 1}^T c_t S^+_t}{\sum_{t = 1}^T c^2_t} \right)^2  = 0.
\end{align*}
Collecting the terms gives
\begin{align*}
     E\left( D_{\vartheta_k} L_m^* \right) &= m \sigma_0^2 \sum_{r = 0}^{\infty} D_{\vartheta_k} \left( \frac{1}{2} \left( \sum_{t = 1}^T g_{t+r}^{(t)} g_{t+r}^{(t)} - \frac{\left( \sum_{t = 1}^T c_t g_{t+r}^{(t)}  \right)^2}{\sum_{t = 1}^T c^2_t} \right) \right) \\
     &\ \ \ + D_{\vartheta_k} m \frac{1}{2} \sum_{t = 1}^T E\left(S_t^{-}  - c_t \frac{\sum_{t = 1}^T c_t S^-_t}{\sum_{t = 1}^T c^2_t} \right)^2
\end{align*}
Simple algebra shows
\begin{align*}
    \frac{1}{2} \sum_{t = 1}^T E\left(S_t^{-}  - c_t \frac{\sum_{t = 1}^T c_t S^-_t}{\sum_{t = 1}^T c^2_t} \right)^2 = \sigma^2_0 \sum_{r = 0}^{\infty} \left( \frac{1}{2} \left( \sum_{t = 1}^T g_{t+r}^{(t)} g_{t+r}^{(t)} - \frac{\left( \sum_{t = 1}^T c_t g_{t+r}^{(t)}  \right)^2}{\sum_{t = 1}^T c^2_t} \right) \right)
\end{align*}
which can be shown to be $O(1)$ by an application of Cauchy-Schwarz inequality and the bound on $g_t$ in Lemma \ref{r12-}. From Lemma \ref{genlemmaa1} and Lemma \ref{genlemmaa1stat}, it holds that $m = 1 + O(\log(T) T^{-1})$ and $ D_{\vartheta_k} m = O(\log(T) T^{-1})$, and we obtain the desired result. 

The bounds for the expectation of the score are obtained as follows. 
For the expectation of $L^*$ in Theorem \ref{genlemmaexpectations}, the first term is treated in Theorem \ref{t52}.  
The second term consists of four components, see \eqref{initbiasterm}. We bound these components one by one.

From Lemma \ref{r12-}, the first component in \eqref{initbiasterm} satisfies
\begin{align*}
    c\sum_{t = 1}^T \sum_{r = t}^{\infty} |g_r^{(t)}| |D_{\vartheta_k} g_r^{(t)}| = O(\sum_{t = 1}^T \sum_{r = t}^{\infty} r^{-2-2\varsigma} \log^2(r)) = O(\sum_{t = 1}^T  t^{-1-2\varsigma} \log^2(t)) = O(1),
\end{align*}
since $1+2\varsigma > 1$.

For the remaining components we distinguish the two cases $d_0>1/2$ and $d_0<1/2$. We first consider $d_0>1/2$.

\medskip\noindent
\textit{Case $d_0>1/2$.} The second component in \eqref{initbiasterm} is bounded above by 
\begin{align*}
  c (\sum_{t = 1}^T c^2_t)^{-1}\sum_{t = 1}^T |c_{\vartheta_k t}| \sum_{s = 1}^T |c_s| \sum_{r = 0}^{\infty} |g_{s+r}^{(s)}| |g_{t+r}^{(t)}|. 
\end{align*}
By Cauchy-Schwarz
\begin{align*}
   \sum_{r = 0}^{\infty} |g_{s+r}^{(s)}| |g_{t+r}^{(t)}| \leq (\sum_{r = 0}^{\infty} (g_{s+r}^{(s)})^2)^{1/2} (\sum_{r = 0}^{\infty} (g_{t+r}^{(t)})^2)^{1/2}.   
\end{align*}
Using Lemma \ref{r12-}, 
$\sum_{r = 0}^{\infty} (g_{t+r}^{(t)})^2 < c t^{-1-2\varsigma}$. Therefore,
\begin{align*}
 &c (\sum_{t = 1}^T c^2_t)^{-1}\sum_{t = 1}^T |c_{\vartheta_k t}| \sum_{s = 1}^T |c_s| \sum_{r = 0}^{\infty} |g_{s+r}^{(s)}| |g_{t+r}^{(t)}| \\
 &\leq  c (\sum_{t = 1}^T c^2_t)^{-1}   \sum_{t = 1}^T  |c_{\vartheta_k t}|  t^{-1/2-\varsigma} \sum_{s = 1}^T |c_s| s^{-1/2-\varsigma}. 
\end{align*}
Applying Cauchy-Schwarz 
\begin{align*}
 &c (\sum_{t = 1}^T c^2_t)^{-1}   \sum_{t = 1}^T  |c_{\vartheta_k t}|  t^{-1/2-\varsigma} \sum_{s = 1}^T |c_s| s^{-1/2-\varsigma} \\
  &\leq  c (\sum_{t = 1}^T c^2_t)^{-1/2}   (\sum_{t = 1}^T  c_{\vartheta_k t}^2)^{1/2}  (\sum_{t = 1}^T t^{-1-2\varsigma}). 
\end{align*}
By Lemma \ref{genlemmaaaa2n}, $(\sum_{t = 1}^T c^2_t)^{-1/2}   (\sum_{t = 1}^T  c_{\vartheta_k t}^2)^{1/2} = O(1)$ and $\sum_{t = 1}^T t^{-1-2\varsigma} = O(1)$ since $1+2\varsigma > 1$. Hence the second component is $O(1)$.

The third component in \eqref{initbiasterm} is bounded above by 
\begin{align*}
  c(\sum_{t = 1}^T c^2_t)^{-1}\sum_{t = 1}^T  |c_{t} |\sum_{s = 1}^T |c_s| \sum_{r = 0}^{\infty} |g_{s+r}^{(s)}||  D_{\vartheta_k} g_{t+r}^{(t)}|
\end{align*}
Cauchy-Schwarz gives
\begin{align*}
    \sum_{r = 0}^{\infty} |g_{s+r}^{(s)}  D_{\vartheta_k} g_{t+r}^{(t)}| \leq (\sum_{r = 0}^{\infty} (g_{s+r}^{(s)})^2)^{1/2} (\sum_{r = 0}^{\infty}   (D_{\vartheta_k} g_{t+r}^{(t)})^2 )^{1/2}.
\end{align*}
From Lemma \ref{r12-}
\begin{align*}
    \sum_{r = 0}^{\infty} (g_{s+r}^{(s)})^2 &\leq c s^{-1-2\varsigma}, \\
    \sum_{r = 0}^{\infty}   (D_{\vartheta_k} g_{t+r}^{(t)})^2 &\leq c t^{-1-2\varsigma} \log^4(t).
\end{align*}
Therefore
\begin{align*}
  &c(\sum_{t = 1}^T c^2_t)^{-1}\sum_{t = 1}^T  |c_{t} |\sum_{s = 1}^T |c_s| \sum_{r = 0}^{\infty} |g_{s+r}^{(s)}||  D_{\vartheta_k} g_{t+r}^{(t)}| \\
  &\leq   c(\sum_{t = 1}^T c^2_t)^{-1}(\sum_{t = 1}^T  |c_{t}| t^{-1/2-\varsigma} \log^2(t))( \sum_{s = 1}^T |c_s| s^{-1/2-\varsigma}) 
  \end{align*}
Applying Cauchy-Schwarz
\begin{align*}
  &c(\sum_{t = 1}^T c^2_t)^{-1}(\sum_{t = 1}^T  |c_{t}| t^{-1/2-\varsigma} \log^2(t))( \sum_{s = 1}^T |c_s| s^{-1/2-\varsigma})
  \\
  &= c (\sum_{t = 1}^T \log^4(t) t^{-1-2\varsigma})^{1/2} (\sum_{s = 1}^T s^{-1-2\varsigma})^{1/2}) \\
  &= O(1)
\end{align*}
since both series converge for $\varsigma > 0$.

Finally, the fourth component in \eqref{initbiasterm} is bounded above by 
\begin{align*}
    c (\sum_{t = 1}^T c^2_t)^{-2} \sum_{r = 0}^{\infty} (\sum_{t = 1}^T c_t g_{t+r}^{(t)})^2       \sum_{t = 1}^T |c_{t}||c_{\vartheta_k t}|.
\end{align*} 
By Cauchy-Schwarz
\begin{align*}
   (\sum_{t = 1}^T c_t g_{t+r}^{(t)})^2 \leq \sum_{t = 1}^T c^2_t \sum_{t = 1}^T (g_{t+r}^{(t)})^2,  
\end{align*}
so
\begin{align*}
 c (\sum_{t = 1}^T c^2_t)^{-2} \sum_{r = 0}^{\infty} (\sum_{t = 1}^T c_t g_{t+r}^{(t)})^2       \sum_{t = 1}^T |c_{t}||c_{\vartheta_k t}|
  \leq c  (\sum_{t = 1}^T c^2_t)^{-1} \sum_{t = 1}^T |c_{t}| |c_{\vartheta_k t}| \sum_{r = 0}^{\infty} \sum_{t = 1}^T (g_{t+r}^{(t)})^2.    
\end{align*}
By Lemma \ref{genlemmaaaa2n}, $(\sum_{t = 1}^T c^2_t)^{-1} \sum_{t = 1}^T |c_{t}| |c_{\vartheta_k t}| = O(1)$, and by Lemma \ref{r12-}, $\sum_{r = 0}^{\infty} \sum_{t = 1}^T (g_{t+r}^{(t)})^2 = O(1)$. Hence the fourth component is also $O(1)$.
So for $d_0 > 1/2$, each of the four components of \eqref{initbiasterm} is $O(1)$.

\medskip\noindent
\textit{Case $d_0<1/2$.} An upper bound for the second component in \eqref{initbiasterm} is 
\begin{align*}
    c (\sum_{t = 1}^T c^2_t)^{-1}\sum_{t = 1}^T |c_{\vartheta_k t}| \sum_{s = 1}^T |c_s| \sum_{r = 0}^{\infty} |g_{s+r}^{(s)}| |g_{t+r}^{(t)}|.
\end{align*}
We first derive a bound for $\sum_{r = 0}^{\infty} |g_{s+r}^{(s)}| |g_{t+r}^{(t)}|$. Let $s \leq t$. Then
\begin{align*}
    \sum_{r = 0}^{\infty} |g_{s+r}^{(s)}| |g_{t+r}^{(t)}| &= \sum_{r = 0}^{t}  |g_{s+r}^{(s)}| |g_{t+r}^{(t)}| + \sum_{r = t+1}^{\infty}  |g_{s+r}^{(s)}| |g_{t+r}^{(t)}|   \\ 
    & \leq c\sum_{r = 0}^{t}  (s+r)^{-1-\varsigma}(t+r)^{-1-\varsigma} + c\sum_{r = t+1}^{\infty}  (s+r)^{-1-\varsigma}(t+r)^{-1-\varsigma} 
\end{align*}
For $0\leq r \leq t$ we use $s+r \geq s$ and $t+r\geq t$, and for $r>t$ we use 
$s+r \geq r$ and $t+r \geq r$. We obtain
\begin{align*}
    c\sum_{r = 0}^{t}  (s+r)^{-1-\varsigma}(t+r)^{-1-\varsigma} + c\sum_{r = t+1}^{\infty}  (s+r)^{-1-\varsigma}(t+r)^{-1-\varsigma} \geq c s^{-1-\varsigma} t^{-\varsigma}.
\end{align*}
By symmetry, for $t \leq s$ we obtain $\sum_{r = 0}^{\infty} |g_{s+r}^{(s)}| |g_{t+r}^{(t)}| \leq c t^{-1-\varsigma}s^{-\varsigma}$. Combining both cases yields the uniform bound
\begin{align}
      \sum_{r = 0}^{\infty} |g_{s+r}^{(s)}| |g_{t+r}^{(t)}| \leq c s^{-1-\varsigma}t^{-\varsigma} + c t^{-1-\varsigma}s^{-\varsigma}. \label{sumofg}
\end{align}
Using $|c_{\vartheta_k t}| = O(\log(t)t^{-d_0})$, $|c_{t}| = O(t^{-d_0})$ together with this bound, we obtain 
\begin{align*}
    \sum_{t = 1}^T |c_{\vartheta_k t}| \sum_{s = 1}^T |c_s| \sum_{r = 0}^{\infty} |g_{s+r}^{(s)}| |g_{t+r}^{(t)}| &\leq c  \sum_{t = 1}^T \log(t)t^{-d_0-\varsigma} \sum_{s = 1}^T s^{-1-d_0-\varsigma} \\
    & \ \ \ + c  \sum_{t = 1}^T \log(t)t^{-1-d_0-\varsigma} \sum_{s = 1}^T s^{-d_0-\varsigma}
\end{align*}
Since $1/2< \varsigma \leq 1$ and $d_0 < 1/2$, we have 
\begin{align*}
    \sum_{t = 1}^T \log(t)t^{-d_0-\varsigma} &= O( \log(T) T^{\max(0,1-d_0-\varsigma)}, \\
     \sum_{s = 1}^T s^{-1-d_0-\varsigma} &= O(T^{\max(0,-d_0-\varsigma)}) \\
     \sum_{t = 1}^T \log(t)t^{-1-d_0-\varsigma} &= O( \log(T) T^{\max(0,-d_0-\varsigma)} \\
    \sum_{s = 1}^T s^{-d_0-\varsigma} &=  O(T^{\max(0,1-d_0-\varsigma}).
\end{align*}
Hence
\begin{align*}
    &\sum_{t = 1}^T |c_{\vartheta_k t}| \sum_{s = 1}^T |c_s| \sum_{r = 0}^{\infty} |g_{s+r}^{(s)}| |g_{t+r}^{(t)}| \\ &= O(\log^2(T) T^{K(d_0,\varsigma)}),
\end{align*}
where 
\begin{align*}
    K(d_0,\varsigma) 
    &= \max(0,-d_0-\varsigma) + \max(0,1-d_0-\varsigma) \\
    &= 
    \begin{cases}
        0, & d_0 \ge 1 - \varsigma, \\[0.3em]
        1 - d_0 - \varsigma, & -\varsigma \le d_0 < 1 - \varsigma, \\
        1 - 2d_0 - 2\varsigma, & d_0 < -\varsigma.
    \end{cases}
\end{align*}
By Lemma \ref{genlemmaaaa2s} we have $\sum_{t = 1}^T c^2_t = O(T^{1-2d_0})$ so
\begin{align*}
    c(\sum_{t = 1}^T c^2_t)^{-1}\sum_{t = 1}^T c_{\vartheta_k t} \sum_{s = 1}^T c_s \sum_{r = 0}^{\infty} g_{s+r}^{(s)} g_{t+r}^{(t)} = O(T^{-1+2d_0 +K(d_0,\varsigma)}\log^2(T)).
\end{align*}
In all three cases $-1+2d_0 +K(d_0,\varsigma) < 0$. Thus, the second component in \eqref{initbiasterm} is $o(1)$ when $d_0 < 1/2$.

The third and fourth components in \eqref{initbiasterm} can be handled in a similar way. In the third component, $D_{\vartheta_k} g_{t+r}^{(t)}$ contributes an additional factor $\log^2(t+r)$, which only produces an extra harmless $\log^2 T$ in the final bound. For the fourth component, the factor
$
(\sum_{t = 1}^T c_t^2)^{-2}
\sum_{t = 1}^T c_t c_{\vartheta_k t}$
is $O(T^{-1+2d_0}\log T)$ by Lemma \ref{genlemmaaaa2s} and the associated quadratic form $\sum_{r = 0}^{\infty} g_{s+r}^{(s)} g_{t+r}^{(t)}$ is the same as in the second component. In both cases, the resulting terms are $o(1)$ for $d_0<1/2$. 

The same type of terms appear in the expectations of $L_{\mu_0}^*$ and $L_{m}^*$ and hence the proof of Theorem \ref{genlemmaexpectations-} is complete.
\end{proof}

This also completes the proof of Theorem \ref{th:corinitial}. Up to this point we
have worked with exact bias expressions. In line with
the main part of the paper, we now 
replace these exact terms by their asymptotic counterparts, referred to as approximate biases, in order to obtain more transparent and
interpretable formulae. The following lemma provides the approximate misspecification biases, i.e.\ the asymptotic counterparts of the exact misspecification biases $S_T^{\mathrm{init},\mu}(\vartheta_0)$ and 
$S_T^{\mathrm{init},\mu_0}(\vartheta_0)$ defined in \eqref{eq:init-mu}-\eqref{eq:init-mu0} of Theorem \ref{th:corinitial}.

\begin{lemma}\label{apprmb}
Let the model for the data $x_t$, t = 1,$\ldots$,T, be given by \eqref{genq1} and let Assumptions \ref{A2}-\ref{A1} be satisfied.
Define 
$T\mathcal{S}_T^{\mathrm{init},\mu}(\vartheta_0) = \lim_{T \rightarrow \infty} TS_T^{\mathrm{init},\mu}(\vartheta_0)$ and $T\mathcal{S}_T^{\mathrm{init},\mu_0}(\vartheta_0) = \lim_{T \rightarrow \infty} TS_T^{\mathrm{init},\mu_0}(\vartheta_0)$.
    Then: If $d_0<1/2$
      \begin{align*}
      T \mathcal{S}^{init,\mu}_T(\vartheta_0) &= - \lim_{T \rightarrow \infty} A^{-1}  \sum_{r = 0}^{\infty} D_{\vartheta} \left( \frac{1}{2} \left( \sum_{t = 1}^{T} (g_{t+r}^{(t)}(\vartheta))^2 \right) \right)\Bigg|_{\vartheta=\vartheta_0}, \\
       T \mathcal{S}^{init,\mu_0}_T(\vartheta_0) &=- \lim_{T \rightarrow \infty} A^{-1}  \sum_{r = 0}^{\infty} D_{\vartheta} \left( \frac{1}{2} \left( \sum_{t = 1}^{T} (g_{t+r}^{(t)}(\vartheta))^2 \right) \right)\Bigg|_{\vartheta=\vartheta_0}, 
  \end{align*}
  and if $d_0>1/2$
      \begin{align*}
      T \mathcal{S}^{init,\mu}_T(\vartheta_0) &= - \lim_{T \rightarrow \infty} A^{-1}  \sum_{r = 0}^{\infty} D_{\vartheta} \left( \frac{1}{2} \left( \sum_{t = 1}^{T} (g_{t+r}^{(t)}(\vartheta))^2  - \frac{\left( \sum_{t = 1}^{T} c_t(\vartheta) g_{t+r}^{(t)}(\vartheta)  \right)^2}{\sum_{t = 1}^{T} c^2_t(\vartheta)} \right) \right) \Bigg|_{\vartheta=\vartheta_0} , \\
       T \mathcal{S}^{init,\mu_0}_T(\vartheta_0) &=- \lim_{T \rightarrow \infty} A^{-1} \sum_{r = 0}^{T} D_{\vartheta} \left( \frac{1}{2} \left( \sum_{t = 1}^{T} (g_{t+r}^{(t)}(\vartheta))^2 \right) \right)\Bigg|_{\vartheta=\vartheta_0}.
  \end{align*}
\end{lemma}
\begin{proof}
    The result follows by taking limits in $T S_T^{\mathrm{init},\mu_0}(\vartheta_0)$ 
and $T S_T^{\mathrm{init},\mu}(\vartheta_0)$ as given in 
\eqref{eq:init-mu}-\eqref{eq:init-mu0} and using the bounds established
in the proof of Theorem \ref{genlemmaexpectations-}. 
\end{proof}

The interpretation of Lemma \ref{apprmb} is instructive. It shows how the bias caused by ignoring the infinite past (i.e.\ treating the process as if it started at $t = 1$) contributes to the bias of our estimators. This misspecification bias depends on whether the process is stationary ($d_0 < 1/2$) or nonstationary ($d_0 > 1/2$) and on whether $\mu$ is estimated. In the stationary case, the approximate misspecification bias is the same whether $\mu$ is known or not, whereas in the nonstationary case there is an additional contribution when $\mu$ is estimated.

The key intuition is that $\hat\mu$ is a GLS-type (or feasible GLS) estimator: it is a weighted average of the observations with weights that depend on $d$ and $\varphi$. For highly persistent (nonstationary) processes, e.g.\ $d = 1$, $\hat\mu$ puts most weight on the first observations. These early observations are precisely those most affected by the fact that the true process starts in the infinite past and this is reflected in the additional component of $T \mathcal{S}^{init,\mu}_T(\vartheta_0)$. On the other hand, for stationary processes (e.g.\ $d = 0$) the GLS weights are spread much more evenly across all observations. Because only a small number of initial observations are meaningfully distorted, their influence is washed out asymptotically. As a result, the contribution of the mean estimation to the misspecification bias disappears for $d_0 < 1/2$.

We now turn to the extensions of the two particular model specifications analysed in Sections \ref{arfima1d} and \ref{bshortm}, namely the ARFIMA(1,$d$,0) model and short-memory models, respectively. Again, the extensions are for an error process $u_t$ that starts in the infinite past.

The following theorem is an extension of Theorem \ref{t54} on  ARFIMA(1,$d$,0) models.

\begin{theorem}\label{th:ExtARFIMA1d0}
     Let $x_t$, $t$ = 1,$\ldots$,$T$, be given by \eqref{genq1} and let $u_t = \varphi u_{t-1} + \epsilon_t$. Let Assumptions \ref{A2} to \ref{A1} be satisfied with $\varphi_0 \neq 0$. The approximate biases of $\hat \vartheta$, $\hat \vartheta_{\mu_0}$ and $\vartheta_{\mu}$ are as in \eqref{init_1}, \eqref{init_2} and \eqref{init_3}, respectively. Specifically, the intrinsic bias $\mathcal{B}_T(\varphi_0)$ and the unknown-level bias $\mathcal{S}_T(\vartheta_0)$ are given in Theorem \ref{t54}. The approximate model-misspecification bias is given by 
    \begin{align*}
        \mathcal{S}^{init,\mu}_T(\vartheta_0) &= \begin{cases}
0, & d_0 < \frac{1}{2},\\
 -A^{-1} Q(\vartheta_0), & d_0 > \frac{1}{2},
\end{cases} \\  
        \mathcal{S}^{init,\mu_0}_T(\vartheta_0) &= 0  .
    \end{align*}
   where $A$ is given in Theorem \ref{t54}, $\vartheta_0 = (d_0,\varphi_0)'$,
    and
    \begin{align*}
      Q(\vartheta_0) 
      = \frac{1}{1-\varphi_0^2}
        \begin{pmatrix}
          Q_d(\vartheta_0) \\
          Q_\varphi(\vartheta_0)
        \end{pmatrix},
    \end{align*}
    with components
    \begin{align*}
      Q_d(\vartheta_0)
      &= - \frac{S_2(\vartheta_0)}{C_0(\vartheta_0)}
         + \frac{\varphi_0^2}{C_0(\vartheta_0)^2} R_d(\vartheta_0), \\
      Q_\varphi(\vartheta_0)
      &= - \frac{S_1(\vartheta_0)}{C_0(\vartheta_0)}
         + \frac{\varphi_0^2}{C_0(\vartheta_0)^2} R_\varphi(\vartheta_0), 
    \end{align*}
    and the limiting quantities
    \begin{align*}
      C_0(\vartheta_0)
      &= \sum_{t = 1}^{\infty} c_t(\vartheta_0)^2
       = (1-\varphi_0)^2 \binom{2d_0-2}{d_0-1}
         + \varphi_0 \binom{2d_0}{d_0}, \\
      R_\varphi(\vartheta_0)
      &= \sum_{t = 1}^{\infty} c_t(\vartheta_0) D_{\varphi} c_t(\vartheta_0)
       = (\varphi_0-1)\binom{2d_0-2}{d_0-1}
         + \tfrac{1}{2}\binom{2d_0}{d_0}, \\
      R_d(\vartheta_0)
      &= \sum_{t = 1}^{\infty} c_t(\vartheta_0) D_{d} c_t(\vartheta_0)\\
      &= (1-\varphi_0)^2 \binom{2d_0-2}{d_0-1}\bigl[\Psi(2d_0-1)-\Psi(d_0)\bigr]
       + \varphi_0 \binom{2d_0}{d_0}\bigl[\Psi(2d_0+1)-\Psi(d_0+1)\bigr], 
     \\
      S_1(\vartheta_0)
      &= \sum_{t = 2}^{\infty} c_t(\vartheta_0) \varphi_0^{t}
       = \varphi_0\Bigl[(1+\varphi_0)(1-\varphi_0)^{d_0}-1\Bigr],\\
      S_2(\vartheta_0)
      &= \sum_{t = 2}^{\infty} \frac{1}{t-1} c_t(\vartheta_0) \varphi_0^{t+1} \\
      &= \varphi_0^2 \sum_{k = 1}^{\infty} (-1)^k  \binom{d_0-1}{k}\frac{\varphi_0^k}{k}
         - \varphi_0^3 \frac{1-(1-\varphi_0)^{d_0}}{d_0}. 
    \end{align*}
    Here
   \begin{align*}
      c_t(\vartheta)
      = \sum_{j = 0}^{t-1} \phi_j(\varphi) \kappa_{0(t-j)}(d)
      = \kappa_{0t}(d) - \varphi\,\kappa_{0(t-1)}(d) I(t \ge 2), 
   \end{align*}
   where $\kappa_{0t}(d) = \pi_{t-1}(1-d)$.
\end{theorem}
The theorem delivers two main insights.

First, when $\mu$ is known, the approximate misspecification bias is zero for
all values of $d_0$, i.e.\ $\mathcal{S}_T^{\mathrm{init},\mu_0}(\vartheta_0)=0$. Thus,
the conditional nature of the CSS objective function does not affect the approximate bias of
$\hat\vartheta$. This reflects the fact that under AR(1) short-run
dynamics, $u_t$ depends only on $u_{t-1}$.
Writing down the CSS objective function, one can see that the first residual is essentially given by $x_1$ and hence
does not depend on the unknown parameters, while the remaining residuals are
correctly specified (filtered). This first (misspecified) residual therefore only adds a
constant to the CSS objective and does not alter the optimisation problem. Consequently, the approximate misspecification bias is zero.

Second, when $\mu$ is estimated, the approximate misspecification bias is zero in the
stationary region $d_0 < 1/2$, but becomes non-zero for $d_0 > 1/2$, where it
is given by $-A^{-1} Q(\vartheta_0)$, as explained following
Lemma \ref{apprmb}.

\begin{proof}
    Recall from the proof of Theorem \ref{t54} that $\omega_j(\varphi) = \varphi^j$, $j\geq 0$, and $\phi_0(\varphi) = 1$ and $\phi_1(\varphi) = -\varphi$, $\phi_s(\varphi) = 0$, $s \geq 2$, and  $\partial \phi_0(\varphi)/\partial \varphi = 0$, $\partial \phi_1(\varphi)/\partial \varphi = -1$, and $\partial \phi_s(\varphi)/\partial \varphi = 0$, $s \geq 2$. Plugging these into \eqref{defc_tr}, \eqref{defl_tr} and \eqref{defb_tr} yields for $r \geq t$,  
\begin{align*}
g_r^{(t)}(\vartheta_0)
&=\sum_{i=0}^{t-1}\phi_i(\varphi_0) \omega_{r-i}(\varphi_0)
=\begin{cases}
\varphi_0^{r}, & t=1,\\
0, & t\ge 2,
\end{cases} \\
D_{d} g_r^{(t)}(\vartheta_0)&=\begin{cases}
0, & t=1,\\
-\frac{1}{t-1} \varphi_0^r, & t\ge 2,
\end{cases} \\
D_{\varphi} g_r^{(t)}(\vartheta_0)
&=\begin{cases}
0, & t=1,\\
-\varphi_0^{r-1}, & t\ge 2.
\end{cases}
\end{align*}
We now derive the expressions for 
$\mathcal{S}_T^{init,\mu_0}$ and $\mathcal{S}_T^{init,\mu}$ in Lemma \ref{apprmb}, using the decomposition in \eqref{initbiasterm}. For $d_0 < 1/2$ the result follows from Lemma \ref{apprmb} and that the first term of \eqref{initbiasterm} (see below) is equal to zero, so we focus on $d_0 > 1/2$. 

The first term in \eqref{initbiasterm} is
\begin{align*}
    \sum_{t = 1}^T \sum_{r = t}^{\infty} g_r^{(t)}(\vartheta_0) D_{\vartheta_k} g_r^{(t)}(\vartheta_0), 
\end{align*}
which is zero for $k = 1,2$ and hence $S^{init,\mu_0}_T(\vartheta_0) = 0$, 

The second term in \eqref{initbiasterm}
\begin{align*}
    \frac{1}{\sum_{t = 1}^T c^2_t} \sum_{t = 1}^T  c_{\vartheta_k t} \sum_{s = 1}^T c_s \sum_{r = 0}^{\infty} g_{s+r}^{(s)} g_{t+r}^{(t)}.
\end{align*}
Only $s,t = 1$ contributed to the inner sum, $c_{\vartheta_k 1} = 0$, so the second term is zero. 

The third term of \eqref{initbiasterm},
\begin{align*}
    \frac{1}{\sum_{t = 1}^T c^2_t} \sum_{t = 1}^T  c_{t} \sum_{s = 1}^T c_s \sum_{r = 0}^{\infty} g_{s+r}^{(s)}  D_{\vartheta_k} g_{t+r}^{(t)} =   \frac{1}{\sum_{t = 1}^T c^2_t} \sum_{t = 2}^T  c_{t} \sum_{r = 0}^{\infty} g_{1+r}^{(1)}  D_{\vartheta_k} g_{t+r}^{(t)} \\
\end{align*}
since only $s = 1$ and $t \geq 2$ can contribute to the inner sum. 
For $k = 1$,
\begin{align*}
     \frac{1}{\sum_{t = 1}^T c^2_t}\sum_{t = 2}^T  c_{t} \sum_{r = 0}^{\infty} g_{1+r}^{(1)}  D_{\vartheta_1} g_{t+r}^{(t)} &= - \frac{1}{\sum_{t = 1}^T c^2_t}\sum_{t = 2}^T  c_{t} \sum_{r = 0}^{\infty} \varphi_0^{r+1}  \frac{1}{t-1} \varphi_0^{r+t} \\
    &= - \frac{1}{\sum_{t = 1}^T c^2_t}\sum_{t = 2}^T  \frac{1}{t-1}  c_{t}\varphi_0^{t+1} \sum_{r = 0}^{\infty} \varphi_0^{2r}  \\
    &= - (1-\varphi_0^2)^{-1}  \frac{1}{\sum_{t = 1}^T c^2_t}\sum_{t = 2}^T  \frac{1}{t-1}  c_{t}\varphi_0^{t+1}.  
\end{align*}
For $k  = 2$, 
\begin{align*}
     \frac{1}{\sum_{t = 1}^T c^2_t}\sum_{t = 2}^T  c_{t} \sum_{r = 0}^{\infty} g_{1+r}^{(1)}  D_{\vartheta_2} g_{t+r}^{(t)} &= - \frac{1}{\sum_{t = 1}^T c^2_t}\sum_{t = 2}^T  c_{t} \sum_{r = 0}^{\infty} \varphi_0^{r+1}  \varphi_0^{r+t-1} \\
    &= - \frac{1}{\sum_{t = 1}^T c^2_t}\sum_{t = 2}^T   c_{t}\varphi_0^{t} \sum_{r = 0}^{\infty} \varphi_0^{2r}  \\
    &= - (1-\varphi_0^2)^{-1} \frac{1}{\sum_{t = 1}^T c^2_t}\sum_{t = 2}^T  c_{t}\varphi_0^{t}. 
\end{align*}

The fourth term in \eqref{initbiasterm} becomes
\begin{align*}
    \frac{1}{(\sum_{t = 1}^T c^2_t)^2} \sum_{r = 0}^{\infty} (\sum_{t = 1}^T c_t g_{t+r}^{(t)})^2       \sum_{t = 1}^T c_{t}c_{\vartheta_k t} &=  \frac{1}{(\sum_{t = 1}^T c^2_t)^2} \sum_{r = 0}^{\infty} (g_{1+r}^{(1)})^2       \sum_{t = 1}^T c_{t}c_{\vartheta_k t} \\
    &= \frac{1}{(\sum_{t = 1}^T c^2_t)^2} \sum_{r = 0}^{\infty} \varphi_0^{2r+2}  \sum_{t = 1}^T c_{t}c_{\vartheta_k t} \\
    &= \frac{1}{(\sum_{t = 1}^T c^2_t)^2} \frac{\varphi^2_0}{1-\varphi^2_0}  \sum_{t = 1}^T c_{t}c_{\vartheta_k t}. \\
\end{align*}

The asymptotic limits are found as follows. From the proof of Theorem \ref{t54} we have 
\begin{align*}
    \sum_{t = 1}^{\infty} c^2_t &= (1-\varphi)^2 \binom{2d-2}{d-1} +  \varphi  \binom{2d}{d}, \\
    \sum_{t = 1}^{\infty} c_t(d,\varphi) D_{\varphi} c_t(d,\varphi) &= (\varphi- 1)   \binom{2d-2}{d-1}  + 0.5 \binom{2d}{d},\\
     \sum_{t = 1}^{\infty} c_t(d,\varphi) D_{d} c_t(d,\varphi) &= (1-\varphi)^2  \binom{2d-2}{d-1} \left( \Psi(2d-1)-\Psi(d) \right)  \\
     &\ \ \ + \varphi  \binom{2d}{d} \left( \Psi(2d+1)-\Psi(d+1) \right).
\end{align*}
Moreover,
\begin{align}
    \lim_{T  \rightarrow \infty} \sum_{t = 2}^{T}  c_{t}\varphi_0^{t} &= \varphi_0( (1+\varphi_0)(1-\varphi_0)^{d_0}-1), \label{sum1a}\\
    \lim_{T  \rightarrow \infty} \sum_{t = 2}^{T}  \frac{1}{t-1}  c_{t}\varphi_0^{t+1}  &= \varphi_0^2 \sum_{k = 1}^{\infty} (-1)^k  \binom{d_0-1}{k}\frac{\varphi_0^k}{k} - \varphi_0^3 \frac{1-(1-\varphi_0)^{d_0}}{d_0} \label{sum1b},
\end{align}
where  $c_t(\vartheta) = \sum_{j = 0}^{t-1} \phi_j(\varphi) \kappa_{0(t-j)}(d) =  \kappa_{0t}(d) - \varphi \kappa_{0(t-1)}(d) I(t \geq 2)$ and $\kappa_{0t}(d) = \pi_{t-1}(1-d)$. The limit in \eqref{sum1a} follows directly from the binomial series $\sum_{t = 1}^{\infty} \kappa_{0t}(d_0) z^{t-1} = (1-z)^{d_0-1}$. For \eqref{sum1b}, note that $(t-1)^{-1} \varphi_0^{t-1} = \int_{0}^{\varphi_0} u^{t-2} du$ and use this together with the binomial series. Letting $T\to\infty$ in the expressions for the third and fourth terms and using the limits above gives $Q(\vartheta_0)$. \end{proof}

The following theorem is an extension of Theorem \ref{t55} and Corollary \ref{t56} on short-memory models. The proofs are omitted.

\begin{theorem} \label{th:ExtShortMemory}
  Let $x_t$, $t$ = 1,$\ldots$,$T$, be given by \eqref{genq1} with $d_0 = 0$ and let Assumptions \ref{A2} to \ref{A1} be satisfied. Furthermore, when $d$ is set to zero in the respective objective functions, the approximate biases
  of $\hat{\varphi}$, $\hat{\varphi}_{\mu_0}$ and $\hat{\varphi}_{m}$ are as in \eqref{init_1}, \eqref{init_2} and \eqref{init_3}, respectively. Specifically, the intrinsic bias $\mathcal{B}_T(\varphi_0)$ and the unknown-level bias $\mathcal{S}_T(\varphi_0))$ are given in Theorem \ref{t55}.  The model misspecification bias terms $S^{init,\mu}_T(\varphi_0)$ and $S^{init,\mu_0}_T(\varphi_0)$ are given by 
  \begin{align}
      T S^{init,\mu}_T(\varphi_0) &= - \tilde{A}^{-1}  \sum_{r = 0}^{\infty} D_{\varphi} \left( \frac{1}{2} \left( \sum_{t = 1}^T (g_{t+r}^{(t)}(\varphi))^2  - \frac{\left( \sum_{t = 1}^T c_t(\varphi) g_{t+r}^{(t)}(\varphi)  \right)^2}{\sum_{t = 1}^T c^2_t(\varphi)} \right) \right) \Bigg|_{\varphi=\varphi_0}, \\
       T S^{init,\mu_0}_T(\varphi_0) &=- \tilde{A}^{-1} \sum_{r = 0}^{\infty} D_{\varphi} \left( \frac{1}{2} \left( \sum_{t = 1}^T (g_{t+r}^{(t)}(\varphi))^2 \right) \right)\Bigg|_{\varphi=\varphi_0},  
  \end{align}
  where $\tilde{A}$ is given in Theorem \ref{t55} and $c_t(\varphi) = \sum_{j = 0}^{t-1} \phi_j(\varphi)$ and $g_{t+r}^{(t)}(\varphi) =  \sum_{i=0}^{t-1}\phi_i(\varphi) \omega_{r-i}(\varphi_0)$. 
\end{theorem}

Based on this theorem, we can deduce the approximate model-specification bias of an AR(1) model as an illustration, extending the result in Corollary \ref{t56}.
\begin{corollary} \label{cr:ExtShortMemory}
  Let $x_t$, $t$ = 1,$\ldots$,$T$, be given by \eqref{genq1} with $d_0 = 0$ and let $u_t = \varphi u_{t-1} + \epsilon_t$. Let Assumptions \ref{A2} to \ref{A1} be satisfied. Furthermore, when $d$ is set to zero in the respective
  objective functions, the approximate biases of $\hat{\varphi}$, $\hat{\varphi}_{\mu_0}$ and $\hat{\varphi}_{m}$ are as in \eqref{init_1}, \eqref{init_2} and \eqref{init_3}, respectively. Specifically, the intrinsic bias  $T \mathcal{B}_T(\varphi_0)$ and the unknown-level score bias $T\mathcal{S}_T(\varphi_0)$ are given in Corollary \ref{t56}. The approximate misspecification biases are equal to zero, i.e.\ $ \mathcal{S}^{init,\mu}_T(\varphi_0) =  \mathcal{S}^{init,\mu_0}_T(\varphi_0) = 0$ 
\end{corollary}

In the pure AR(1) case with $d_0 = 0$ and known $\mu_0$ (for simplicity set $\mu_0 = 0$), the CSS objective
function is
\begin{align*}
L^*(\varphi) = x_1^2 + (x_2-\varphi x_1)^2 + \cdots + (x_T-\varphi x_{T-1})^2 .
\end{align*}
If the process in fact starts in the infinite past, then only the first residual
$x_1$ is misspecified, while the remaining residuals are correctly specified.
Since this first term does not depend on the unknown parameter $\varphi$, it only contributes an additive constant to the CSS objective and therefore does not affect the optimisation problem for $\varphi$. Consequently, the approximate misspecification bias is zero. When $\mu$ is estimated, the same conclusion holds in this stationary AR(1) case, see the discussion following
Lemma \ref{apprmb}.

\end{document}